\documentclass[11pt, a4paper, oneside]{Thesis} 

\graphicspath{{Pictures/}} 

\usepackage[square,sort,compress,numbers]{natbib}
\usepackage{cite}
\usepackage{amsmath,latexsym} 
\usepackage{float}

\title{\ttitle} 

\begin{document}

\frontmatter 

\setstretch{1.3} 

\fancyhead{} 
\rhead{\thepage} 
\lhead{} 

\pagestyle{fancy} 

%

\thesistitle{Study of Accelerated Expansion of Universe in the framework of $f(R,T)$ Gravity}
\documenttype{Thesis}
\supervisor{\textbf{Dr. Pradyumn Kumar Sahoo}}
\supervisorposition{Associate Professor}
\supervisorinstitute{BITS-Pilani, Hyderabad Campus}
\cosupervisor{\textbf{Dr. Bivudutta Mishra}}
\cosupervisorposition{Associate Professor}
\cosupervisorinstitute{BITS-Pilani, Hyderabad Campus}
\examiner{}
\degree{Ph.D. Research Scholar}
\coursecode{DOCTOR OF PHILOSOPHY}
\coursename{Thesis}
\authors{\textbf{PARBATI SAHOO}}
\IDNumber{2015PHXF0008H}
\addresses{}
\subject{}
\university{\texorpdfstring{\href{http://www.bits-pilani.ac.in/} 
                {Birla Institute of Technology and Science, Pilani}} 
                {Birla Institute of Technology and Science, Pilani}}
\UNIVERSITY{\texorpdfstring{\href{http://www.bits-pilani.ac.in/} 
                {BIRLA INSTITUTE OF TECHNOLOGY AND SCIENCE, PILANI}} 
                {BIRLA INSTITUTE OF TECHNOLOGY AND SCIENCE, PILANI}}



\department{\texorpdfstring{\href{http://www.bits-pilani.ac.in/pilani/Mathematics/Mathematics} 
                {Mathematics}} 
                {Mathematics}}
\DEPARTMENT{\texorpdfstring{\href{http://www.bits-pilani.ac.in/pilani/Mathematics/Mathematics} 
                {Mathematics}} 
                {Mathematics}}
\group{\texorpdfstring{\href{Research Group Web Site URL Here (include http://)}
                {Research Group Name}} 
                {Research Group Name}}
\GROUP{\texorpdfstring{\href{Research Group Web Site URL Here (include http://)}
                {RESEARCH GROUP NAME (IN BLOCK CAPITALS)}}
                {RESEARCH GROUP NAME (IN BLOCK CAPITALS)}}
\faculty{\texorpdfstring{\href{Faculty Web Site URL Here (include http://)}
                {Faculty Name}}
                {Faculty Name}}
\FACULTY{\texorpdfstring{\href{Faculty Web Site URL Here (include http://)}
                {FACULTY NAME (IN BLOCK CAPITALS)}}
                {FACULTY NAME (IN BLOCK CAPITALS)}}

\maketitle
\Certificate
\begin{abstract}
The main purpose of the thesis is to provide a coherent description of current accelerated expansion of the universe within $f(R,T)$ gravity theory, at an appropriate level of understanding. In recent cosmic scenario, various cosmological observations have reported the accelerated expansion of the universe. This phenomenon can be addressed by invoking the exotic source of energy, the so called dark energy which dominates the total energy content of the universe. On the other hand, we may need to modify the Einstein's general relativity (GR) as possible gravitational scenario for the early and the late time universe.\\
Einstein's theory of general relativity has been accepted as the theory of gravity for nearly a century as it passed several cosmological tests. Despite this, it cannot account for some existing mysteries. Therefore, the investigations on accelerating models have led to modifications of gravity theories involving a number of interesting features. These modifications can prompt a theoretical study of accelerated universe within modified gravity theories in a large-scale structure. It also has substantial impact on structure formation and its observable predictions. In the context of gravitational modification in GR, a number of modified theories are presented from time to time viz. scalar-tensor theories, $f(R)$ gravity, $f(\mathcal{G})$ gravity, $f(R,\mathcal{G})$ gravity,  $f(\mathcal{G},T)$ gravity, $f(\mathcal{T})$ gravity, and $f(R,T)$ gravity. In particular, we have focused on details of the last one mentioned above i.e. $f(R,T)$ gravity theory proposed by Harko et al. in 2011 \cite{Harko11}, while the other modified theories are briefly discussed in chapter \ref{Chapter1}.
\clearpage
In addition, we have focused on some parametrization scheme to obtain the exact solutions of each cosmological model within the $f(R,T)$ formalism. There are two cosmological parameters known as Hubble's parameter ($H$) and deceleration parameter ($q$). The evolution of the universe can be defined through the parametrization of these two parameters, on the other hand  scale factor parameter has very important role in understanding the present evolution of the universe. This scale factor determines the rate of expansion of the universe. However, in this thesis, one can find different types of parametrizations of deceleration parameters used in order to obtain exact solutions of the field equations. This is because from mathematical point of view, the exact solutions of field equations and its physical behaviors are more realistic. Therefore, in this thesis from chapter \ref{Chapter2} to chapter \ref{Chapter5}, we have employed several types of time varying deceleration parameters along with different matter sources. Thereafter, the first part of chapter \ref{Chapter6} consists the hybrid law of scale factor. 
The chapterization details are discussed in the following section.\\
In chapter \ref{Chapter2}, the formalism of first case of $f(R,T)$ gravity (i.e. $f(R,T)=R+2f(T)$) where $R$ is the Ricci scalar and $T$ is the trace of the stress energy momentum tensor is studied for the Bianchi universe  in the context of accelerating expansion of the universe as suggested by the present observations. The matter considered in this model is filled with string fluid. \\
Chapter \ref{Chapter3} focused on the study of spatially homogeneous anisotropic Bianchi type I universe in the frame work of $f(R,T)$ gravity with two different cases viz. $f(R,T)=R+2f(T)$ and $f(R,T)=f_1(R)+f_2(T)$. The exact solution of the field equations embedded with bulk viscus matter content is obtained by employing a time varying deceleration parameter, which generates an accelerating universe in both cases. Furthermore, in this chapter we have explored the nature of weak energy condition (WEC), dominant energy condition (DEC), strong energy condition (SEC), null energy condition (NEC) and the compatibility of cosmic jerk parameter with three kinematical data sets. \\
Cosmic parametrization has an important role in cosmological model, as it describes the behavior of model solutions. Parametrization of deceleration parameter is one of the best approach among them to obtain the exact solution as well as to elaborate the physical behavior of models. The bilinear and special form of time varying deceleration parameter for Bianchi type I universe filled with magnetized strange quark matter (MSQM) distribution have been discussed in chapter  \ref{Chapter4}. \\
Chapter \ref{Chapter5} serves as an introduction to the idea of the finite cosmic singularity called Big Rip singularity within the scope of both anisotropic and isotropic cosmological models. In order to study Big Rip cosmological model we have considered a linearly varying deceleration parameter (LVDP) in previous MSQM $f(R,T)$ model. The details of transitional behavior of LVDP are discussed in the first model of this chapter. In addition, the second model also has showed the Big Rip singularity in the presence of perfect fluid matter content. In this model, a periodic varying deceleration parameter plays a vital role to study the oscillating behavior as well as the future singularity in isotropic flat Friedmann-Lema\^itre-Robertson-Walker (FLRW) universe.\\
In chapter \ref{Chapter6}, we have explored the modification of geometry in left hand side of Einstein's field equation. In details, we have focused on higher order curvatures as well as different space-time geometries. That means the first model of this chapter deals with FLRW metric with different forms of curvature scalar, while in the second model, we concentrate on one of the recent development in astrophysics named as wormhole geometry. It has been studied extensively within $f(R,T)$ formalism and provides ideas and tools which turned out to be useful to propose new models. Finally, in chapter  \ref{Chapter7} we have studied more details of the wormhole geometry through energy conditions in $f(R,T)$ gravity formalism. In this chapter the late time evolution of universe has been studied extensively, which is an important step in establishing the viability of such modified gravity theories.    
\end{abstract}

\begin{acknowledgements}
It gives me great pleasure, that I have an opportunity to remember and express my sincere gratitude to all the people who contributed some or the other way to the completion of this wonderful journey in obtaining my doctoral degree and made my stay in BITS memorable. \\
First and foremost, I wish to express my special appreciation and thanks to my supervisor \textbf{Dr. Pradyumn Kumar Sahoo}, Associate Professor, Department of Mathematics, BITS-Pilani, Hyderabad Campus, Hyderabad, Telangana. A person with outstanding commitment and dedication, he has always made himself available with immense care and support in my professional as well as personal life throughout my Ph.D. tenure. Without his guidance, encouragement and moral support this journey would not have been achievable. The work environment given to me under him, the experience gained from him and his creative working culture with perfect time management are treasured and will be remembered throughout my life.\\
I express my profound gratitude and sincere thanks to my co-supervisor \textbf{Dr. Bivudutta Mishra}, Associate Professor, Department of Mathematics, BITS-Pilani, Hyderabad Campus, for his valuable suggestions, guidance, constant encouragement and intent supervision at every stage of my Ph.D. tenure.\\
I am extremely grateful to my Doctoral Advisory Committee (DAC) members \textbf{Dr. Suresh Kumar} and \textbf{Dr. T. S. L. Radhika} for simultaneously encouraging, guiding, and supporting my research work.\\
I extend my warm gratitude to HOD and all the faculty members, Department of Mathematics, BITS-Pilani, Hyderabad Campus, for their support and encouragement to carry out my research work. I would also like to thank the Doctoral Research Committee (DRC) members for providing valuable support and constant feedback during this research work.\\
I take this opportunity to thank my co-authors \textbf{Dr. P.H.R.S. Moraes}, \textbf{Dr. S. Ayg\"{u}n}, \textbf{Dr. S. K. Tripathy}, \textbf{Dr. S. K. J. Pacif}, \textbf{Dr. Binaya K. Bishi}, and \textbf{Dr. K. L. Mahanta} for their continuous support and motivations. \\    
I am very much grateful to all my friends for their co-operations and constant support. I would also like to acknowledge some special individuals \textbf{Pratik}, \textbf{Sanjay}, \textbf{Tapas}, \textbf{Revathi}, \textbf{Sirisa Akka}, \textbf{Basu}, \textbf{Tusar}, \textbf{Banchha}, \textbf{Sankarsan}, \textbf{Karthik}, \textbf{Faiz} for their valuable help during this course. It's my fortune to  gratefully acknowledge all my research colleague and friends for the time they had spent for me and making my stay at campus a memorable one.\\
Last but not least, I express my gratitude to my late uncle, my parents, my brothers, my sister-in-law, and my dear best friend Deepak Samal, for their constant support, love, care and encouragement during my work.

Parbati Sahoo\\
ID-2015PHXF0008H           
\end{acknowledgements}



\pagestyle{fancy}

\lhead{\emph{Contents}} 
\tableofcontents 

\lhead{\emph{List of Figures}}
\listoffigures 

\lhead{\emph{List of Tables}}
\listoftables 


\clearpage 

\setstretch{1.5}

\lhead{\emph{Abbreviations}} 
{
\textbf{\Large Abbreviations}\\
\textbf{B}\\
\textbf{BAO}:  \textbf{B}aryon \textbf{A}coustic \textbf{O}scillations\\
\textbf{BD}: \textbf{B}rans-\textbf{D}icke \\
\textbf{C}\\
\textbf{COBE}: \textbf{C}osmic \textbf{B}ackground \textbf{E}xplorer \\
\textbf{CMB}: \textbf{C}osmic \textbf{M}icrowave \textbf{B}ackground\\
\textbf{CMBR}: \textbf{C}osmic \textbf{M}icrowave \textbf{B}ackground \textbf{R}adiation\\
\textbf{D}\\
\textbf{DE}:  \textbf{D}ark \textbf{E}nergy\\
\textbf{DEC}:  \textbf{D}ominant \textbf{E}nergy \textbf{C}ondition\\
\textbf{DM}:  \textbf{D}ark \textbf{M}atter\\
\textbf{DP}:  \textbf{D}eceleration \textbf{P}arameter\\
\textbf{E}\\
\textbf{EH}:  \textbf{E}instein- \textbf{H}ilbert\\
\textbf{EC}:  \textbf{E}nergy \textbf{C}ondition\\
\textbf{EoS}:  \textbf{E}quation of \textbf{S}tate\\
\textbf{F}\\
\textbf{FLRW}: \textbf{F}riedmann-\textbf{L}ema\^itre-\textbf{R}obertson-\textbf{W}alker\\
\textbf{G}\\
\textbf{GR}:  \textbf{G}eneral \textbf{R}elativity\\
\textbf{GUT}: \textbf{G}rand \textbf{U}nified \textbf{T}heory\\
\textbf{H}\\
\textbf{HP}:  \textbf{H}ubble \textbf{P}arameter\\
\textbf{HSST}: \textbf{H}igh \textbf{R}edshift \textbf{S}upernovae \textbf{S}earch \textbf{T}eam\\
\textbf{J}\\
\textbf{JLA}: \textbf{J}oint \textbf{L}ight-Curve \textbf{A}nalysis\\
\textbf{L}\\
\textbf{$\Lambda$CDM}: \textbf{L}ambda \textbf{C}old \textbf{D}ark \textbf{M}atter\\
\textbf{LSS}: \textbf{L}arge \textbf{S}cale \textbf{S}tructures\\
\textbf{LVDP}:  \textbf{L}inearly \textbf{V}aring \textbf{D}eceleration \textbf{P}arameter\\
\textbf{M}\\
\textbf{MGT}: \textbf{M}odified \textbf{G}ravity \textbf{T}heories\\
\textbf{MSQM}:  \textbf{M}agnetised \textbf{S}trange \textbf{Q}uark \textbf{M}atter\\
\textbf{N}\\
\textbf{NEC}:  \textbf{N}ull \textbf{E}nergy \textbf{C}ondition\\
\textbf{P}\\
\textbf{PVDP}:  \textbf{P}eriodic \textbf{V}aring \textbf{D}eceleration \textbf{P}arameter\\
\textbf{Q}\\
\textbf{QGP}: \textbf{Q}uark \textbf{G}luon \textbf{P}lasma\\
\textbf{S}\\
\textbf{SCP}:  \textbf{S}upernovae \textbf{C}osmology \textbf{P}roject\\
\textbf{SEC}:  \textbf{S}trong \textbf{E}nergy \textbf{C}ondition\\
\textbf{SEP}: \textbf{S}trong \textbf{E}quivalence \textbf{P}rinciple\\
\textbf{SR}: \textbf{S}pecial \textbf{R}elativity\\
\textbf{SNeIa}:  \textbf{T}ype Ia \textbf{S}upernova\\
\textbf{SNLS3}: \textbf{S}upernova \textbf{L}egacy \textbf{S}urvey three year sample\\
\textbf{W}\\
\textbf{WEC}:  \textbf{W}eak \textbf{E}nergy \textbf{C}ondition\\
\textbf{WEP}: \textbf{W}eak \textbf{E}quivalence \textbf{P}rinciple\\
\textbf{WH}:  \textbf{W}ormhole \\
\textbf{WMAP}: \textbf{W}ilkinson \textbf{M}icrowave \textbf{A}nisotropy \textbf{P}robe
}


%
%


\clearpage 

\lhead{\emph{Glossary}} 

{
The standard nomenclatures used in this thesis are listed below. \\
 
$\mu$ \& $\nu$: Indicices varies from 0(1)3 for co-moving coordiates $(t, x, y, z)$ and $(t, r, \theta, \phi)$\\
$c$: Speed of light\\
$^.$:  Differentiation with respect to $t$\\
$'$: Differentiation with respect to radial coordinate $r$\\
$g_{\mu \nu}$: Metric potential\\
$\Gamma_{\mu \nu}^{\lambda}$: Christoffel symbol\\
$R_{\mu \nu}$: Ricci tensor\\
$R$: Ricci scalar\\
$q$: Deceleration parameter\\
$a(t)$: Scale factor\\ 
$H$:  Hubble's parameter\\
$G_{\mu \nu}$: Einstein tensor\\
$G$: Gravitational constant\\
$\mathcal{L}_m$: Matter Lagrangian\\
$S$: Einstein- Hilbert action\\
$T_{\mu \nu}$: Stress energy-momentum tensor\\
$T$: Trace of stress energy-momentum tensor\\
$\mathcal{T}$: Torsion scalar\\
$\mathcal{G}$: Gauss- Bonnet invariant\\
$p$: Pressure \\ 
$\rho$: Energy density\\
$p_r$: Radial pressure of WH\\
$p_t$: Tangential pressure of WH \\
$\omega$: EoS parameter\\
$t_0$: Present time\\
$H_0$: Hubble constant at present epoch\\
$a_0$: Scale factor at present epoch\\
$h^2$: Magnetic flux\\
$\Lambda$: Cosmological constant
}
\clearpage
\setstretch{1.3} 

\pagestyle{empty} 

\Dedicatory{\bf Dedicated to my loving parents \\and\\ beloved uncle Late K. C. Sahoo.}

\addtocontents{toc}{\vspace{2em}} 


\mainmatter 

\pagestyle{fancy} 



\chapter{Introduction} 
\label{Chapter1}

\lhead{Chapter 1. \emph{Introduction}} 
%

This chapter deals with an overview of modern cosmology, addressing its most important observational phenomenon, expansion of universe. In addition, some physical aspects to provide a theoretical prove to the expansion of universe.\\
Cosmology is a branch of physical science that studies the ultimate fate of the universe. It reveals the origin, evolution, structure as well as nature and laws of the universe. It is not only based on the theoretical facts and some fundamental principles but also confronted by various observations. The birth of modern cosmology started with the discovery that almost everything in the universe appears to be moving away from us. Mostly, the galaxies appears to be receding from us at speeds proportional to their distance. These types of large scale features are commonly interpreted as expansion of the entire universe. 
\section{Expansion of the universe}\label{ch1EU}
Expansion of the universe is a known fact since 1929.  The first study of expansion of the universe by Hubble through galaxy observation is considered to be one of the most important discoveries ever made. Redshift of galaxies and distance moduli have played a vital role in the determination of Hubble's law. Edwin Hubble discovered a simple proportionality relation between the redshift in the light coming from nearby galaxies and their distances. Hubble found the recession speeds of nearby galaxies were linearly related to their radial distance with a proportionality constant $H_0$. \\
The dynamics of this expansion can be defined by universe and dimensionless expansion function $a(t)$, known as cosmic scale factor. So, the distance $d$ between any two galaxies is given as\\
\begin{equation}
 d(t)=a(t)d_0
\end{equation}
where $d_0=d(t_0)$ is the present distance and $a(t_0)=1$ (at present time $t_0$). The relative velocity can be obtained as 
\begin{eqnarray}
v=\dot{d}=\dot{a} d_0 \Rightarrow \dot{d}=\dot{a} \frac{d}{a}\\
\Rightarrow \dot{d}=\frac{\dot{a}}{a}d.
\end{eqnarray}

Here, the rate of expansion denoted as $\frac{\dot{a}}{a}=H$, known as Hubble parameter (HP) named after the eponymous astronomer Edwin Hubble. He proposed the following relation for galaxies at cosmological distance and it is called Hubble's law.
\begin{equation}
v=\frac{\dot{a}}{a}d=Hd
\end{equation}
 As per the observational evidence for large number of galaxies, the value of Hubble constant $H_0$ (where $H(t_0)=H_0$, HP at present time $t_0$) can be estimated as $H_0\approx 72\pm 5$ km Sec$^{-1}$ Mpc$^{-1}$, or $\approx 0.074$ Gyr$^{-1}$. \\
In the universal evolution, Big Bang theory constitutes the current cosmological paradigm. It explains the early phase evolution of the universe, hot phase of the entire universe, which is known as the beginning of the universe. This theory indicates that the universe starts in a very hot, dense state from which it expands and cools. If the rate of expansion is constant with $a(0)=0$ at an initial point, then the expansion function would evolve as $a(t)=\dot{a}t$ which implies $H=\frac{\dot{a}}{a}=\frac{1}{t}$, and the current age would be the Hubble time, $t_H=\frac{1}{H_0}$. This is known as time scale in an expanding universe. According to the observational estimation, the value of $H_0=0.074$ Gyr$^{-1}$ and the present age of the universe is $t_0=t_H \approx 13.5$ Gyr. It is similar to the oldest stars in our galaxy, and suggests that our galaxy formed early in the universe's expansion. But gravitation and dark energy will modify the expansion rate. So, such age estimates can only be approximate estimations. \\
The dynamics of the expanding universe is inherited from Big Bang theory, and its matter energy content, which is the source of gravitation. But, the evolving history of universal dynamics is dominated by an unknown form of energy, named as dark energy (DE), which constitutes 95\% of the universe's gravitating contents. Observations of the effects of these unobserved, and mysterious DE and dark matter (DM) on the visible universe and reasonable extrapolations from known physics give a possibility to understand the fate of the universe. Also, one can model the dynamics of these forms of matter/energy in order to understand them theoretically. The search for the nature of these dark components is a major challenge of modern cosmological research. However, these physical models that based largely on explicit form of the expansion function $a(t)$ are one of the best attempt to explain large-scale features of the universe. On the other hand, the observational consequences of the models, such as the Hubble relation between distance and recession velocity are tested against actual observations in order to assess their validity. On the cosmological scales Einstein's general theory of relativity (GR) is the most accepted theory to address the dominant force of gravitation. 
\section{General Relativity}\label{GR}
Albert Einstein published `General theory of relativity' as geometric theory of gravitation in the year 1915 for describing the universe. It has turned out to be the underlying theory of every cosmological model of the universe, providing a unified description of gravity as a geometric property of space-time through matter energy distribution. The evolution of the universe is described by the EFEs in GR.\\
In the first step of Einstein's principle of special relativity; the laws of physics were the same in all non-accelerating reference frames. Further development of his goal of relativistic physics came with his general theory of relativity, in which the laws of physics to be identical in all reference frames, including accelerating ones. This extension of the relativity principle leads to the GR. Later on, it is employed as a successful theory in cosmology to describe large scale structure and evolution of the universe. The fundamental concept of  Einstein's GR are
\begin{itemize}
\item \textbf{General co-variance}- It express the relativity principle that is, laws of physics take the same form in all coordinate systems.
\item \textbf{Equivalence principle}- It defines the equality of gravitational and inertial mass.
\item \textbf{Space-time curvature}- It provides the mass by which, gravitation controls the dynamics.  
\end{itemize}  
This co-variance of inertial coordinate systems allowed Einstein to write the laws of mechanics and of electrodynamics in different ways that revealed new aspects (e.g. $E=mc^2$ ) and extended their validity to reference frames moving at high velocity. \\
The main purpose of GR was to extend this principle of \textbf{general co-variance} that the equations of physics were invariant to change the coordinate system to all reference frames including accelerating one. For this purpose, some tools of mathematical physics also have developed in ways that revealed new aspects. A coordinate-independent mathematical quantities associated with geometry has been developed by mathematicians. So that equality in any coordinate system  led to equality in all coordinate system. These were different forms of \textbf{tensors}, which have since been employed in many areas of advanced physics.\\
\textbf{\textit{Metric tensor}}\\
In GR, geometrically related objects play a central role to represent physical quantities of interest. In general, the incremental distance in any Riemannian curved space is given as
\begin{equation}
ds^2=\sum_{\mu,\nu}g_{\mu \nu}dx^{\mu}dx^{\nu}
\end{equation}
where $x^\mu$ is the $\mu$th coordinate (e.g. $x, y,z$ or $r$, $\theta$, $\phi$ etc.) and $g_{\mu \nu}$ represents the components of metric tensor. It express as the curvatures of space in which the path length is being measured. The metric tensor \textbf{g} (where $g_{\mu \nu}=(\textbf{g})_{\mu \nu}$)  characterizes the space and coordinate system. It is used in GR such that, gravitating matter and energy establish the form of this tensor through the curvature they induce, and that in turn determines the paths of freely falling objects.\\
In general, the metric tensor characterizes the geometry underlying the coordinate system and is thus a fundamental quantity in GR as well as for all cosmological model formulations.\\
 The second fundamental principle of GR, namely, \textbf{equivalence principle}, constitutes the connection between curvature and gravitation. In this principle, the observed equality of gravitational and inertial mass lead all objects to experience the same acceleration in a gravitational field. There are two types of equivalence principle; weak equivalence principle (WEP) and strong equivalence principle (SEP). In both cases, acceleration and gravitation are equivalent, but they assert different senses and consequences.\\
\textbf{\textit{Weak equivalence principle}}\\
This principle states that, all objects in a gravitational field experience the same acceleration. As a consequence, the accelerating effects of gravitation can be transformed away by going over to a coordinate system falling freely with the gravitational field. Thus, WEP can be rewritten as
``the dynamical effects of a gravitational field can be transformed away by moving to a reference frame that is freely falling in the gravitational field". The effects of changing coordinate systems are entirely expressed in the term called ``\textbf{affine connection} $\Gamma$" (commonly known as Christoffel symbol). It can be expressed directly in terms of the metric tensor components as
  \begin{equation}
  \Gamma_{\mu \nu}^{\lambda}=\frac{1}{2}g^{\lambda k}\biggl(\frac{\partial g_{\mu k}}{\partial x^{\nu}}+\frac{\partial g_{\nu k}}{\partial x^{\mu}}-\frac{\partial g_{\mu \nu}}{\partial x^{k}}\biggr).
  \end{equation}
This affine connection incorporates metric tensor components, is a consequence of role played by curvature in the GR formulations of gravitation. It can be noted that, when all the metric tensor components are constant then $\Gamma_{\mu \nu}^{\lambda}=0$, for all indices which lead to zero acceleration. In this sense, the affine connection describes departures from inertial reference frames.\\
\textbf{\textit{Strong equivalence principle}}\\
The WEP deals only with gravitational forces. The SEP includes all fundamental forces and claims that by no means internal to a reference frames one can distinguish between acceleration of that frame, and the presence of a gravitational field.\\
From a geometrical point of view, a locally inertial reference frame is the equivalent of a locally flat space-time and the above statement of SEP is evocative of one characteristics of the Riemannian geometries that underlie GR: any sufficiently small region in such a geometry must be locally flat. Thus Einstein chooses Riemannnian geometry for modeling curvature of space-time.\\
 According to Einstein's concept of universes; the matter and energy causes space-time to be curved and that curvature determines the path of freely falling objects. This curvature can be reflected in terms of affine connections as well as in terms of metric tensor. It is necessary to define appropriate measures of curvature in terms of the metric tensor components.\\
\textbf{\textit{Ricci tensor}}\\
As we have focused Riemannian geometry for curved space-time, it is a pre-requisite to analyse the curvature in Riemannian space-time. Riemann has chosen to characterise curvature in terms of measurable quantities on curved surfaces. One can find out an easiest way to visualize the characteristic of curvatures and movement of vectors along geodesics through ``\textit{parallel transport}". Thus, Riemann has taken as a measure of curvature of any space to the amount of vector rotation $\Delta V$ upon parallel transport about an infinitesimal closed path $\Delta f^{\mu k}$, bound by coordinate, $\mu$ and $k$. Then the change in vector component can be written as
\begin{equation}
\Delta V_\nu=R_{\mu \nu k}^{\lambda} V_{\lambda} \Delta f^{\mu k}
\end{equation} 
where $R_{\mu \nu k}^{\lambda}$ is known as Riemannian curvature tensor and is given by
 \begin{equation}
 R_{\mu \nu k}^{\lambda}=\frac{\partial \Gamma_{\mu \nu}^{\lambda}}{\partial x^k}-\frac{\partial \Gamma_{\mu k}^{\lambda}}{\partial x^\nu}+\Gamma_{\mu \nu}^{\eta} \Gamma_{k \eta}^{\lambda}-\Gamma_{\mu k}^{\eta}\Gamma_{\nu \eta}^{\lambda}.
 \end{equation}
This tensor fully characterizes curvature. For flat space-time all components of Riemannian curvature are zero in all coordinate system and vice verse by its rank two contraction. The Riemannian curvature is termed as \textbf{Ricci tensor} 
\begin{equation}
R_{\mu k}=R_{\mu \lambda k}^{\lambda}\\
         =   \frac{\partial \Gamma_{\mu \lambda}^{\lambda}}{\partial x^k}-\frac{\partial \Gamma_{\mu k}^{\lambda}}{\partial x^\lambda}+\Gamma_{\mu \lambda}^{\eta} \Gamma_{k \eta}^{\lambda}-\Gamma_{\mu k}^{\eta}\Gamma_{\lambda \eta}^{\lambda}.
\end{equation}
This can be further contracted to the \textbf{Ricci scalar/ curvature scalar}
\begin{equation}
R=g^{\mu \nu}R_{\mu \nu}.
\end{equation} 
These two quantities are the only useful contractions of Riemann tensor: all others are equivalent to them, or identically zero.\\
\textbf{\textit{Einstein field equation}}\\
Space-time curvature is established by the density of gravitating matter and energy, and is reflected in the metric tensor components. The equation relating the resulting metric tensor to matter/energy is known as the EFE of gravitation. It is in the form $G(\textbf{g})=T(\rho, \epsilon)$, where $T$ stands for source of gravitation and $G(\textbf{g})$ termed as Einstein tensor, a curvature tensor  containing metric tensor components and their derivatives.\\
\textbf{\textit{Source of Gravitation}}- Possible sources of gravitation include all forms of mass/energy, including such things as kinetic energy, pressure (which carries dimensions of energy density), stress, electromagnetic field energies etc.; in addition to normal matter and radiation. One of the  consequences is that the tensor describing the densities of sources of gravitation is known by many names: mass energy tensor, energy momentum tensor, and stress energy tensor. One will stick to the most commonly used name the energy momentum tensor and that is denoted by $T^{\mu \nu}$. It is a second rank tensor quantity and describes the density and flux of energy and momentum in space-time. It contains all the essential physics of a system pertinent to gravitation and space-time.  So $T^{\mu \nu}$ can be formulated as
 \begin{itemize}
 \item $T^{00}$= energy density,
 \item $T^{\mu0}$=$c\times$ density of $\mu$th component of momentum $(\mu=1,2,3)$,
 \item $T^{0\nu}$=$c^{-1}\times$ energy flux in the $\nu$th direction $(\nu=1,2,3)$, 
 \item $T^{\mu \nu}$= flux in the $\mu$th direction of $\nu$th component of momentum $(\mu,\nu=1,2,3)$.
 \end{itemize}
It is normally assumed that one has local energy momentum conservation, which translates into saying that energy momentum tensor satisfies the covariant form of the \textit{continuity equation}.
\begin{equation}
T^{\mu \nu}_{;\nu}=0
\end{equation} 
That means, the energy momentum tensor is divergent free, and its covariant divergence vanishes. According to special relativity (SR), it strictly holds in flat space-time. But in curved space-time this does not imply mass/energy conservation. As a consequence, the concept of conservation is poorly defined in curved space-time, which is a dynamical partner to matter and energy (as opposed to the situation in SR where space-time is a non-interacting background).\\
\textbf{\textit{Einstein tensor}}\\
One side of EFE contains energy momentum tensor where the other side must be a second-rank, divergent-free tensor containing space-time curvature. The simplest choice for the curvature tensor are the metric tensor $g_{\mu \nu}$ and Ricci tensor $R_{\mu \nu}$. But the Ricci tensor is not divergent-free i.e. it does not satisfy the continuity equation. Metric tensor yields a form for the EFE that does not reduce to Newtonian gravitation in static, weak-field limit. So Einstein choose the curvature side of the field equations, the simplest possible combination of these two that satisfies both requirements.
\begin{equation}
G_{\mu \nu}=R_{\mu \nu}-\frac{1}{2}g_{\mu \nu} R
\end{equation} 
where $R= g^{\mu \nu} R_{\mu \nu}$, is the curvature scalar. This is known as the \textbf{Einstein tensor}. In order to connect the geometry (curvature) of space-time with the matter/energy components of the universe, the simplest way is to combine Einstein tensor and energy momentum tensor, which formed EFE. So that, it can be reduced to Newtonian gravitation in the static, weak field limit. Hence the EFE is given as
 \begin{equation}\label{eqEFE}
 G_{\mu \nu}=R_{\mu \nu}-\frac{1}{2}R g_{\mu \nu}=\frac{8 \pi G}{c^4} T_{\mu \nu}
\end{equation}  
where $T_{\mu \nu} =g_{\mu k}g_{\lambda \nu } T^{k \lambda}$ is the covariant form of the energy momentum tensor.
\section{Cosmological model}\label{ch1cm}
Relativistic cosmology is based on two basic principles:\\
\textbf{Cosmological principle}- It states that the universe is homogeneous and isotropic, the same in all places and in all directions. In other words, isotropy means that there is no preferred direction in the universe; it looks same from each point. Homogeneity means that there are no preferred locations in the universe. It would be interesting to note that, homogeneity and isotropy do not imply each other. For example, a universe with a uniform magnetic field is homogeneous, but it fails to be isotropic due to its perpendicular directional field lines. At the same point, a spherically-symmetric distribution, is viewed from its central point, is isotropic but not necessarily homogeneous. The principle implies that the universe is maximally symmetric, i.e. it has the maximum number of symmetries. Mathematically, symmetric manifolds are spacial with constant curvature. Observations coming from radio wave, cosmic X-rays  \cite{Allen/2003,Bahcall/2003,Voevodkin/2004,Allen/2008} and specially, the cosmic microwave back ground radiation point out towards the fact that universe is very homogeneous.\\
\textbf{Weyl's postulate}- On cosmological scale, matter behaves as perfect fluid whose components move along temporal geodesics. These geodesics do not intersect, except (possibly) at one point in the past. The peculiar velocities produced by the gravitational interactions are usually negligible with respect to the velocities generated by the evolution of the universe. It is also possible to define a co-moving time, measured by the co-moving observer.\\
In the envision of the expanding universe, we can transform the expanding universe into an inertial reference frame by adopting a freely falling coordinate system in which the spatial coordinate of a galaxy does not change as a consequence of expansion. This is termed as \textit{co-moving coordinate system}. We can figure out the coordinates in terms of galaxy world lines. When the geometry of the bundle of world lines correspond to a set of galaxies evolving over a period of time, that system is defined in terms of hyper surfaces connecting the bundle of world lines and commonly called \textbf{cosmic} or \textbf{coordinate} time. \\
\textbf{\textit{Robertson-Walker metric}}\\
The form of space-time metric can be determined by these two aforesaid principle. Meanwhile, Weyl's postulate implies the space-time can be foliated in spatial hyper surfaces, while cosmological principle implies that such spatial hyper surfaces are maximally symmetric. \\
By considering these two premises the resulting metric can be formed as 
\begin{equation}\label{m1}
ds^2=-c^2dt^2+a(t)^2 \biggl[\frac{1}{1-kr^2}dr^2+r^2(d\theta^2+\sin^2(\theta)d\phi^2)\biggr].
\end{equation}
This is commonly known as the Robertson-Walker metric of cosmology. It casts the expanding universe of galaxies into the form of an inertial reference frame and is the basis for homogeneous and isotropic model of the expanding universe. The parameter $k$ in eqn. (\ref{m1}) is known as curvature parameter. It can take three values that represents the three hyper surface of constant curvature:
\begin{itemize}
\item Positive curvature, $k=1$, the sphere $S^3$
\item Null curvature, $k=0$, the plane $R^3$
\item Negative curvature, $k=-1$, the hyperboloid.
\end{itemize}
The shapes of universe referred as closed, flat and open corresponding to positive, null and negative curvature respectively.\\
The universal expansion function $a(t)$ describes the spatial evolution of the universe. At current time $t=t_0$, it can be normalized as $a(t_0)=1$. All the cosmological models can be expressed entirely as a functional form of $a(t)$ determined by solving the EFE with metric eqn. (\ref{m1}) and the curvature $k$.\\
In 1922 \cite{FRIED/1922}, Alexander Friedmann first proposed this metric given in eqn. (\ref{m1}). Later, it was developed by Lem$\hat{a}$itre independently as most general one in describing homogeneous and isotropic universe. Furthermore, Friedmann has not proposed a cosmological model based on the use of EFEs. In 1917, Einstein realized that his equations, as he wrote them in 1916, resulted in a non-statistical universe when it was supposed to be a normal content of matter for the universe. The idea of a non-statical universe seemed senseless to Einstein and irritated him, according to letter addressed to the astronomer Willem de Sitter. When de Sitter deduced the equation of an empty universe which could be expanding. This fact persuaded Einstein to modify his field equations by introducing a new term proportional to a constant $\Lambda$, the so-called cosmological constant. It was interpreted as the energy density of the vacuum. The new field equation took the following form
\begin{equation}
R_{\mu \nu}-\frac{1}{2}R g_{\mu \nu}+\Lambda g_{\mu \nu}=8 \pi G T_{\mu \nu}.
\end{equation}
The positive cosmological constant $\Lambda$ generates a repulsive cosmic force, while if it is negative the new force is attractive. After several years, when Hubble had demonstrated that the universe was expanding, Einstein declared that the introduction of the cosmological constant had been the worst error of his scientific carrier. He quoted it as ``Biggest blunder of my life".\\
Nevertheless, at the end of twentieth century the cosmological constant appears again in the scientific community for the discovery of accelerated expansion of the universe. Cosmological solutions of EFE are equations relating the two adjustable elements of space-time metric expansion function $a(t)$ and curvature $k$ to the mass/energy density of universe as expressed in the energy momentum tensor $T_{\mu \nu}$. The energy momentum tensor for perfect fluid matter is given as
\begin{equation}
T_{\mu \nu}=(\epsilon+p)u_{\mu}u_{\nu}+p g_{\mu \nu},
\end{equation}
where $u_{\mu}=(-1,0,0,0)$ is the four-velocity vector, $\epsilon=\rho c^2$ is the energy/matter density and $p$ is the pressure. Energy momentum tensor components are read as:
\begin{eqnarray}
T_{00}=(\epsilon+p)(-1)^2-p=\epsilon,\\
T_{\mu \mu}=(\epsilon+p)(0)^2+pg_{\mu \mu}=pg_{\mu \mu}.
\end{eqnarray}
The basic solutions to the EFE of cosmology are commonly known as Friedmann equations, given as:
\begin{eqnarray}
\frac{\dot{a}^2}{a^2}=\frac{8\pi G}{3c^2}\epsilon-\frac{kc^2}{a^2},\\
\frac{\ddot{a}}{a}=\frac{-4\pi G}{3c^2}(\epsilon+3p).
\end{eqnarray}
Here, we have two equations and three unknowns viz. $a, p$ and $\rho$. In order to get a solution for Friedmann equation in terms of desired $a(t)$, we require some additional information relating these functions.\\
In a cosmological context, the contribution of mass/energy densities to gravitation are given as: radiation, matter, and cosmological constant.\\
\textbf{\textit{Radiation}}- The radiation of energy density represented in term of $\epsilon_\text{r}$. It is dominated by isotropic thermal radiation and primordial neutrinos, which are remnants of the hot, early stages of universal evolution. Black body spectrum also referred as cosmic microwave background radiation (a.k.a. CMB) describes the photon context of this mass/energy density radiation.\\
\textbf{\textit{Matter}}- The energy density of matter content can be written as $\epsilon=\rho c^2$. Usually, non-relativistic baryonic and non-baryonic (dark matter) components dominate matter density while relativistic matter is either in the form of low mass neutrinos and is included in the radiation budget or is existed so early in the universe that it influences the expansion.\\
\textbf{\textit{Cosmological constant}}- The energy density corresponding to cosmological constant ($\Lambda$) is defined as
\begin{equation}
\epsilon_{\Lambda}=\frac{c^4}{8\pi G}\Lambda.
\end{equation}
With these definitions, the field equations can be rewritten as 
\begin{eqnarray}
\frac{\dot{a}^2}{a^2}=\frac{8\pi G}{3c^2}(\epsilon_\text{r}+\epsilon_\text{m}+\epsilon_{\Lambda})-\frac{kc^2}{a^2}\\
\frac{\ddot{a}}{a}=\frac{-4\pi G}{3c^2}(\epsilon_\text{r}+\epsilon_\text{m}+\epsilon_{\Lambda}+3p).
\end{eqnarray}
Another relation between mass/energy density evolution and expansion function is also useful to get solutions which can be directly derived from divergent-free property of the energy momentum tensor (i.e. conservation equation) or from two Friedmann equations themselves. From the first law of thermodynamics, we can obtain an \textit{energy equation},
\begin{equation}\label{eqe}
  \dot{\epsilon}+3\frac{\dot{a}}{a}(\epsilon+p)=0.
\end{equation}
This equation is an equal footing with the two Friedmann equations and referred as the third differential equation for $a, \epsilon$ and $p$.\\
\textbf{\textit{Equation of State}}\\
A parameter, which defines a relation between energy density and pressure, termed as equation of state a.k.a. EoS parameter, which is an independent variable that plays an important role in each cosmological models. It relates the pressure and energy density in the way as\\
 \begin{equation}
 p=\omega \epsilon.
 \end{equation}
The values of EoS parameter $\omega$ for the known energy components of the universe are as follows:
\begin{itemize}
\item Radiation: $p_\text{r}=\frac{\epsilon_\text{r}}{3}$, standard thermodynamics relation also applies to highly relativistic mater.
\item For non-relativistic matter: $\omega \approx 0$
\item For DE:\\
The DE arising from the cosmological constant is more interesting than others. It can be noted that $\epsilon_\Lambda$ is constant for constant $\Lambda$. So $\dot{\epsilon_\Lambda}=0$, then from the energy equation, we obtained $p_{\Lambda}=-\epsilon_{\Lambda}$ and $\omega_\Lambda=-1$. The pressure is obtained from the $\Lambda$ is negative, that means $\Lambda$ corresponds to a force of expansion. In short, the EoS parameter for standard energy components are
\end{itemize}
\begin{equation}
   \omega= 
\begin{cases}
    \frac{1}{3},& \text{radiation } \\
    0,              & \text{non-relativistic matter}\\
    -1,      & \Lambda- \text{DE}
\end{cases}
\end{equation}
By using this EoS, the Friedmann equations reduce to two equations with two unknowns which become consistent system of equations to evaluate exact solution of cosmological model. On the other hand, evolution of the energy density by the use of EoS can be obtained from energy equation, that means we can get some extra relation between mass energy and expansion function which provides the necessary additional constraints that allow the Friedmann equations to be fully solved. So energy eqn. (\ref{eqe}) can be rewritten in terms of EoS as
\begin{eqnarray}
\dot{\epsilon}+3\frac{\dot{a}}{a}\epsilon(1+\omega)=0\\
\Rightarrow \frac{\text{d} \epsilon}{\epsilon}=-3(1+\omega(a))\frac{\text{d}a}{a}\\
\Rightarrow \epsilon(a)=\epsilon_0 \exp \biggl(-3\int_{1}^{a}\frac{1+\omega(\alpha)}{\alpha}\text{d}\alpha\biggr)
\end{eqnarray}
where $\epsilon_0$ is the current value of energy density. If $\omega$ is a constant, it reduces to $\epsilon(a)=\epsilon_0 a^{-3(1+\omega)}$ for constant $\omega$. Then the corresponding energy densities vary with the universal expansion function as
\begin{itemize}
\item radiation: $\epsilon_{\text{r}}(a)=\epsilon_{\text{r},0}a^{-4}$,
\item matter: $\epsilon_{\text{m}} (a)=\epsilon_{\text{m},0} a^{-3}$,
\item cosmological constant: $\epsilon_{\Lambda}(a)=\epsilon_{\Lambda,0}$. 
\end{itemize}
Here, $\epsilon_{\text{r},0}, \epsilon_{\text{m},0}$ and $\epsilon_{\Lambda,0}$ are the energy densities in the current epoch. With the help of above energy density and EoS relations, we can rewrite the Friedmann equation in terms of the expansion function alone,
\begin{equation}
\text{Expansion:}\,\,\,\ \frac{\dot{a}^2}{a^2}=\frac{8\pi G}{3c^2}\biggl(\frac{\epsilon_{\text{r},0}}{a^4}+\frac{\epsilon_{\text{m},0}}{a^3}+\epsilon_{\Lambda,0}\biggr)-\frac{k_0c^2}{a^2},
\end{equation}
\begin{equation}
\text{Acceleration:} \, \, \ \frac{\ddot{a}}{a}=\frac{-4\pi G}{3c^2}\biggl(2\frac{\epsilon_{\text{r},0}}{a^4}+\frac{\epsilon_{\text{m},0}}{a^3}-\epsilon_{\Lambda,0}\biggr).
\end{equation}
The solutions to the Friedmann equation in terms of scale factor $a(t)$ are the models for expansion of the universe. From the expansion equation we can constitute a four-dimensional family of expansion function by the parametrized energy densities and curvature.\\
Hence, the expansion equation with the parametrization $\epsilon_{\text{r},0}=\epsilon_{\text{m},0}=k_0=0$ is given as
\begin{equation}
\dot{a}=\biggl(\frac{8\pi G}{3c^2}\epsilon_{\Lambda,0}\biggr)^{\frac{1}{2}}(t-t_0).
\end{equation}

The aforesaid model is called \textbf{de Sitter} model. It is first proposed by Willem de Sitter in 1917 \cite{SITTER/1917}. This model is characterized as a geometrically flat model and describes the exponential expansion of the universe.\\
On the other hand, soon after discoveries of Hubble's universal expansion in 1932 Einstein and de Sitter jointly proposed a model with flat geometry along with only matter (i.e. $\epsilon_{\text{r},0}=\epsilon_{\Lambda,0}=k_0=0$). The corresponding expansion equation is
\begin{eqnarray}
\dot{a}=\biggl(\frac{8\pi G}{3c^2}\epsilon_{\text{m},0}\biggr)^{\frac{1}{2}}a^{\frac{-1}{2}}\\
\Rightarrow a(t)=\biggl(\frac{6\pi G}{c^2}\epsilon_{\text{m},0}\biggr)^{\frac{1}{3}}t^{\frac{2}{3}}
\end{eqnarray}
This \textbf{flat}, \textbf{matter}-only model known as \textbf{Einstein-de Sitter} model.
\section{Cosmological model parameters}\label{ch1cmp}
This part deals with the development of some tools to describe the expansion of universe in terms of observational and cosmological parameters. Some alternative parameters are introduced to use in Friedmann equations which are mostly related to observation. Thereafter, we need to derive the connection between model parameters and the basic properties of expansion: time, distance, redshift. Finally, we need to check the dynamical properties of expansion in relation to observations. \\
\textbf{\textit{Hubble parameter}}\\
It is defined by the relative rate $\frac{\dot{a}}{a}$, such as 
\begin{equation}
 H=\frac{\dot{a}}{a}
\end{equation}

The dimension of $H$ is the inverse of time. The time and distances corresponding to HP are given as: Hubble time-$t_H=\frac{1}{H}$ and distance- $d_H=ct_H=\frac{c}{H}$. Roughly speaking that Hubble time is the time required for the universe to double in size. The current value of $H$ is called as Hubble constant \[H_0=\biggl(\frac{\dot{a}}{a}\biggr)_{t=t_0}=\dot{a}(t_0),\]
which is a fundamental observable characteristic of the expanding universe.\\
\textbf{\textit{Deceleration parameter}}\\
A non-dimensional parameter, to relate the observable feature to the expansion function is defined as 
\begin{equation}
q=-\frac{\ddot{a}a}{\dot{a}^2}
\end{equation}
where $a$ is the scale factor of the universe by which all length scale $\dot{a}$ is the first time derivative of $a$ and $\ddot{a}$ is the second derivative of $a$. In this notation $\frac{\dot{a}}{a}$ is equivalent to the HP $H$ and its present value is $H_0$ i.e. the Hubble constant. Recent observations have suggested that the rate of expansion of the universe is currently accelerating due to the effects of DE. This yields negative values for the DP.
In terms of $q_0$ (DP at present epoch) the Taylor's series representation of $d_{\text{L}}(z)$ (discussed in later) is 
\begin{equation}
d_{\text{L}}=\frac{c}{H_0}\biggl[z+\frac{1}{2}(1-q_0)z^2+....\biggr].
\end{equation}
Higher order terms are entitled as \textit{jerk}, \textit{snap}, \textit{crackle}, \textit{pop} etc. for more details one can refer \cite{WEINBERG/2008}.\\
\textbf{\textit{Energy density parameter}}\\
We can rewrite the expansion equation in terms of HP as
\begin{equation}
H^2=\frac{8\pi G}{3c^2}\epsilon-\frac{kc^2}{a^2},
\end{equation}

where $\epsilon=\epsilon_\text{r}+\epsilon_\text{m}+\epsilon_\Lambda$. If the universe is to be flat ($k=0$), then the energy density takes the form 
\begin{equation}
\epsilon \rightarrow\epsilon_\text{c}=\frac{3c^2H^2}{8\pi G},
\end{equation}

 which is known as \textbf{critical energy density}. It is required for a flat universe expanding at rate $H$.
For present value of $H$, $H_0=72$ km Sec $^{-1}$ Mpc$^{-1}$ $=0.074$ Gpc$^{-1}$, the corresponding estimated value of $\epsilon_\text{c}$ is 
\begin{equation}
\epsilon_{\text{c},0}=\frac{3c^2 H_0}{8\pi G}\approx 8.8\times 10^{-10} \text{J m$^{-3}$} \approx 5500  \text{Mev m$^{-3}$.}
\end{equation}

Then the expansion equation can be rewritten as 
\begin{equation}
\frac{\dot{a}^2}{a^2}=\frac{H_0^2}{\epsilon_{\text{c},0}}\biggl(\frac{\epsilon_{\text{r},0}}{a^4}+\frac{\epsilon_{\text{m},0}}{a^3}+\epsilon_{\Lambda,0}\biggr)-\frac{k_0c^2}{a^2}.
\end{equation}
The dimensionless forms of energy densities are then defined as the energy density parameters
\begin{equation}
\Omega_{\text{x}}= \frac{\epsilon_\text{x}}{\epsilon_\text{c}}\\
\Rightarrow \Omega_{\text{x},0}= \frac{\epsilon_{\text{x},0}}{\epsilon_{\text{c},0}}=\frac{8\pi G}{3c^2 H_0^2}\epsilon_{\text{x},0}.
\end{equation}
Here, $x=(r, m, \Lambda)$, (note: for $\epsilon_\text{x}=\epsilon_\text{c}$, $\Omega=1$). The expansion equation can be written in terms of these parameters as
\begin{equation}
\frac{\dot{a}^2}{a^2}=H_0^2\biggl(\frac{\Omega_{\text{r},0}}{a^4}+\frac{\Omega_{\text{m},0}}{a^3}+\Omega_{\Lambda,0}\biggr)-\frac{k_0c^2}{a^2}.
\end{equation}
The total energy density parameter is denoted by $\Omega$ such that $\Omega=\Omega_\text{r}+\Omega_\text{m}+\Omega_{\Lambda}$ and the current value is given as $\Omega_0=\Omega_{\text{r},0}+\Omega_{\text{m},0}+\Omega_{\Lambda,0}= \frac{1}{\epsilon_{\text{c},0}}(\epsilon_{\text{r},0}+\epsilon_{\text{m},0}+\epsilon_{\Lambda,0})$. 
Present estimated values for the energy density parameters are $\epsilon_{\text{r},0} \approx 0, \epsilon_{\text{m},0} \approx 0.27, \Omega_{\Lambda,0} \approx 0.73, \,\,\,\text{so that,}\ \Omega_0 \approx 1.$
The cosmological curvature is closely related with energy density such as, $k=\frac{H_0^2(\Omega_0-1)}{c^2 a^2}$. $k$ carries the same sign as  $k(<,=,>)0$ resulting $\Omega_0 (<,=, >)1$. That means, the curvature decreases monotonically, but does not change sign, as the universe expands. Open universe stays open, closed universe stays closed, and flat universe stays flat respectively. The universe is geometrically flat for $\Omega_0=1$. Moreover, the Friedmann equations can be parametrized with observable parameters as 
\begin{equation}\label{expeqn}
\text{Expansion:}\,\,\,\ \dot{a}^2=H_0^2\biggl[\frac{\Omega_{\text{r},0}}{a^4}+\frac{\Omega_{\text{m},0}}{a}+a^2 \Omega_{\Lambda,0}-(\Omega_0-1)\biggr],
\end{equation}
\begin{equation}\label{acleqn}
\text{Acceleration:} \, \, \ \ddot{a}=H_0^2\biggl(-\frac{\Omega_{\text{r},0}}{a^3}-\frac{\Omega_{\text{m},0}}{2a^2}+a\Omega_{\Lambda,0}\biggr).
\end{equation}
Here, it can be noted that, positive cosmological constant implies $\Omega_{\Lambda,0}>0$. The acceleration eqn. (\ref{acleqn}) yields the possibility of accelerating ($\ddot{a}>0$) expansion at sufficiently large value of $a$. \\
In case of Einstein-de Sitter model (i.e. flat, matter only model), $\Omega_{\text{m},0}=\Omega_0=1$ and $\Omega_{\text{r},0}=\Omega_{\Lambda,0}=0$, then expansion equation takes the form 
\begin{equation}
\dot{a}=H_0 a^{\frac{-1}{2}},
\end{equation}
which gives a solution for $a(0)=0$ as 
\begin{equation}
 a(t)=\biggl(\frac{3}{2}H_0 t\biggr)^{\frac{2}{3}}=\biggl(\frac{t}{t_0}\biggr)^{\frac{2}{3}},
\end{equation}
 
where $t_0=\frac{2}{3H_0}$ known as the \textbf{present time} of the universe.\\
\textbf{\textit{Redshift}}\\
According to the position of spectral lines of astronomical objects, it has been noted by astronomers in their analysis that the observed spectral lines are all shifted to longer (redder) wavelengths. This phenomenon termed as cosmological redshift. However, it becomes a fundamental object of distance galaxies and defined in terms of wavelength as
\begin{equation}
z=\frac{\lambda_\text{o}-\lambda_\text{e}}{\lambda_\text{e}}=\frac{\lambda_\text{o}}{\lambda_\text{e}}-1,
\end{equation} 
where $\lambda_\text{e}$ stands for emitted wavelength from a distant galaxy and $\lambda_\text{o}$ denotes as observed wavelength. For objects in relative radial motion in an inertial reference frame this quantity is given simply by the usual \textit{Doppler shift} formula, but in a co-moving coordinate system galaxies are not moving with respect to each other. In Doppler shift, the wavelength of the emitted radiation depends on the motion of the objects at the instant the photons are emitted. If the object is traveling towards us, the wavelength is shifted towards the blue end of the spectrum, which is known as \textit{blueshift}. If the object is traveling away from us, the wavelength is shifted towards the red end and the effect is known as \textit{redshift}. This technique was first used by Vesto Slipher around 1912 and was applied systematically by one of the most famous cosmologist Edwin Hubble in the following decades- it turns out that almost all galaxies are receding from us. The redshift of light from cosmological objects thus maps out history of the universe's expansion between emission and reception times but in an incomplete manner, only the expansion end points are reflected in the redshift. Thus, a set of redshifts spanning the history of the universe's expansion is required to fully explicate that expansion. In effect, a galaxy's redshift serves as an observable proxy for its unobservable expansion factor. Simply, we can write it as $z+1=\frac{1}{a}$ with the understanding that $z$ is the redshift observed at the current time and $a$ is for the time of emission of currently observed photons.\\
\textbf{\textit{Cosmic time}}\\
The cosmic time associated with a value of the expansion function is given by integrating $\frac{\text{d}t}{\text{d}a}$ as per expansion equation (\ref{expeqn})
\begin{multline}
t_\text{a}(a)=\int_{0}^{a} \frac{\text{d}t}{\text{d}x}\text{d}x=\int_{0}^{a} \frac{\text{d}x}{\text{d}x/\text{d}t}
= \frac{1}{H_0}\int_{0}^{a}\biggl[\frac{\Omega_{\text{r},0}}{x^2}+\frac{\Omega_{\text{m},0}}{x}+x^2 \Omega_{\Lambda, 0}-(\Omega_0-1)\biggr]^{\frac{-1}{2}}\text{d}x.
\end{multline}
The current time is given as 
\begin{equation}
t_0 \equiv t_\text{a}(1)=\frac{1}{H_0}\int_{0}^{1}\biggl[\frac{\Omega_{\text{r},0}}{a^2}+\frac{\Omega_{\text{m},0}}{a}+a^2 \Omega_{\Lambda, 0}-(\Omega_0-1)\biggr]^{\frac{-1}{2}}\text{d}a.
\end{equation}
The cosmic time of emission of photons from a galaxy of redshift $z$ is 
\begin{equation}
t_\text{e} (z)=\frac{1}{H_0}\int_{0}^{(z+1)^{-1}}\biggl[\frac{\Omega_{\text{r},0}}{a^2}+\frac{\Omega_{\text{m},0}}{a}+a^2 \Omega_{\Lambda, 0}-(\Omega_0-1)\biggr]^{\frac{-1}{2}}\text{d}a.
\end{equation}
The corresponding look-back time (the time elapsed since currently observed photons were emitted from a galaxy of redshift $z$) is 
\begin{equation}
t_{\text{lookback}}(z)=t_0-t_\text{e}(z)= \frac{1}{H_0}\int_{(z+1)^{-1}}^{1}\biggl[\frac{\Omega_{\text{r},0}}{a^2}+\frac{\Omega_{\text{m},0}}{a}+a^2 \Omega_{\Lambda, 0}-(\Omega_0-1)\biggr]^{\frac{-1}{2}}\text{d}a.
\end{equation}
For our flat, matter only model, $t=t_0 a^{\frac{3}{2}}=t_0(z+1)^{\frac{-3}{2}}$ and $t_0=\frac{2}{3H_0}$, so the time relations for flat, matter model only are 
\begin{eqnarray}
t_0=t_\text{a}(1)=\frac{2}{3H_0}\\
t_\text{e}(z)=t_0 (z+1)^{-3/2}\\
t_{\text{lookback}}=t_0[1-(z+1)^{-3/2}]
\end{eqnarray}
\textbf{\textit{Cosmic time dilation}}\\
In an expanding universe, enlargement of time interval is one of the fundamental phenomenon called as time dilation. The cosmological redshift is also interpreted as time dilation due to the expansion of the universe. In this phenomenon, both the duration and wavelength of the emitted light from a distance object at the redshift $z$ will be dilated by a factor of $(1+z)$ at the observer i.e. the observed time intervals systematically change with redshift according to $\vartriangle t_0=(1+z) \vartriangle t_\text{e}$. Roughly, we can say an event that span a time interval multiplied by $\frac{1}{a}=1+z$ i.e. cosmologically distant clocks appear to run slowly by this factor. This time dilation applies to chronometric wave periods as well as in type Ia supernovae observation. So that supernovae light curve decay times are observed to be systematically longer in galaxies of larger redshift, by a factor of `$1+z$'. In literature, Leibundut et al. \cite{Leibundgut/1996}
have claimed that the apparent time dilation observed in the light curves of a large-redshift supernovae is uniquely indicative of the ``expansion of our Universe". A similar analysis  is made by Goldhaber et al. \cite{GOLD/1996}, who describes their findings as the ``first clear observation of the cosmological time dilation for macroscopic objects". Moreover, one can expect the observed duration to be stretched by a factor $(1+z)$; exactly the same factor by which wavelength are redshifted for light observed from the same events. This method was proposed by Wilson \cite{Wilson/1939}, as a test of the expanding universe theory rather than the ``gradual dissipation of photonic energy" (a.k.a tried light).
\section{Cosmological distances}\label{ch1cd}
\textbf{\textit{Proper distance}}\\
The cosmological distance defined in our reference frame at a specific time is known as proper distance. This proper distance analogues to proper time, because the proper distance is defined between two space-like separated events, while the proper time is defined between two time-like separated events.
Therefore, for a galaxy of co-moving radial coordinate $r_g$ at time $t$ the proper distance $d_p$ defined for FLRW metric for a purely radial path $d\phi=d\theta=0$ is given as
\begin{equation}
d_\text{p}(r_\text{g},t)=\biggl(\int \text{d}s\biggr)_{\text{d}t=0}=a(t)\int_{0}^{r_g}\frac{\text{d}r}{\sqrt{1-kr^2}}.
\end{equation}
Since $ds=0$ for a photon traveling from a galaxy to ours,
\begin{equation}
c\int _{t_\text{e}}^{t}\frac{\text{d}t}{a(t)}=\int_0^{r_\text{g}}\frac{\text{d}r}{\sqrt{1-kr^2}}.
\end{equation}

Here, $t_\text{e}$ is the emission time of photon. Then the proper distance in terms of $t_\text{e}$ is given as
\begin{equation}
d_\text{p}(t_\text{e},t)=c a(t)\int_{t_\text{e}}^{t}\frac{\text{d}\tau}{a(\tau)}.
\end{equation}
It is more important to discuss the current proper distance (denotes as $d_0(z)$) for present time ($t=t_0$) with redshift parameter. To express this as a function of redshift, $\text{d}t$ can be replaced as $\text{d}a/\dot{a}$ with $a=(1+z)^{-1}$ and also at $t=t_0$, $a(t_0)=1$,
\begin{equation}
\text{d}_0(z)=c\int_{(1+z)^{-1}}^{1}\frac{1}{a}\frac{\text{d}a}{\text{d}a/\text{d}t}.
\end{equation} 
Using expansion equation (\ref{expeqn}), the current proper distance to a galaxy of redshift $z$ can be written as 
\begin{equation}\label{eqd0}
d_0(z)=\frac{c}{H_0}\int_{(1+z)^{-1}}^{1}\frac{\text{d}a}{\sqrt{\Omega_{\text{r},0}+a\Omega_{\text{m},0}+a^4 \Omega_{\Lambda,0}-a^2(\Omega_0-1)}}
\end{equation}
for $a\approx 1$ and $z<<1$
\begin{equation}
d_0(z)\approx \biggl(\frac{c}{H_0}a\biggr)_{(1+z)^{-1}}^{1}=\frac{c}{H_0}[1-(1+z)^{-1}]
\end{equation}
\begin{equation}
\Rightarrow \lim_{z\rightarrow 0}d_0(z)=\frac{c}{H_0}z.
\end{equation}
This holds for all cosmological model. For flat, matter only model, $d_0(z)=\frac{2c}{H_0}[1-(1+z)^{-1/2}]$. In all expanding models, $d_p$ is an increasing function of redshift with a negative second derivative.\\
\textbf{\textit{Luminosity distance}}\\
A classical way of measuring distances in astronomy is to measure the flux from an object of known luminosity ($\therefore$ the total amount of energy emitted per unit of time by a star, galaxy, or other astronomical objects \cite{Hopkins/1980}). An observed flux $F$ at a distance $d_{\text{L}}$ from a source of known luminosity $L$ is given as $F=\frac{L}{4 \pi d_{\text{L}}^2}$, which comes from the fact that, in flat space, the variation with distance of brightness of an isotropic emitter is $F=L/A(d)$, where $A(d)$ is the surface area. That means for a source at distance $d$, the flux $F$ over the luminosity $L$ is just the inverse of the area of a sphere centered around the source. In an expanding universe, the flux is diluted by two additional effects: the cosmological redshift which decreases photon energies by a factor of $1/a=1+z$, and the cosmological time dilation which decreases the photon arrival rate by the same factor. Hence $F\rightarrow \frac{F_{\text{static}}}{(1+z)^2}$. We can neatly incorporate all these matters into the flux computation by defining the luminosity distance as
\begin{equation}\label{eqd}
d_{\text{L}}=(1+z)R_0 \sin (d_0(z)/R_0).
\end{equation}  
Here, $R_0$ is the radius of curvature given by 
\begin{equation}\label{eqR0}
R_0=\frac{c}{H_0}\frac{1}{\sqrt{\Omega_0-1}}
\end{equation}
and the surface area at present time is given as $A_0(z)=4\pi [R_0\sin (d_0(z)/R_0)]^2$. The observed flux in an expanding curved universe is given as
\begin{equation}\label{eqF}
F(z)=\frac{L}{4\pi d_{\text{L}}(z)^2}
\end{equation}
\begin{equation}
\Rightarrow d_{\text{L}}=\biggl(\frac{L}{4 \pi F}\biggr)^{1/2}.
\end{equation}
Combining both equation (\ref{eqd}) and (\ref{eqF}), we obtained 
\begin{equation}
F(z)=\frac{1}{4\pi}\frac{L}{(1+z)^2[R_0 \sin(d_0(z)/R_0)]^2}.
\end{equation}
This is known as the basic form of Hubble's relation. Furthermore, to measure the distant of an astronomical object, astronomers usually prefer to deal with apparent ($m$) and absolute ($M$) magnitude rather than fluxes and luminosities. Therefore, apparent (observed) and absolute (assumed) magnitudes are then related to distance in the form as
\begin{equation}\label{eqdL}
m(z)-M=5\log d_{\text{L}}(z)+\mu_{dm 0},
\end{equation} 
where $\mu_0$ is a calibrating distance modulus; $\mu_0=(25,40)$ for $d_{\text{L}}$ in units of (Mpc, Gpc) respectively. The quantity $\mu_{dm}=m-M$ is commonly known as the \textbf{distance modulus}. Then the Hubble relations can be rewritten as
\begin{equation}
\text{d}_{\text{L}}(z)\equiv \biggl(\frac{l}{4\pi F}\biggr)^{1/2}=(1+z)R_0 \sin(d_0(z)/R_0),
\end{equation}
\begin{equation}
\mu_{dm}(z)\equiv=m(z)-M=5\log[(1+z)R_0\sin(d_0(z)/R_0)]+\mu_{dm 0}.
\end{equation}
Here, the values of $R_0$ and $d_0$ can be computed for any model by using the formula given by (\ref{eqR0}) and (\ref{eqd0}). For nearby galaxies, Hubble relations can be obtained as 
\begin{equation}
\lim_{z\rightarrow 0}d_{\text{L}}=\frac{c}{H_0}z, \,\,\,\ \lim_{z\rightarrow 0}\mu_{dm}=5\log \biggl(\frac{c}{H_0}z\biggr)+\mu_{dm0}.
\end{equation}
\section{Observational constraints}\label{ch1OB}
\textbf{\textit{Type Ia Supernovae Observation}}\\
The late time cosmic acceleration observed by distance Supernovae of Type Ia (SNeIa) has been announced separately by two teams; Supernova Cosmology Project (SCP) \cite{Perlmutter/1999}, and High-redshift Supernova Search Team (HSST) \cite{Riess/1998}. 
Till 1998, in the redshift range 42 Supernovae and 16 high redshift supernovae near by 34 supernovae have been discovered by Perlmutter et al. \cite{Perlmutter/1999} and Riess et al. \cite{Riess/1998} respectively.  The history of supernovae is extremely bright and causes a blast of radiations. Scientists classify different types of supernovae according to their light curves and the absorption lines of different chemical elements (e.g. hydrogen, helium etc.) that appears in their spectra. They are classified as type I if they have no hydrogen lines in their spectra, otherwise it is type II Supernovae. Then each of the two types are subcategorised according to the presence of other element's absorption line. The subclass type Ia refers to those which have a strong silicon line at 615 nm in their spectra, type Ib refers to the presence of strong helium lines, and Ic categorized as if they do not have helium lines. Type Ia supernovae have become very important as the most reliable distance measurement at cosmological distances, useful at distances in excess of 1000 Mpc. Astronomers believe that the origin of type Ia is from white dwarf (the carbon-oxygen remnant of a sun-like star). When the mass of white dwarf in binary system overtakes the Chandrasekhar limit of 1.4 solar masses \cite{Chandrasekhar/1931} by absorbing gas from other star then explosion of type Ia developed. That is why, the peak of brightness occur when the absolute luminosity of Type Ia is almost constant. The distance to a SNeIa can be calculated by its apparent luminosity. In this manner, the SNeIa plays major role of standard candle and we can measure the luminosity distances practically. Since the 1990s, it has become apparent that type Ia supernovae offer a unique opportunity for the consistent measurement of distance out to perhaps 1000 Mpc. Measurement at these great distances have provided the first data to suggest that the expansion rate of the universe is actually accelerating. That acceleration implies an energy density that acts in opposition to gravity which would cause the expansion to accelerate. This is an energy density which we have not directly detected observationally and it has been given the name ``DE".\\
\textbf{\textit{Cosmic Microwave Background Radiation}}\\
According to Big Bang theory, the CMB is electromagnetic radiation as a remnant of the hot, dense early universe. It can be thought as the leftover radiation from the Big Bang, or the time when the universe began. Also, the CMB discovery is one of the landmark evidences of the Big Bang origin of the universe. The origin of the CMB appears to be a thermal equilibrium between radiation and matter during the period when the universe's contents were ionized and photons scattered readily off free electrons, producing a nearly uniform and isotropic radiation field of the same temperature as baryonic matter. Also, the detection of CMB by Penzias and Wilson \cite{Penzias/1965} in 1965, is one of the most important pillars of modern cosmology today providing an incredibly rich source of information about parameters describing our Universe \cite{White/1994,Hu/2002}. The CMB photons detected today originated from the last scattering surface (LSS), when the universe was roughly 380000 years old and due to its expansion had cooled down sufficiently ($\sim$ 3000K) to allow the formation of neutral atoms. The temperature of the universe would have fallen about ten thousand million degrees just one second after the Big Bang. At that time, universe would have contained particles like photons, electrons and neutrinos (the particles which are affected by the weak force and gravity) and their antiparticles, also protons and neutrons. However, in the standard picture of the CMB photons were born at much earlier epochs, very close to the Big Bang, and therefore represents the oldest detectable electromagnetic relics in the universe, with roughly 410 photons per cm$^3$ today.  In 1992, a satellite named Cosmic Background Explorer (COBE) measured the spectrum of the CMB radiation and detected slight fluctuations of the temperature of CMB \cite{Bennett/2003}. It revealed that the energy spectrum of the CMB is extremely close to a perfect black body of temperature $T_0=2.725 \pm 0.001$K, with spectral deviations of not more than $10^{-4}$ \cite{Fixsen/1996,Fixsen/2002,Mather/1994} and fluctuations of the radiation temperature at a level of $\frac{\triangle T}{T}\sim 10^{-5}$ on angular scales smaller than $7^0$ \cite{Smoot/1992}. Also, the fluctuation results from WMAP \cite{Hinshw/2013} reveals that temperature variations in CMB follow a distinctive pattern predicted by cosmological theory. The anisotropies in the CMB are confirmed from the recent observations and this behavior is helpful to tell us about the present and past history of the universe. The discovery of CMB was the strongest evidence in the favor of hot Big Bang.

\textbf{\textit{Planck's observation}}\\
Planck’s high-precision cosmic microwave background map \cite{Ade/2014} has offers scientists to select the most clear value of the cosmos ingredients. The simple matter forms galaxies and stars is just contribute 4.9\% of the cosmos. Dark matter, which is
marked indirectly by its gravitational impact on closely contributes 26.8\%
of the cosmos and the remaining part of the cosmos is 68.3\% which is responsible of accelerated expansion of cosmos known as DE.\\
Beyond the standard cosmological scenario, 
PLANCK Collaboration \cite{plank/2016} has investigated the
implications of cosmological data for models of DE and modified gravity. They have been improved the present constraints and have found that the early estimated DE density has to be below $\approx$ 2\% (at 95\% confidence) of the critical density. They also have consider the general parametrizations of the DE or modified gravity perturbations that encompass both effective field theories and the phenomenology of gravitational potentials in modified gravity models. They have tested a range of specific models, such as $\textbf{k}$-essence, $f(R)$ theories and coupled DE. It is an important tool in order to test cosmological models.
\section{Dark energy}\label{ch1DE}
One of the most remarkable discoveries of our time is the late time cosmic acceleration of the universe. But what causes this phenomenon in the puzzle of modern cosmology that there is no convincing answer to this question at present. However, the current best fit in the Hubble's diagram seems to imply a preference for a universe with more than 70\% of the energy in the form of DE \cite{Dodelson/2003}, for this reason, it appears more important to investigate the scenario of DE dominated universe. Usually, normal matter (cold dark matter/ radiation or baryonic matter) is gravitationally attractive. We need an exotic matter with repulsive in character which can account for late time acceleration. The hypothetical matter with the said unusual property is known as DE \cite{Weinberg/2013,Sahni/2006,Padmanabhan/2006,Bamba/2012,Copeland/2006,Frieman/2008,Sami/2009}. Since it has never been detected or created in laboratory, treated as hypothetical \cite{Liddle/2003}. The recognition that DE appears to exist has completely altered the landscape of theoretical physics and driving most of astrophysicists to launch new cosmic probes to detect its nature. \\
One of the simplest form of this DE is `cosmological constant' ($\Lambda$) or vacuum energy density, with negative pressure, whose equation of state can be written as $\omega=\frac{p}{\rho}=-1$. $\Lambda$ is called the vacuum energy density, since, in particle physics, it naturally arises as the energy density of the vacuum. Originally, the cosmological constant was introduced by Einstein and included in his field equations of GR to keep the universe static. However, it later turned out that the cosmological constant itself can be regarded as a form of DE that driving the late time acceleration of the universe.\\
The standard model of cosmology known as $\Lambda$CDM model, is a very good model in agreement with observational data. However, there exists several fine-tuning problems, one of which is the value of $\Lambda$ in many orders of magnitude smaller than that of the vacuum energy predicted in quantom field theories. It is severly fine-tuned and is the order of about $10^{121}$ wrong. The observational value of DE is expected to be about $10^{74}$ GeV while the vacuum energy is approximatly $10^{-47}$ GeV. This problem is known as the cosmological constant problem \cite{Carroll/2001,Peebles/2003}. It has not been resolved satisfactorily until today.\\
It is belived that if DE evolves with time, the cosmological coincidence problem may be resolved. For this solution purpose many aspects are introduced in literature. For example, quintessence scalar field model \cite{Wetterich/1988,Ratra/1988}, phantom fields \cite{Caldwell/2002}, $\textbf{k}$-essence \cite{Chiba/2000, Armendariz/2000, Armendariz/2001}, tachyons \cite{Padmanabhan/2002,Padmanabhan/20002}, Chaplygin gases \cite{Kamenshchik/2001, Bilic/2002, Bento/2002}, etc.\\
\textbf{\textit{Quintessence}}\\
In order to solve the fine-tuning problem of cosmological constant $\Lambda$ at present era, quintessence scalar field model has been introduced by Wetterich \cite{Wetterich/1988}, Caldwell \cite{Caldwell/1998}, and Ratra and Peebles \cite{Ratra/1988}. In this model, a dynamical scalar field which can explain the role of DE in an accelerating universe. Usually the cosmological constant is attributed to the vacuum energy with constant energy density $\rho$, pressure $p$ and EoS $\omega=-1$, whereas quintessence is a time varying inhomogeneous field with an EoS $-1<\omega<0$ \cite{Wetterich/1988, Caldwell/1998, Ratra/1988}. In this case, DE dominates the cosmic acceleration of the universe for future evolution. The dominance of quintessence field increases with the increase in $\omega$.\\
\textbf{\textit{Phantom}}\\
It is another hypothetical form of DE with EoS $\omega<-1$. The difference between the DE with $\omega>-1$ and $\omega<-1$ becomes apparent if we consider expansion of the universe. Phantom energy \cite{Caldwell/2002} violates NEC \cite{Hawking/973}, which yields the existence of wormholes. For phantom energy the energy density emerge and becomes unbounded in a bounded time. Phantom energy increases the gravitational repulsion that will destroy the galaxies and then any bound system including elementary particles \cite{Caldwell/2002,Caldwell/2003}. Expansion factor of the universe dominated by the phantom energy diverges in a finite time to approach the future singularity \cite{Caldwell/2003, Mcinner/2002}. This situation is also termed as cosmic doomsday when all the objects from galaxies to nucleons will be ripped apart. According to literature \cite{Baushev/2010}, phantom energy is not enough to produce Big Rip because $\omega$ does not seem to be constant throughout the evolution of the universe. \\
\textbf{\textit{Quintom}}\\
Quintom is a unified DE model with EoS parameter getting across the cosmological constant boundary $\omega=-1$ from either side. Feng et al. \cite{Feng/2005} have considered the effects of cosmic age and Supernovae Ia limits on the variation of the EoS parameter $\omega$. They have found that age limits can lower the variation of amplitude on the EoS parameter. Current Supernovae Ia data favors the transition of $\omega$ from quintessence to phantom and the quintom model predicts some interesting features related to the evolution and fate of the universe. In quintom scenario, the universe would avoid the singularities such as Big Bang, Big Rip \cite{Guo/2005, Feng/2006}.\\
Since existence of DE has not yet been proved, it may be possible to find alternative theories that can explain the observed accelerated expansion of the universe, at the same time solving the cosmological constant problem.
\section{Modified Gravity Theories}\label{ch1MGT}
At the end of the previous section, we discussed about the expansion of the universe and one of the main causes of this expansion is the presence of DE. We can explain accelerated expansion of the universe through cosmological models without invoking DE. It yields the possibilities of modifications/extensions of GR. It is well known that Einstein's theory of gravitation has great implications in describing the gravitational phenomenon through mathematical elegance and experimental validation.  But some theoretical arguments indicate that GR suffers from shortcomings such as:
\begin{itemize}
\item	GR fails to represent local energy momentum tensor \cite{Carmeli/1990}: GR theory doesn't assign a definite stress-energy tensor to the gravitational field. This property of GR is not satisfactory as all the fundamental interactions in the universe follow the principle of local conservation of energy-momentum tensor.
\item	GR fails to be quantized:  The possibility of formulating gravity as quantum field theory is essential for unification of all fundamental interactions. However, GR has failed all attempts to find a consistent quantum gauge field theory. 
\item GR predicts space-time singularities: Space-time
singularities and event horizons are a consequence of general relativity, appearing in the
solutions of the gravitational field. Although the ``Big Bang" singularity and ``black holes" have been an topic of intensive study in theoretical astrophysics, one can seriously doubt that such mathematical monsters should really represent physical objects. In fact, in order to predict black holes, one has to
extrapolate the theory of general relativity far beyond observationally known gravity strengths. 
Albert Einstein has showed that, he was being aware of this conceptual problem:\\
``For large densities of field and of matter, the field equations and even the field variables which enter into them will have no real significance. One may not therefore assume the validity of the equations for very high density of field and of matter, and one may not
conclude that the `beginning of the expansion' [of the universe] must mean a singularity in the mathematical sense. All we have to realize is that the equations may not be continued over such regions" \cite{Einstein/1956}.\\
Many physicists would prefer a gravity theory without mathematical anomalies in its field solutions.
\end{itemize}
In order to resolve these issues several MGTs have been proposed as alternatives to Einstein's theory from time to time. There are two ways to approach this modification; firstly, modification in matter part of EFE, treated as modified matter theories. These theories are also introduced to solve cosmological constant problems as discussed in the previous section (\ref{ch1DE}). In addition, these models produce the acceleration introducing new energy components alongside matter and radiation in matter part. Second approach contains the modification in gravitational part of Einstein–Hilbert (EH) action. \\
There are several gravitational theories, that have been proposed as alternatives to GR. Now, we will concentrate on a few of specific gravitational theories. Many interesting theories of gravity have been proposed by deriving from EH action. So we have discussed the derivation of the field equations through the application of a suitable variational principle and analyzed the basic characteristic of the theory, as expressed through the field equations. Here, one can focus on the derivation of field equation from an action by variational principle, and later by modifying  the action various MGTs can be obtained as an alternative to GR. In contrast, we will focus on the cosmological models of expanding universe within $f(R,T)$ gravity formalism. \\
The universe is described by EFE given by eqn. (\ref{eqEFE}).
It is necessary to introduce the derivation of this EFE from Lagrangian through the least action principle. Let us consider the Einstein Lagrangian with matter $\mathcal{L}_\text{m}$, and assuming speed of light $c=1$, (hereafter $c=1$) the EH action is given as 
\begin{equation}\label{ch1eqactn}
S=\frac{1}{16 \pi G}\int \text{d}^4x\sqrt{-g}R+\int \text{d}^4x\sqrt{-g}\mathcal{L}_\text{m},
\end{equation} 
where $G$ is Gravitational constant.
We require the variation of action $\delta S$ to vanish. Therefore 
\begin{equation}\label{ch1eqvar}
\delta S=\frac{1}{16 \pi G}\int \text{d}^4x \delta(\sqrt{-g}R)+\int \text{d}^4x\sqrt{-g} \delta\mathcal{L}_\text{m}=0.
\end{equation}
Variation of the eqn. (\ref{ch1eqvar}) reads, 
\begin{equation}
\delta S =\frac{1}{16\pi G}\int \text{d}^4x [\delta \sqrt{-g} g^{\mu \nu}R_{\mu \nu}+\\sqrt{-g}(\delta g^{\mu \nu}R_{\mu \nu}+g^{\mu \nu}\delta R_{\mu \nu})]+\int \text{d}^4x\sqrt{-g} \delta\mathcal{L}_\text{m}=0.
\end{equation}
It is possible to show that 
\begin{eqnarray}
\delta \sqrt{-g}=\frac{1}{2} \sqrt{-g}g_{\mu \nu}\delta g^{\mu \nu}, \\
\delta R_{\mu \nu}=\delta \Gamma_{\mu \nu;\lambda}^{\lambda}-\delta \Gamma_{\mu \lambda;\nu}^{\lambda}.
\end{eqnarray}
Thus the variation of the action can be expressed as 
\begin{multline}
\delta S =\frac{1}{16\pi G}\int \text{d}^4x \biggl[ \sqrt{-g}\biggl(R_{\mu \nu}-\frac{1}{2}g_{\mu \nu} R\biggr)\delta g^{\mu \nu}+\sqrt{-g} g^{\mu \nu}\biggl(\delta \Gamma_{\mu \nu;\lambda}^{\lambda}-\delta \Gamma_{\mu \lambda;\nu}^{\lambda}\biggr)\biggr]\\
+\int \text{d}^4x\sqrt{-g} \biggl(\frac{\delta \mathcal{L}_\text{m}}{\delta g^{\mu \nu}}-\frac{1}{2}g^{\mu \nu} \mathcal{L}_{\text{m}}\biggr)\delta g^{\mu \nu}=0.
\end{multline}
This equation can be further simplified by noticing that the term containing the Christoffel symbols behaves like a vector divergence and they vanishes when the equations are integrated. The term containing $\mathcal{L}_\text{m}$ can be expressed as 
\begin{equation}
\frac{\delta \mathcal{L}_\text{m}}{\delta g^{\mu \nu}}-\frac{1}{2}g^{\mu \nu} \mathcal{L}_{\text{m}}=-\frac{1}{2}T_{\mu \nu}.
\end{equation}

Therefore, we obtain 
\begin{equation}
\delta S =\frac{1}{16\pi G}\int \text{d}^4x \biggl[ \sqrt{-g}\delta g^{\mu \nu}\biggl(R_{\mu \nu}-\frac{1}{2}g_{\mu \nu}R-8\pi G T_{\mu \nu}\biggr)\biggr]=0.
\end{equation}
This equation must be valid for every $\delta g^{\mu \nu}$. Thus we obtained the EFE in the form of eqn. (\ref{eqEFE}).
Henceforth, one can obtain several MGTs by using arbitrary choice of functions in EH action (\ref{ch1eqactn}). In GR, the gravity is described by using only one tensor field as the metric $g_{\mu \nu}$. The most popular MGTs are those in which new degrees of freedom are introduced apart from the metric. Scalar, vectors and tensors are introduced in order to produce specific features in different theories. There are several theories in this context from which some are discussed briefly in the next section.
\subsection{Scalar tensor theories}\label{ch1ScT}
Scalar tensor theories are one of the widely studied
modifications of GR \cite{Bergmann/1968,Nordtvedt/1970,Fujii/2003}. In theoretical point of view, the earliest and most famous version of scalar tensor theory is the Brans-Dicke (BD) theory, one of the paradigmatic alternative theory of GR.
In 1961, it was introduced by R. H. Dicke and his student Carl H. Brans, which is later termed as Brans-Dicke theory \cite{Brans/1961}. In this theory a scalar field included in the gravitational part, apart from metric and it was based on the earlier work of P. Jordan \cite{Jordan/1955}.\\
The action for BD theory is given as
\begin{equation}\label{eqactn1}
S=\frac{1}{16 \pi G}\int \text{d}^4x\sqrt{-g}\biggl[\phi R-\frac{\omega_0}{\phi}(\delta_\mu \phi \delta^\mu \phi)\biggr]+\int \text{d}^4x\sqrt{-g}\mathcal{L}_\text{m},
\end{equation}
where $\phi$ is a scalar field and $\omega_0$ is known as BD parameter. Note that, it is a non-minimally coupled gravity, but the scalar field $\phi$ is not directly coupled with matter. Hence matter responds only to the metric. In addition, this theory is a theory with varying gravitational constant where one can always define an effective gravitational constant $G_{eff}=\frac{G}{\phi}$. Therefore, the theory can indeed be thought as a manifestation of Dicke's formulation of Mach's principle. \\
Furthermore, BD theory generalized to a scalar tensor theory with the action,
\begin{equation}\label{eqactn2}
S=\frac{1}{16 \pi G}\int \text{d}^4x\sqrt{-g}\biggl[\phi R-\frac{\omega}{\phi}(\delta_\mu \phi \delta^\mu \phi)-V(\phi)\biggr]+\int \text{d}^4x\sqrt{-g}\mathcal{L}_\text{m},
\end{equation}
where $V(\phi)$ is the scalar field potential and $\omega(\phi)$ is an arbitrary function of $\phi$. If we set $\omega(\phi\omega_0)$ and exclude the potential term $V(\phi)$, it will reduces to previous one.
The simplicity of their formulation allowed to obtain exact field equations in many interesting physical situations or addition of single scalar field also attempts to unify gravity with other fundamental forces such as string theory. It is worth clarifying that BD theory or its generalized scalar tensor theories are metric theories of gravity. There are no direct connections between matter and scalar field. 
\subsection{$f(R)$ gravity}\label{ch1f(R)}
Adding scalar to the EH action is not the only possible ways to modify GR and its field equations. There are several ways to approach the modification of gravity instead of adding scalar field into the EH action with a potential form of a non-minimal coupling between the scalar field and gravitational sector. So that one can prefer a simplest way that directly modify the EH action by replacing the Ricci scalar $R$ to an arbitrary function of it i.e. $f(R)$. Such kind of modification is termed as \textit{$f(R)$ gravity} theory, and it is one of the most well studied modified gravity in literature \cite{Sotiriou/2010,Felice/2010,Capozziello/2011}. Thus, the action for $f(R)$ gravity theory is given as 
\begin{equation}\label{eqactn3}
S=\frac{1}{16 \pi G}\int \text{d}^4x\sqrt{-g}f(R)+\int \text{d}^4x\sqrt{-g}\mathcal{L}_\text{m}.
\end{equation}
It is worth to recall that, the field equations can be obtained by varying the action. One could vary the action with respect to the metric or alternatively with respect to the metric and connection. The variation with respect to metric by variational principle called as \textit{metric $f(R)$ gravity}. Similarly, variation with respect to both metric and connection yielding the \textit{palatini $f(R)$ gravity}, where the connection is independent of metric and vice versa.  
These two approaches lead the same field equation in case of usual EH action. The third and most general way is the \textit{metric-affine $f(R)$ gravity}, in which we use the palatini formalism abandoning the assumption that the matter action is independent of the connection. There are some theoretical difficulties in the palatini version of $f(R)$ gravity, for example it appears to be in conflict with the standard model \cite{Flanagan/2004,Iglesias/2007}, and suffers from the existence of singularities appearing in stars \cite{Barausse/2008}. We thus consider metric $f(R)$ gravity. Performing the variation of action the field equation is
\begin{equation}
FR_{\mu \nu}-\frac{1}{2}f g_{\mu \nu}=g_{\mu \nu} \Box F-\nabla_{\mu} \nabla_{\nu}F=8\pi G T_{\mu \nu},
\end{equation}
where $\Box=\nabla_{\mu} \nabla^{\nu}$ and $F=f'(R)$. The above theory has been extensively studied in a variety of applications. Surprisingly, such a simple replacement can account for many observed phenomena. Since, the function $f(R)$ can be expressed as a series expansion, and contains terms which are of a phenomenological interest. The appealing feature of $f(R)$ gravity  action is that it combines mathematical simplicity with a fair amount of generality. For instance if we take a series expansion of $f(R)$ i.e.
\begin{equation}
f(R)=...+\frac{a_2}{R^2}+\frac{a_1}{R}-2\Lambda +R+b_2 R^2+b_3R^3+....
\end{equation}
where $a_i$ and $b_j$ coefficients have the appropriate dimensions. We see that the action includes a number of
phenomenologically interesting terms. For example, the Starobinsky model of $f(R)$ gravity \cite{starobinsky/1980,Starobinsky/2007}, where $f(R)=R+\alpha R^2$, is a strong candidate to describe inflation. The functions of the form $f(R)=R^n$ have been shown interesting properties in analyzing the galaxy rotation curves \cite{Martins/2007}. They are also of great interest for applications in late time acceleration for a dynamical system analysis of the cosmology in general $f(R)$ gravity, see \cite{Amendola/2007}.\\
In last two decades,  $f(R)$ theories have been studied extensively with the general function $f(R)$ and with some particular choices like,
 $R^n$, $\ln(\lambda R)$, $e^{\lambda R}$, $R-a(R-\Lambda_1)^{-1}+b(R-\Lambda_2)^n$ etc.\\
Mostly, it is considered as an interesting toy model, but some researchers find an appropriate expression for the function $f(R)$ which could satisfy all the constraints in order to describe the real nature. Also, as a viable theory it should confirm to the galaxy clustering and CMB anisotropy spectra. It allows the existence of matter dominated era and stable stars. But it should not contain ghost or tachyons. In addition, it should agree with the small scale results of Einstein's gravity. To fulfil all these necessary requirements, the function $f(R)$ has to satisfy some important conditions as follows \cite{Felice/2010,Sotiriou/2010}
\begin{itemize}
\item  In order to avoid the ghost states, $f_R>0$ for $R\geq R_0$, where $R_0$ is the present value of Ricci scalar.
\item $f_{RR}>0$ for $R\geq R_0$. This is necessary to avoid the existence of scalar degree of freedom with negative mass, i.e. tachyons.
\item $f(R)\rightarrow R-2\Lambda$ for $R\geq R_0$. This needs to be true for the presence of the matter dominated era and for agreement with the local gravity constraints.
\item $0< \frac{Rf_{RR}}{f_R}<1$ when $\frac{Rf_R}{f}=2$. This condition is important for stability and late de Sitter limit of the universe. Also, the matter epoch is followed by an accelerated phase where the global EoS parameter goes asymptotically to $-1$.
\end{itemize}    
These conditions rule out many $f(R)$ models, as the simple $f(R)=R-\alpha R^{-n}$. This model seems to have good properties, such as the importance of the modification at low curvatures. Also there is a special function designed by Starobinsky in \cite{Starobinsky/2007} to pass all such requirements as 
\begin{equation}
 f(R)=R+\lambda R_0\left[\biggl(1+\frac{R^2}{R_0^2}\biggr)^{-n}-1\right]
\end{equation} 
where $n$, $\lambda$ are positive constants and $R_0$ is of the order of $H_0^2$.
Henceforth, various exact solutions have been found in $f(R)$ theories. It can be shown that any vacuum
solution to Einstein's gravity is also a solution of $f(R)$ gravity except for some pathological choices of the function $f$. This includes the usual black hole solutions, e.g. Schwarzschild
space-time. However, since the Birkhoff theorem does not hold here, other spherically symmetric solutions exist in the $f(R)$ gravity. Propagating scalar degree of freedom implies the existence of an additional types of longitudinally polarized wave-like solution. Since the scalaron is massive, these waves would travel with the speed lower than that of light.
\subsection{Gauss-Bonnet gravity}\label{ch1GBT}
Other modified gravity models can be built by replacing the Ricci scalar $R$ with other quantities coming from the Ricci tensor $R_{\mu \nu}$ in the Lagrangian. We can allow any combination of the quadratic curvature terms, namely, $R^2$, $R_{\mu \nu}R^{\mu \nu}$, and $R_{\mu \nu \alpha \beta}R^{\mu \nu \alpha \beta}$ etc.  in the Lagrangian to derive the field equations of second order. One of the special case of quadratic gravity theories is Gauss-Bonnet gravity. The action of this theory is formulated as \cite{Nojiri/05}
\begin{equation}\label{eqactn4}
S=\int \text{d}^4x\sqrt{-g}\biggl[\frac{1}{16 \pi G}\biggl(R-f(\Phi)\mathcal{G} \biggr)-\frac{1}{2} g^{\mu \nu}\delta _{\mu}\Phi \delta_{\nu} \Phi-V(\Phi)\biggr]+\int \text{d}^4x\sqrt{-g}\mathcal{L}_\text{m}.
\end{equation} 
In this action a scalar field is non-minimally coupled with a Gauss invariant $\mathcal{G}=R^2-4R_{\mu \nu}R^{\mu \nu}+R_{\mu \nu \alpha \beta}R^{\mu \nu \alpha \beta}$ in 4 dimensional space-time. Here, $\Phi$ is the scalar field. The non-topological character of the Gauss-Bonnet term in the above action is ensured by the coupling function between the scalar field and the Gauss-Bonnet term, symbolized by $f(\Phi)$.\\
One of the interesting properties of this theory is it does not lead to any ghost on many background space-time after quantization \cite{Chiba/0005}.
\subsection{$f(\mathcal{G})$, $f(R,\mathcal{G})$, $f(\mathcal{G},T)$, $f(\mathcal{T})$ gravity}\label{ch1FRGT}
Some specific MGTs are investigated with the Gauss-Bonnet invariant, namely, $f(\mathcal{G})$, $f(R,\mathcal{G})$, and $f(\mathcal{G},T)$ etc. This section introduce briefly these theories.\\
In EH action, the Lagrangian is modified with an addition of arbitrary function $f(\mathcal{G})$ \cite{Odintsov/2006,Amendola/20007}. The latter approach is introduced by Nojiri and Odintsov, known as $f(\mathcal{G})$ gravity \cite{Nojiri/0005}. Like other modified theories, this theory is an alternative to study DE and is consistent with solar system constraints \cite{Felice/2009}. The action of this theory is written as 
\begin{equation}\label{eqactn5}
S=\frac{1}{16 \pi G}\int \text{d}^4x\sqrt{-g}(R+f(\mathcal{G}))+\int \text{d}^4x\sqrt{-g}\mathcal{L}_\text{m}.
\end{equation}
Another specific modified gravity that yields a general class of non-linear gravity model having the action in the following form
\begin{equation}\label{eqactn5}
S=\frac{1}{16 \pi G}\int \text{d}^4x\sqrt{-g}f(R,\mathcal{G})+\int \text{d}^4x\sqrt{-g}\mathcal{L}_\text{m},
\end{equation}
where $R$ and $\mathcal{G}$ are the Ricci scalar and Gauss-Bonnet scalar respectively.\\ 
Similarly, the general action for $f(\mathcal{G},T)$ gravity can be written as \cite{Sharif/2016,Shamir/2018}
\begin{equation}\label{eqactn7}
S=\frac{1}{16 \pi G}\int \text{d}^4x\sqrt{-g}(R+f(\mathcal{G},T))+\int \text{d}^4x\sqrt{-g}\mathcal{L}_\text{m},
\end{equation}
where $\mathcal{G}$ and $T$ are the Gauss-Bonnet invariant and trace of stress energy momentum tensor respectively.\\
We have introduced various kind of MGTs. One can consider a straightforward modification of the teleparallel gravity action with an arbitrary function of torsion scalar $\mathcal{T}$. Such a theory is named as $f(\mathcal{T})$ gravity theory, and the action of this is given by \cite{Bengochea/2009}
\begin{equation}\label{eqactn8}
S=\frac{1}{16 \pi G}\int \text{d}^4x e f(\mathcal{T})+\int \text{d}^4x e\mathcal{L}_\text{m}.
\end{equation}
Since the torsion scalar $\mathcal{T}$ only depends on the first derivative of the tetrads, this theory is a second order theory.
\subsection{$f(R,T)$ gravity}\label{ch1f(R,T)}
The modified $f(R)$ theory provides several viable models, passing solar system tests, satisfying local tests and also unifying inflation with DE era \cite{Nojiri007,Nojiri08,Cognola/2008}. Therefore, it is a successful modified theory in account of late time acceleration. Furthermore, it can be generalized into several theories for better theoretical results of late time evolution of universe. In this context, the gravitational Lagrangian of EH action is modified by an arbitrary function of the Ricci scalar $R$ and of the matter Lagrangian $\mathcal{L}_\text{m}$ and this maximal extended theory was proposed by Harko et al. \cite{Harko/2010}. The gravitational field equations and equation of motion have been obtained through metric formalism and covariant divergence of the stress energy tensor. In this way, Harko et al. in \cite{Harko11} have  proposed another extension of $f(R)$ gravity, a successively alternative to GR, is called the $f(R,T)$ gravity theory (where the gravitational Lagrangian is given by an arbitrary function of Ricci scalar $R$ and trace of the stress energy tensor $T$). In the frame work of this theory covariant of stress energy is obtained and the cosmological models depend on a source term, which represent the variation of the matter stress energy tensor with respect to metric. Since, the source term is expressed as a function of the matter Lagrangian $\mathcal{L}_\text{m}$, then several set of field equations can be obtained with each choice of $\mathcal{L}_\text{m}$. In addition, the non-zero value of covariant divergence of stress energy tensor also provides a non-geodesic motion of massive test particles due to the matter energy coupling, it always represents an extra acceleration of expansion. In this thesis, all the cosmological models in each chapter are constructed within the framework of $f(R,T)$ gravity. \\
The action for $f(R,T)$ gravity is proposed as 
\begin{equation}\label{eqactn9}
S=\int \sqrt{-g}\biggl(\frac{1}{16\pi G}f(R,T)+\mathcal{L}_\text{m}\biggr)\text{d}^{4}x,
\end{equation}%
where the gravitational Lagrangian consists of an arbitrary function of Ricci scalar $R$ and the trace $T$ of the energy-momentum tensor $T_{\mu \nu}$ of the matter source. $\mathcal{L}_\text{m}$ is the usual matter Lagrangian density, and the gravitational constant is considered as $G=1$ (hereafter $G=1$).  For matter source, the stress energy tensor $T_{\mu \nu}$ is given by
\begin{equation}
T_{\mu \nu}=-\frac{2}{\sqrt{-g}}\frac{\delta (\sqrt{-g}\mathcal{L}_\text{m})}{\delta g^{\mu \nu}},
\end{equation}%
and its trace is $T=g^{\mu \nu}T_{\mu \nu}$.\\
Here, we have assumed that the matter Lagrangian $\mathcal{L}_\text{m}$ depends only on the metric tensor component $g_{\mu 
\nu}$ rather than its derivatives. Hence, we obtain
\begin{equation}
T_{\mu \nu}=g_{\mu \nu}\mathcal{L}_\text{m}-\frac{\partial \mathcal{L}_\text{m}}{\partial g^{\mu \nu}}.
\end{equation}%
By varying the action $S$ in eqn. (\ref{eqactn9}) with respect to $g_{\mu \nu}$, the $f(R,T)$ gravity field equations are obtained as
\begin{equation}\label{eqnfld1}
f_{R}(R,T)R_{\mu \nu}-\frac{1}{2}f(R,T)g_{\mu \nu}+(g_{\mu \nu}\Box -\nabla _{\mu}\nabla
_{\nu})f_{R}(R,T)=8\pi T_{\mu \nu}-f_{T}(R,T)T_{\mu \nu}-f_{T}(R,T)\Theta _{\mu \nu},
\end{equation}%
where
\begin{equation}
\Theta _{\mu \nu}=-2T_{\mu \nu}+g_{\mu \nu}\mathcal{L}_\text{m}-2g^{l\eta}\frac{\partial ^{2}\mathcal{L}_\text{m}}{\partial
g^{\mu \nu}\partial g^{l\eta}}.
\end{equation}%
Here, $f_{R}(R,T)=\frac{\partial f(R,T)}{\partial R}$, $f_{T}(R,T)=\frac{%
\partial f(R,T)}{\partial T}$, $\Box \equiv \nabla ^{\mu}\nabla _{\mu}$ where $\nabla _{\mu}$ is the covariant derivative.\\
Contracting eqn. (\ref{eqnfld1}), we get
\begin{equation}\label{eqnfld2}
f_{R}(R,T)R+3\Box f_{R}(R,T)-2f(R,T)=(8\pi -f_{T}(R,T))T-f_{T}(R,T)\Theta,
\end{equation}%
where $\Theta =g^{\mu \nu}\Theta _{\mu \nu}$.\\
From eqns. (\ref{eqnfld1}) and (\ref{eqnfld2}), the $f(R,T)$ gravity field equations take the form
\begin{multline}
f_{R}(R,T)\biggl(R_{\mu \nu}-\frac{1}{3}Rg_{\mu \nu}\biggr)+\frac{1}{6}f(R,T)g_{\mu \nu}= \\
8\pi -f_{T}(R,T)\biggl(T_{\mu \nu}-\frac{1}{3}Tg_{\mu \nu}\biggr)-f_{T}(R,T)\biggl(%
\Theta _{\mu \nu}-\frac{1}{3}\Theta g_{\mu \nu}\biggr)+\nabla _{\mu}\nabla _{\nu}f_{R}(R,T).
\end{multline}%
Since different cosmological models of $f(R,T)$ gravity are possible depending on the nature of the matter source. Harko et al. \cite{Harko11} constructed
three types of frames of $f(R,T)$ gravity and their corresponding field equations are as follows:
\begin{itemize}
\item $f(R,T)=R+2f(T)$\\
For an arbitrary choice of function $f(T)$, the field equation from general equation (\ref{eqnfld1}) takes the form
\begin{equation}\label{eqnfld3}
G_{\mu \nu}= 8\pi T_{\mu \nu}-2f'(T)T_{\mu \nu}-2f'(T)\Theta_{\mu \nu}+f(T)g_{\mu \nu}.
\end{equation}
For a perfect fluid matter source $\Theta_{\mu \nu}=-2T_{\mu \nu}-pg_{\mu \nu}$, the field equation can be written as 
\begin{equation}\label{eqnfld4}
G_{\mu \nu}= 8\pi T_{\mu \nu}+2f'(T)T_{\mu \nu}+(2pf'(T)+f(T))g_{\mu \nu},
\end{equation}
where the prime stand for the derivative with respect to the argument.
\item $f(R,T)=f_{1}(R)+f_{2}(T)$\\
With the choice of both arbitrary function of $R$, $f_{1}(R)$, and function of $T$, $f_{2}(T)$, the general eqn. (\ref{eqnfld1}) rewritten as
\begin{equation}\label{eqnfld5}
f_{1}'(R)R_{\mu \nu}-\frac{1}{2}f_1(R)g_{\mu \nu}+(g_{\mu \nu}\Box -\nabla _{\mu}\nabla
_{\nu})f_{1}(R)=8\pi T_{\mu \nu}-f_{2}'(T)T_{\mu \nu}-f_{2}'(T)\Theta _{\mu \nu}+\frac{1}{2}f_2(T)g_{\mu \nu}.
\end{equation}
Similarly, for perfect fluid source, 
\begin{equation}\label{eqnfld6}
f_{1}'(R)R_{\mu \nu}-\frac{1}{2}f_1(R)g_{\mu \nu}+(g_{\mu \nu}\Box -\nabla _{\mu}\nabla
_{\nu})f_{1}(R)=8\pi T_{\mu \nu}+f_{2}'(T)T_{\mu \nu}+\biggl(f_{2}'(T)p+\frac{1}{2}f_2(T)\biggr)g_{\mu \nu}.
\end{equation} 
\item $f(R,T)=f_{1}(R)+f_{2}(R)f_{3}(T)$\\
The field equation for the third generalized $f(R,T)$ gravity for an arbitrary matter source is 
\begin{multline}\label{eqnfld7}
(f_{1}'(R)+f_2'(R)f_3(T))R_{\mu \nu}-\frac{1}{2}f_1(R)g_{\mu \nu}+(g_{\mu \nu}\Box -\nabla _{\mu}\nabla
_{\nu})(f_{1}'(R)+f_2'(R)f_3(T))=8\pi T_{\mu \nu}\\+f_{2}(R)f_3'(T)T_{\mu \nu}-f_{2}(R)f_3'(T)\Theta _{\mu \nu}+\frac{1}{2}f_2(R)f_3(T)g_{\mu \nu}.
\end{multline}
Similarly, for perfect fluid source, 
\begin{multline}\label{eqnfld8}
(f_{1}'(R)+f_2'(R)f_3(T))R_{\mu \nu}-\frac{1}{2}f_1(R)g_{\mu \nu}+(g_{\mu \nu}\Box -\nabla _{\mu}\nabla
_{\nu})(f_{1}'(R)+f_2'(R)f_3(T))=8\pi T_{\mu \nu}\\+f_{2}(R)f_3'(T)T_{\mu \nu}+f_{2}(R)\biggl(f_3'(T)p+\frac{1}{2}f_3(T)\biggr)g_{\mu \nu}.
\end{multline}
\end{itemize}
\section{Bianchi universes}\label{ch1Bianchi}
In recent studies, it has been proposed that present isotropic universe may have evolved from an early anisotropic phase. This has led to the study of a wide range of cosmological scenario based on a spatially homogeneous, but anisotropic manifold. It can be possible with a consideration of Bianchi type space-time which is globally hyperbolic spatially homogeneous and ansotropic.\\
Precise measurements of the temperature anisotropies of the CMB and the Wilkinson Microwave Anisotropy Probe (WMAP) \cite{Bennett/2003,Hinshaw/2009} provide a universe dominated by cold dark matter (CDM) and cosmological constant ($\Lambda$) \cite{Coles/2005}, which termed as a $\Lambda$CDM model. The most necessary part of this model is the primordial metric perturbations that give rise to the galaxies and Large-scale structure we observed around us today should be Gaussian and statistically homogeneous \cite{Guth/1982,Starobinsky/1982,Bardeen/1983} and the observed temperature fluctuations in the CMB should be Gaussian and statistically isotropic. Some anomalous behavior of universe has been reported in \cite{Yadav/2008}. Also, there is no clear evidence of primordial non-Gaussianity, but there are several indications of statistical anisotropy in the CMB sky. 
 Moreover, the further possibility of the WAMP temperature fluctuation may be affected by other systematic problems suggested in literature \cite{Chiang/2007,Chiang/20007,Short/2010}. Even slight effects of this type could seriously hamper our attempts to uncover evidence of physics beyond the standard model. From the evidence of global asymmetry and other analyses, it can be confirmed the possibility that we may live in universe which is described by a background cosmology that globally anisotropic. The approach we have followed in this thesis is to study some Bianchi cosmological model within $f(R,T)$ gravity theory, based on the exact solutions to modified EFE.\\
The Bianchi classification groups all possible spatially homogeneous but anisotropic relativistic cosmological models into types depending on the symmetry properties of their spatial hyper surfaces \cite{Ellis/1969}. In literature, it has been known for some time that interesting localized features in the CMB temperature pattern can occur in Bianchi models with negative spatial curvature \cite{Collins/1973,Dautcourt/1978,Tolman/1984,Matzner/1982,Barrow/1985}. The physical origin of such features lies in the focusing effect of space on the geodesics that squeezes the pattern of the small region of the sky. More recently, attention has shifted to the possibility of using the additional parameters available in such models to reproduce a cold spot such as that claimed to exist in the WMAP data. Since we know that our present universe is close to isotropic, attention has focused on the subset of the Bianchi type that contains the FLRW model as a limiting case. \\
From mathematical point of view, there must be some symmetry that relates what the universe looks like as seen by observer `A' to what is seen in a coordinate system centered on any other observer `B'. The possible space-time consistent with this requirement possess symmetries that can be classified into the Bianchi types. The Bianchi classification is based on the construction of space like hyper surfaces. The set of killing vectors will have some $L$-dimensional group structure, read $G_{\text{L}}$, which depends on the equivalence classes of the structure constant $C_{\mu \nu}^{\gamma}$. This is used to classify all spatially homogeneous cosmological models. For any particular space like hyper surface, the killing vector basis can be chosen so that the structure constants can be decomposed as 
\begin{equation}
C_{\mu \nu}^{\gamma}=\epsilon_{\mu \nu l}n^{l\gamma}+\delta_{\nu}^{\gamma}a_\mu-\delta_{\mu}^{\gamma}a_\nu,
\end{equation}
where $\epsilon_{\mu \nu l}$ is the total antisymmetric term and $\delta_{\mu}^{\nu}$ is the Kronecker delta. The tetrad basis can be chosen to diagonalise the tensor, $n^{\mu \nu}=\text{diag}(n_1, n_2,n_3)$ and to set the vector $a_\mu=(a,0,0)$, then the Jacobi identity are simply $n_1 a=0$ (from $n^{\mu \nu}a_\nu=0$, for tetrad basis). The possible combination of $n_\mu$ and $a$ then fix the different Bianchi types as shown in the given table \ref{tblebianchi1}.
\begin{table}[]
\begin{center}
\begin{tabular}{|c|c|c|c|c|c|c| } 
 \hline
Bianchi Type & Class & $a$ & $n_1$ & $n_2$ & $n_3$ \\ 
\hline
 I & A & 0 & 0 & 0 & 0 \\
 \hline
II & A & 0 & + & 0 & 0 \\
\hline
VI$_0$ & A & 0 & 0 & + & - \\
\hline
VII$_0$ & A & 0 & 0 & + & + \\
\hline
VIII & A & 0 & - & + & + \\
\hline
IX & A & 0 & + & + & + \\ 
 \hline
IV & B & + & 0 & 0 & +\\ 
 \hline 
V & B & + & 0 & 0 & 0 \\ 
 \hline
VI$_h$ & B & + & 0 & + & - \\ 
 \hline
VII$_h$ & B & + & 0 & + & + \\ 
 \hline 
\end{tabular}
\end{center} 
\caption{The Bianchi types shown in terms of whether the various parameters used to construct the classification are zero, positive or negative. The designation of Class A or Class B depends on whether $a = 0$, or not. The parameter $h$ is defined by $h = a^2/(n_1 n_2)$. In particular, the spaces I, V, VII$_0$, VII$_h$ and IX are contain the isotropic FLRW spaces as limiting cases.}\label{tblebianchi1}
\end{table} 
The isotropic spaces that feature in the Friedmann models have $G_6$ symmetry groups with $G_3$ subgroup. So that the zero curvature ($k=0$) FLRW model can be thought of as a special case of Bianchi type I on VII$_0$. Like-wise the open ($k<0$) FLRW model is a special case of type V or VII$_h$. The closed FLRW case ($k>0$) is special case of Bianchi type IX. Since, Bianchi I and VII$_0$ are spatially flat, Bianchi IX is positively curved and Bianchi V and  VII$_h$ have negative spatial curvature. Thus the scalar curvature can be defined in terms of Bianchi parameters
\begin{equation}
R=-\frac{1}{2}\biggl[(n_1-n_2)^2+(n_1-n_3)^2+(n_2-n_3)^2\biggr]+\frac{1}{2}(n_1^2+n_2^2+n_3^2)-6a^2
\end{equation}
where $n_1=n_2=n_3=0$, $R=-6a^2$, it yields Bianchi V.
For VIII$_h$, we have $n_1=0$, $n_2\neq 0$, and $n_3 \neq 0$; the parameter $h$ is defined as $h=\frac{a^2}{n_2 n_3}$. Details are depicted in the following table
 (\ref{tblebianchi2})\\
\begin{table}[]
\begin{center}
\begin{tabular}{|c|c|c|} 
 \hline
Bianchi Type & $R$ & $k$  \\ 
\hline
 I & 0 & 0 flat \\
 \hline
V & $-6 a^2$ & $<0$ open \\
\hline
VII$_0$ & $-\frac{1}{2}(n_2-n_3)^2 $ & 0 flat \\
\hline
VII$_h$ & $-6a^2-\frac{1}{2}(n_2-n_3)^2 $ & $<0$ open   \\
\hline
IX & $n_1n_2+n_1n_3+n_2n_3-\frac{1}{2}(n_1^2+n_2^2+n_3^2) $ & $>0$ closed  \\ 
 \hline
\end{tabular}
\end{center} 
\caption{The Bianchi classification  shown in terms of curvature scalar.}\label{tblebianchi2}
\end{table}
One of the special cases- Bianchi type I is \textbf{Kasner solution}, which demonstrates the difficulty of finding meaningful exact solutions in situations of restricted symmetry. In 1922 Kasner has found the following solution for the EFE in vacuum \cite{Kasner/1921} as
\begin{equation}
ds^2=dt^2-t^{2p_1}dx^2+t^{2p_2}dy^2+t^{2p_3}dz^2
\end{equation}
with \begin{eqnarray}
p_1+p_2+p_3=1 \label{k1}\\
p_1^2+p_2^2+p_3^2=1\label{k2}
\end{eqnarray}
where $p_1$, $p_2$, and $p_3$ are Kasner indexes, that can be distributed in the following ways \cite{Belinksi/1970}.\\
$\frac{-1}{3}\leq p_1\leq 0$, $0\leq p_2 \leq \frac{2}{3}$, $\frac{2}{3} \leq p_3 \leq 1.$\\
Thus, we may conclude that the three scale factors of Bianchi type I metric referred to the spatial axes with two of them that increase with time and one of which conversely decreases. It is worthy to note that, for (\ref{k1}) and (\ref{k2}) the spatial volume of a Bianchi type I space-time grows with time. Thus in the limit $t\rightarrow 0$ we have a Big Bang like singularity. 
\section{Past and future singularities in cosmological model}\label{singularity}
Another important physical property of cosmological models in the behavior of universe, which we have covered in this thesis is called \textit{cosmic singularity}. In this section, we have covered some important definitions of what a singularity stands for and what it involves qualitatively. In many space-time, points of infinite coordinate is known as `existence of singularity'. Coordinates can not be extended beyond the singular point and hence for geodesics are incomplete in such space-time that singularity. If we consider a manifold $M$, there can be two points which are not connected by any causal curve. Then, it is suggested that the geodesics faces a singularity, which can be thought of as the edge of manifold. Manifolds which have such singularities are known as geodesically incomplete manifold. In fact the singularity theorem states that for a reasonable matter content (positive energies), space-times are almost guaranteed to be geodesically incomplete in GR \cite{Wald/1984}. However, the initial singularity (Big Bang) occurs when the scale factor of the universe $a(t)\rightarrow 0$ in a finite time. This particular singularity results from a homogeneous contraction of space down to ``zero size", but does not represent an explosion of matter concentrated at a point of pre-existing non-singular space-time. Similarly, curvature singularity occurs, when the curvature becomes infinite. The curvature is measured by the Riemann tensor from which it can be constructed various scalar quantities like Ricci scalar $R$, or higher order scalars $R_{ab}R^{ab}$, $R_{abcd}R^{abcd}$, etc. If any one of those scalars tend to infinity at some point then there is a curvature singularity at that point. Any curvature singularity which is not surrounded by an event horizon is called naked singularity. Since there is no event horizon, there is no obstruction to an observer traveling to the singularity and returning to report on what was observed.\\
Other than the initial singularities there also appears another type of singularity in future at finite time, which is known as ``future singularities". It is a serious problem that face a large number of gravitational theories in its way to explain the accelerated expansion. Among these theories some models like phantom models, some quintessence models and other modified gravity models suffer this singularity problem. Also, it has an important fact that it can produce instabilities in black holes and in stellar physics. Nevertheless, this problem can be understood and/or solved only from the perspective given by a quantum theory of gravity that, we do not have yet.\\ 
In general FLRW universe, the singularities appear during cosmological evolution when the HP is expressed as
\begin{equation}
H=\frac{h_\text{s}}{(t_\text{s}-t)^\beta},
\end{equation}
where $h_\text{s}$, and $t_\text{s}$ are positive constants, and $t_\text{s}$ is the time when singularity appears. $\beta$ is a constant. We can seen that if $\beta>0$, $H$ becomes singular in the limit $t\rightarrow t_\text{s}$. On the other hand, if $\beta<0$, even for non-integer values of $\beta$ some derivatives of $H$ and therefore the curvature or some combination of curvature invariants, become singular. Moreover, for an expanding universe, $t<t_\text{s}$. In order to get the singularities properties $\beta$ must be $\beta\neq 0$, because $\beta=0$ corresponds to de Sitter space, which has no singularity.\\
The finite-time future singularities appears in cosmological models can be classified into several types, depending on the divergent magnitude. In \cite{Nojiri/2005}, the authors propose the classifications for the different types of singularities in the following way:
\begin{itemize}
\item \textbf{Type I (Big Rip)}: for $t\rightarrow t_\text{s}$ and for $\beta=1$ or $>0$, $a(t)\rightarrow \infty$, $\rho \rightarrow \infty$, and $|p|\rightarrow \infty$. A wide range of literature are available for this singularity \cite{Barboza/2006,Caldwell/2003,Cognola/2007,Elizalde/2003,Elizalde/2005,Lobo/20005,Nojiri/0003,Sami/20004,Singh/2003,Wu/2005}. $p$ and $\rho$ are finite at $t=t_\text{s}$.
\item \textbf{Type II (Sudden singularity)}: for $t\rightarrow t_\text{s}$, $a(t)\rightarrow a_\text{s}$, $\rho \rightarrow \rho_\text{s}$, and $|p|\rightarrow \infty$ \cite{Barrow/1990}. It corresponds to $-1<\beta<0$. 
\item \textbf{Type III}: for $t\rightarrow t_\text{s}$, $a(t)\rightarrow a_\text{s}$, $\rho \rightarrow \infty$, and $|p|\rightarrow \infty$. It corresponds to $0<\beta<1$.
\item \textbf{Type IV}: for $t\rightarrow t_\text{s}$, $a(t)\rightarrow a_\text{s}$, $\rho \rightarrow \rho_\text{s}$, $|p|\rightarrow p_\text{s}$, and higher derivatives of $H$ diverge. It corresponds to $\beta<-1$ but $\beta$ is not any integer number. 
\end{itemize}
Here $t_\text{s}$, $a_\text{s}$, $\rho_\text{s}$, $p_\text{s}$ are constants, with $a_\text{s} \neq 0$, while $a$, $\rho$, $p$ are the scale factor, energy density and the pressure respectively.
\section{Energy conditions}\label{ch1ECs}
According to GR, matter and the energy density are always related with the geometry of the space-time through EFE eqn. (\ref{eqEFE}). In which the left side Einstein tensor $G_{\mu \nu}$ describes space-time geometry, $T_{\mu \nu}$ is the energy momentum tensor, which describes the matter and the energy in this space-time.\\
In principle, one can consider any metric $g_{\mu \nu}$ imaginable (for example, traversable wormhole discussed below) and as long as its second derivative exists-plug that metric into the LHS of field eqn. (\ref{eqEFE}) to allow the stress energy tensor responsible to that metric. In this way, we will obtain the exact solutions easily, but the stress energy tensor will not necessarily be physically reasonable. In order to accept the stress energy tensor as a source of gravitational field, it is useful to impose certain conditions, called ``energy conditions" (ECs), which serves precisely certain ideas about what is physically reasonable. In addition, one key generic feature of every matter is that, energy densities (almost) always seem to be positive. It is only possible through so called ECs of GR by making this notion of locally positive energy density more precise.\\
Moreover, these ECs are coordinate invariant restrictions on the stress-energy tensor. This invariance is enforced by scalar quantities, which are usually considered are contractions of $T_{\mu \nu}$ with time like or null vector. There are various ways to formulate all the energy conditions. Here, we have followed on \textit{geometric}, the \textbf{physical} and the \textbf{effective} ways. One can write down formal conditions expressed by using only the value of the stress-energy tensor itself, but it is required to stand in relation to a fixed family of vectors or other tensors. In every case, the physical formulation is logically equivalent to the geometric formulation if the EFE is assumed to hold. According to a useful classification of GR \cite{Hawking/973}, a stress-energy tensor (for a perfect fluid) is given by 
\begin{equation}
T_{\mu \nu}=(\rho+p)u_{\mu} u_{\nu}+pg_{\mu \nu},
\end{equation} 
where $u_{\mu}$ is the fluid four velocity, $g_{\mu \nu}$ is the space-time metric, $\rho$ is the energy density and $p$ is the pressure in the $x^i$ direction (i=1,2,3). \\
Several types of ECs in classical GR \cite{Carroll/2004} are given as follows:
\begin{itemize}
\item the null energy condition (NEC):\\
\textbf{geometric}:  $R_{\mu \nu} k^{\mu}k^{\nu}\geq 0$, $\forall$ null vector $k^{\mu}$, where $R_{\mu \nu}$ is Ricci tensor.\\
\textbf{physical}: $T_{\mu \nu} k^{\mu}k^{\nu}\geq 0$, $\forall$ null vector $k^{\mu}$.\\
\textbf{effective}: for each $i$, $\rho+p_i \geq 0$. 

\item  the weak energy condition (WEC):\\
It must be positive and the geometric definition refers to the EFEs.
 \textbf{geometric}:  $G_{\mu \nu} t^{\mu}t^{\nu}\geq 0$, $\forall$ timelike  vector $t^{\mu}$, where $G_{\mu \nu}$ is Einstein tensor.\\
\textbf{physical}: $T_{\mu \nu} t^{\mu}t^{\nu}\geq 0$, $\forall$ timelike vector $t^{\mu}$.\\
\textbf{effective}: for each $i$, $\rho \geq 0$ $\rho+p_i \geq 0$ .
\item  the strong energy condition (SEC):\\
\textbf{geometric}:$R_{\mu \nu} k^{\mu}k^{\nu}\geq 0$, $\forall$ null vector $k^{\mu}$.\\
\textbf{physical}: for any timelike vector $k^{\mu}$, $(T_{\mu \nu}-\frac{1}{2}Tg_{\mu \nu})k^{\mu}k^{\nu}\geq 0$.\\
\textbf{effective}: $\rho+\sum_{i}p_i \geq 0$, and for each $i$, $\rho+p_i\geq 0$.
\item  the dominant energy condition (DEC):\\
It states that matter flows along timelike or null world lines.\\
\textbf{geometric}:  $G_{\mu \nu} t^{\mu}t^{\nu}\geq 0$, $\forall$ timelike vector $t^{\mu}$, where $G_{\mu \nu}$ is Einstein tensor.\\
for any two co-oriented timelike vectors $t^{\mu}$  and $v^{\mu}$,$G_{\mu \nu} t^{\mu}v^{\nu}\geq 0$.\\
\textbf{physical}: $T_{\mu \nu} t^{\mu}t^{\nu}\geq 0$, $\forall$ timelike vector $t^{\mu}$.\\
for any two co-oriented timelike vectors $t^{\mu}$  and $v^{\mu}$,$T_{\mu \nu} t^{\mu}v^{\nu}\geq 0$.\\
\textbf{effective}: for each $i$, $\rho \geq 0$ $\rho \geq |p_i|$. 
\end{itemize}
It can be noted from all the above descriptions that NEC implies WEC.  NEC, WEC and the SEC are mathematically independent assumptions. In particular, the SEC does not imply the WEC. Violating the NEC implies violating the DEC, SEC and WEC as well. Hawking area theorem for black hole horizon relies on the NEC, and hence evaporation of a black hole must violate the NEC. All these considerations are related to standard matter which satisfies regular equations of state and is minimally coupled to the geometry. They can be generalized to other theories of gravity assuming that at least causal structure is preserved. In this thesis, we have considered these ECs in modified gravity cosmological models in both validation and violation aspects. In contrast, the null, weak, and dominant energy conditions are still extensively used in the GR community. The weakest of these is the NEC, and it is in many cases also the easiest to work with and analyse. In addition, the refinement of the ECs are applied in the development of some powerful mathematical theorems such that the singularity theorem \cite{Hawking/973,Wald/1984}, positive energy theorem, the non-existence of traversable wormholes, and limits on the extent to which light cones can trip over in strong gravitational fields. Hence, some form of ECs and some notion of positivity of the stress energy tensor as an input of hypothesis, are the necessary requirements of these theorems.\\
Despite this, there are some classical systems and field theories these are compatible with experimental results, and  quantum field theory. It has been cleared that the violation of all ECs are possible. In fact, the WEC prohibits observers from seeing negative energy densities \cite{Hawking/973,Poisson/2004}. It may seem reasonable to postulate that the WEC always holds, experiments have shown that it is violated by certain phenomena such as the Casimir effect. However, the current evidence suggests that there are strong limits on how severe such violations can be, globally (e.g.\cite{Poisson/2004} page 32). These classical violation of ECs can be made an huge changes and provide some new features in wired physics. For instance, it is possible to demonstrate that Lorentzian signature traversable wormhole arises as classical solution of field equations \cite{Barce/1999}.
\section{Wormhole geometry}\label{ch1WH}
Wormhole (WH) is a hypothetical connection between two widely separated regions of space-time \cite{Morris/1988}. WH solutions were firstly considered from physics standpoint by Einstein and Rosen (ER) in 1935, which is known today as ER bridges connecting two identical sheets \cite{Einstein/1935}. Then Wheeler added the term ``wormhole" to the physics literature, however he defined it at the quantum scale. After that, first traversable WH was proposed by Morris-Thorne in 1988 \cite{Morris/1988}. Then Morris, Thorne, and
Yurtsever investigated the requirements of the EC for WHs \cite{Yurtsever/1988}. After that, many research works have been done in literature to support this idea \cite{Visser/1989a,Visser/1989b,Visser/1995a,Visser/2003}.\\
A WH is any compact region on the space-time that contains two mouth and a throat connecting the two each at separate points in space-time like a tunnel and can act either as a passage in space or in time \cite{Lobo/2005a,Lobo/2016}. Also, it provides solutions to field equations in a  reverse manner. Firstly, one considers an interesting exotic space-time metric and solves the EFE, then finds the exotic matter needed as a source responsible for the respective geometry. It is needed because the exotic matter violates the NEC. It also violates the causality by allowing closed time-like curves. Furthermore, the other interesting outcome is that time travel is possible without excess the speed of light. \\
Now, lets take an example of a traversable WH solution which is characterized by the given line element  
\begin{equation}\label{eqnwh}
ds^2=e^{a(r)}dt^2-\frac{dr^2}{1-\frac{b(r)}{r}}-r^2(d\theta^2+\sin^2 \theta d\phi^2).
\end{equation}
Here, $a(r)$ and $b(r)$ are arbitrary functions of radial coordinate $r$. The function $a(r)$ is called redshift function as it describes about gravitational redshift. Similarly $b(r)$ is termed as shape function and it determines the shape of the wormhole throat. The radial coordinate $r$, is non-monotonic, i.e. it decreases from $+\infty$ to a minimum value $r_0$ and then increases from $r_0$ to $+\infty$. The point at $r=r_0$, represents the throat of wormhole, where the shape function must satisfy a condition,  i.e. $b(r_0)=r_0$, called ``\textit{throat condition}".\\ 
Although the metric coefficient $g_{\text{rr}}$ becomes divergent at the throat, it is signaled by the coordinate singularity, the proper distance,
\begin{equation}
l(r)=\pm \int_{r_0}^{r}\biggl(\frac{1}{b(r)}-1\biggr)^{-1/2}\text{d}r
\end{equation} 
must be finite everywhere. For an embedded surface to have an equation $z=z(r)$, the line element eqn. (\ref{eqnwh}) can be rewritten as 
\begin{equation}\label{eqnwh1}
ds^2=\biggl[1+\biggl(\frac{\text{d}z}{\text{d}r}\biggr)^2\biggr]dr^2+r^2d\phi^2.
\end{equation}
From eqns. (\ref{eqnwh}) and (\ref{eqnwh1}), it can be obtained as
\begin{equation}\label{eqnwh2}
\frac{\text{d}z}{\text{d}r}=\pm\biggl(\frac{1}{b(r)}-1\biggr)^{-1/2}.
\end{equation}
For the surface to be vertical, i.e. $\frac{\text{d}z}{\text{d}r} \rightarrow \infty$, we must have a minimum radius at $r=r_0$. Similarly, for ``\textit{asymptotic flatness}" we need $\frac{\text{d}z}{\text{d}r} \rightarrow 0$ as $r\rightarrow \infty$ which implies $\frac{b(r)}{r}\rightarrow 1$ as $r\rightarrow \infty$.\\
In order to find the wormhole solution, the surface must ``flare out", that means the inverse of the embedding function $r=r(z)$, must satisfy $\frac{\text{d}^2r}{\text{d}z^2}>0$ near the throat $r_0$. From eqn. (\ref{eqnwh2}), we find the \textit{flare out} condition as 
\begin{equation}\label{ch1flareout}
\frac{\text{d}^2r}{\text{d}z^2}=\frac{b-b'r}{2b^2}>0,
\end{equation}
where $'$ denotes the differentiation with respect to radial coordinate $r$. This ``flaring-out" condition is a fundamental ingredient of wormhole physics, and
plays a fundamental role in the analysis of the violation of the ECs. At the throat we verify that the form function satisfies the condition $b'(r_0)<1$. With this condition, ensuring that the two spherical volumes on each side of the wormhole throat are smoothly joined together. One must verify the absence of horizons,
in order for the wormhole to be traversable. This \textit{traversability condition} in the absence of event horizon must imply that $g_{tt}=e^{a(r)} \neq 0$ so that $a(r)$ must be finite everywhere. This finite characteristic of redshift function insists the violation of NEC. In fact, it implies the violation of all the point-wise energy condition.\\
A dedicated discussion on spherically symmetric traversable wormhole solutions in MGTs are covered in chapter \ref{Chapter6} and chapter \ref{Chapter7} respectively.


\chapter{Bianchi type string cosmological models in $f(R,T)$ gravity} 

\label{Chapter2} 

\lhead{Chapter 2. \emph{Bianchi type string cosmological models in $f(R,T)$ gravity}} 

\justify
This chapter  \blfootnote{The work, in this chapter, is covered by the following publication: \\ 
\textit{Bianchi type string cosmological models in $f(R,T)$ gravity}, Eur. Phys. J. Plus, \textbf{131} (2016) 333.} contains two cosmological models (Bianchi III and $VI_0$) with string fluid source in the framework of $f(R,T)$ gravity in the context of late time accelerating expansion of the universe as suggested by the present observations. The exact solutions of the field equations are obtained by using a time varying DP. The obtained models are anisotropic and free from initial singularity. The models initially show acceleration for a certain period of time and then decelerates consequently. Several dynamical and physical behaviors of the models are discussed in detail.
\section{Introduction}
\justify
Presently, standard models of cosmology are the most accepted models to study the origin and evolution of the current universe. 
All the alternatives for the physics beyond the standard model represented up to now are inspired by the principle of naturalness and are in the search for the unification of forces, but they do not offer a new conceptual framework in which gravity may be conciliated with quantum physics. 
To date, the foremost promising candidate as a theory of quantum gravity is \textit{string theory}. The substantial theoretical progress in string theory continues to own variety of challenges to deal with, if it is to be made experimentally verifiable. Since the string cosmological model plays a significant role within the description of our early universe, an exciting opportunity is offered by modern cosmology to explore such challenges by constructing a concrete string model in the background that is compatible with our understanding of the early universe. In fact, string theory unites all the matter and forces in a single theoretical framework, which provides a unified description of the fundamental structure and nature of the early universe. Therefore, the presence of the strings in the early universe can be explained using grand unified theories (GUTs) \cite{Kibble/1976,Everett/1981}. Moreover, it is necessary to allow  the early universe to go through an accelerated expansion phase, known as inflation, to overcome some present conceptual and observational problems (e.g. flatness and horizon problem). This is because, it can be associated to a phase transition of a scalar field. On the other hand, one important consequence of such transition is to provide a mechanism for the formation of topological defects, like domain walls, monopoles, and cosmic strings \cite{Hind/1995,Vilen/2000,Dzhunu/2007,Bezerra/2003}. Therefore, cosmic strings are one dimensional topological defect, which are formed during the transition of phase due to the broken axial symmetry when the temperature cooled down below some critical temperature in the early stage of the universe after the Big Bang.  \\
Thereafter, many cosmologists have been inspired by the study of string cosmological models in GR and alternative theories of gravitation. In GR, the importance of cosmic string and its gravitational effects are extensively studied by Kibble and Turok in  \cite{Kibble82}. Various aspects of cosmic strings coupled with perfect fluid and electromagnetic field are also investigated in \cite{Mahanta01, Hawking/1987, Bhatta01}. In fact, the cosmic string model constructed by Letelier \cite{Letel83} is used as a source for many cosmological models such as Bianchi I and Kantowski-Sach type. In the same context, Krori et al. \cite{k7} have studied the spatially homogeneous models of Bianchi types II, VI$_{0},$ VII and IX in the presence of strings. By introducing the 
stress energy tensor matter for a perfect dust along with cosmic strings, some of the cosmological models can be generalized to null strings and to perfect fluid strings. Equations of motion for such strings and conservation laws of dust can be derived by eliminating divergence equation of the stress energy tensor  \cite{Stachel80}. Furthermore, cosmic string cosmological models are widely studied in literature by several authors in various aspects, for example: the Bianchi II magnetic string cosmological model in addition to loop quantum cosmology are investigated in \cite{Rikhvitsky13}, and five dimensional Bianchi I cosmological models in the framework of the Lyra manifold \cite{Samanta12}.\\
 Our Universe is sustaining an expansion at an increasing rate which is confirmed in \cite{Riess/1998,Perlmutter/1999,Gold09}. This late time cosmic acceleration is an observed phenomenon at present \cite{Riess/1998} whereas a similar conformation for inflation is still awaited.
Even if the hot Big Bang model is sandwiched between inflation \cite{Guth81, Linde83} and late time cosmic acceleration \cite{Sahni2000, Copeland2006}. In order to address this expansion, MGTs, which disagree with GR in the context of low or high curvature scalar are being extensively examined in the past few years. The reconstruction of curvature scalar is very essential for cosmological models. In the same context, at low curvature regime, the accelerating expansion of the universe is described by $f(R)$ theory of gravity \cite{Carr05} and $f(R,T)$ gravity \cite{Harko11}. In this chapter, we have focused on the modified $f(R,T)$ gravity (for details one can check in previous chapter). In literature, several cosmological models are discussed within the $f(R,T)$ gravity formalism with various aspects \cite{Harko11,Hound12}. Then,  Jamil et al. \cite{Jamil12} have reconstructed the minimally coupled scalar field model with the Chaplygin gas and have demonstrated them for a specific form of $f(R,T)=R^{2}+f(T)$, and also Sharif et al. \cite{Sharif13} studied
the energy conditions for FLRW universe with perfect fluid in $f(R,T)$ gravity. For more details about various cosmological models within $f(R,T)$ gravity, one can refer to these refs.\cite{Houndjo14,Ahmed14,Moraes15,Sahoo15,Sahoo16}. \\ 
Despite this, many string cosmological models are investigated in different scenarios of MGTs  \cite{Hendi13, Reddy13, Wang06, Bali07, Sahoo08, Sahoo09, Sahoo10, Rao13, Bishi15}. For instance, viscous string cosmic models for Bianchi type V metric have been discussed with the help of barotropic EoS in the framework of $f(R,T)$ gravity \cite {Naidu13}. Sharma and Singh \cite{Sharma14} have presented a Bianchi type II string cosmological model in magnetic field epoch within
the $f(R,T)$ gravity. Then the massive string dominated by the Bianchi type-V universe is addressed by Yadav \cite{Yadav14} which no longer survived in $f(R,T)$ gravity. In addition, various dynamical properties of Bianchi type cosmological models in $f(R,T)$ gravity has been extensively studied by Zubair and Hassan \cite{Zub16}, Singh et al. \citep{GPS16}. The anisotropic behavior of Bianchi type III cosmological model in $f(R,T)$ gravity have been discussed by considering a simple power law form of the scale factor in \cite{PKS16b}. Recently, Sahoo \cite{PK16} has constructed a LRS Bianchi type I one dimensional string cosmological model in $f(R,T)$ theory of gravity.\newline 
Since, the present universe is homogeneous and isotropic on a larger scale, it is generally believed that the early universe was highly anisotropic and was isotropized later, with the cosmic expansion \cite{Ellis69}. In fact, the universe was not completely symmetric, that is indicated by WMAP data \cite{Camci2001, Pradhan2005}. Thus, the homogeneous and anisotropic Bianchi type space-times are more important in constructing models to analyze the anisotropy level of early universe. For the details about the Bianchi type space-time, one can refer previous chapter section \ref{ch1Bianchi}.\\ 
With the motivation of above discussions, we have explored two (Bianchi type III and $VI_{0}$) string cosmological models in the $f(R,T)$ formalism with the particular choice of $f(R,T)$ gravity i.e. $f(R,T)=R+2f(T)$. This chapter is organized in the following manner. The exact solution of first Bianchi type III model in presence of cosmic fluid is derived in section  \ref{ch2m1}. Then the solution and model for Bianchi type $VI_{0}$ metric are presented in section \ref{ch2m2}. Finally, in section \ref{ch2conclusion}, we have discussed the conclusions for both the models.\\
\textbf{\textit{The model field equation}}\\
Here, we have considered the standard stress energy tensor for string matter Lagrangian as
\begin{equation}\label{ch2ent}
T_{\mu \nu}=(\rho +p)u_{\mu}u_{\nu}+pg_{\mu \nu}-\lambda x_{\mu}x_{\nu}
\end{equation}%
where the four velocity vector $u^{\mu}$ satisfies the relation $u_{\mu}u_{\nu}=-x_{\mu}x_{\nu}=-1,\ \ \ u_{\mu}x_{\mu}=0$, $x^{\mu}$ is the direction of the string and $u^{\mu}\nabla _{\nu}u_{\mu}=0$. Here, 
 $\lambda $ is the string tension density. The particle density $\rho_{p}$ is defined as
\begin{equation}
\rho =\rho_{\text{p}}+\lambda
\end{equation}%
$\lambda$ may be positive or negative \cite{Letel83}.\newline
Since, the matter Lagrangian has no unique representation. So, the source term
is described as a function of Lagrangian matter through different choices of it. Choosing the perfect fluid matter as $\mathcal{L}_{\text{m}}=-p$, (hereafter $\mathcal{L}_{\text{m}}=-p$) $\Theta _{\mu \nu}$ can be written as 
\begin{equation}
\Theta _{\mu \nu}=-2T_{\mu \nu}- pg_{\mu \nu},
\end{equation}
which describes the physical nature of the matter field and used for the field equations of $f(R,T)$ gravity.
According to the nature of matter source, several cosmological models of $f(R,T)$ gravity are possible. 
Hence, Harko et al. \cite{Harko11} have constructed three different frames of $f(R,T)$ gravity (see section \ref{ch1f(R,T)} in chapter \ref{Chapter1}). Here, we have chosen $f(R,T)=f_{1}(R)+f_{2}(T)$ with $f(R)=\gamma R$ and $f(T)=\gamma T$ where $\gamma$ is an arbitrary constant. For a perfect fluid matter source the field equations of $f(R,T)$ gravity given in chapter \ref{Chapter1} (section \ref{ch1f(R,T)}) becomes
\begin{equation}\label{ch2fldeqn}
R_{\mu \nu}-\frac{1}{2}Rg_{\mu \nu}=\biggl(\frac{8\pi +\gamma }{\gamma }\biggr)T_{\mu \nu}+\biggl(p+\frac{1}{2}T\biggr)g_{\mu \nu}
\end{equation}%
By using this field equation, we have obtained two different cosmological models are as follows:

\section{Bianchi type III model}\label{ch2m1}
In this section the spatially homogeneous Bianchi type III metric is considered as
\begin{equation}\label{ch2met1}
\text{d}s^{2}=-\text{d}t^{2}+a_{1}^{2}\text{d}x^{2}+a_{2}^{2}e^{-2x}\text{d}y^{2}+a_{3}^{2}\text{d}z^{2}
\end{equation}%
where $a_{1},a_{2},a_{3}$ are functions of cosmic time $t$ only. \newline
The field eqn. (\ref{ch2fldeqn}) with energy momentum tensor eqn. (\ref{ch2ent}) for the metric  in eqn. (\ref{ch2met1}) take the form
\begin{eqnarray}
\frac{\dot{a_{1}}\dot{a_{2}}}{a_{1}a_{2}}+\frac{\dot{a_{2}}\dot{a_{3}}}{a_{2}a_{3}}+\frac{\dot{a_{1}}\dot{a_{3}}}{a_{1}a_{3}}-\frac{1}{a_{1}^{2}} &=&%
\frac{(16\pi +3\gamma )}{2\gamma }\rho +\frac{p}{2}+\frac{\lambda }{2},\label{ch2en1} \\
\frac{\ddot{a_{2}}}{a_{2}}+\frac{\ddot{a_{3}}}{a_{3}}+\frac{\dot{a_{2}}\dot{a_{3}}}{a_{2}a_{3}} &=&-\frac{(16\pi +3\gamma )}{2\gamma }p-\frac{(16\pi +3\gamma )}{2\gamma }\lambda +\frac{\rho }{2},\label{ch2en2}\\
\frac{\ddot{a_{1}}}{a_{1}}+\frac{\ddot{a_{3}}}{a_{3}}+\frac{\dot{a_{1}}\dot{a_{3}}}{a_{1}a_{3}} &=&-\frac{(16\pi +3\gamma)}{2\gamma }p-\frac{\lambda }{2}+\frac{\rho }{2},\label{ch2en3} \\
\frac{\ddot{a_{1}}}{a_{1}}+\frac{\ddot{a_{2}}}{a_{2}}+\frac{\dot{a_{1}}\dot{a_{2}}}{a_{1}a_{2}}-\frac{1}{a_{1}^{2}} &=&-\frac{(16\pi +3\gamma )}{2\gamma }p-%
\frac{\lambda }{2}+\frac{\rho }{2},\label{ch2en4}\\
\frac{\dot{a_{1}}}{a_{1}}-\frac{\dot{a_{2}}}{a_{2}} &=&0.\label{ch2en5}
\end{eqnarray}
Here, the overhead dot denotes derivative with respect to time $t$.\\
From eqn. (\ref{ch2en5}), we obtain
\begin{equation}
a_{1}=c_{1}a_{2},
\end{equation}%
where $c_{1}$ is an integration constant. By assuming $c_{1}=1$ the above equation reduces,
\begin{equation}
a_{1}=a_{2}.
\end{equation}%
Using this, the field eqns. (\ref{ch2en1} - \ref{ch2en4}) can be rewritten as
\begin{eqnarray}
\frac{\dot{a_{2}^{2}}}{a_{2}^{2}}+2\frac{\dot{a_{2}}\dot{a_{3}}}{a_{2}a_{3}}-\frac{1}{a_{2}^{2}} &=&\biggl(\frac{16\pi +3\gamma}{2\gamma }\biggr)\rho +\frac{p}{2}+\frac{\lambda}{2},\\
\frac{\ddot{a_{2}}}{a_{2}}+\frac{\ddot{a_{3}}}{a_{3}}+\frac{\dot{a_{2}}\dot{a_{3}}}{a_{2}a_{3}} &=&-\biggl(\frac{16\pi +3\gamma }{2\gamma }\biggr)p-\biggl(\frac{16\pi +3\gamma }{2\gamma }\biggr)\lambda +\frac{\rho }{2}, \\
\frac{\ddot{a_{2}}}{a_{2}}+\frac{\ddot{a_{3}}}{a_{3}}+\frac{\dot{a_{2}}\dot{a_{3}}}{a_{2}a_{3}} &=&-\biggl(\frac{16\pi +3\gamma }{2\gamma}\biggr)p-\frac{\lambda }{2}+\frac{\rho }{2}, \\
2\frac{\ddot{a_{2}}}{a_{2}}+\frac{\dot{a_{2}}^{2}}{a_{2}^{2}}-\frac{1}{a_{2}^{2}} &=&-\biggl(\frac{16\pi +3\gamma }{2\gamma}\biggr)p-\frac{\lambda }{2}+\frac{\rho }{2}.
\end{eqnarray}

Here, we have five unknowns $a_{2}$, $a_{3}$, $p$, $\rho$,
 and $\lambda $ in a set of four equations. Therefore, we require a valid assumption to find a consistent solution to these equations  In order to obtain exact solutions of the above non-linear system of equations we have assumed a well-motivated ansatz considered by Abdussattar and S. R. Prajapati \cite{Abdus2011}, which puts a constraint on the functional form of the DP $q$  as
\begin{equation}\label{ch2dec}
q=-\frac{\alpha }{t^{2}}+(\beta -1),
\end{equation}%
where $\alpha >0$ (dimension of square of time) and $\beta >1$ (dimensionless) are constants, also known as model parameters through which the dynamics of model can be extracted.\\
The use of a time dependent DP $q$ becomes an well motivated assumption due to the fact
that the universe exhibits phase transitions from the past decelerating expansion to the recent accelerating one as revealed by the recent observations of SNe Ia \cite{Riess/1998} and CMB anisotropies \cite{Bennet03}. In this context the DP describes the acceleration or deceleration behavior of the universe depending on its negative or positive value. Hence, the choice of variable $q$ is physically acceptable.\\
From eqn. (\ref{ch2dec}) it can be observed that the $q\rightarrow \infty $ at $t=0$ and it reduces to zero at $t=\sqrt{\frac{\alpha }{\beta -1}}$. The period of acceleration depends on $\alpha $ and $\beta $.\newline
Usually, DP $q$ can be defined as $q=-\frac{a\ddot{a}}{\dot{a}^{2}}=\frac{\text{d}}{\text{d}t}\biggl(\frac{1}{H}\biggr)-1$. By integrating this expression the scale factor can be written as  
\begin{equation}
a(t)=e^{\eta}\exp \int \frac{\text{d}t}{\int (1+q)\text{d}t+\delta},
\end{equation}
where $\delta, \eta $ are integrating constants. This can not be solved for different values of constants. Thus, by assuming $\delta=\eta=0$ and with help of eqn. (\ref{ch2dec}), we obtain 
\begin{equation}
a(t)=\biggl(t^{2}+\frac{\alpha }{\beta }\biggr)^{\frac{1}{2\beta }}.
\end{equation}
Note: for $\alpha=0$ in the above expression, it becomes $a(t)=t^{\frac{1}{\beta }}$ which yields a constant DP $q=\beta -1$ throughout the universe \cite{Berm83}.\\
\begin{figure}[H]
\begin{minipage}[t]{.45\textwidth}
\centering
      \includegraphics[width=68mm]{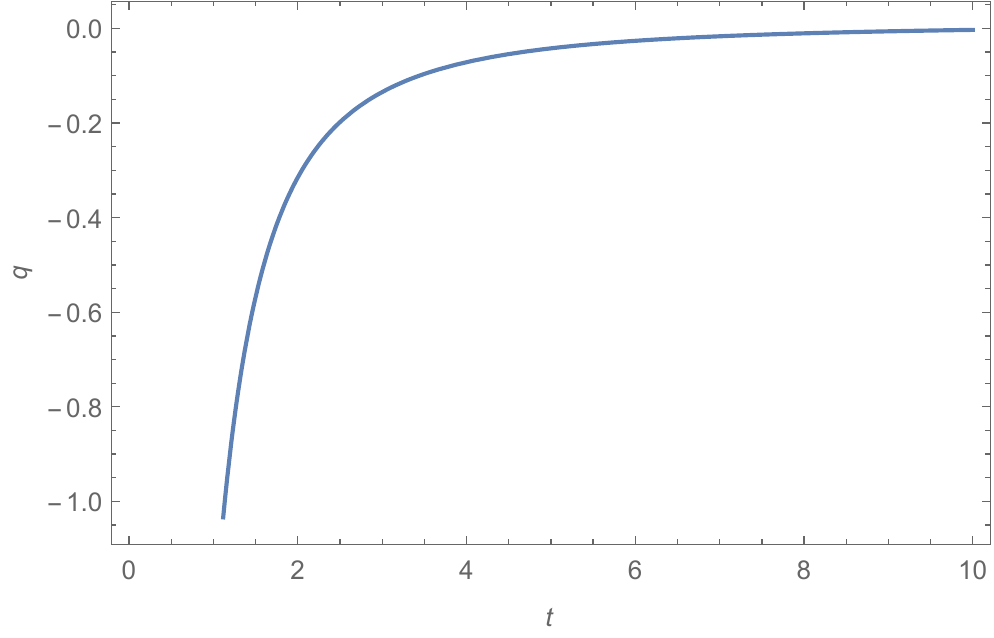}
\caption{The plot of $q$ versus $t$ with $\alpha=1.3$ and $\beta=1.01$.}\label{ch2fig1}
  \end{minipage}
\begin{minipage}[t]{.45\textwidth}
\centering
      \includegraphics[width=68mm]{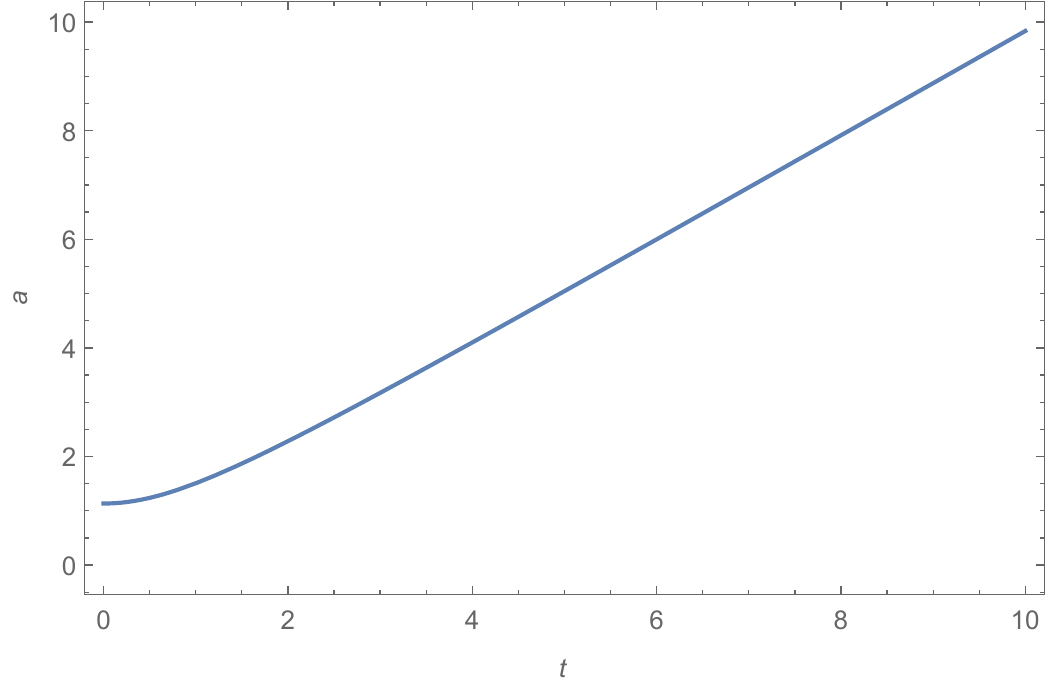}
\caption{The plot of $a$ versus $t$ with $\alpha=1.3$ and $\beta=1.01$.}\label{ch2fig2}
  \end{minipage}
\end{figure}
The behavior of DP and average scale factor with respect to cosmic time are depicted  in Fig.   \ref{ch2fig1} and Fig. \ref{ch2fig2} respectively. The negative DP represents that the model is accelerating and the scale factor is a positive increasing function. By using the relation between spatial volume and scale factor i.e. $a_3=V^b$, where $b$ is any constant, we obtain the  values of directional scale factors are 
\begin{eqnarray}
a_1=a_2=\biggl(t^2+\frac{\alpha}{\beta}\biggr)^\frac{3-3b}{4\beta},\\
a_3=\biggl(t^2+\frac{\alpha}{\beta}\biggr)^\frac{3b}{2\beta}.
\end{eqnarray}
The values of energy density for the cloud of strings $\rho$, fluid pressure $p$
and the string tension density $\lambda$ of the Bianchi type III model with respect to the above scale factors are given by
\begin{multline}
\rho= \frac{2}{1+4\zeta^2}\biggl[t^2 \times\biggl(\frac{6\zeta(1-3b)+3(b-3)%
}{2\beta}+\frac{27b^2(1-2\zeta)-9b(3-10\zeta)+18}{4\beta^2}\biggr)\times%
\biggl(t^2+\frac{\alpha}{\beta}\biggr)^{-2} \\
+\frac{6\zeta(3b-1)-3(b-3)}{4\beta}\biggl(t^2+\frac{\alpha}{\beta}\biggr)%
^{-1}-\frac{2\zeta+1}{2}\biggl(t^2+\frac{\alpha}{\beta}\biggr)^\frac{3b-3}{%
2\beta}\biggr],
\end{multline}
\begin{multline}
p= \frac{2}{1+4\zeta^2}\biggl[t^2 \times\biggl(\frac{6\zeta(3-b)-3(3b-1)}{%
2\beta}+\frac{9b^2(1-6\zeta)+9b(1+6\zeta)-36\zeta}{4\beta^2}\biggr)\times%
\biggl(t^2+\frac{\alpha}{\beta}\biggr)^{-2} \\
+\frac{6\zeta(2b-3)+3(2b-1)}{4\beta}\biggl(t^2+\frac{\alpha}{\beta}\biggr)%
^{-1}+\frac{2\zeta-1}{2}\biggl(t^2+\frac{\alpha}{\beta}\biggr)^\frac{3b-3}{%
2\beta}\biggr],
\end{multline}
\begin{equation}
\lambda=0,
\end{equation}
where $\zeta=\frac{16\pi+3\gamma}{2\gamma}$. \newline

\begin{figure}[H]
\begin{minipage}[t]{.45\textwidth}
\centering
      \includegraphics[width=68mm]{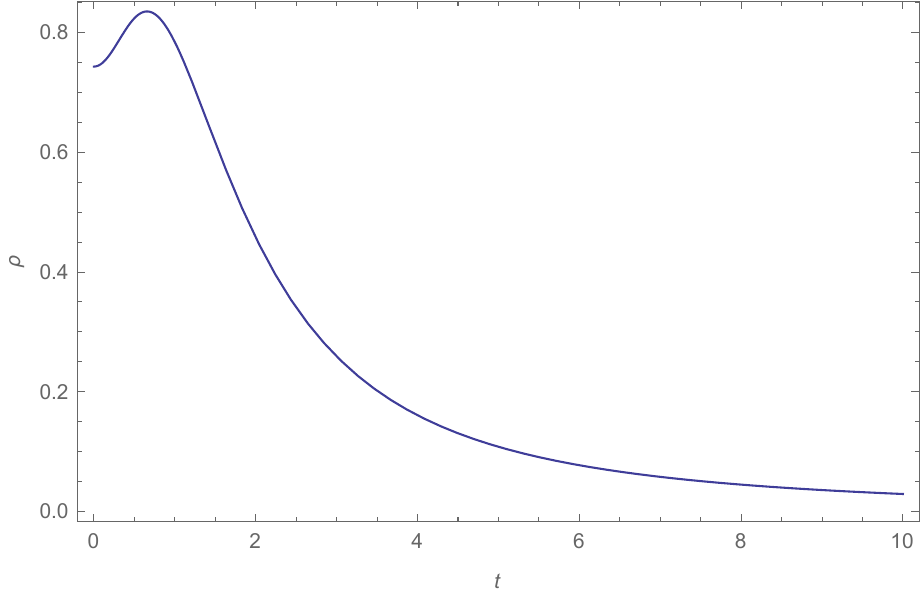}
\caption{The plot of $\rho$ versus $t$ with $b=0.3$, $\gamma=-25$, $\alpha=1.3$ and $\beta=1.01$.}\label{ch2fig3}
  \end{minipage}
\begin{minipage}[t]{.45\textwidth}
\centering
      \includegraphics[width=68mm]{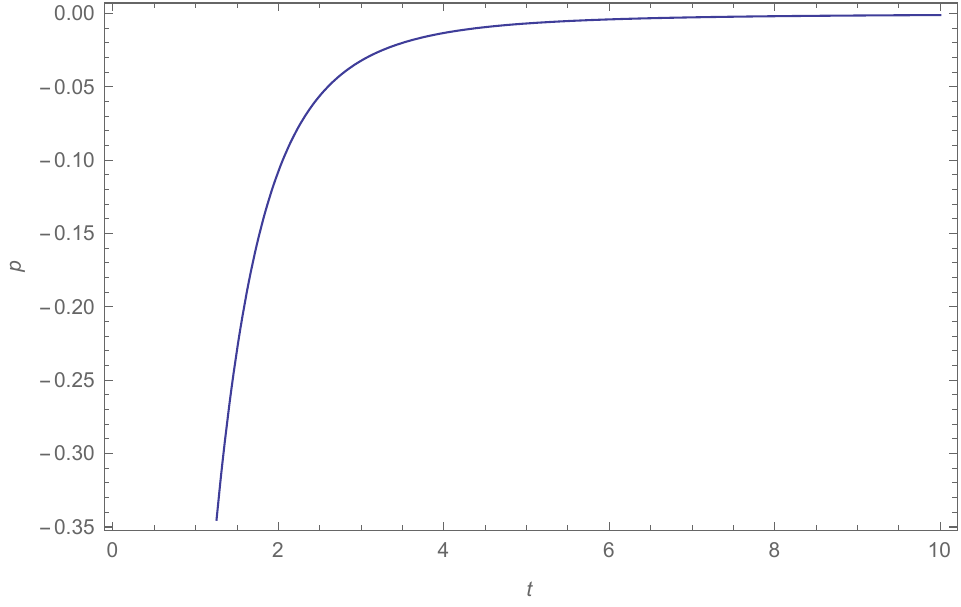}
\caption{The plot of $p$ versus $t$ with $b=0.3$, $\gamma=-25$, $\alpha=1.3$ and $\beta=1.01$.}\label{ch2fig4}
  \end{minipage}
\end{figure}
\begin{figure}[H]
\centering
\includegraphics[width=0.5\textwidth]{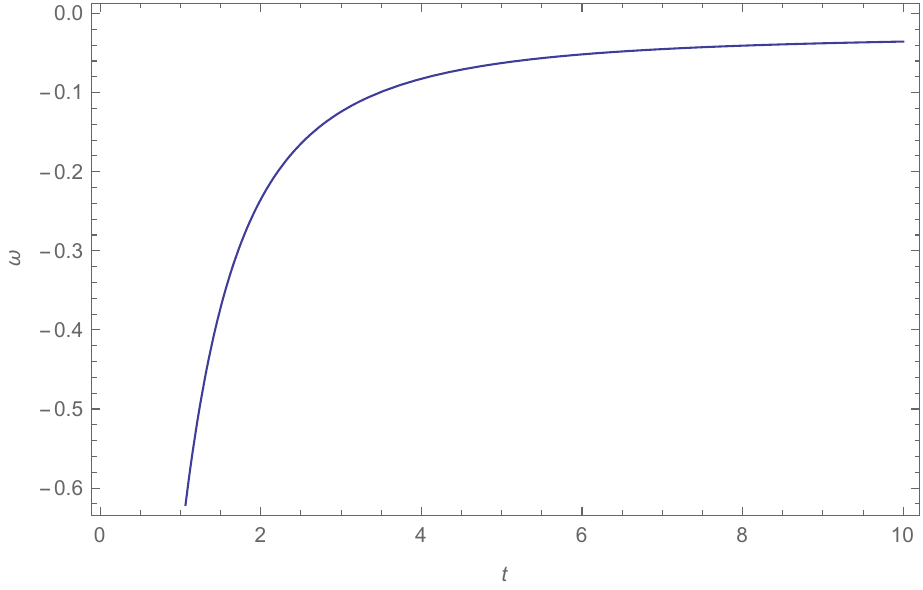}
\caption{The plot of EoS $\omega$ versus $t$
with $b=0.3$, $\gamma=-25$, $\alpha=1.3$ and $\beta=1.01$.}\label{ch2fig5}
\end{figure}
In this model the string tension density vanishes. 
From Fig. \ref{ch2fig3}, it can be observed that the large value of $\rho$ in the beginning indicates that the density dominates in the early time but it is negligible later. Similarly, in Fig. \ref{ch2fig4} the pressure is negative at initial time and later it vanishes. The relation between pressure and energy termed as EoS parameter  $\omega (t)=\frac{p}{\rho}$, which specifies the evolution of the expansion rate in GR. Therefore, its measurement for DE is one of the biggest efforts in observational cosmology today. The DE model has been described in a conventional manner by the EoS parameter, which is not necessarily constant. The present observational data seem to slightly favor an evolving DE with EoS $\omega < -1$ at the present epoch and $\omega > -1$ in the recent past. Fig. \ref{ch2fig5} shows the behavior of the EoS parameter for the Bianchi type III model. We can observe that the model represents a quintessence $(\omega > -1)$ in the present epoch. 

The HP, expansion scalar, shear scalar and mean anisotropic parameters are obtained as
\begin{equation}
H=\frac{1}{3}\biggl(\frac{\dot{a_{1}}}{a_{1}}+\frac{\dot{a_{2}}}{a_{2}}+%
\frac{\dot{a_{3}}}{a_{3}}\biggr)= \frac{t}{\beta}\biggl(t^2+\frac{\alpha}{\beta}\biggr)^{-1},
\end{equation}
\begin{equation}
\theta=3H=\frac{3t}{\beta}\biggl(t^2+\frac{\alpha}{\beta}\biggr)^{-1},
\end{equation}
\begin{equation}
\sigma^2=\frac{1}{2}\biggl(\sum H_{i}^{2}-\frac{1}{3}\theta ^{2}\biggr)=t^2 \times\frac{27b^2-18b+3}{4\beta^2}\times\biggl(t^2+\frac{\alpha%
}{\beta}\biggr)^{-2},
\end{equation}
and
\begin{equation}
\Delta=\frac{1}{3}\sum_{i=1}^{3}\biggl(\frac{H_{i}-H}{H}\biggr)^{2}=6\biggl(%
\frac{\sigma }{\theta }\biggr)^{2}= \frac{9b^2-6b+1}{2}.
\end{equation}
Here, $H_{i}(i=1,2,3)$ represent the directional Hubble parameters in the directions of $x,y$ and $z$ respectively.
\begin{figure}[H]
\centering
\includegraphics[width=0.5\textwidth]{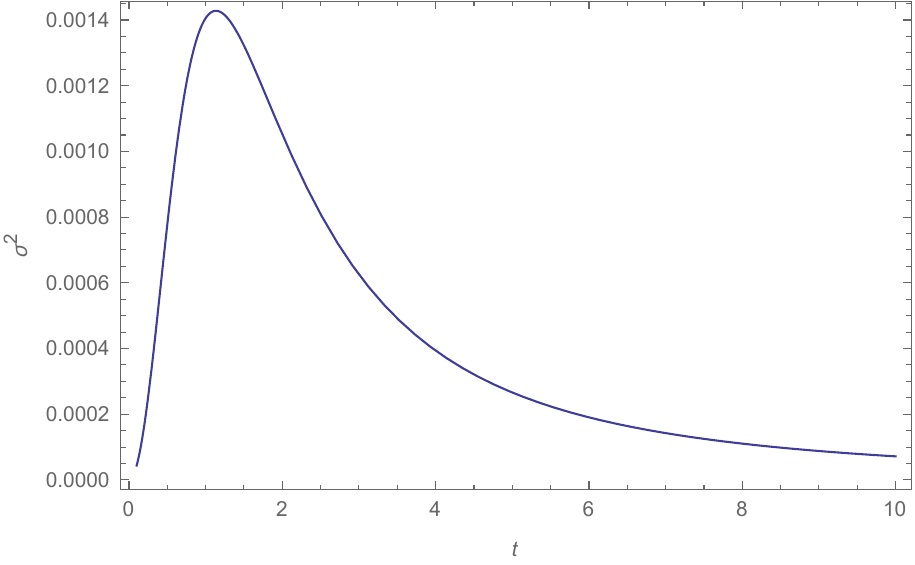}
\caption{The plot of $\protect\sigma$ versus $t$
with $\protect\alpha=1.3$, $\protect\beta=1.01$ and $b=0.3$}.\label{ch2fig6}
\end{figure}
From the above values, we can observe that the present model is free from initial singularity. The values of $H, \theta$ and $\sigma$ are finite at $t=0$. All the above parameters are decreasing function of cosmic time and tend to zero
for later time. Fig. \ref{ch2fig6} shows the behavior of the shear expansion for the Bianchi type III model. The mean anisotropic parameter is constant. Hence, the obtained model is
anisotropic throughout the universe as $\frac{\sigma^{2}}{\theta^2}\neq 0$.
\section{Bianchi type $VI_0$ model}\label{ch2m2}

\begin{figure}[H]
\begin{minipage}[t]{.45\textwidth}
\centering
      \includegraphics[width=65mm]{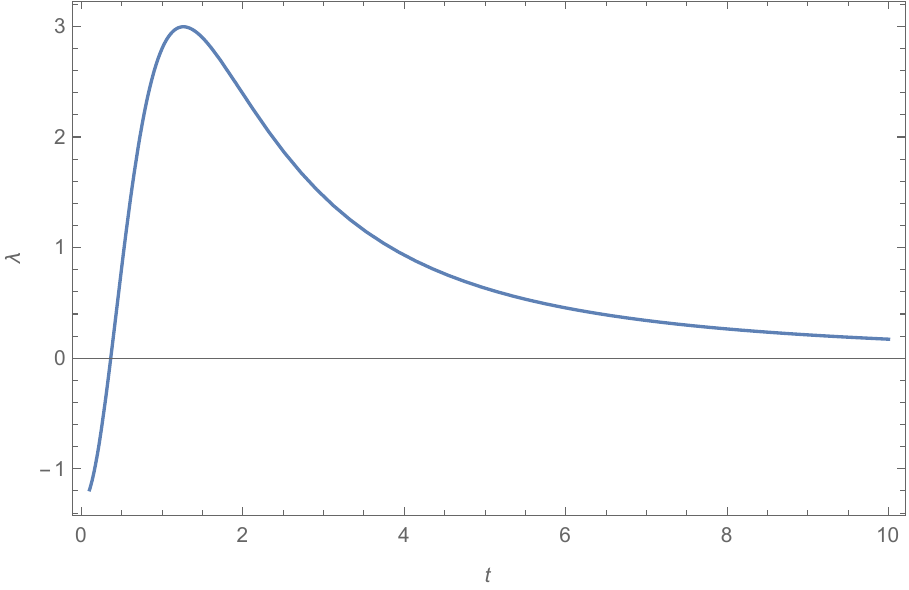}
\caption{The plot of  $\lambda$ versus  $t$ with $b=0.1$, $\gamma=-18$, $\alpha=1.3$ and $\beta=1.01$.}\label{ch2fig10}
  \end{minipage}
\begin{minipage}[t]{.45\textwidth}
\centering
      \includegraphics[width=65mm]{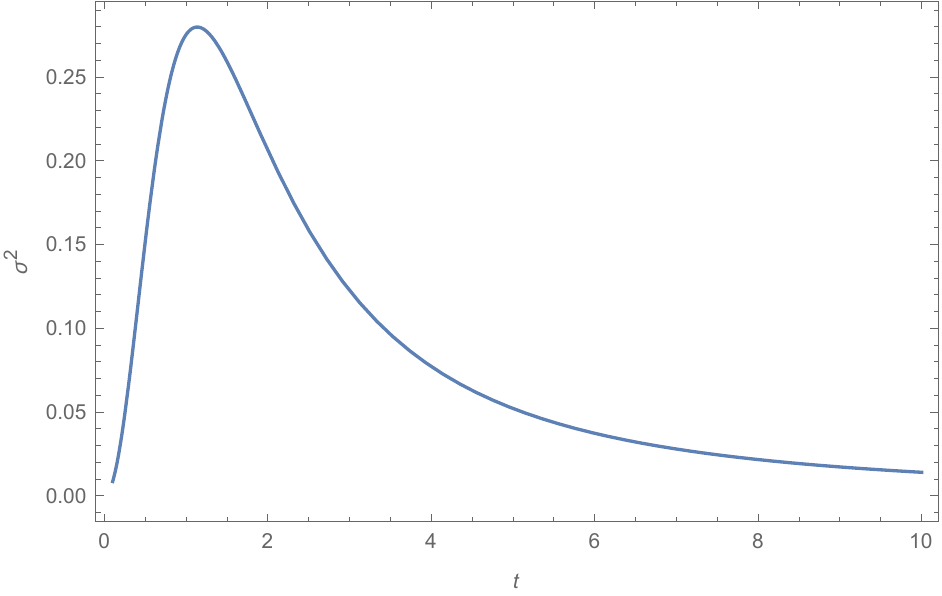}
\caption{The plot of $\sigma$ versus $t$ with $\alpha=1.3$, $\beta=1.01$ and $b=0.1$.}\label{ch2fig11}
  \end{minipage}
\end{figure}

In this model we have considered the spatially homogeneous Bianchi type $VI_0$ line element as
\begin{equation}\label{ch2met2}
ds^{2}=-dt^{2}+a_1^{2}dx^{2}+a_2^2e^{-2x}dy^{2}+a_3^{2}e^{2mx}dz^{2},
\end{equation}
where $a_1, a_2, a_3$ are functions of cosmic time $t$ and $m$ is a constant. 
Using eqns. (\ref{ch2ent}) and (\ref{ch2met2}) in eqn. (\ref{ch2fldeqn}), we obtain the following set of field equations,
\begin{eqnarray}
\frac{\dot{a_1}\dot{a_2}}{a_1a_2}+\frac{\dot{a_2} \dot{a_3}}{a_2a_3}+\frac{\dot{a_1}\dot{a_3}}{a_1a_3}-\frac{1}{a_1^2}=\biggl(\frac{16\pi+3\gamma}{2\gamma}%
\biggr)\rho+ \frac{p}{2}+\frac{\lambda}{2}, \label{ch2en6}\\
\frac{\ddot{a_{2}}}{a_{2}}+\frac{\ddot{a_{3}}}{a_{3}}+\frac{\dot{a_2} \dot{a_3}}{a_2a_3}+\frac{1}{a_1^2}=-\biggl(\frac{16\pi+3\gamma}{2\gamma}\biggr)p-\biggl(%
\frac{16\pi+3\gamma}{2\gamma}\biggr)\lambda+\frac{\rho}{2}, \label{ch2en7} \\
\frac{\ddot{a_1}}{a_1}+\frac{\ddot{a_3}}{a_3}+\frac{\dot{a_1}\dot{a_3}}{a_1a_3}-\frac{1}{a_1^2}=-\biggl(\frac{16\pi+3\gamma}{2\gamma}\biggr)p-\frac{\lambda}{2}+\frac{\rho}{2},\label{ch2en8}\\
\frac{\ddot{a_1}}{a_1}+\frac{\ddot{a_2}}{a_2}+\frac{\dot{a_1}\dot{a_2}}{a_1a_2}-\frac{1}{a_1^2}=-\biggl(\frac{16\pi+3\gamma}{2\gamma}\biggr)p-\frac{\lambda}{2}+\frac{\rho}{2}
,\label{ch2en9} \\
\frac{\dot{a_3}}{a_3}-\frac{\dot{a_2}}{a_2}=0.\label{ch2en10}
\end{eqnarray}
From eqn. (\ref{ch2en10}) we have
\begin{equation}
a_2=c_2a_3,
\end{equation}
where $c_2$ an integration constant. Assuming $c_2=1$, we obtain the following relation
\begin{equation}
a_2=a_3.
\end{equation}
Using this, the field eqns. (\ref{ch2en6} - \ref{ch2en9}) are reduced to
\begin{eqnarray}
\frac{\dot{a_3}^2}{a_3^2}+2\frac{\dot{a_1}\dot{a_3}}{a_1a_3}-\frac{1}{a_1^2}=\biggl(\frac{16\pi+3\gamma}{2\gamma}\biggr)\rho+ \frac{p}{2}+\frac{\lambda}{2},\\
2\frac{\ddot{a_{3}}}{a_{3}}+\frac{\dot{a_3}^2}{a_3^2}+\frac{1}{a_1^2}=-\biggl(\frac{16\pi+3\gamma}{2\gamma}\biggr)p-\biggl(\frac{16\pi+3\gamma}{2\gamma}\biggr)%
\lambda+\frac{\rho}{2}, \\
\frac{\ddot{a_1}}{a_1}+\frac{\ddot{a_3}}{a_3}+\frac{\dot{a_1}\dot{a_3}}{a_1a_3}-\frac{1}{a_1^2}=-\biggl(\frac{16\pi+3\gamma}{2\gamma}\biggr)p-\frac{\lambda}{2}+\frac{\rho}{2}.
\end{eqnarray}
By using the method as given in previous section, we obtain the scale
functions as
\begin{equation}
a_1=\biggl(t^2+\frac{\alpha}{\beta}\biggr)^\frac{3-6b}{2\beta}
\end{equation}
\begin{equation}
a_2=a_3=\biggl(t^2+\frac{\alpha}{\beta}\biggr)^\frac{3b}{2\beta}
\end{equation}
The energy density, fluid pressure and string tension density for Bianchi type $VI_0$ model are
\begin{multline}
\rho=\frac{4\zeta}{1+4\zeta^2}\biggl[t^2 \times\biggl(\frac{%
18b-6b\beta-45b^2}{\beta^2}-\frac{9b^2+6b\beta}{2\zeta\beta^2}\biggr)\times%
\biggl(t^2+\frac{\alpha}{\beta}\biggr)^{-2} \\
+\frac{3b(\zeta+1)}{\zeta\beta}\biggl(t^2+\frac{\alpha}{\beta}\biggr)^{-1}+%
\frac{1-2\zeta}{2\zeta}\biggl(t^2+\frac{\alpha}{\beta}\biggr)^\frac{6b-3}{%
\beta}\biggr]
\end{multline}
\begin{multline}
p=t^2\frac{\zeta^3[(48\beta+360b)(b-1)]+72b\zeta^2(3b+\beta-1)+\zeta%
[12\beta(4b-1)-198b^2+126b-18]+6b\beta}{\zeta(1-2\zeta)(1+4\zeta^2)\beta^2%
}\\
\times\biggl(t^2+\frac{\alpha}{\beta}\biggr)^{-2} 
+ \frac{24\zeta^3(1-b)-36b\zeta^2+(6-12b)\zeta}{\zeta(1-2\zeta)(1+4%
\zeta^2)\beta}\biggl(t^2+\frac{\alpha}{\beta}\biggr)^{-1}-\frac{8\zeta^2+6%
}{(1-2\zeta)(1+4\zeta^2)}\biggl(t^2+\frac{\alpha}{\beta}\biggr)^\frac{6b-3%
}{\beta}
\end{multline}
\begin{equation}
\lambda=\frac{2}{1-2\zeta}\biggl[t^2 \times\biggl(\frac{6-18b}{\beta}+\frac{%
9(1-2b)(1-3b)}{\beta^2}\biggr)\times\biggl(t^2+\frac{\alpha}{\beta}\biggr)%
^{-2}+\frac{9b-3}{\beta}\biggl(t^2+\frac{\alpha}{\beta}\biggr)^{-1}+2\biggl(%
t^2+\frac{\alpha}{\beta}\biggr)^{\frac{6b-3}{\beta}}\biggr]
\end{equation}
\begin{figure}[H]
\begin{minipage}[t]{.45\textwidth}
\centering
     \includegraphics[width=68mm]{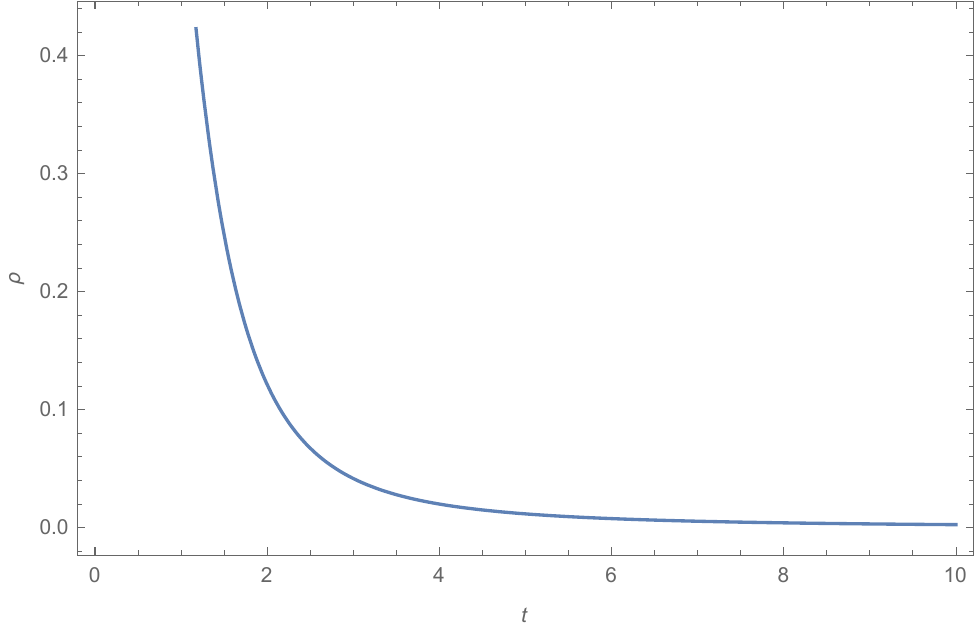}
\caption{The plot of $\rho$ versus $t$ with $b=0.1$, $\gamma=-18$, $\alpha=1.3$ and $\beta=1.01$. }\label{ch2fig7}
  \end{minipage}
\begin{minipage}[t]{.45\textwidth}
\centering
      \includegraphics[width=68mm]{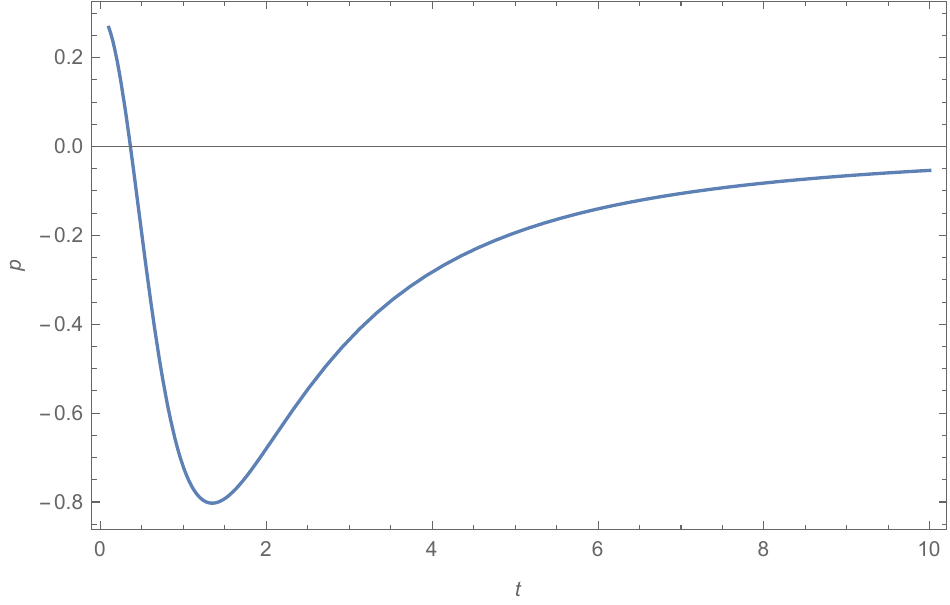}
\caption{The plot of  $p$ versus $t$ with $b=0.1$, $\gamma=-18$, $\alpha=1.3$ and $\beta=1.01$.}\label{ch2fig8}
  \end{minipage}
\end{figure}
\begin{figure}[H]
\centering
\includegraphics[width=0.5\textwidth]{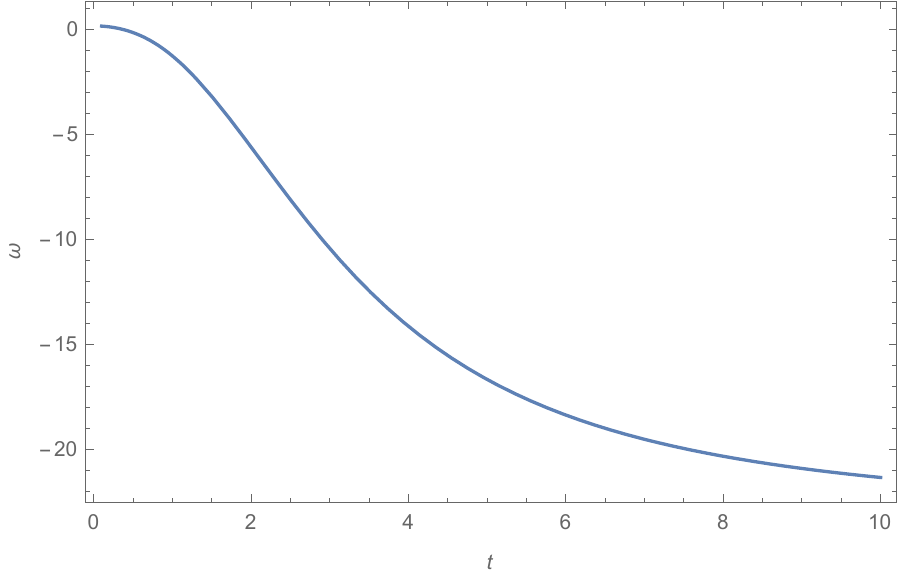}
\caption{The plot of $\omega$ versus $t$
 with $b=0.1$, $\gamma=-18$, $\alpha=1.3$ and $\beta=1.01$.}\label{ch2fig9}
\end{figure}
In this model, the behavior of energy density $\rho$, pressure $p$ and EoS parameter $\omega$ with respect to time are depicted in Fig. \ref{ch2fig7} to Fig. \ref{ch2fig9}. It can be observed that this model is free from initial singularity like previous model. The volume of this model increases with time showing the accelerated expansion of the universe. Moreover, the values of $\rho$ and $p$ are finite at the initial epoch $t=0$ and then vanishes for large time i.e. $t\rightarrow \infty$. Fig. \ref{ch2fig9} depicts the variation of the EoS $\omega$ versus time $t$ as a representative case with appropriate choice of physical parameters using reasonably well known situations. It shows the negative value $\omega$ throughout the universe which is a good agreement with recent observational data. The string tension density $\lambda$ for this model is presented in Fig. \ref{ch2fig10} in which it can be observed that the strings disappear from the universe at larger times.\\  
The HP, expansion scalar, shear scalar and mean anisotropic parameters for the model are obtained as
\begin{equation}
H=\frac{t}{\beta}\biggl(t^2+\frac{\alpha}{\beta}\biggr)^{-1},
\end{equation}
\begin{equation}
\theta=\frac{3t}{\beta}\biggl(t^2+\frac{\alpha}{\beta}\biggr)^{-1},
\end{equation}
\begin{equation}
\sigma^2=t^2 \times\frac{27b^2-18b+3}{\beta^2}\times\biggl(t^2+\frac{\alpha}{%
\beta}\biggr)^{-2},
\end{equation}
and
\begin{equation}
\Delta= 18b^2-12b+2.
\end{equation}

In this model, the values of $H, \theta$ and $\sigma$ are also finite at initial time $t=0$ and they vanish when $t\rightarrow \infty$. The behavior of the shear scalar is depicted in Fig. \ref{ch2fig11}. The mean anisotropy parameter is constant for this model. Since $\frac{\sigma^{2}}{\theta^2}\neq 0$, the present model indicates that it is not approaching isotropy at later time.
\section{Conclusion}\label{ch2conclusion}
In this chapter, it is shown that in $f(R,T)$ gravity the two Bianchi models representing string fluid as source of matter are free from initial singularity and start expanding with finite acceleration. That means 
at $t=0$, it shows that $a(0)\neq 0,\dot{a}(0)=0$ but $\ddot{a}(0)=$ constant. In addition, the assumed DP $q$ approaches to $-\infty $ at $t=0$ and reduces to zero at $t=\sqrt{\alpha /(\beta -1)}$. Since the period of accelerated expansion depends on the values of $\alpha $ and $\beta $, the model shows deceleration when DP $q$ approaching to $\beta -1$ for large values of $t$. Therefore it puts a restriction on $\beta $ as $1\leq \beta \leq 2$. The string tension density
vanishes for Bianchi III and exists for Bianchi $VI_{0}$ model. In both the cases, the energy density is positive and decreasing function of time and pressure is an increasing function of time. In the Bianchi type III model the pressure $p$ starts from a large negative value in starting and later approaches to zero, whereas, in Bianchi type $VI_0$ model it starts with a small positive value at $t=0$ and it becomes negative at later time. Since the causes of accelerated expansion of the universe assume that some kind of unknown energy that is DE reveals, pressure must be negative, the nature of pressure in the model is also in a good agreement with the current observation about accelerated expansion. The physical parameters $H, \theta ,\sigma ^{2}$ are decreasing function of time and tends to zero at $t\rightarrow -\infty $ for both the models. Since $\frac{\sigma ^{2}}{\theta ^{2}}\neq 0$, both the models are anisotropic throughout the evolution of the universe. 



\chapter{Anisotropic cosmological models in $f(R,T)$ gravity with variable deceleration parameter} 

\label{Chapter3} 

\lhead{Chapter 3. \emph{Anisotropic cosmological models in $f(R,T)$ gravity with variable deceleration parameter}} 


\blfootnote{The work, in this chapter, is covered by the following publication: \\
\textit{Anisotropic cosmological models in $f(R,T)$ gravity
with variable deceleration parameter}, Int. J. Geom. Methods Mod. Phys., \textbf{14} (2017) 1750097.}
According to the models reported in the previous chapter, Bianchi models in $f(R,T)$ gravity theory provide a comprehensive and coherent description of space-time, gravity, and matter at a macroscopic level. In understanding the large scale structures and realizing the picture of early stages of universe, these cosmological models studied in $f(R,T)$ gravity are very essential. After inspired by the outcomes of Bianchi models, we have explored in this chapter a new feature of spatially homogeneous anisotropic Bianchi type I model with bulk viscous matter content. The model is constructed in $f(R,T)$ gravity with two different cases viz. $f(R,T)=R+2f(T)$ and $f(R,T)=f_1(R)+f_2(T)$. 
Also, a time varying DP is employed to obtain the exact solution of the field equations. Moreover, we have developed some tools like the nature of WEC, DEC, SEC, by which the physical and kinematical properties of the model can be discussed to understand the future evolution of the universe.
       
\section{Introduction}\label{ch3sec1}

In the last two decades, modern cosmology has reached a new vision to establish considerable advancements in the account of  expanding universe. The relevant observational evidence for the same are provided by these cosmic observations in ref. \cite{Riess/1998,Perlmutter/1999,Plank15,ACTPol14,Garnavich998,
Garnavich98,Perlmutter97,Spergel03,Spergel07,Tegmark04,
Daniel08,Caldwell004,Huang06}. In this context, some MGTs become most attractive aspirant to observe the accelerated expansion of the universe as well as the effective causes related to DE.
One of the MGTs, $f(R)$ gravity (see details in \ref{ch1f(R)}) in which the matter Lagrangian is replaced by an arbitrary function of $R$. This $f(R)$ gravity becomes an adequate theory to provide the gravitational alternative for DE and about the early inflation and the late-time cosmic acceleration of the universe \cite{Capozziello02,Caroll004,Dolgov03,
 Nojiri03,Nojiri04,Abdalaa05,Mena06,Bamba08}. After that Harko et al. \cite{Harko11} have extended that $f(R)$ theory to a new modified theory named as $f(R,T)$ theory, where the gravitational part of the action still depends on the Ricci scalar $R$ and trace of stress energy momentum tensor $T$. It is suggested that due to the matter-energy coupling, the leading model of this theory depends on source term representing the variation of energy-momentum tensor. For this purpose, the matter content of the model in this chapter is considered as Bulk viscous fluid. \\
Several models are investigated with perfect fluid matter to analyze the accelerated expansion of the universe. As per the recent observations an unknown form of energy (DE) having negative pressure is treated as the main candidate for accelerated expansion of the universe. On this account, we need to construct a cosmological model for expanding universe without invoking DE, while choosing most reliable matter component. In this context, one can approach the cosmic viscosity which may act as the DE candidate and can play an important role in causing accelerated expansion of the universe by consuming negative effective pressure \cite{Zimdahl01}. At present, the cosmological model with dissipative fluid matter becomes more acceptable than with dust (a pressure-less distribution) or with a perfect fluid matter. It gives more realistic model than others and is most effective in paying attention to the dynamical background of homogeneous and isotropic universe. If we focus on the background origin of viscous fluid, it is commonly accepted that, during the neutrino decoupling in radiation era, the early phase matter content of the universe behaved like a viscous fluid \cite{Misner67,Misner68,Klimerk76}. The dissipative mechanism of viscous fluid helps to modify the nature of singularity occurred for perfect fluid matter. On the basis of GUTs, it can be suggested that the phase transition and string creation are also involved in viscous effects. Moreover, such kind of viscous fluid cosmological model helps to explain the matter distribution on the large entropy per baryon in the present universe. In fact, the mixture of minimally coupled self-interacting scalar field with viscous fluid can successfully derive an accelerated expansion of the universe, while the same mixture with perfect fluid is unable to do so \cite{Chimento02}. Hence, the models with viscous fluid are widely investigated in literature \cite{Singh11,Yadav12,Jamil12,Singh07,Hu06}.
  
In the present chapter, we take in account a Bianchi type I space-time, which is known as the immediate generalization of the FLRW flat metric with different directional scale factors. In some special cases, the simplest spatially homogeneous and anisotropic Bianchi type I model include \textit{Kasner metric}, which helps to govern the dynamics near the singularity (see details in section \ref{ch1Bianchi}). The nature of Bianchi type I cosmological model in the context of a viscous fluid can cause a qualitative behavior of solutions near the singularity without removing the total initial Big Bang singularity \cite{Belinskii76}. Thereafter, the bulk viscous fluid matter within Bianchi type I universe was studied with the assumption of constant DP in $f(R,T)$ gravity theory \cite{Singh09,Fabris06,Saha07,Sharif012,Sharif 013,Sharif14}.\\ 
The chapter is organized in the following ways, section \ref{ch3sec1} deals with the basic idea about the whole concept. Then the  details of exact solutions in both the cases of $f(R,T)$ gravity ($f(R,T)=R+2f(T)$ and $f(R,T)=f_1(R)+f_2(T)$) for the Bianchi type I space-time with the help of time varying DP are obtained in the section \ref{ch3sec2}. The detail discussion of the figures and the physical properties of both models are presented in the section \ref{ch3sec3}, which is followed by the concluding remarks in section \ref{ch3sec4}.
\section{Field equations and solutions}\label{ch3sec2}
The spatially homogeneous LRS Bianchi type I metric  reads,
\begin{equation}\label{ch3met}
ds^{2}=dt^{2}-A^{2}dx^{2}-B^2(dy^{2}+dz^{2}),
\end{equation}
where $A$ and $B$ are functions of cosmic time $t$ only.\\
The energy momentum tensor for bulk viscous fluid is considered in the following form
\begin{equation}
T_{\mu \nu}=(\rho+\overline{p})u_{\mu}u_{\nu}-\overline{p}g_{\mu \nu},
\end{equation}
where the four velocity vector $u^\mu$ satisfying $u_\mu u^\nu=1$. The bulk viscous pressure which satisfies the linear equation of state $p=\omega \rho, \ \ 0\leq \omega \leq 1$ can be expressed as 
\begin{equation}
\overline{p}=p-3\xi H.
\end{equation}
where $\xi$ is the bulk viscous coefficient. The trace of energy momentum tensor is given as $T=\rho-3\overline{p}$.


\subsection{Case I: $f(R,T)=R+2f(T)$}
\begin{figure}[H]
\centering
\includegraphics[width=78mm]{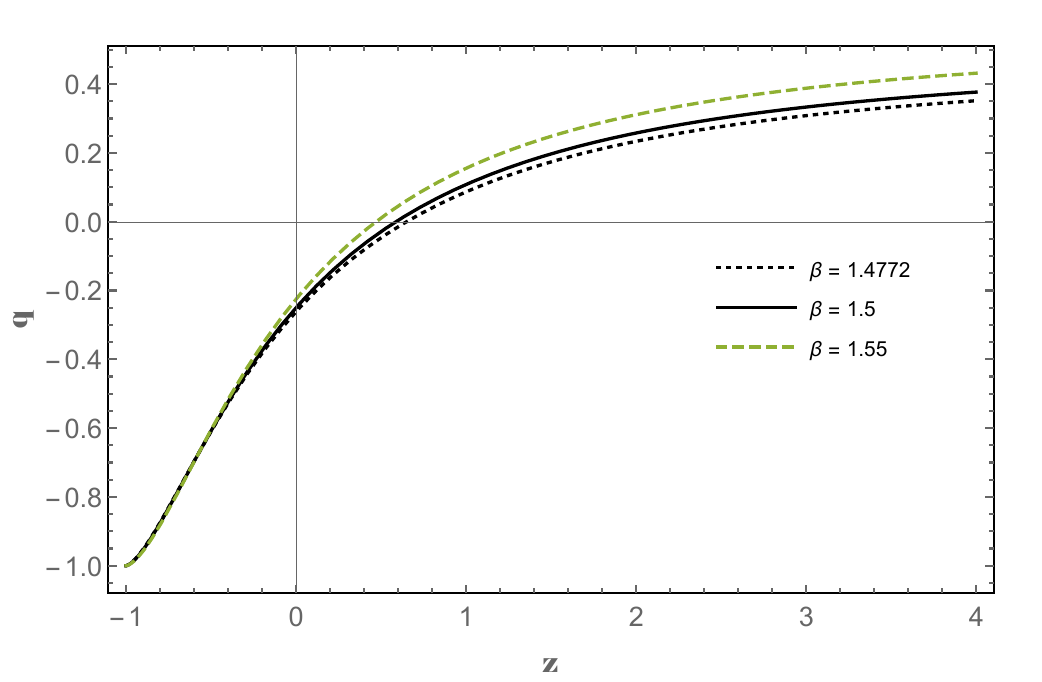}
\caption{$q$  vs. $z$ with different $\beta$.}\label{ch3fig1}
\end{figure}

\begin{figure}[H]
\centering
\minipage{0.50\textwidth}
  \includegraphics[width=68mm]{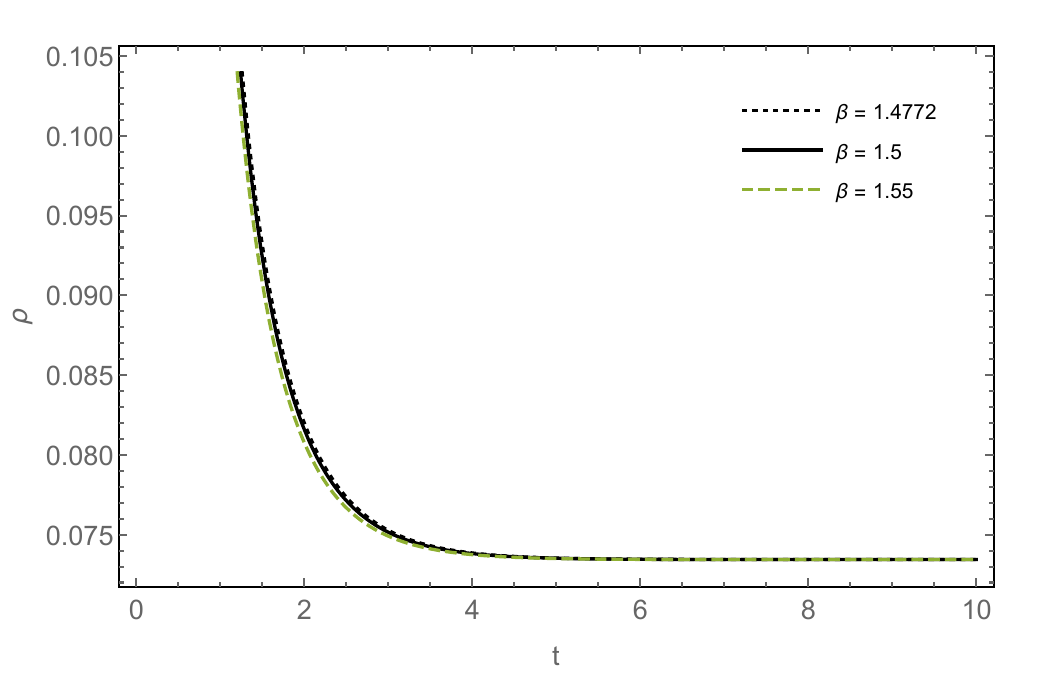}
  \caption{ $\rho$  vs. $t$ with $\alpha=1$ and different $\beta$.}\label{ch3fig2}
\endminipage\hfill
\minipage{0.50\textwidth}
  \includegraphics[width=68mm]{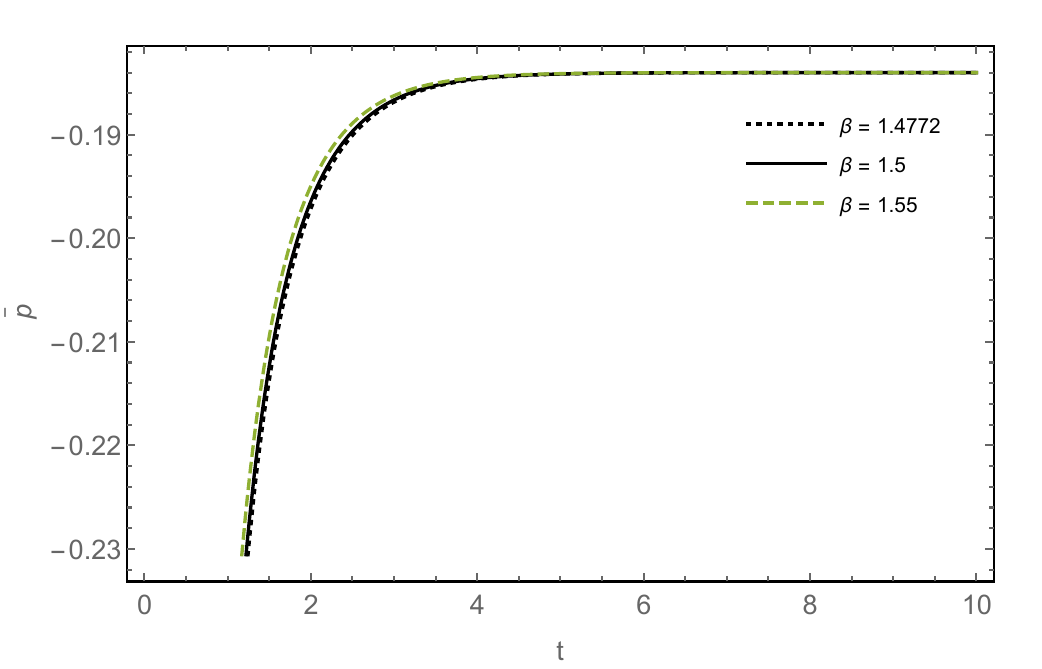}
  \caption{ $\overline{p}$  vs. $t$ with $\alpha=1$ and different $\beta$.}\label{ch3fig3}
\endminipage
\end{figure}

\begin{figure}[H]
\centering
\minipage{0.48\textwidth}
\includegraphics[width=68mm]{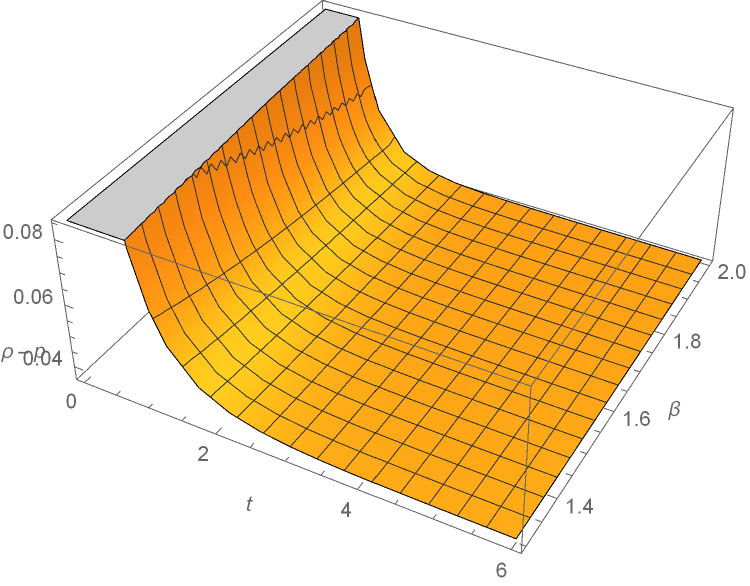}
  \caption{ Behavior of WEC versus $t$ and $\beta$ with $\alpha=1, \omega=0.5 $.}\label{ch3fig5}
\endminipage\hfill
\minipage{0.48\textwidth}
  \includegraphics[width=68mm]{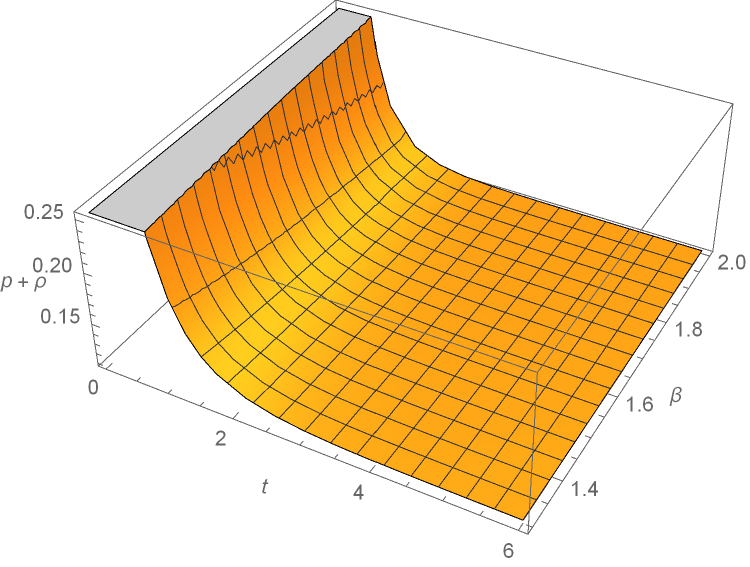}
  \caption{ Behavior of DEC versus $t$ and $\beta$ with $\alpha=1, \omega=0.5 $.}\label{ch3fig6}
\endminipage
\end{figure}

\begin{figure}[H]
\centering
  \includegraphics[width=78mm]{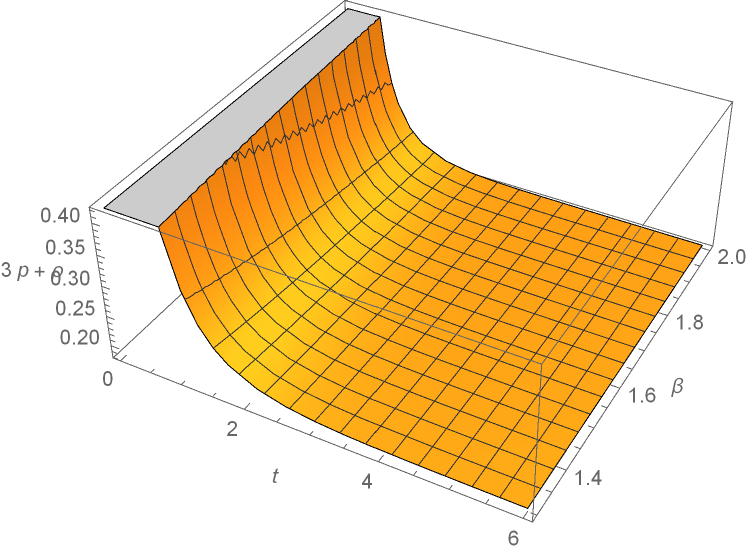}
  \caption{ Behavior of SEC versus $t$ and $\beta$ with $\alpha=1, \omega=0.5$.}\label{ch3fig7}
\end{figure}

The $f(R,T)$ gravity field equations  of the section (\ref{ch1f(R,T)}) for linear case with $f(T)=\alpha T$, where $\alpha$ is an arbitrary constant for the metric (\ref{ch3met}) are obtained as
\begin{eqnarray}
2\frac{\dot{A}\dot{B}}{AB}+\frac{\dot{B}^{2}}{B^{2}}=(8+3\alpha) \rho-\alpha \overline{p},\label{ch3eqn1} \\
-2\frac{\ddot{B}}{B}-\frac{\dot{B}^{2}}{B^{2}}=(8\pi+3\alpha)\overline{p}-\alpha \rho, \label{ch3eqn2}\\
-\frac{\ddot{A}}{A}-\frac{\ddot{B}}{B}-\frac{\dot{A}\dot{B}}{AB}=(8\pi+3\alpha)\overline{p}-\alpha \rho.\label{ch3eqn3}
\end{eqnarray}
Here, we have three eqns. (\ref{ch3eqn1} - \ref{ch3eqn3}) including four parameters as $A, B, \overline{p} \ \&\ \rho$.
In order to get an exact solution, we consider a time varying DP as
\begin{equation}\label{ch3DP}
q=-1+\frac{\beta}{1+a^{\beta}},
\end{equation}
where, $\beta>0$ is a constant.\\
The HP we have obtained from the given DP 
\begin{equation}\label{ch3HP}
H=\frac{\dot{a}}{a} =1+a^{-\beta}.
\end{equation}
By assuming the integrating constant as unity and integrating the above one, we have found 
\begin{equation}\label{ch3sf}
a= (e^{\beta t}-1)^{\frac{1}{\beta}}.
\end{equation}
Setting $a(t)=\frac{1}{1+z}$, where $z$ is the redshift, leads to relation
\begin{equation}
t=\frac{\log \left[\left(\frac{1}{z+1}\right)^\beta+1\right]}{\beta}.
\end{equation}
The corresponding $q(z)$ is obtained as
\begin{equation}\label{ch3q(z)}
q=\frac{\beta}{\left(\frac{1}{z+1}\right)^\beta+1}-1.
\end{equation}
The evolution of the present universe is described by the cosmic DP, and the cosmological models are classified on the basis of the time dependence of DP. Therefore, it is more adequate to focus on the behavior of this parameter in the present model. The recent observations like SNe Ia \cite{Riess/1998} and CMB anisotropy \cite{Bennet03} have confirmed the present accelerated phase of expansion and the DP specifies its range value in between $-1\leq q\leq 0$ for describing the expansion phase. Moreover, the behavior of DP with respect to redshift  lies in the specified range of accelerating phase, which can be observed from Fig. \ref{ch3fig1}. The transition from deceleration to acceleration occurs at some transition redshift $z_{tr}$ is completely dependent upon the choice of $\beta$ values. In this model, we have considered three representative values of $\beta$ i.e. $1.4772$, $1.5$ and $1.55$ corresponding to $z_{tr} =0.65$, $0.5874$ and $0.4706$ respectively. The values of transition redshift $z_{tr}$ for the model are agreeing with the observational data \cite{Capozziello14,Capozziello15,Farooq17}.
From the definition of volume in terms scale factor (i.e. $V=a^3=AB^2$) and by using (\ref{ch3sf}) the values of the metric potentials $A, B$ are obtained as
\begin{eqnarray}
A= (e^{\beta t}-1)^{\frac{2}{\beta}},\label{ch3A} \\
B= (e^{\beta t}-1)^{\frac{1}{2 \beta}}.\label{ch3B}
\end{eqnarray}
Solving the field eqns. (\ref{ch3eqn1} - \ref{ch3eqn3}), the values of $\rho$ and $\overline{p}$ are obtained as
\begin{equation}\label{ch3rho1}
\rho=\frac{1}{(8\pi+3\alpha)^2-\alpha^2}\biggl[\biggl(18\pi+\frac{3\alpha}{2}+\frac{5\alpha\beta}{2}\biggr)e^{2\beta t} (e^{\beta t}-1)^{-2}-\frac{5\alpha \beta}{2}e^{\beta t} (e^{\beta t}-1)^{-1}\biggr],
\end{equation}
\begin{multline}\label{ch3p1}
\overline{p}= \frac{-1}{(8 \pi +3\alpha)^2-\alpha^2}\biggl[\biggl(42\pi+\frac{27\alpha}{2}-\frac{5(8\pi+3\alpha)\beta}{2}\biggr)e^{2\beta t} (e^{\beta t}-1)^{-2}\\+\frac{5(8\pi+3\alpha) \beta}{2}e^{\beta t} (e^{\beta t}-1)^{-1}\biggr].
\end{multline}
From eqn. (\ref{ch3rho1}) and Fig. \ref{ch3fig2}, it can be observed that the energy density $\rho$ is a decreasing function of time (e.g. it starts with a positive value and later approaches to zero when $t\rightarrow \infty$) and remains positive throughout the evolution of the universe. At the same time, the bulk viscous pressure $\overline{p}$ given in eqn. (\ref{ch3p1}) and represented graphically in Fig. \ref{ch3fig3} is also an increasing function of time. 
Since it begins from a large negative value and tends to zero at present epoch, which shows observational compatibility that the pressure is negative due to DE in the context of accelerated expansion of the universe. Hence, the behavior of bulk viscous pressure in the model is agreed with the current observation.\\
In this way the coefficient of bulk viscosity $\xi$ and the pressure are also expressed as
\begin{multline}\label{ch3chi1}
\xi=\frac{1}{(8 \pi +3\alpha)^2-\alpha^2}\biggl[\biggl(2\pi(3\omega+7)+\frac{\alpha(\omega+9)}{2}-\frac{5\beta(\alpha\omega-8\pi-3\alpha)}{6}\biggr)e^{\beta t} (e^{\beta t}-1)^{-1}\\-\frac{5\beta(\alpha\omega-8\pi-3\alpha)}{6}\biggr],
\end{multline}
\begin{equation}
p=\omega\rho
=\frac{\omega}{(8\pi+3\alpha)^2-\alpha^2}\biggl[\biggl(18\pi+\frac{3\alpha}{2}+\frac{5\alpha\beta}{2}\biggr)e^{2\beta t} (e^{\beta t}-1)^{-2}-\frac{5\alpha \beta}{2}e^{\beta t} (e^{\beta t}-1)^{-1}\biggr].
\end{equation}
\begin{figure}[H]
\centering
  \includegraphics[width=78mm]{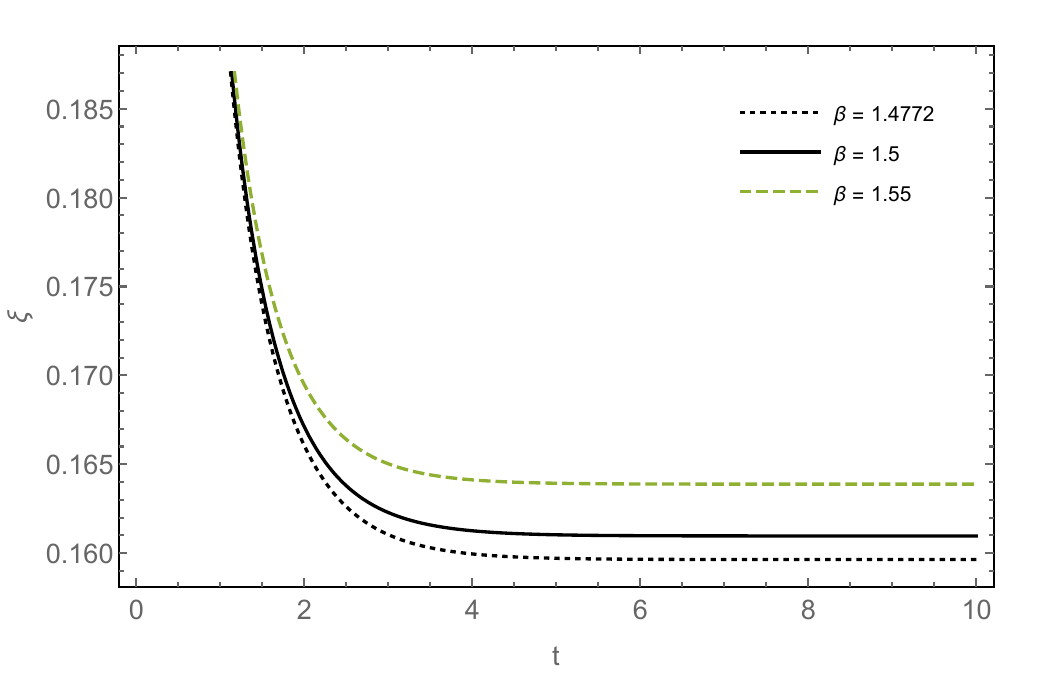}
  \caption{ $\xi$  vs. $t$ with $\alpha=1,\omega=0.5$ and different $\beta$.}\label{ch3fig4}
\end{figure}
Now the behavior of bulk viscous coefficient $\xi$ in this case can be observed from Fig. \ref{ch3fig5}. The bulk viscous coefficient is positive through out the universe and becomes finite when  $t\rightarrow \infty$ for the model.\\
\textbf{\textit{Energy Conditions}}\\
In addition, the alternative conditions for matter content of any theory can be studied by adopting the ECs. In GR, the role of these ECs are widely accepted to prove the existence of space-time singularity theorem and black holes  \cite{Wald/1984}. The ECs are used in many approaches to understand the evolution of the universe as we have already mentioned in previous chapter (\ref{ch1ECs}). Here, we have discussed those ECs for this model and their importance in various aspects. For example, the stability of matter source in a model can be studied by invoking the DEC and also it imposes the DE along with EoS parameter $\omega$ for lower bound $\omega \geq -1$, which is considered as one of the cause for Big Rip singularity \cite{Carroll03}. Similarly, the violation of SEC provides an idea about late time acceleration and also it considered as a typical trait of a positive  $\Lambda$ \cite{Lake04}. Finally, WEC shows that the matter-energy is always non-negative. In literature, the nature of solution for FLRW model with perfect fluid matter is studied through ECs \cite{Zubair16,Sharif13}. According to the predefined literature, one can consider the ECs to be useful to analyze the behavior of cosmological solutions throughout the universe. Therefore, we have dealt with some well-known ECs like WEC, DEC and SEC to observe the solutions in both the cases of this chapter. The behavior of WEC, DEC, and SEC  with the proper choice of constants are depicted in Fig. \ref{ch3fig5} to Fig. \ref{ch3fig7} respectively. Also, it is observed that all the ECs are respected for this model.
The values of Ricci scalar $R$ and the trace of matter source $T$ are obtained as
\begin{equation}
R=\biggl(6\beta-\frac{27}{2}\biggr)e^{2\beta t}(e^{\beta t}-1)^{-2}-6\beta e^{\beta t}(e^{\beta t}-1)^{-1},
\end{equation}
\begin{multline}
T=\frac{1}{(8\pi+3\alpha)^2-\alpha^2}\biggl[(144\pi+42\alpha-(60\pi+20\alpha)\beta)e^{2\beta t}(e^{\beta t}-1)^{-2}\\+(60\pi+20\alpha)\beta e^{\beta t}(e^{\beta t}-1)^{-1}\biggr].
\end{multline}
Using the above equations, the function $f(R,T)$ is obtained as
\begin{multline}\label{ch3frt}
f(R,T)=\biggl(6\beta-\frac{27}{2}+\frac{(288-120\beta)\alpha \pi+48\alpha^2-40\alpha^2 \beta}{(8\pi+3\alpha)^2-\alpha^2}\biggr)e^{2\beta t}(e^{\beta t}-1)^{-2}\\+\biggl(\frac{120\pi \alpha+40\alpha^2\beta}{(8\pi+3\alpha)^2-\alpha^2}-6\beta \biggr)e^{\beta t}(e^{\beta t}-1)^{-1}.
\end{multline}
\begin{figure}[H]
\centering
\includegraphics[width=78mm]{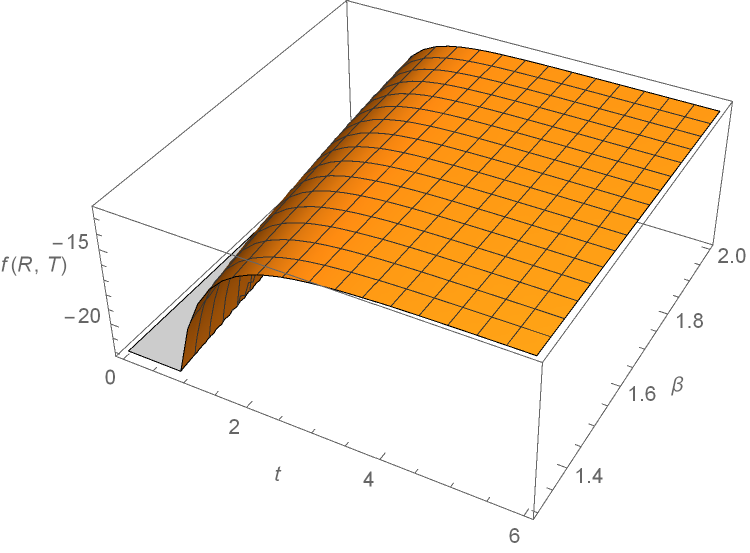}
\caption{ Behavior of $f(R,T)$ versus $t$ and $\beta$ with $\alpha=1$.}\label{ch3fig8}
\end{figure}
Fig. \ref{ch3fig8} shows the behavior of the function $f(R,T)$ for this first model of this chapter.
\subsection{Case II: $f(R,T)=f_1(R)+f_2(T)$}
\begin{figure}[H]
\centering
\minipage{0.50\textwidth}
  \includegraphics[width=68mm]{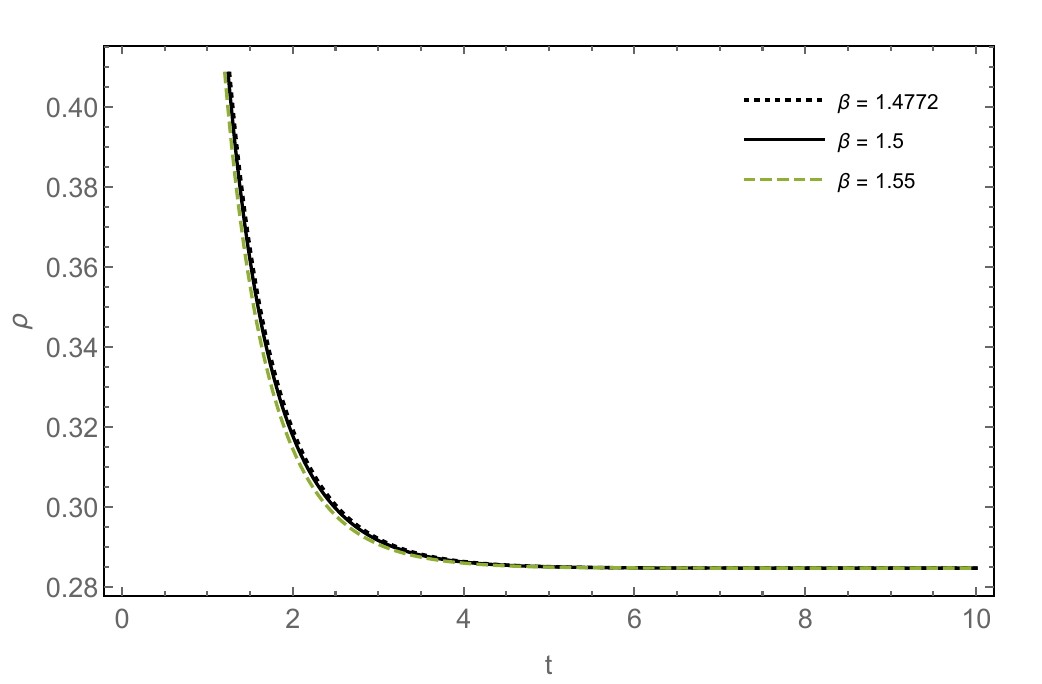}
  \caption{ $\rho$  vs. time with  $\gamma=5$ and different $\beta$.}\label{ch3fig9}
\endminipage\hfill
\minipage{0.50\textwidth}
  \includegraphics[width=68mm]{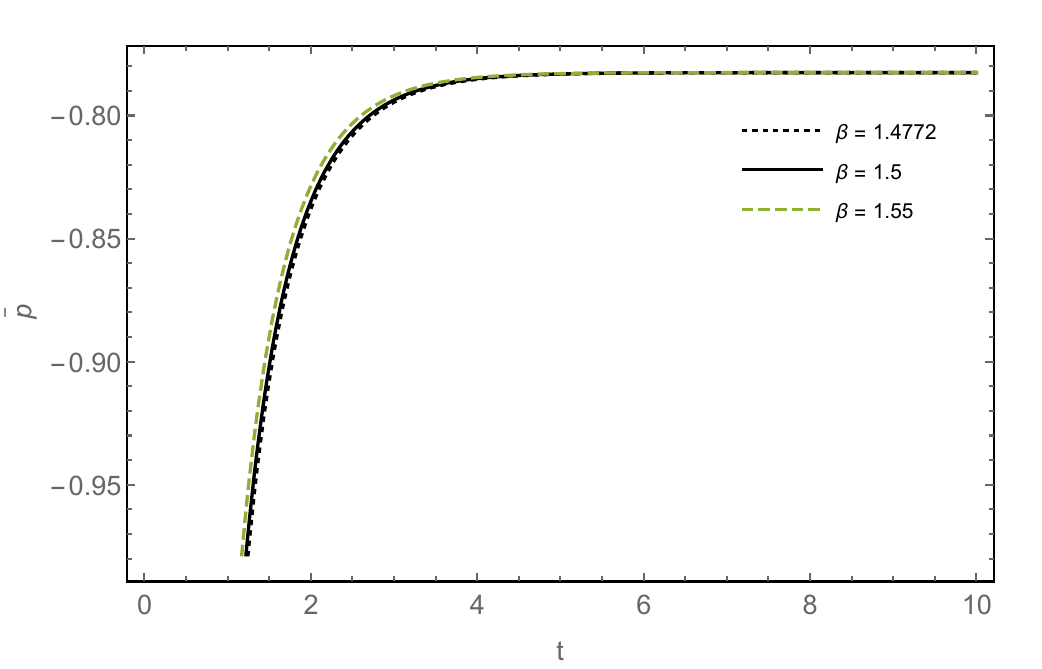}
  \caption{ $\overline{p}$  vs. time with $\gamma=5$ and different $\beta$.}\label{ch3fig10}
\endminipage
\end{figure}

In this case we have assumed $f_1(R)=\gamma R$ and $f_2(T)=\gamma T$, where $\gamma$ is an arbitrary constant. The corresponding field equations from section \ref{ch1f(R,T)} take the form as
\begin{equation}
 R_{\mu \nu}-\frac{1}{2}Rg_{\mu \nu}=\biggl(\frac{8\pi +\gamma}{\gamma}\biggr)T_{\mu \nu}+\biggl(p+\frac{1}{2}T\biggr)g_{\mu \nu}=\chi T_{\mu \nu}+\biggl(p+\frac{1}{2}T\biggr)g_{\mu \nu},
\end{equation}
where $\chi=\biggl(\frac{8\pi +\gamma}{\gamma}\biggr)$. The set of field equations for the metric in eqn. (\ref{ch3met}) are
\begin{eqnarray}
2\frac{\dot{A}\dot{B}}{AB}+\frac{\dot{B}^{2}}{B^{2}}=\biggl(\chi+\frac{1}{2}\biggr)\rho-\frac{1}{2}\overline{p},\label{ch3eqn4}\\
-2\frac{\ddot{B}}{B}-\frac{\dot{B}^{2}}{B^{2}}=\biggl(\chi+\frac{1}{2}\biggr)\overline{p}-\frac{1}{2} \rho,\label{ch3eqn5} \\
-\frac{\ddot{A}}{A}-\frac{\ddot{B}}{B}-\frac{\dot{A}\dot{B}}{AB}=\biggl(\chi+\frac{1}{2}\biggr)\overline{p}-\frac{1}{2}\rho. \label{ch3eqn6}
\end{eqnarray}
In this case, we have obtained the same metric potential eqn. (\ref{ch3A}) and eqn. (\ref{ch3B}) from the above set of field eqns.  (\ref{ch3eqn4} - \ref{ch3eqn6}) as obtained in the previous case. By using these metric potentials eqn. (\ref{ch3A}) and eqn. (\ref{ch3B}), the values of energy density $\rho$ and bulk viscous pressure $\overline{p}$ are expressed as
\begin{equation}\label{ch3rho2}
\rho=\frac{1}{\biggl(\chi+\frac{1}{2}\biggr)^2-\frac{1}{4}}\biggl[\biggl(\frac{9\chi-6}{4}+\frac{5\beta}{4}\biggr)e^{2\beta t} (e^{\beta t}-1)^{-2}-\frac{5\beta}{4}e^{\beta t} (e^{\beta t}-1)^{-1}\biggr],
\end{equation}
\begin{multline}\label{ch3p2}
\overline{p}= \frac{-1}{\biggl(\chi+\frac{1}{2}\biggr)^2-\frac{1}{4}}\biggl[\biggl(\frac{21\chi+6}{4}-\frac{5\beta}{2}\biggl(\chi+\frac{1}{2}\biggr)\biggr)e^{2\beta t} (e^{\beta t}-1)^{-2}\\+\frac{5\beta}{2}\biggl(\chi+\frac{1}{2}\biggr)e^{\beta t} (e^{\beta t}-1)^{-1}\biggr].
\end{multline}
Here, the energy density $\rho$ is a positive decreasing function of time and it converges to zero as $t\rightarrow \infty$ (see Fig. \ref{ch3fig9}). Also the Fig. \ref{ch3fig10} specify the similar behavior of $\overline{p}$ versus cosmic time $t$ as obtained in previous case.\\
For this case, the values of bulk viscosity coefficient $\xi$ and pressure $p$ are also given as
\begin{multline}\label{ch3chi2}
\xi=\frac{\omega \rho-\overline{p}}{3H}
= \frac{1}{\biggl(\chi+\frac{1}{2}\biggr)^2-\frac{1}{4}}\biggl[\biggl(\frac{9\omega+21-10\beta}{12}\chi+\frac{(5\beta-6)(\omega-1)}{12}\biggr)e^{\beta t} (e^{\beta t}-1)^{-1}\\-\frac{5\omega \beta}{12}+\frac{5\beta}{6}\biggl(\chi+\frac{1}{2}\biggr)\biggr],
\end{multline}
\begin{equation}
 p =\omega\rho=\frac{\omega}{\biggl(\chi+\frac{1}{2}\biggr)^2-\frac{1}{4}}\biggl[\biggl(\frac{9\chi-6}{4}+\frac{5\beta}{4}\biggr)e^{2\beta t} (e^{\beta t}-1)^{-2}-\frac{5\beta}{4}e^{\beta t} (e^{\beta t}-1)^{-1}\biggr].
\end{equation}
\begin{figure}[H]
\centering
  \includegraphics[width=78mm]{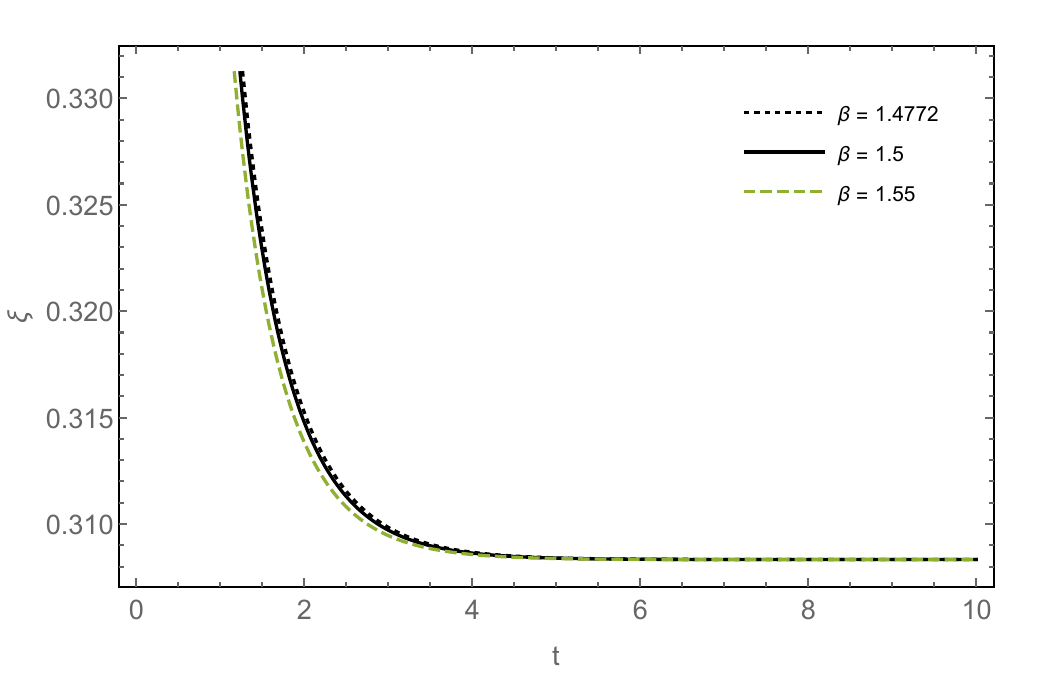}
  \caption{ $\xi$  vs. time with $\gamma=5$ and different $\beta$}\label{ch3fig11}
\end{figure}
The bulk viscosity coefficient is constant throughout the universe as required and presented in Fig. \ref{ch3fig11}.\\
The various ECs for this model are plotted below.
\begin{figure}[H]
\centering
\minipage{0.48\textwidth}
  \includegraphics[width=68mm]{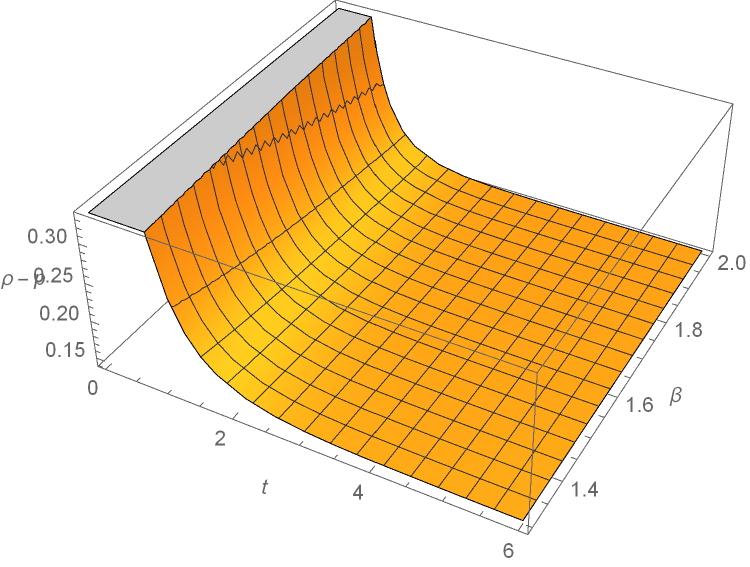}
  \caption{ Behavior of WEC versus $t$ and $\beta$  with  $\gamma=5$.}\label{ch3fig12}
\endminipage\hfill
\minipage{0.48\textwidth}
  \includegraphics[width=68mm]{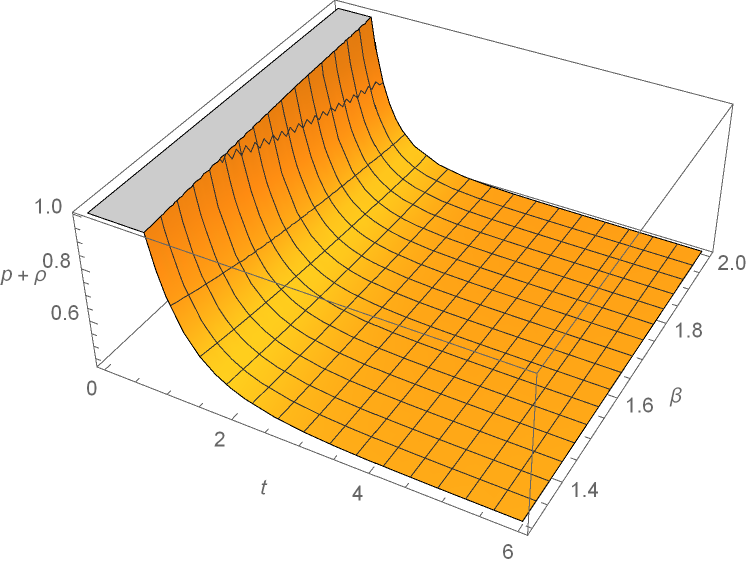}
  \caption{ Behavior of DEC versus $t$ and $\beta$ with $\gamma=5$.}\label{ch3fig13}
\endminipage
\end{figure}
\begin{figure}[H]
\centering
  \includegraphics[width=78mm]{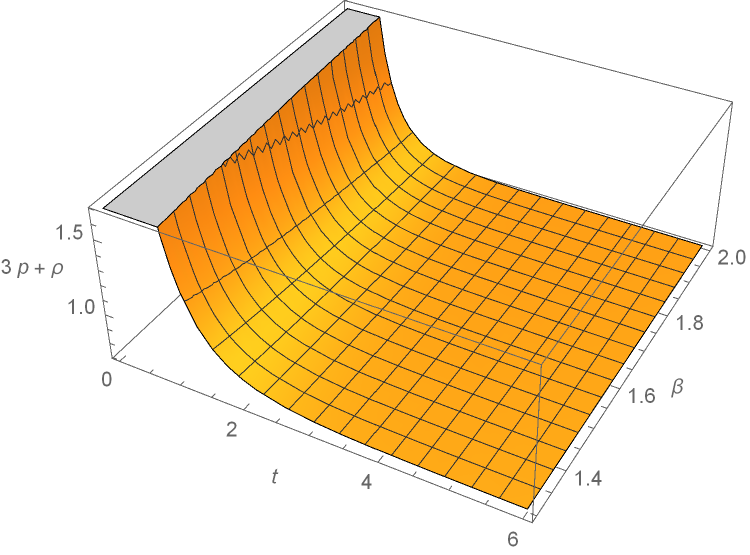}
  \caption{ Behavior of SEC versus $t$ and $\beta$ with $\gamma=5$.}\label{ch3fig14}
\end{figure}
Fig. \ref{ch3fig12} to Fig. \ref{ch3fig14}, show that the ECs are completely agreed with GR.\\
We can obtain the trace of matter $T$ for this model as
\begin{multline}
T=\rho-3\overline{p}=\frac{1}{(\chi+\frac{1}{2})^2-\frac{1}{4}}\biggl[\biggl(\biggl(\frac{36-15\beta}{2}\biggr)\chi-\frac{12-10\beta}{4}\biggr)e^{\beta t}(e^{\beta t}-1)^{-2}\\+\biggl(\frac{15\beta \chi}{2}-\frac{10\beta}{4}\biggr)\beta e^{\beta t}(e^{\beta t}-1)^{-1}\biggr].
\end{multline}
The relation $f(R,T)$ for the above case is obtained in the form
\begin{multline}
f(R,T)=\biggl(6\beta-13.5+\frac{(18-7.5\beta)\chi-2.5\beta+3}{(\chi+\frac{1}{2})^2-\frac{1}{4}}\biggr)\mu e^{2\beta t}(e^{ \beta t}-1)^{-2}\\+\biggl(\frac{7.5\beta \chi-2.5\beta}{(\chi+\frac{1}{2})^2-\frac{1}{4}}-6\beta \biggr)\mu e^{\beta t}(e^{\beta t}-1)^{-1}.
\end{multline}
\begin{figure}[H]
\centering
\includegraphics[width=78mm]{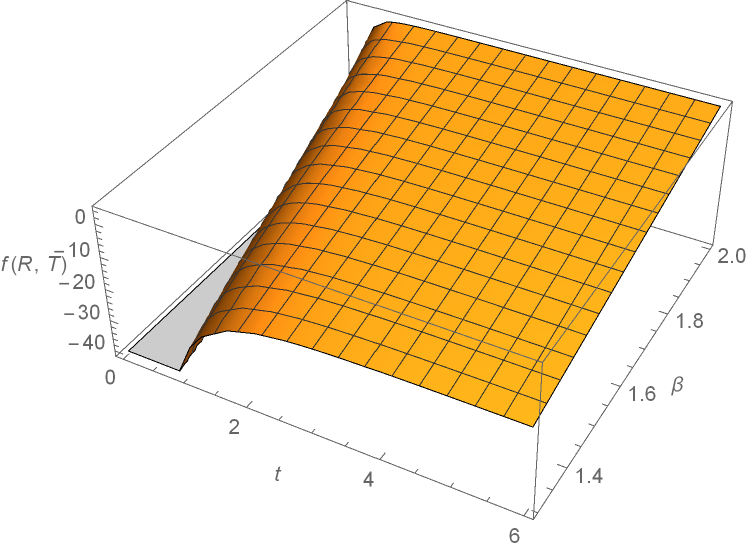}
\caption{ Behavior of $f(R,T)$ versus $t$ and $\beta$ with  $\gamma=5$.}\label{ch3fig15}
\end{figure}
Fig. \ref{ch3fig15} shows the behavior of $f(R,T)$ for $f(R,T)=f_1(R)+f_2(T)$ model.
\section{Physical properties of the models}\label{ch3sec3}
\begin{figure}[H]
\centering
\includegraphics[width=78mm]{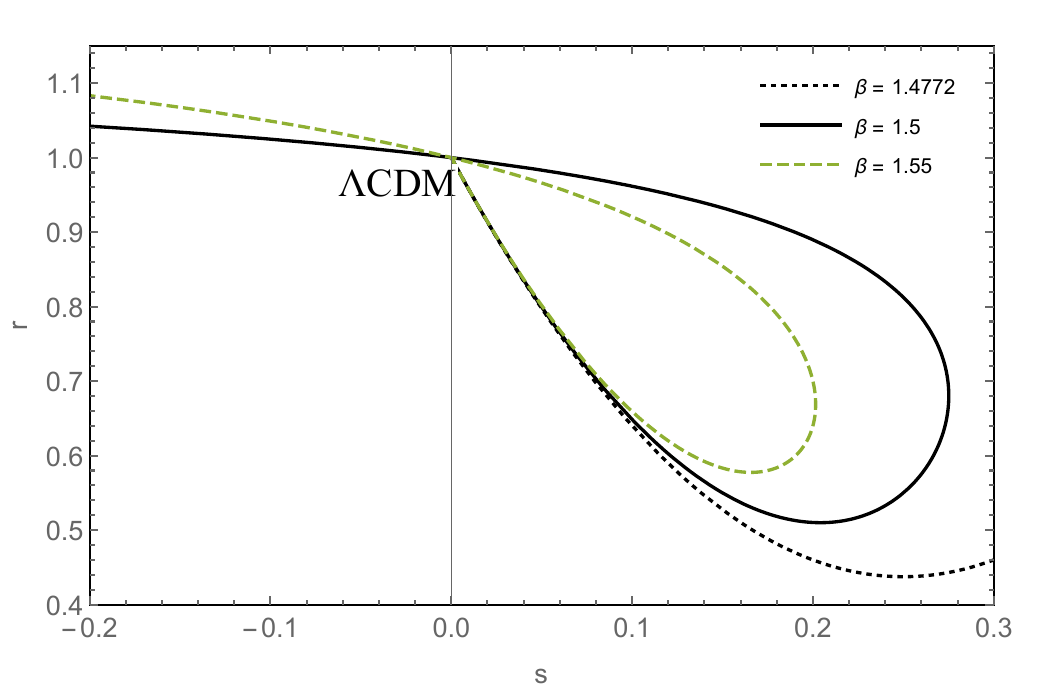}
\caption{ $r$  vs. $s$.}\label{ch3fig17}
\end{figure}

It is well known fact that HP as well as DP describe the rate expansion of the universe with respect to time. But for detail kinematical descriptions of the cosmological expansions, one has to consider some extended set of parameters having higher order time derivatives of the scale factor.\\
The spatial volume for both the models of this chapter turns out to be
\begin{equation}\label{ch3vol}
V=AB^2=(e^{\beta t}-1)^{\frac{3}{\beta}},
\end{equation}
which indicates that in both the models, it is zero at initial time $t=0$ and then increases with time. That means the evolution of our Universe starts with Big Bang scenario for these models. It is further noted that from eqn. (\ref{ch3sf}), the average scale factor becomes zero at the initial epoch. Hence, both the models have a point type singularity \cite{Maclum71}.\\
The HP $H$, expansion scalar $\theta$, shear scalar $\sigma^2$, and anisotropy parameter become
\begin{equation}
H=e^{\beta t} (e^{\beta t}-1)^{-1},
\end{equation}
\begin{equation}
\theta=3e^{\beta t} (e^{\beta t}-1)^{-1},
\end{equation}
\begin{equation}
\sigma^2= \frac{3}{4} e^{2\beta t} (e^{\beta t}-1)^{-2}.
\end{equation}
\begin{equation}
\Delta =\frac{1}{2}.
\end{equation}
All above parameters are diverge at $t=0$ and they become finite as $t\rightarrow \infty$ for both the models of this chapter. It is noted here that the isotropic condition $\frac{\sigma^2}{\theta^2}$ becomes constant (from early to late time), which shows that the model does not approach isotropy throughout the evolution of the universe. Therefore, the aniostropic parameter becomes constant for the models. Hence, from the above mentioned equations it can be observed that the models represents an expanding and accelerating universe which starts at a Big Bang singularity.\\
\textbf{\textit{Jerk parameter}}\\
The third derivative of the scale factor with respect to time and the third-order term in the Taylor series expansion of the Hubble's law is termed as the cosmological jerk parameter. This dimensionless parameter is one of the most important quantities for describing the dynamics of the universe. It plays an important role to describe models close to $\Lambda$CDM through a convenient method \cite{Sahni02, Visser05}. Cosmic transition from decelerating phase to accelerating phase occurs for models with a positive value of $j$ and negative $q$. For flat $\Lambda$CDM model the value of jerk is constant $j=1$ \cite{Rapetti06}. The jerk parameter in terms of scale factor and DP is
$$j=\frac{a^2}{\dot{a}^3}\frac{d^3 a}{dt^3},\,\,\,\,\ 
j=q+2q^2-\frac{\dot{q}}{H}.$$
Thus, the jerk parameter for the models is
\begin{equation}
j=1-3\beta e^{-\beta t}+2\beta^2 e^{-2\beta t}+\beta^2e^{-2\beta t}(e^{\beta t}-1).
\end{equation}
\begin{figure}[H]
\centering
\includegraphics[width=78mm]{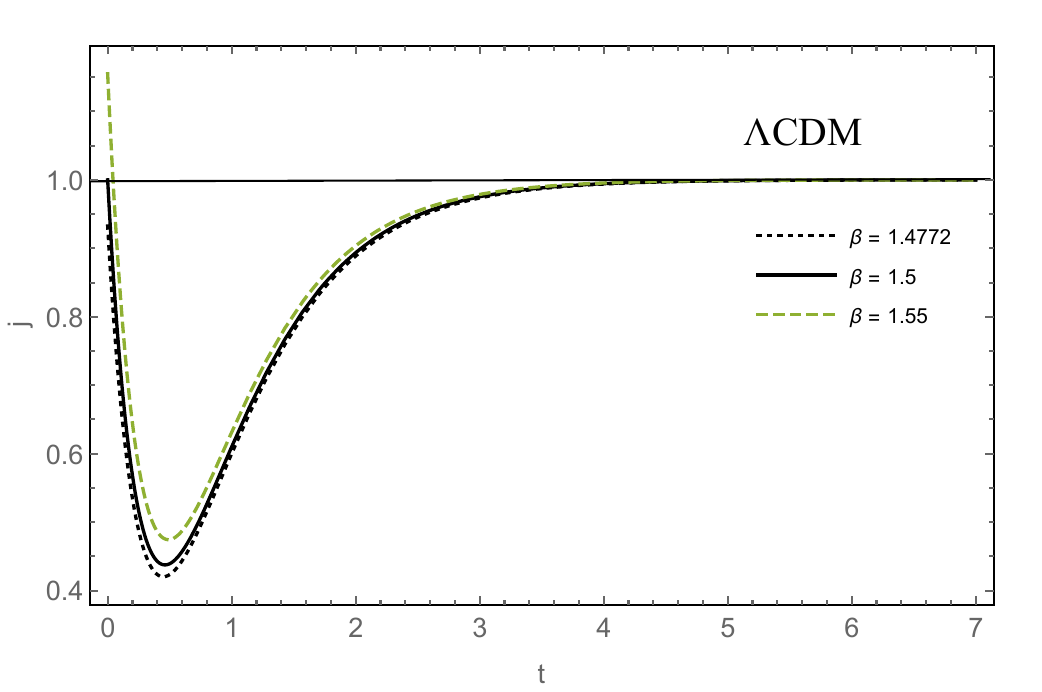}
\caption{ Behavior of Jerk parameter versus $t$ with different $\beta$.}\label{ch3fig16}
\end{figure}
From the Fig. \ref{ch3fig16}, it is clear that our value does not overlap with the value $j=2.16^{+0.81}_{-0.75}$ obtained from combination of three kinematical data sets: the gold sample data of type Ia supernovae \cite{Riess04}, the SNIa data obtained from the SNLS project \cite{Astier06}, and the X-ray galaxy cluster distance measurements \cite{Rapetti06}. In Fig. \ref{ch3fig16}, it can be observed that the jerk parameter remains positive through out the universe and  is equal to the $\Lambda$CDM model at $t\geq5.5$ for the considered values of $\beta$.  It is interesting to note that the model is close to  $\Lambda$CDM model for the following set of values as represented in table \ref{ch3tbl1}.\\
\textbf{$r-s$ parameter:}\\
The state-finder pair $\{r,s\}$ is defined as \cite{Sahni03}
\begin{equation}
r=\frac{\dddot{a}}{aH^3}, \ \ \ \ s=\frac{r-1}{3(q-\frac{1}{2})}
\end{equation}
The state-finder pair is a geometrical diagnostic parameter, that can be constructed from a space-time metric directly. It is more general compared to physical variables, that depends on the properties of physical fields describing DE. This is because the physical variables are model dependent. For the flat $\Lambda$CDM model the state-finder pair obtained as $\{r,s\}=\{1,0\}$ \cite{Feng08}. The values of the state-finder parameter for the models are
\begin{equation}
r=1-3\beta e^{-\beta t}+2\beta^2 e^{-2\beta t}+\beta^2e^{-2\beta t}(e^{\beta t}-1)
\end{equation}
\begin{equation}
s=\frac{1}{6 \beta-9 e^{\beta t}}\biggl[2\beta^2 e^{-\beta t}(e^{\beta t}-1)+4\beta^2 e^{-\beta t}-6\beta\biggr]
\end{equation}
From the expressions of $r$ and $s$ parameters, we found that $\{r,s\}=\{1,0\}$ only when $t=\frac{1}{\beta}\ln\bigl(\frac{\beta}{3-\beta}\bigr)$. The variation of $\beta$ and $t$ for $\{r,s\}=\{1,0\}$  is presented in Table- \ref{ch3tbl1}. For the set of values of $(\beta,t)$ the models represents $\Lambda$CDM models, which are presented in table \ref{ch3tbl1}. The $r-s$ trajectory of the models is presented in the Fig. \ref{ch3fig17}.
\begin{table}[H]
\centering
\begin{tabular}{|c|c|c|c|}
  \hline
  $\beta$ & $t=\frac{1}{\beta}\ln\bigl(\frac{\beta}{3-\beta}\bigr)$ & $r$  &  $s$\\\hline
  1.5 & 0 & 1 & $\infty$ \\
  1.6 & 0.08345 & 1 & $-1.458333324\times 10^{-9}\approx 0$ \\
  1.7 & 0.15780 & 1 & 0 \\
  1.8 & 0.22525 & 1 & 0 \\
  1.9& 0.28765 & $1.000000002\approx 1$ & 0 \\
  2 & 0.5$\ln(2)$ & 1 & 0\\
  \hline
\end{tabular}
\caption{Variation of $\beta$ and $t$ for $\{r,s\}$}\label{ch3tbl1}
\end{table}
From the above table \ref{ch3tbl1}, it is observed that at initial epoch $t=0$ the parameter $r$ becomes unity and $s$ becomes finite and diverges for $\beta=1.5$.
\section{Conclusion}\label{ch3sec4}
It is well known that the currently available observational data are verified by considering the standard $\Lambda$CDM ( $\Lambda$+Cold Dark Matter) as it is treated as the best model. In fact, the matter density parameter $\Omega_m=0.3089$ in the $\Lambda$CDM model, is being confirmed by the most recent data from Planck+BAO+JLA+$H_0$ \cite{Plank15}. In this sense, the model provides the transition of $\Lambda$CDM universe from deceleration to acceleration at the redshift $z_{tr}=0.65$, which corresponds to $\beta=1.4772$ as seen in Fig. \ref{ch3fig1}. Similar behavior is given in the variation of DP vs redshift at $\beta=1.4772$. Therefore, we can conclude that the present models of this chapter have a transition from deceleration to acceleration with transition redshift $z_{tr}$ satisfying the observational data in both the ways. Moreover, the energy density for both cases is positive valued, decreasing function of time and approaching zero with the evolution of time (see Fig. \ref{ch3fig2} and Fig. \ref{ch3fig9}). Similarly, the variation of bulk viscous pressure and coefficient of bulk viscosity are presented in the Fig. \ref{ch3fig3} and Fig. \ref{ch3fig10} for case I and in the Fig. \ref{ch3fig4} and Fig. \ref{ch3fig11} for case II respectively.  
From the figures, we notice that bulk viscous pressure and coefficient of bulk viscosity are negative and positive valued function of time respectively. Also, the coefficient of bulk viscosity decreases with the evolution of time and maintains a constant rate after $t>4$. In this way, the evolution of ECs against time is satisfied in both the cases, which are presented in Fig. \ref{ch3fig5} to Fig. \ref{ch3fig7} and Fig. \ref{ch3fig12} to Fig. \ref{ch3fig14} respectively. All the physical parameters presented in both the cases follow the same quantitative behavior as that of observational data. Hence, we conclude this chapter by summarizing the obtained results for both the cases as
\begin{itemize}
  \item The models of the universe obtained here are accelerating and expanding with an exponential expansion.
  \item Energy density and coefficient of bulk viscosity are positively valued and decreasing function of time in both the cases and also $\rho \rightarrow 0$ when $t\rightarrow\infty$.
  \item Bulk viscosity pressure $(\bar{p})$ is negatively valued in both the cases.
  \item Energy conditions (SEC, WEC, DEC) are satisfied for both cases.
  \item Jerk parameter and state-finder trajectory in the $r-s$ plane are close to $\Lambda$CDM model.
\end{itemize}



\chapter{MSQM model in $f(R, T)$ gravity with time varying deceleration parameters} 

\label{Chapter4} 

\lhead{Chapter 4. \emph{MSQM model in $f(R, T)$ gravity with time varying deceleration parameters}} 

This chapter \blfootnote{The work, in this chapter, is covered by the following publication: \\ 
\textit{Magnetized strange quark matter in $f(R, T)$ gravity with bilinear and special form of time varying deceleration parameter}, New Astronomy, \textbf{60} (2018) 80.} provides more details about the different parametrization of geometric DP in such a way that it can produce an accelerated cosmological model in $f(R,T)$ gravity. We have already established the vital role of DP in the context of obtaining exact solutions of cosmological model embedded with different matter fluid in our previous chapter. Based on the importance of DP, we have explored here some cosmological models with MSQM distribution and $\Lambda$ in $f(R,T)$ gravity. We have concentrated on various aspects of bilinear and special form of time varying DP in order to achieve exact solution of the field equations. Further we have verified the observational compatibility of the model for the MSQM source of matter. The models presented in this chapter, with the SQM along with magnetic epoch gives an idea of accelerated expansion of the universe as per the observations indicated by type Ia Supernovae. 
\section{Introduction}\label{ch4intro}
A geometric dimensionless DP is one of the most accountable candidate to describe and understand the evolution of the universe in present scenario. For that reason, the present modern cosmology is defined by the search for two numbers: the present universe expansion rate $H_0$ and DP $q_0$, which allows testing the coincidence between the cosmological model with cosmological principle \cite{Sandage70}. From the previous chapter, it has been derived that $\Lambda$CDM model is the best fit and most suitable model to specify the accelerated expansion of the universe. This model predicts the value of DP $q_0 \sim \frac{1}{2}$, while the other CDM models predict $q_0 = 0$. However, the value of $q_0$ is determined to a precision of $\pm 0.2$ by the group of Supernovae cosmology project and the high $z$ supernovae team in which the distant type Ia Supernovae ($z \sim 0.3-0.7$) is used as standard candles. In order to predict the fate of the evolving universe we need not require only the current numeric values of these parameters but also need their time dependence. Thus,  the Taylor series expansion of scale factor at the present time $t_0$ is given as
\begin{equation}
a(t)=a(t_0)+\dot{a}(t_0)(t-t_0)+\frac{1}{2}\ddot{a}(t_0)(t-t_0)^2+......
\end{equation}
or 
\begin{equation}
\frac{a(t)}{a(t_0)}=1+H_0 (t-t_0)+\frac{q_0}{2}H_{0}^2(t-t_0)^2+.....
\end{equation}
It is well known that the difference between the actual age of the universe and Hubble time, deceleration or acceleration are all determined by the sign of DP. Such as
\begin{itemize}
\item Constant DP: $q>0$, the
age of the universe will be less than the Hubble time and decelerating.
\item For $q=0$, expansion occurs at constant rate and the age equals to Hubble time.
\item For $-1<q<0$, the acceleration of the universe exhibits power-law expansion.
\item  For $q=-1$, the acceleration of the universe is exponential  and super-exponential if $q<-1$.
\end{itemize} 
The constant DP is commonly used by cosmologist in literature with various aspects (see for details  \cite{Berm83, berman88}). But for detailed description of the kinematics of cosmological expansion, it is necessary to consider various parametrized forms of time dependence DP. For instance,  Akarsu et al.  \cite{Akarsu14} have described the fate of the universe through parametrization $q = q_0+q_1\bigl(1- \frac{t}{t_0}\bigr)$, which is linear in cosmic time $t$, along with two well-known  additional parametrization of the DP $q = q_0+q_1 z$ and $q = q_0+q_1\bigl(1-\frac{a}{a_0}\bigr)$, where $z$ and $a$ are the redshift parameter and scale factor respectively. Furthermore, they have studied the dynamics of the universe in comparison with the standard $\Lambda$CDM model. \\
Moreover, in this chapter we have considered the MSQM distribution for LRS Bianchi type I universe in the framework of $f(R,T)$ gravity with  $(\Lambda)$.
This SQM is one of the most relevant matter, containing a large quantity of deconfined quark in $\beta$-equilibrium, with electric charge neutrality \cite{Alford07, BA sad07, Aktas11}. It is composed of an equal number of deconfined $u,d,$ and $s$ quarks, and treated as the ground state of matter as well as a strongly interacting matter, well described in \cite{Bodmer71, Witten84, Farhi84}. The creation of this SQM is approaching in two ways; firstly, during phase transition of early universe, another transition also occurred called quark-hadron phase transition in which Quark Gluon Plasma (QGP) got transformed into hadron gas at temp $T\sim 200$ MeV. It is considered as one of the first proof of origin for SQM, while the second approach is the strange matter made from the neutron star at ultra-high density \cite{Bodmer71, Itoh70}. In the medium-dependent quark mass scale, the components of quark mass function are its chemical potential and temperature. These finite chemical potentials presently encounter serious problems. To resolve this, some effective phenomenological models are commonly used like: MIT Bag model \cite{Chodos74} and the Nambu-Jona-Lasinio (NJL) model \cite{nambu, nambu61}. In bag model, ad-hoc bag function is introduced to make all corrections of pressure and energy functions of SQM. This model has been used in literature by Farhi and Jaffe \cite{Farhi84}, in which they have studied the SQM with its EoS. Usually, in this bag model the broken physical vacuum take place inside the hadrons on the basis of strong interaction theories. It gives essentially different vacuum energy densities inside and outside of hadron, and on the bag wall, the pressure of quarks are equilibrated through vacuum pressure and stabilizes the system. Thus, the EoS of
the SQM depends upon the system pressure, various chemical potential components and the degenerate Fermi gasses. Later on these Fermi gases are referred to as the replacement of quarks, which can survive only in a region with a vacuum energy density  $B_\text{c}$ (Bag constant). The unit of bag constant is MeV (fm)$^{-3}$ and it lies in the range of 60-80 MeV (fm)$^{-3}$  \cite{Chakraborty2014}. In the present chapter, we have considered the value of $B_\text{c}$ to be 60 MeV (fm)$^{-3}$. In a simplified bag model, the quarks are considered as massless and non-interacting, and pressure can be defined as  $P_q=\frac{\rho_q}{3}$, where  $\rho_q$ is the quark energy density. The total energy density and pressure given as
\begin{eqnarray}
\rho_\text{m}=\rho_\text{q}+B_\text{c}, \\
   P_\text{m}=P_\text{q}-B_\text{c}.
\end{eqnarray}
Hence the EoS for SQM reads,  \cite{Sotani2004} 
\begin{equation}\label{ch4EoS}
P_\text{m}=\frac{\rho-4B_\text{c}}{3}.
\end{equation}
Recently, the effects of the magnetic field on the stability and on interacting properties of SQM have attracted much attention \cite{Miransky15}. This widespread component of the universe is possessed by Milky way galaxy and many other spiral galaxies along with some common properties of galaxy clusters \cite{Klein88, kronberg92, wolfe92, kronberg94, beck96, vallee04, carilli04}. In the present cosmic scenario, the main focus of research is on the impact of strong magnetic field on the special properties of dense quark matter, neutron star matter \cite{Ducan92} and on the stability of SQM \cite{fukusima13, kojo13, chakrabarty96}. The quark matter has been studied in strong magnetic field with phenomenological bag model, in which the stability of SQM gets stronger when the order of the strength of the magnetic field is greater than some critical value \cite{chakrabarty96}. Thus, it is commonly accepted that the presence of magnetic field causes an anisotropy in pressure and the bag model can be considered as the best satisfactory approach for studying MSQM \cite{mart05,mart07}.\\
The theoretical arguments for the late-time cosmic acceleration are being confirmed by observations through type Ia Supernovae  \cite{Riess/1998, Perlmutter/1999, Garnavich98,Perlmutter97,Letel83}, CMB \cite{Spergel03, Spergel07}, baryon acoustic oscillation (BAO) in galaxy clustering \cite{Eisenstein05, Percival07, Kazin14} and WMAP \cite{permutter03} etc.  In order to investigate the accelerated expansion of the universe, the $f(R,T)$ gravity theory proposed by Harko et al. \cite{Harko11} triggers as one of the best ways to examine the current accelerated behavior of the universe. The magnetized models have also started to be studied frequently. In particular, Agrawal and Pawar \cite{Agrawal2017} have studied Bianchi type V universe model with magnetized domain walls in $f(R,T)$ theory. 
Thereafter, Ayg\"{u}n et al. \cite{Aygun2016} have studied SQM distribution for Marder type anisotropic universe model in $f(R,T)$ theory with $\Lambda$. Later Akta\c{s} and Ayg\"{u}n investigated the dynamics of MSQM distribution in FLRW universe with reconstructed $f(R, T)$ gravity \cite{Aktas17}.\\
In the context of anisotropic properties of space-time, Sahoo and Sivakumar \cite{Sahoo15} have studied $f(R,T)$ theory in LRS Bianchi type I universe where they have presented the Big Rip singularity in this theory (\textbf{NB} \textit{Later discussed in the chapter\ref{Chapter5} about Big Rip}). Also, homogeneous and anisotropic Bianchi type II universe model for dark energy with/without a magnetic field in $f(R,T)$ gravitation theory are studied by Mishra et al. \cite{Chand2016}. Moreover, \c{C}aglar and Ayg\"{u}n \cite{Caglar2017} have obtained homogeneous and anisotropic Bianchi type I universe solutions in $f(R,T)$ gravity with quark matter and $\Lambda$.
 This chapter is organized in the following manner; the basic formalism of $f(R,T)$ gravity field equations and its solutions are narrated in section \ref{ch4fldeqn} and section \ref{ch4sol} respectively. In section \ref{ch4sol}, the solutions of the models with graphical representations are described in detail whereas the conclusion with observational behaviors are discussed in section \ref{ch4conclsn}. 
\section{Field equations in $f(R,T)$ gravity}\label{ch4fldeqn}
In this chapter, we have considered the spatially homogeneous and anisotropic LRS Bianchi type I metric given in eqn. (\ref{ch3met}), 
and the energy-momentum tensor for SQM with magnetic field as \cite{CGT1997, JDB2007}
\begin{equation}\label{ch4eng}
T_{\mu \nu}=(\rho+p+h^2)u_{\mu}u_{\nu}+\bigg(\frac{h^2}{2}-p \bigg)g_{\mu \nu}-h_{\mu} h_{\nu},
\end{equation}
where the magnetic flux $h^2$ is considered in the $x$-direction with $h_{\mu}u^{\nu}=0$.\\
The field equations of $f(R,T)$ gravity with a choice of $f(R,T)=R+2f(T)$ in presence of $\Lambda$ and $f(T)=\lambda T$ ($\lambda=$constant) can be written as 
\begin{equation}\label{ch4fldeqn1}
G_{\mu \nu}=[8\pi+2\lambda]T_{\mu \nu}+[\lambda \rho-\lambda p+2\lambda h^2+\Lambda]g_{\mu \nu}.
\end{equation}
The field eqn. (\ref{ch4fldeqn1}) in presence of $\Lambda$ and $f(T)=\lambda T$ ($\lambda=$constant) can be written as
\begin{equation}\label{ch4fldeqn2}
G_{\mu \nu}=[8\pi+2\lambda]T_{\mu \nu}+[\lambda \rho-\lambda p+2\lambda h^2+\Lambda]g_{\mu \nu}.
\end{equation}
 The set of field equations for the metric in eqn. (\ref{ch3met}) with HP are obtained as
\begin{eqnarray}
2H_\text{x} H_\text{y}+H_\text{y}^2=-(12\pi+5\lambda)h^2-(8\pi+3\lambda)\rho-\Lambda+\lambda p,\label{ch4eqn1}\\
2\dot{H_\text{y}}+3H_\text{y}^2= (4\pi-\lambda)h^2+(8\pi+3\lambda)p-\Lambda-\lambda \rho, \label{ch4eqn2}\\
\dot{H_\text{x}}+\dot{H_\text{y}}+H_\text{x}^2+H_\text{y}^2+H_\text{x}H_\text{y}=-(4\pi+3\lambda)h^2+(8\pi+3\lambda)p-\Lambda-\lambda \rho. \label{ch4eqn3}
\end{eqnarray}
Here, the mean HP, $H=\frac{H_\text{x}+2H_\text{y}}{3}$ where $H_\text{x}=\frac{\dot{A}}{A}$, $H_\text{y}=H_\text{z}=\frac{\dot{B}}{B}$ are the directional HPs. The expansion scalar $\theta$ and shear scalar $\sigma$ for the metric in eqn. (\ref{ch3met}) are obtained as
\begin{eqnarray}
\theta=H_\text{x}+2H_\text{y},\label{ch4theta}\\
\sigma^2=\frac{1}{3}(H_\text{x}-H_\text{y})^2.\label{ch4sigma}
\end{eqnarray}
\section{Solutions of field equations}\label{ch4sol}
Field eqns. (\ref{ch4eqn1} - \ref{ch4eqn3}) contains six unknowns $A, B, \rho, p, h^2$ and $\Lambda$ with three equations. To get a physically viable model of the universe with observational consistency, we assumed the following physically feasible relation.
\begin{enumerate}
\item First we have considered the linear relationship between the directional HPs $H_\text{x}$ and $H_\text{y}$ as
\begin{equation}\label{ch415}
H_\text{x}=nH_\text{y},
\end{equation}
where $n\geq 0$ is an arbitrary constant controls the anisotropy nature of the model. This yields a relation between shear scalar $\sigma$ and scalar expansion $\theta$ as $\theta \propto \sigma$. 
\item Secondly, we have employed the EoS for the SQM as 
\begin{equation}\label{ch416}
p=\frac{\rho-4B_\text{c}}{3},
\end{equation}
where $B_\text{c}$ is the bag constant \cite{Sotani2004}.
\item Finally, we have assumed different type of time varying DP $q$. 
\end{enumerate}
Using eqn. (\ref{ch415}) in the field eqns. (\ref{ch4eqn1}-\ref{ch4eqn3}), we have
\begin{eqnarray}
\frac{9(2n+1)}{(n+2)^2}H^2=-(12\pi+5\lambda)h^2-(8\pi+3\lambda)\rho-\Lambda+\lambda p, \label{ch4eqn4}\\
\biggl[\frac{27}{(n+2)^2}-\frac{6(1+q)}{n+2}\biggr]H^2= (4\pi-\lambda)h^2+(8\pi+3\lambda)p-\Lambda-\lambda \rho,\label{ch4eqn5}\\
\biggl[\frac{9(n^2+n+1)}{(n+2)^2}-\frac{3(n+1)(1+q)}{n+2}\biggr]H^2=-(4\pi+3\lambda)h^2+(8\pi+3\lambda)p-\Lambda-\lambda \rho. \label{ch4eqn6}
\end{eqnarray}
Use of the EoS from eqn. (\ref{ch416}) yields
\begin{equation}\label{ch4h}
h^2=\frac{3(n-1)(q-2)}{2(4\pi+\lambda)(n+2)}H^2,
\end{equation}
\begin{equation}\label{ch4rho}
\rho=\frac{-3}{4(4\pi+\lambda)}\biggl[\frac{9(n-1)}{(n+2)^2}+\frac{3(3+qn-2n)}{(n+2)}\biggr]H^2+B_\text{c},
\end{equation}
\begin{equation}\label{ch4p}
p=\frac{-1}{4(4\pi+\lambda)}\biggl[\frac{9(n-1)}{(n+2)^2}+\frac{3(3+qn-2n)}{(n+2)}\biggr]H^2-B_\text{c},
\end{equation}
\begin{equation}\label{ch4Lambda}
\Lambda=\biggl[\frac{3[(12n\pi+3n\lambda-n^2 \lambda+24\pi+10\lambda)q]}{2(4\pi+\lambda)(n+2)^2}+\frac{(-26\lambda+18n\lambda+6n^2 \lambda-76\pi)}{2(4\pi+\lambda)(n+2)^2}\biggr]H^2-(8\pi+4\lambda)B_\text{c}.
\end{equation}
 Here, we have considered a bilinear DP \cite{Mishra16} in two forms i.e. (i) $ q=\frac{\alpha(1-t)}{1+t}$, (ii) $ q=\frac{-\alpha t}{1+t}$ and (iii) $q=-1+\frac{\beta}{1+a^{\beta}}$  \cite{Debnath2009}.
\subsection{Model I: $q(t)=\frac{\alpha(1-t)}{1+t}$}\label{ch4model1}
\begin{figure}[H]
\minipage{0.48\textwidth}
  \includegraphics[width=68mm]{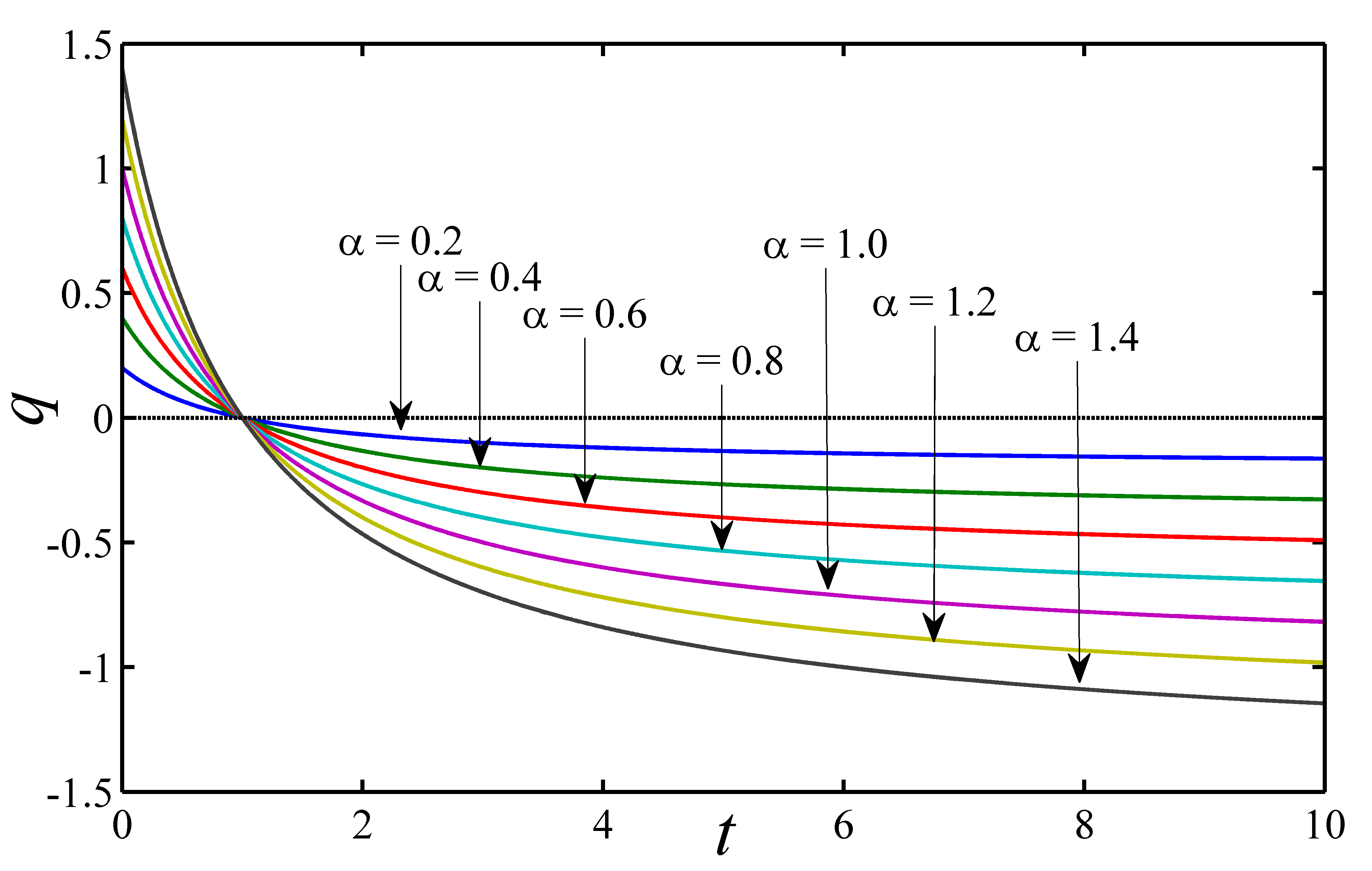}
  \caption{Variation of DP against time for different $\alpha$}\label{ch4fig1}
\endminipage\hfill
\minipage{0.48\textwidth}
  \includegraphics[width=68mm]{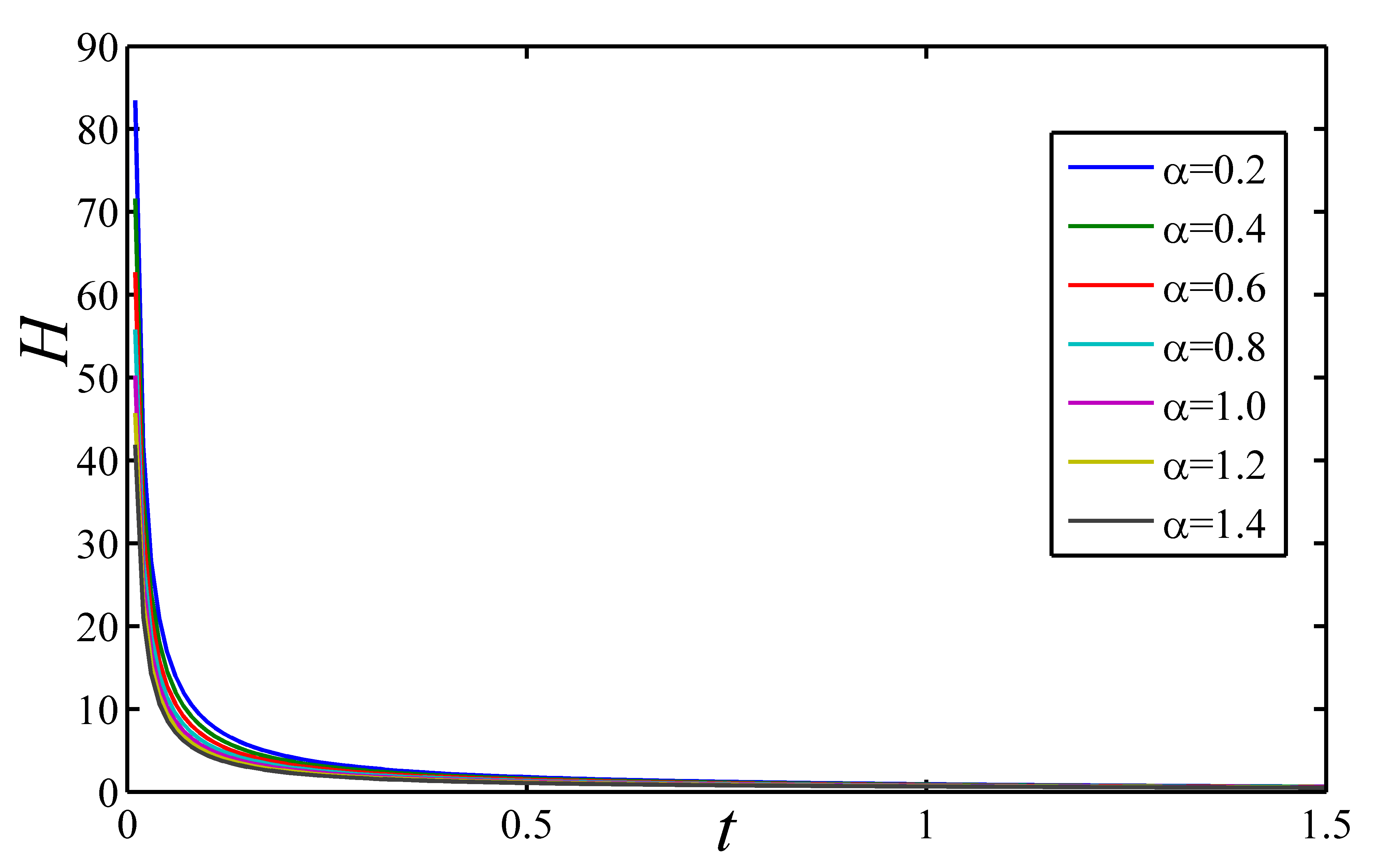}
  \caption{Variation of HP against time for different $\alpha$}\label{ch4fig2}
\endminipage
\end{figure}

\begin{figure}[H]
\minipage{0.48\textwidth}%
  \includegraphics[width=68mm]{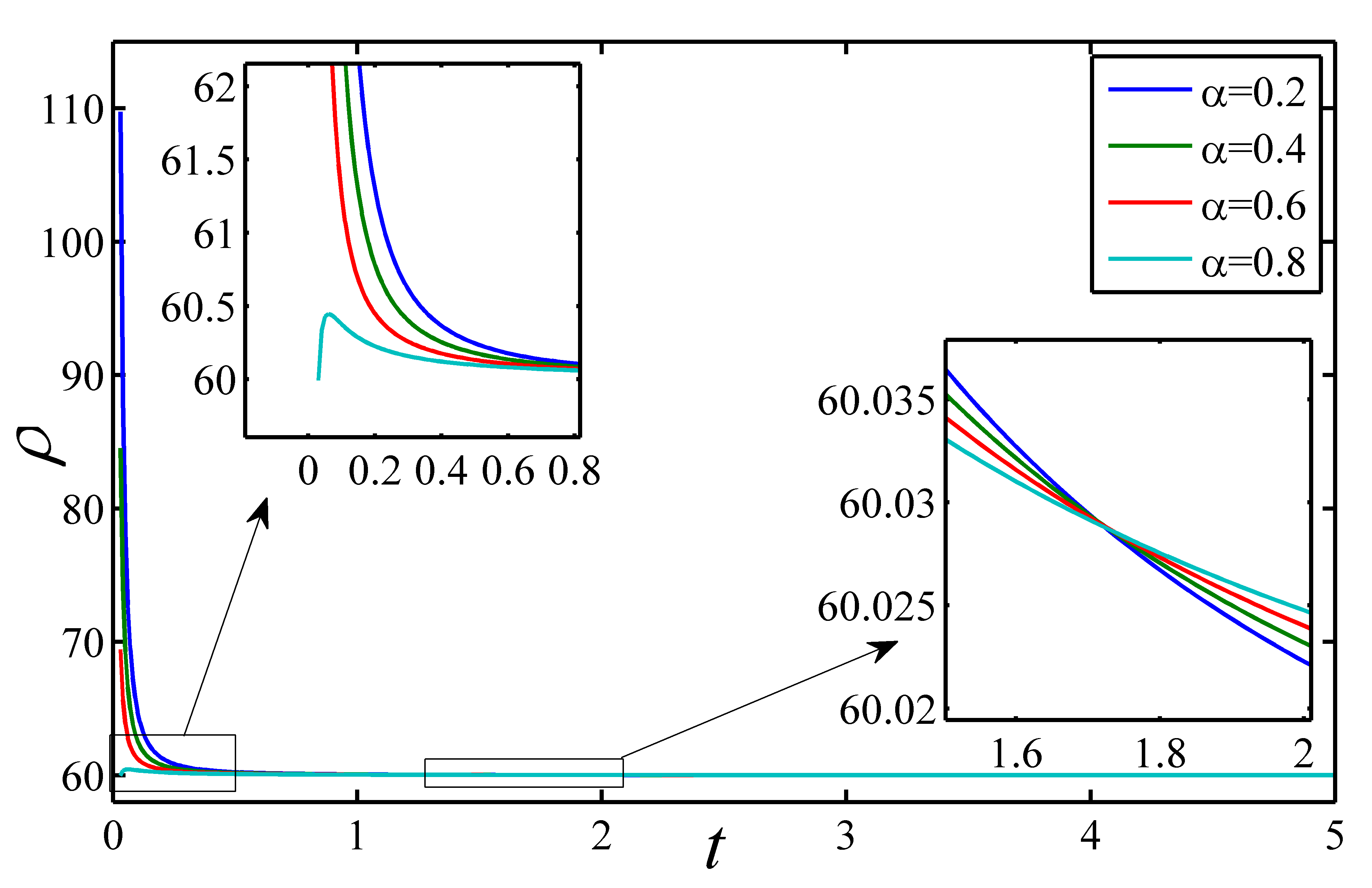}
  \caption{Variation of $\rho$ against $t$ for $\lambda=0.1$, $n=3.5$, $B_\text{c}=60$ and diff. $\alpha$ }\label{ch4fig3}
\endminipage\hfill
\minipage{0.48\textwidth}
\includegraphics[width=68mm]{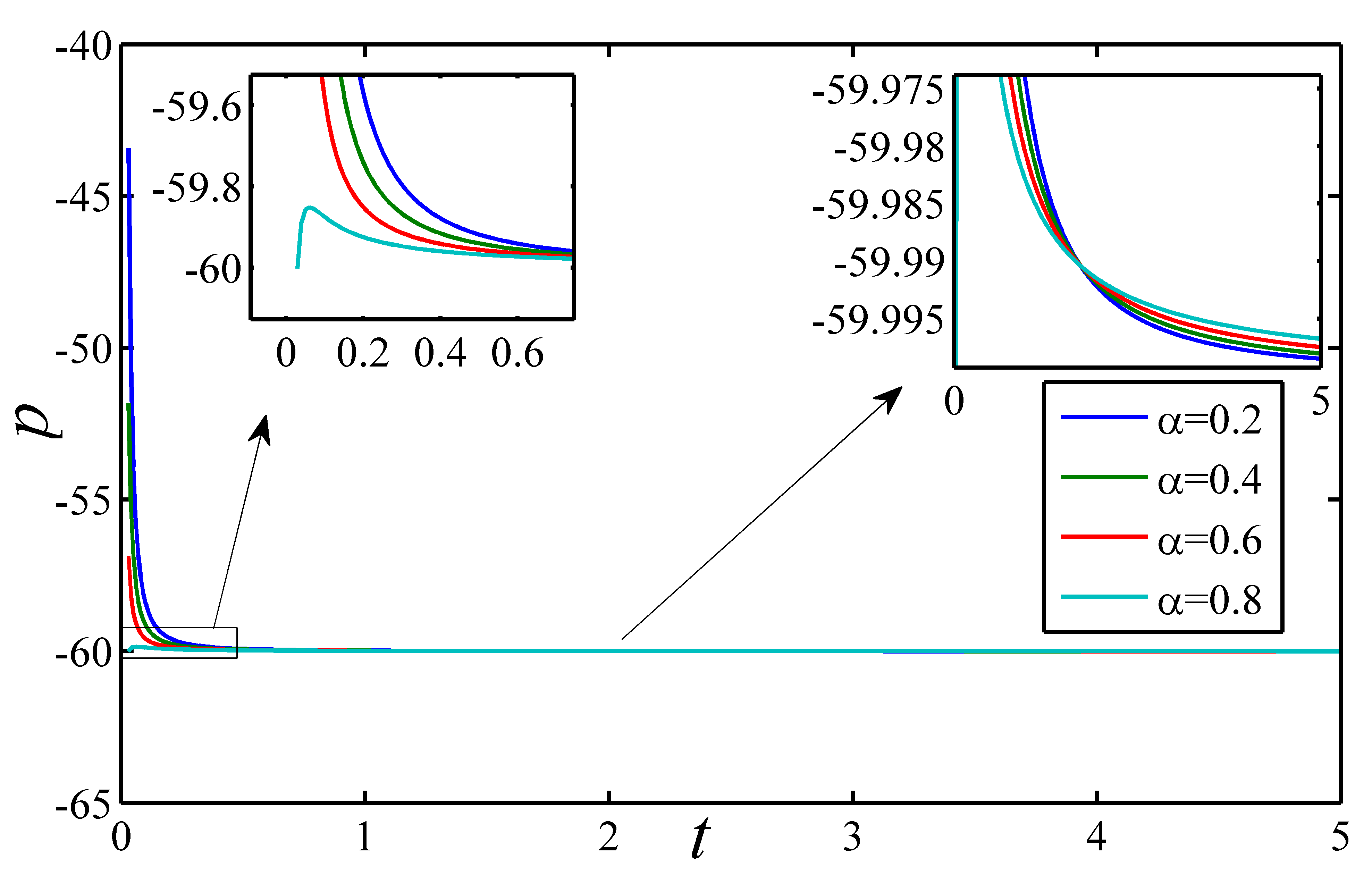}
  \caption{Variation of $p$ against $t$ for $\lambda=0.1$, $n=3.5$, $B_\text{c}=60$ and diff. $\alpha$}\label{ch4fig4}
  \endminipage
\end{figure}

\begin{figure}[H]
\minipage{0.48\textwidth}
  \includegraphics[width=68mm]{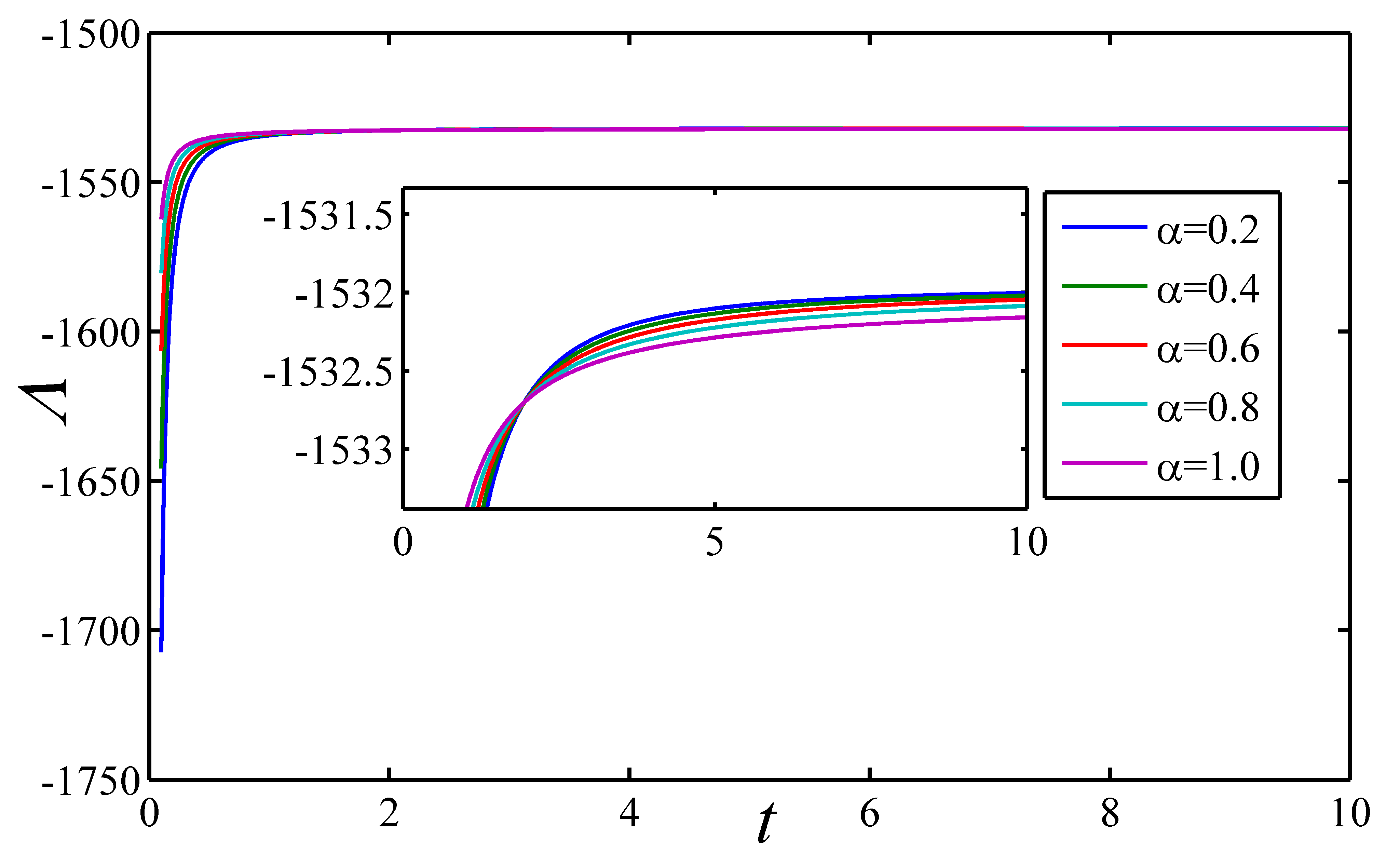}
  \caption{Variation of $\Lambda$ against $t$ for $\lambda=0.1$, $n=0.2$, $B_\text{c}=60$ and diff. $\alpha$ }\label{ch4fig5}
\endminipage\hfill
\minipage{0.48\textwidth}%
  \includegraphics[width=68mm]{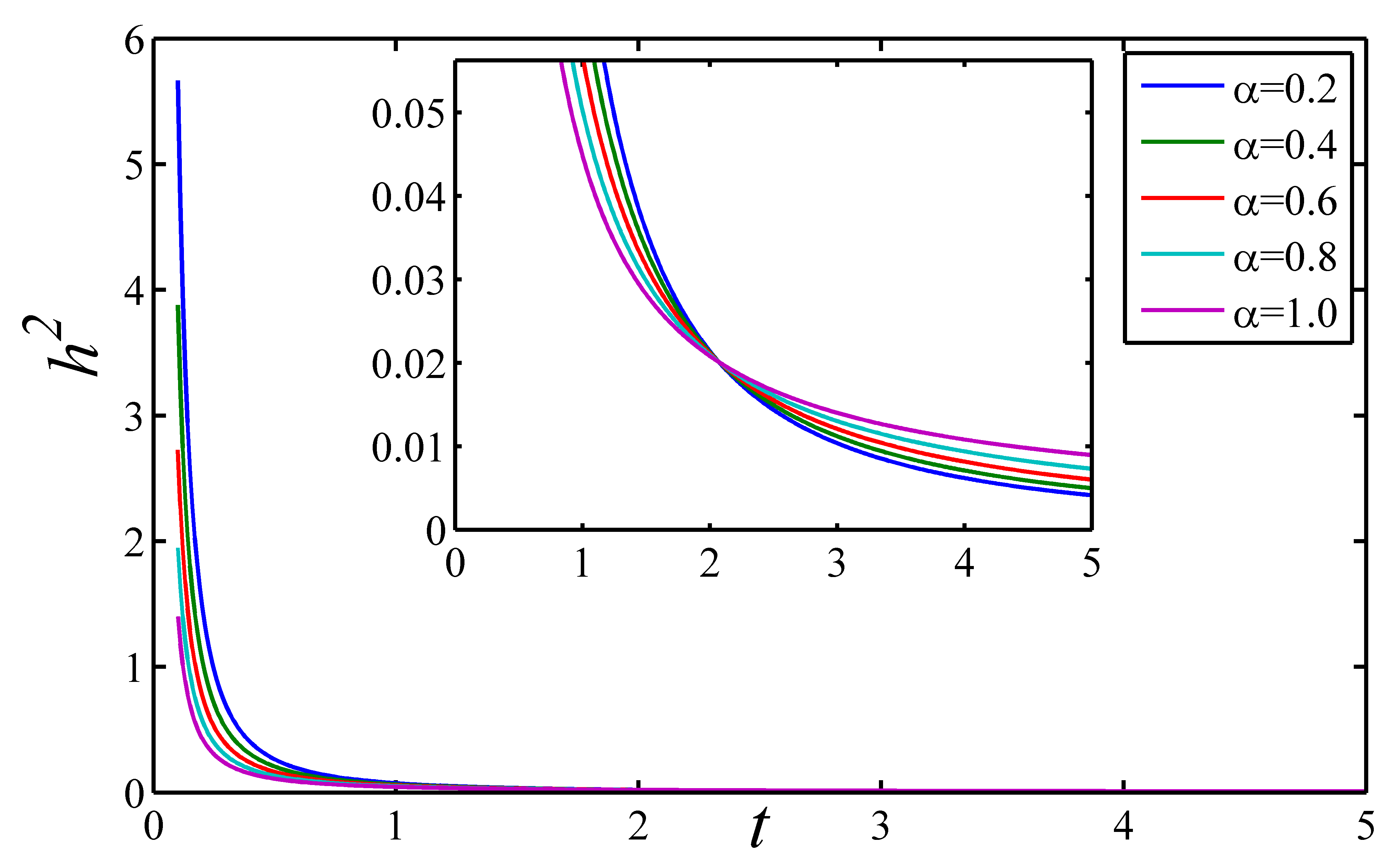}
  \caption{Variation of $h^2$ against $t$ for $\lambda=0.1$, $n=0.2$ and diff. $\alpha$}\label{ch4fig6}
\endminipage
\end{figure}

Here, we have considered the first form of the bilinear DP as \cite{Mishra16}
\begin{equation}\label{ch424}
q=\frac{\alpha(1-t)}{1+t},
\end{equation}
where $\alpha>0$ is a constant. For $0<t<1$, $q>0$ and for $t\geq1$, $q\leq0$ respectively.\\
The HP can be obtained from eqn. (\ref{ch424}) as
\begin{equation}\label{ch425}
H=\frac{1}{(1-\alpha)t+2\alpha log(1+t)}.
\end{equation}
After integrating eqn. (\ref{ch425}) we have
\begin{equation}\label{ch426}
a= a_0 t^{\frac{1}{1+\alpha}}e^{G(t)},
\end{equation}
where
\begin{equation*}
G(t)=\frac{\alpha}{(1+\alpha)^2}t+\frac{-2\alpha+\alpha^2}{6(1+\alpha)^3}t^2+\frac{3\alpha-2\alpha^2+\alpha^3}{18(1+\alpha)^4}t^3
+\frac{-18\alpha+11\alpha^2-14\alpha^3+2\alpha^4}{180(1+\alpha)^5}t^4+O(t^5),
\end{equation*}
and $a_0$ is the constant of integration. The values of $h^2, \rho, p \ \&\ \Lambda$  are obtained as
\begin{equation}\label{ch4h1}
h^2=\frac{3(n-1)[(\alpha-2)-(\alpha+2)t]}{2(1+t)(4\pi+\lambda)(n+2)}H^2
\end{equation}
\begin{equation}\label{ch4rho1}
\rho= \frac{-3}{4(4\pi+\lambda)}\biggl[\frac{9(n-1)}{(n+2)^2}+\frac{3[3+(\alpha-2)n+(3-3n)t]}{(1+t)(n+2)}\biggr]H^2+B_\text{c}
\end{equation}
\begin{equation}\label{ch4p1}
p= \frac{-1}{4(4\pi+\lambda)}\biggl[\frac{9(n-1)}{(n+2)^2}+\frac{3[3+(\alpha-2)n+(3-3n)t]}{(1+t)(n+2)}\biggr]H^2-B_\text{c}
\end{equation}
\begin{multline}\label{ch4Lambda1}
\Lambda=\biggl[\frac{3[(12n\pi+3n\lambda-n^2 \lambda+24\pi+10\lambda)\alpha(1-t)]}{2(1+t)(4\pi+\lambda)(n+2)^2}+\frac{(-26\lambda+18n\lambda+6n^2 \lambda-76\pi)}{2(4\pi+\lambda)(n+2)^2}\biggr]H^2\\
-(8\pi+4\lambda)B_\text{c}
\end{multline}
The expressions of other physical parameters; spatial volume, expansion scalar, shear scalar and anisotropy parameter of this model are given as
\begin{eqnarray}
V=a_0^3 t^{\frac{3}{1+\alpha}}e^{3G(t)}, \label{ch431}\\
\theta=\frac{3}{(1-\alpha)t+2\alpha log(1+t)}, \label{ch432}\\
\sigma^2=\frac{3(n-1)^2}{(n+2)^2}[(1-\alpha)t+2\alpha log(1+t)]^{-2}, \label{ch433}\\
\Delta=\frac{2(n-1)^2}{(n+2)^2}. \label{ch434}
\end{eqnarray}
The quantitative behaviors of $q, H, \rho, p, \Lambda,$ and $h^2$ of this model are depicted in the first set of figures,  Fig. \ref{ch4fig1} to  Fig. \ref{ch4fig6}. In  Fig. \ref{ch4fig1}, phase transition takes place as $q$ is evolving with positive to negative valued for different $\alpha$ against time. Similarly, evolution of HP against time with set of $\alpha$ values is presented in  Fig. \ref{ch4fig2}. The HP posses an initial singularity at $t=0$, later it tends to zero as $t$ tends to infinity.   Fig. \ref{ch4fig3} and  Fig. \ref{ch4fig4} represent the profile of energy density and pressure against time for different $\alpha$ respectively. From eqns. (\ref{ch4rho1}) and (\ref{ch4p1}), one can observe that, $\rho \rightarrow B_\text{c}$ and $p \rightarrow -B_\text{c}$ when $t\rightarrow \infty$. Here, we would like to point out that, for different $n$ the approach of $\rho$ toward $B_\text{c}$ is different. That means for $n\in(0.9,3.5)$, $\rho\rightarrow B_\text{c}$ from the left of $B_\text{c}$ and for $n\geq 3.5$, $\rho\rightarrow B_\text{c}$  from the right of $B_\text{c}$.  As a representative case, we have chosen $n=3.5$ and different $\alpha$ for energy density profile, which is presented in  Fig. \ref{ch4fig3}. Now coming to the pressure profile in  Fig. \ref{ch4fig4}, it is purely negative valued function of time for $n\in(0,16]$. In the interval $(0,2]$ and $[3.5,16]$ of $n$, $p\rightarrow -B_\text{c}$  from the left and right of $B_\text{c}$ respectively. Then the variation of $\Lambda$ against time is presented in  Fig.  \ref{ch4fig5}. From eqn. (\ref{ch4Lambda1}), it is clear that $\Lambda\rightarrow -(8\pi+4\lambda)B_\text{c}$ when $t\rightarrow \infty$. The approach of $\Lambda$ toward $-(8\pi+4\lambda)B_\text{c}$ is differs according to the values of $n$ and $\alpha$ respectively. That means $\Lambda\rightarrow -(8\pi+4\lambda)B_\text{c}$ from either side of the value of $-(8\pi+4\lambda)B_\text{c}$. Here, we have noticed that, $\Lambda$ is a negative quantity. As a representative case we have chosen $n=0.2$ and different $\alpha$ for the profile of $\Lambda$. In this way the magnetic flux $h^2$ is represented in  Fig.  \ref{ch4fig6}. It can be observed that $h^2$ is a positive value for $n\in (0,1)$ and negative value for $n>1$ for given values of $\alpha$. Further, at initial time $t=0$, the spatial volume $V$ is zero and it gradually increases exponentially with time. It is interesting to note that, for $n\neq 1$, the model is anisotropic for late time and not free from shear, whereas for $n=1$, it is isotropic and shear free.
\subsection{Model II: $q(t)=-\frac{\alpha t}{1+t}$}\label{ch4model2}
\begin{figure}[H]
\minipage{0.48\textwidth}
  \includegraphics[width=68mm]{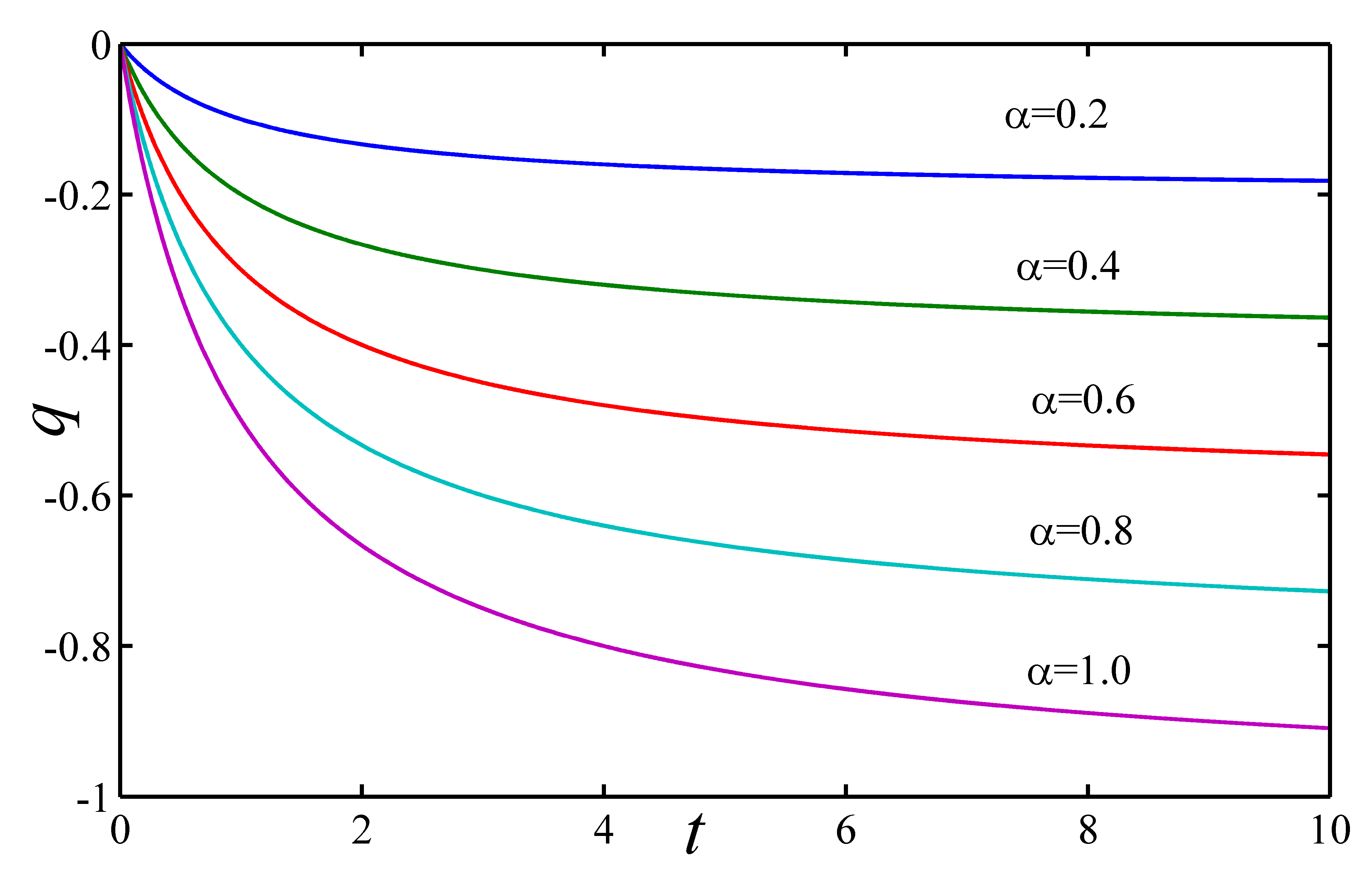}
  \caption{Variation of DP against $t$ for diff. $\alpha$}\label{ch4fig7}
\endminipage\hfill
\minipage{0.48\textwidth}
  \includegraphics[width=68mm]{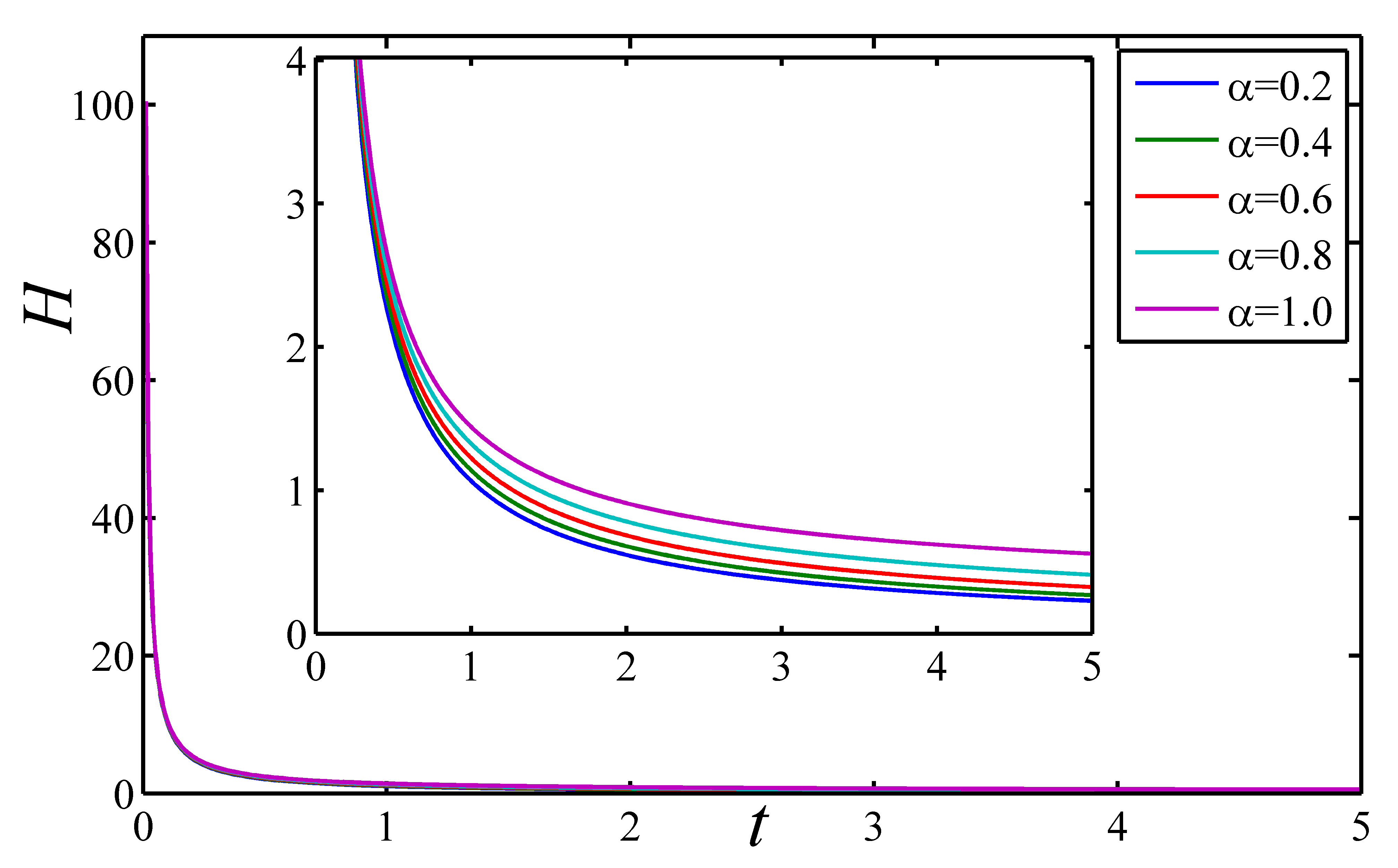}
  \caption{Variation of $H$ against $t$ for diff. $\alpha$}\label{ch4fig8}
\endminipage
\end{figure}

\begin{figure}[H]
\minipage{0.48\textwidth}%
  \includegraphics[width=68mm]{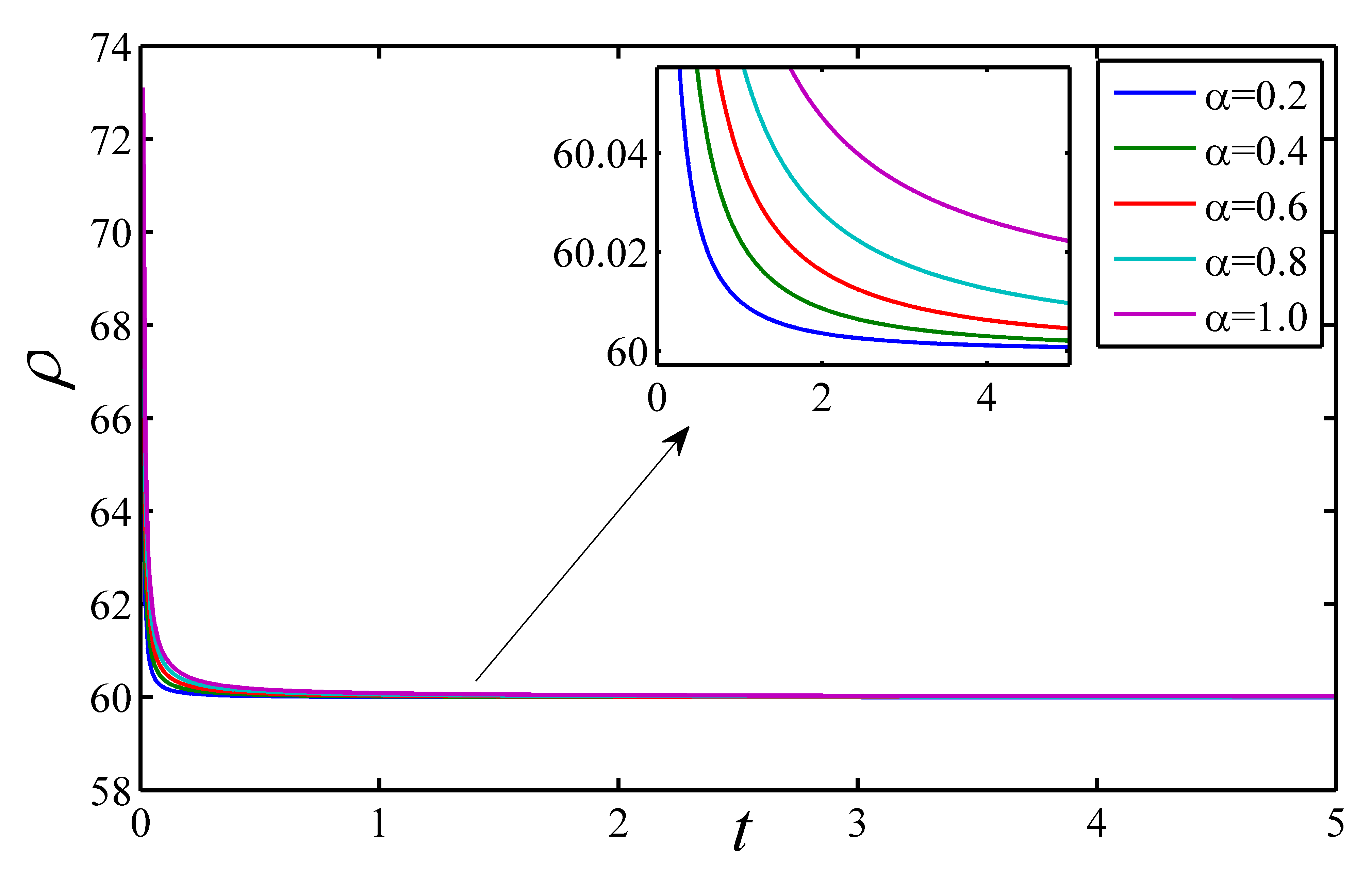}
  \caption{Variation of $\rho$ against $t$ for $\lambda=0.1$, $n=1.83$, $B_\text{c}=60$ and diff. $\alpha$}\label{ch4fig9}
\endminipage\hfill
\minipage{0.48\textwidth}
  \includegraphics[width=68mm]{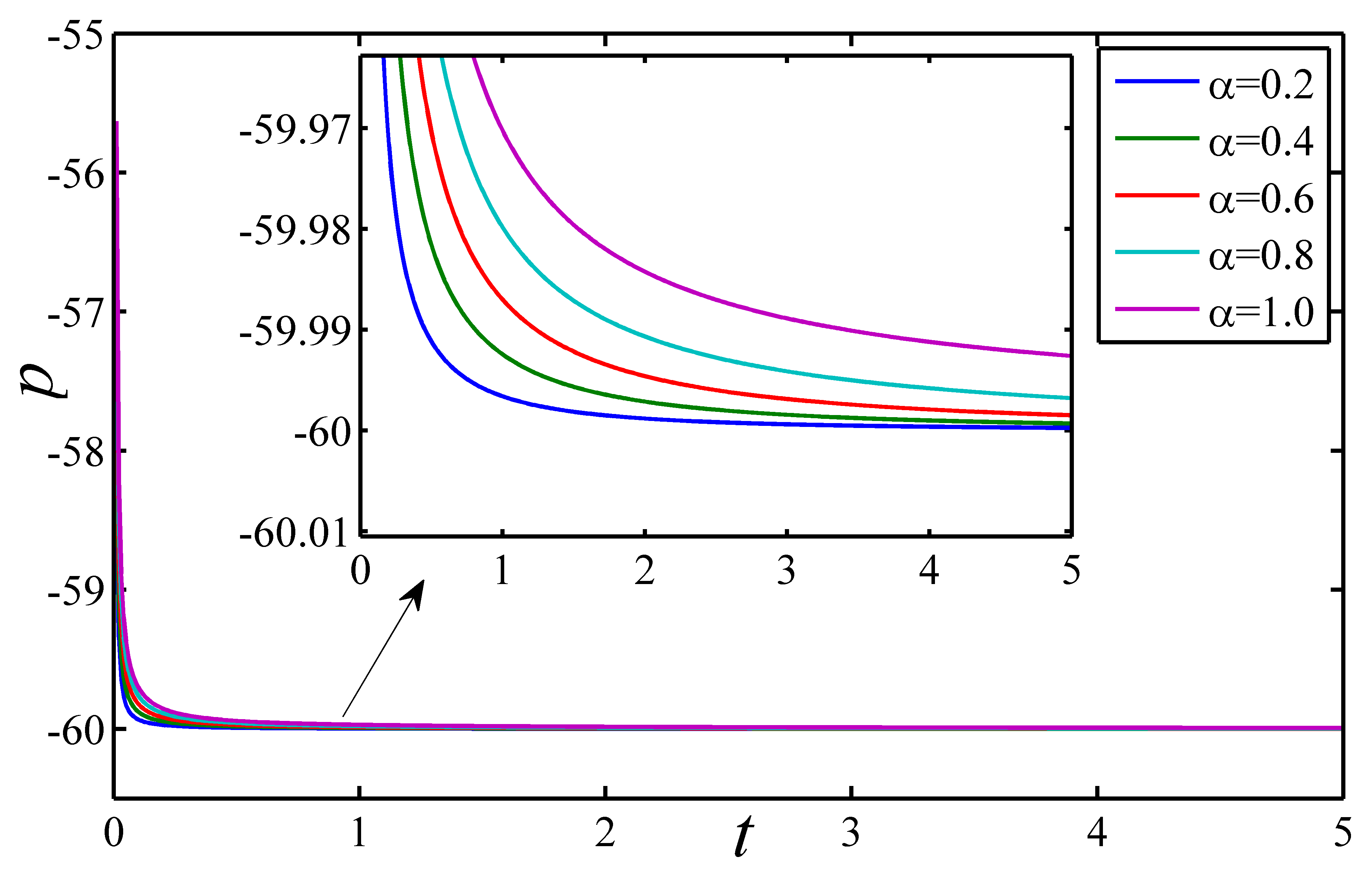}
  \caption{Variation of $p$ against $t$ for $\lambda=0.1$, $n=1.83$, $B_\text{c}=60$ and diff. $\alpha$ }\label{ch4fig10}
\endminipage
\end{figure}

\begin{figure}[H]
\minipage{0.48\textwidth}
  \includegraphics[width=68mm]{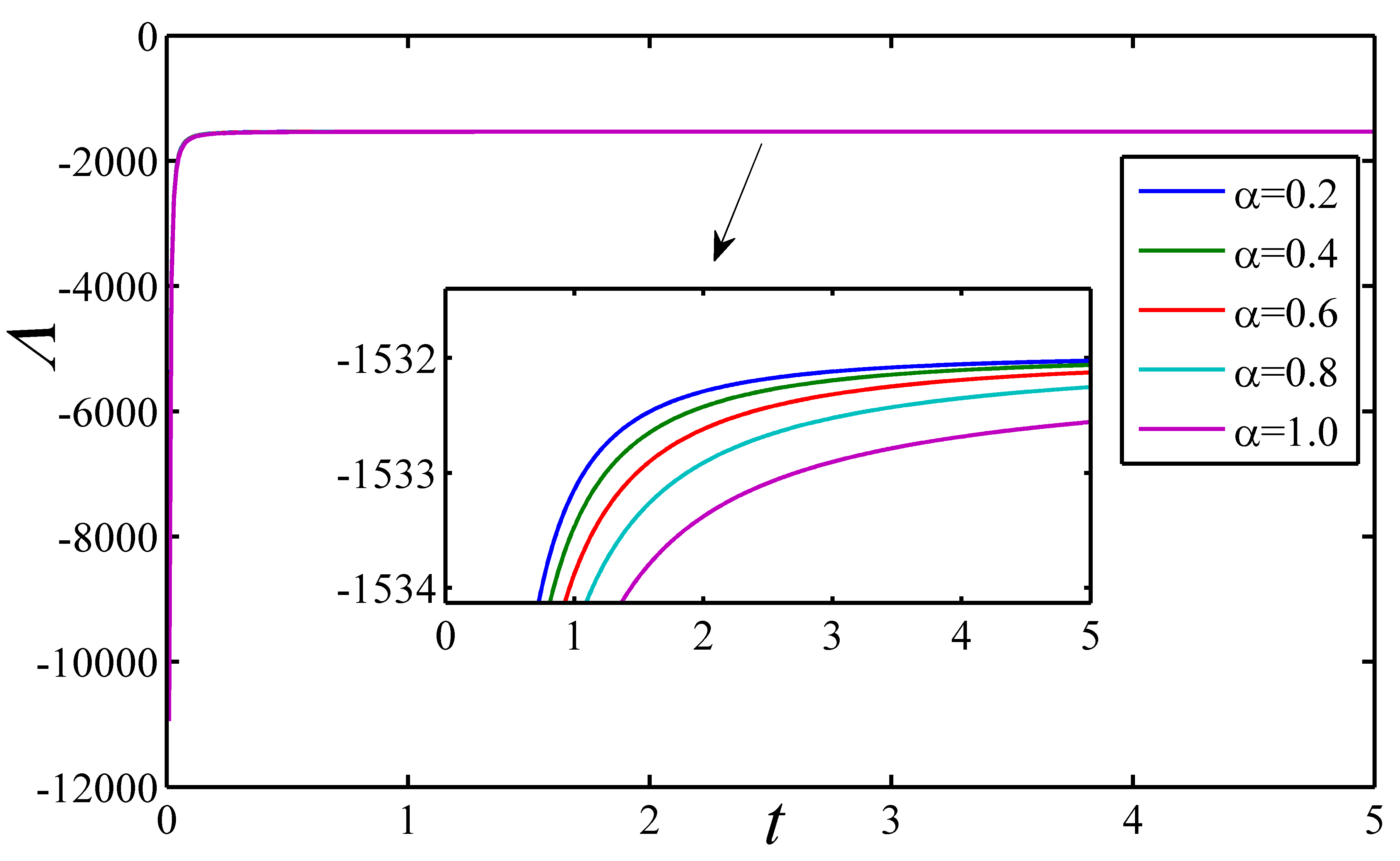}
  \caption{Variation of $\Lambda$ against $t$ for $\lambda=0.1$, $n=1.83$, $B_\text{c}=60$ and diff.$\alpha$  }\label{ch4fig11}
\endminipage\hfill
\minipage{0.48\textwidth}%
  \includegraphics[width=68mm]{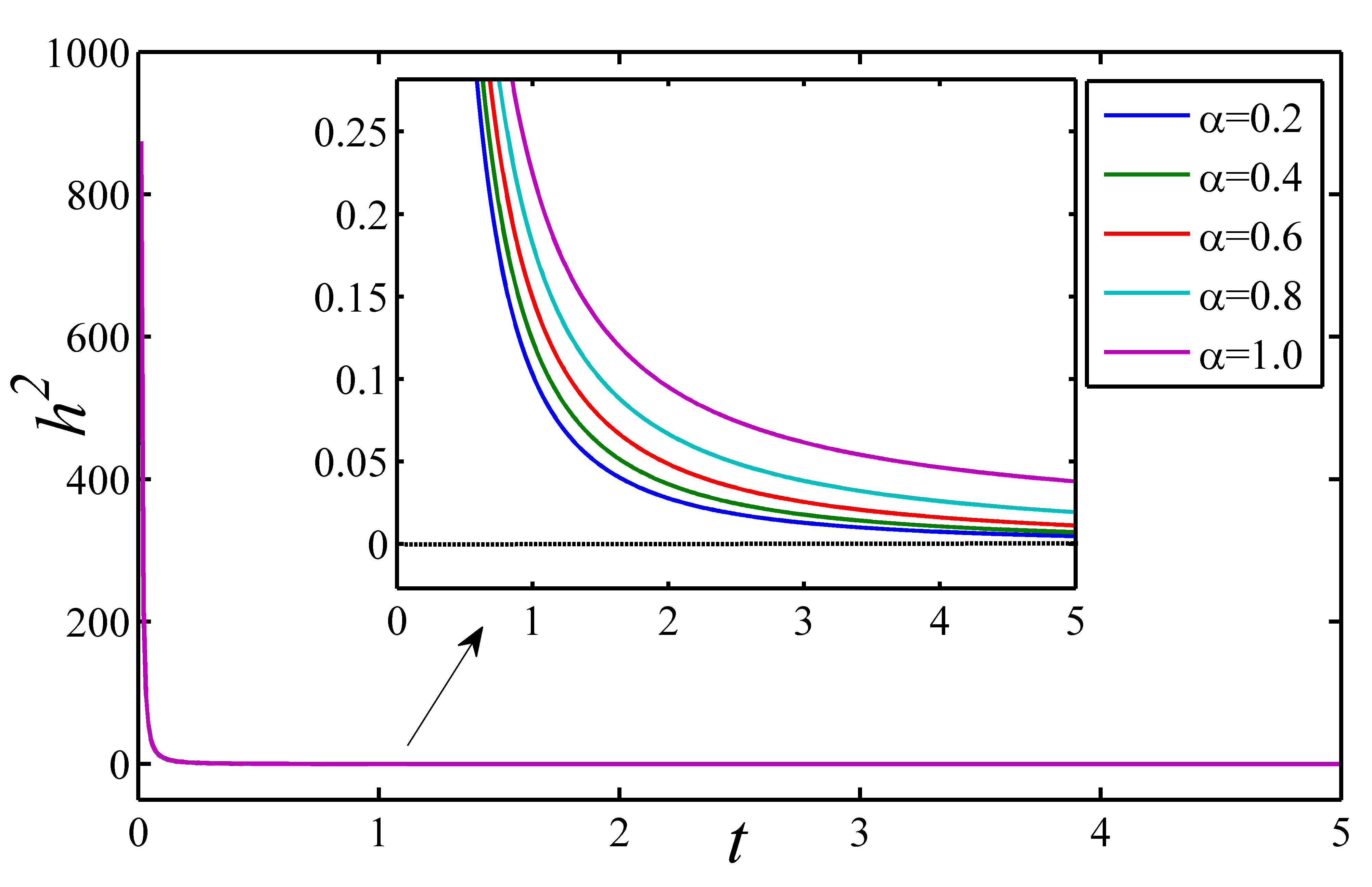}
  \caption{Variation of $h^2$ against $t$ for $\lambda=0.1$, $n=0.2$, $B_\text{c}=60$ and diff. $\alpha$ }\label{ch4fig12}
\endminipage
\end{figure}

In this case we have considered the second form of the bilinear DP \cite{Mishra16}
\begin{equation}\label{ch435}
q(t)=-\frac{\alpha t}{1+t},
\end{equation}
and it yields the HP as
\begin{equation}\label{ch436}
H=\frac{1}{(1-\alpha)t+\alpha log(1+t)}.
\end{equation}
Integrating the above eqn. (\ref{ch436}), we have
\begin{equation}\label{ch437}
a= a_0 t e^{F(t)},
\end{equation}
where
\begin{equation*}
F(t)=\frac{\alpha}{2}t+\frac{-4\alpha+3\alpha^2}{24}t^2+\frac{6\alpha-8\alpha^2+3\alpha^3}{72}t^3
+\frac{-144\alpha+260\alpha^2-180\alpha^3+45\alpha^4}{2880}t^4+O(t^5).
\end{equation*}
The values of $h^2$, $\rho$, $p$, and $\Lambda$  are obtained as
\begin{equation}\label{ch4h2}
h^2=\frac{3(n-1)[-(\alpha+2)t-2]}{2(1+t)(4\pi+\lambda)(n+2)}H^2,
\end{equation}
\begin{equation}\label{ch4rho2}
\rho= \frac{-3}{4(4\pi+\lambda)}\biggl[\frac{9(n-1)}{(n+2)^2}+\frac{3[3-2n+(3-2n-\alpha n)t]}{(1+t)(n+2)}\biggr]H^2+B_\text{c},
\end{equation}
\begin{equation}\label{ch4p2}
p=\frac{-1}{4(4\pi+\lambda)}\biggl[\frac{9(n-1)}{(n+2)^2}+\frac{3[3-2n+(3-2n-\alpha n)t]}{(1+t)(n+2)}\biggr]H^2-B_\text{c},
\end{equation}
\begin{multline}\label{ch4Lambda2}
\Lambda=\biggl[\frac{3[(12n\pi+3n\lambda-n^2 \lambda+24\pi+10\lambda)(-\alpha t)]}{2(1+t)(4\pi+\lambda)(n+2)^2}\\
+\frac{(-26\lambda+18n\lambda+6n^2 \lambda-76\pi)}{2(4\pi+\lambda)(n+2)^2}\biggr]H^2-(8\pi+4\lambda)B_\text{c}.
\end{multline}
The other physical parameters of this model are given as
\begin{eqnarray}
V=a_0^3 t^3 e^{3F(t)}\\\label{42}
\theta=\frac{3}{(1-\alpha)t+\alpha log(1+t)},\label{43}\\
\sigma^2=\frac{3(n-1)^2}{(n+2)^2}[(1-\alpha)t+\alpha log(1+t)]^{-2},\label{44}\\
\Delta=\frac{2(n-1)^2}{(n+2)^2}.\label{45}
\end{eqnarray}
 Fig. \ref{ch4fig7} and  Fig. \ref{ch4fig8} represent the variation of DP and HP with respect to time for different $\alpha$, where $q<0$ for $\alpha>0$ and yield an accelerating universe in this model. Moreover, specifically for $0<\alpha\leq 1$ \& $t>0$ implies $q\in (-1,0)$ and $\alpha>1$ \& $t>0$ implies $q\in (0,-2)$, which indicates that our Universe is accelerating with exponential expansion (see  Fig. \ref{ch4fig7}) and super exponential expansion respectively. Also, HP is decreasing with the time and approaching to zero for $t \rightarrow \infty$. The positivity of energy density $\forall \alpha \in (0,1)$ restricts  $n\geq 1.75$. From  Fig. \ref{ch4fig9}, it can be observed that the energy density $\rho\rightarrow B_\text{c}$ from the left of $B_\text{c}$ in the interval $n\in[1.75,1.82]$ and from the right of $B_\text{c}$ for $n>1.82$. At the same time in case of pressure $p\rightarrow -B_\text{c}$ from the left of $B_\text{c}$ for $n\in[1.75,1.82]$ and $p\rightarrow -B_\text{c}$ from the right of $B_\text{c}$ for $n>1.82$ (see  Fig. \ref{ch4fig10}). Here, pressure is a negative valued function of time. The profile of $\Lambda$ is depicted in  Fig.  \ref{ch4fig11}. Here, we have seen that, for provided value of $\alpha$ and $\forall n>0$, the $\Lambda$ is negative and $\Lambda\rightarrow -(8\pi+4\lambda)B_\text{c}$ (see  Fig.  \ref{ch4fig11} and eqn. (\ref{ch4Lambda2})). The other physical quantities like volume, shear scalar, expansion scalar, magnetic flux have the similar qualitative behavior as that of first model \ref{ch4model1}.
\subsection{Model III: $q(t)=-1+\frac{\beta}{1+a^{\beta}}$}\label{ch4model3}
\begin{figure}[H]
\minipage{0.32\textwidth}
  \includegraphics[width=55mm]{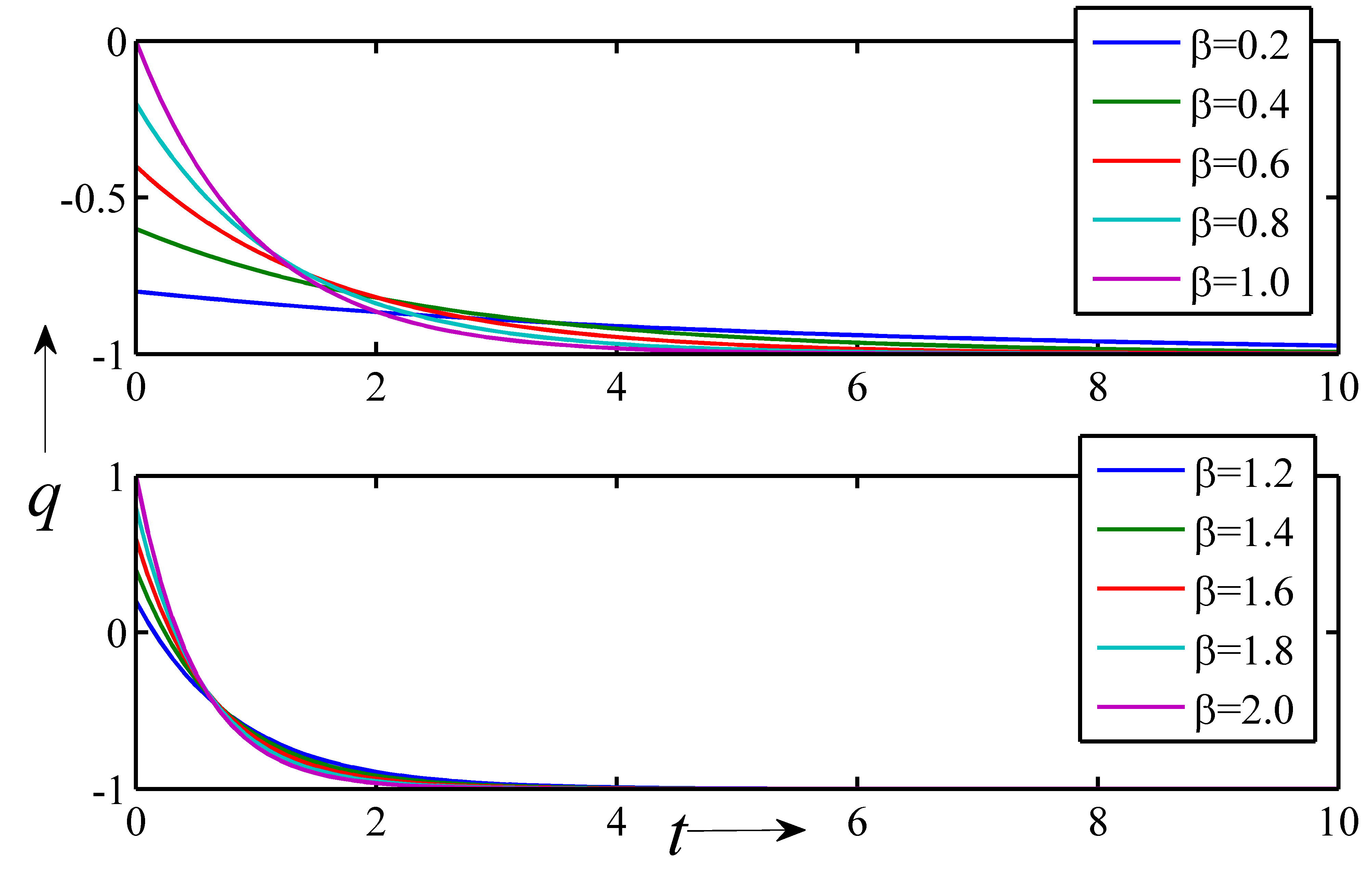}
  \caption{Variation of $q$ against $t$ for diff. $\beta$ }\label{ch4fig13}
\endminipage\hfill
\minipage{0.32\textwidth}
  \includegraphics[width=55mm]{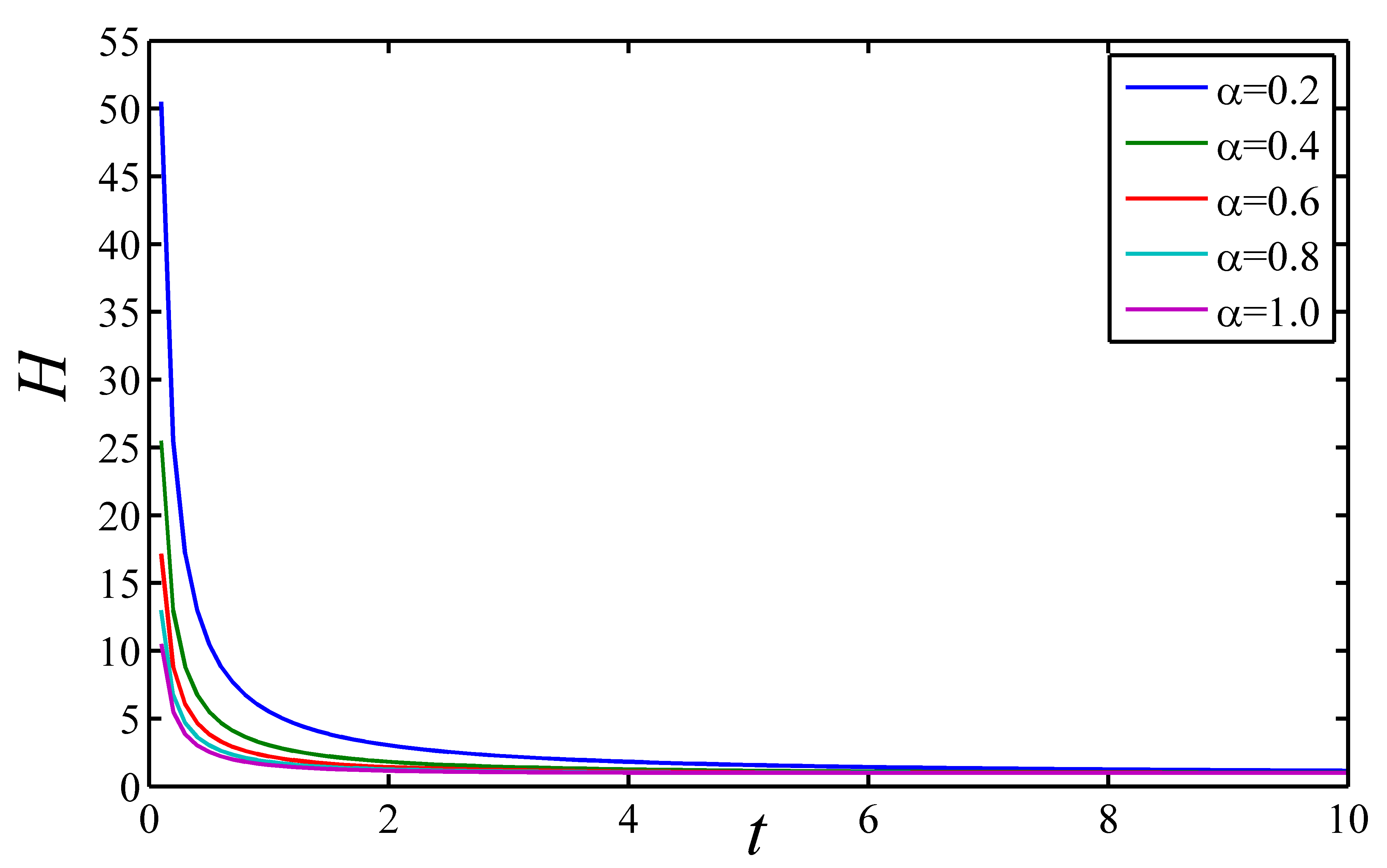}
  \caption{Variation of $H$ against $t$ for diff. $\beta$}\label{ch4fig14}
\endminipage
\hfill
\minipage{0.32\textwidth}%
  \includegraphics[width=53mm]{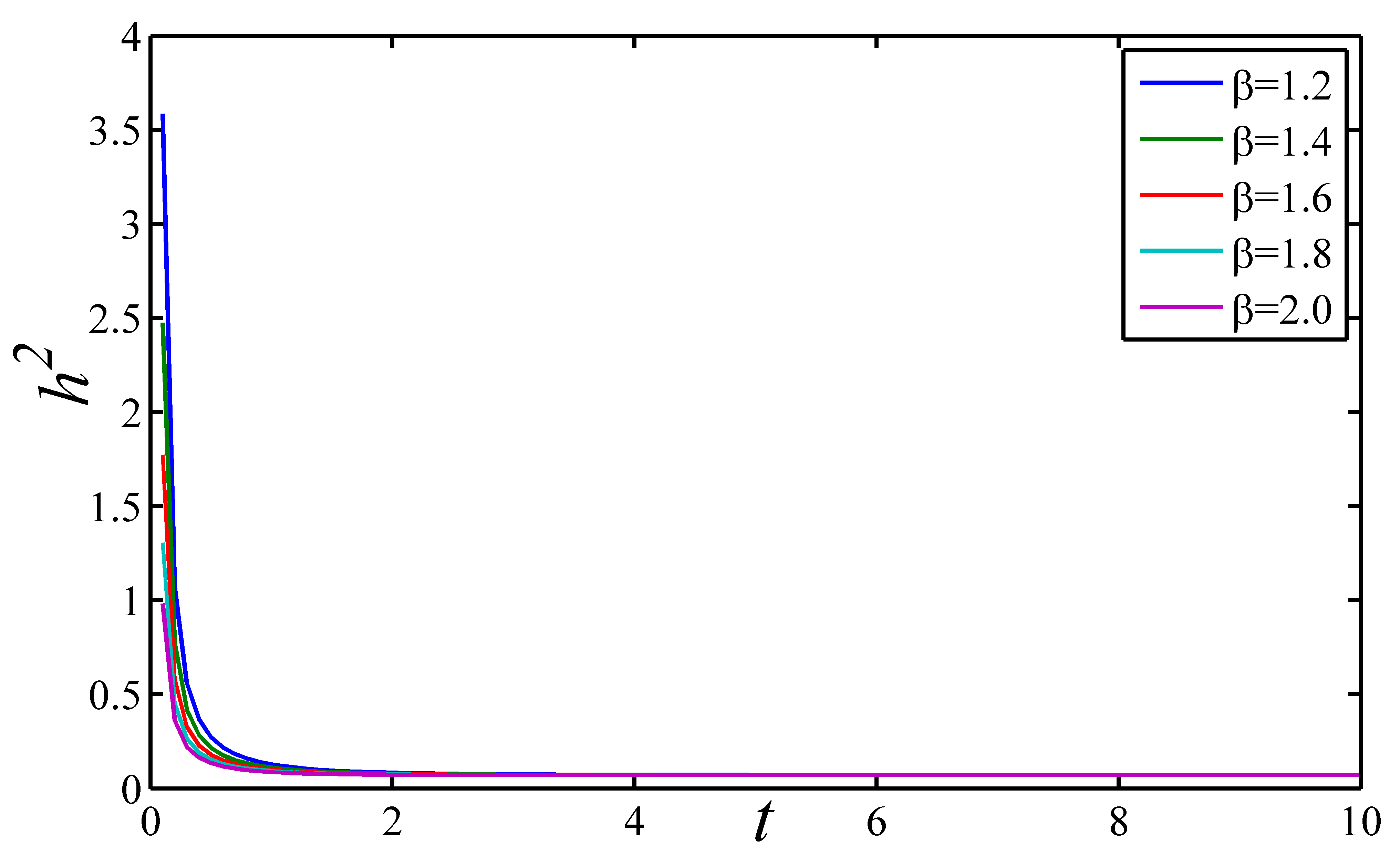}
  \caption{Variation of $h^2$ against $t$ for $\lambda=0.1$, $n=0.5$, $B_\text{c}=60$ and diff. $\beta$}\label{ch4fig15}
\endminipage
\end{figure}

\begin{figure}[H]
\minipage{0.32\textwidth}
  \includegraphics[width=53mm]{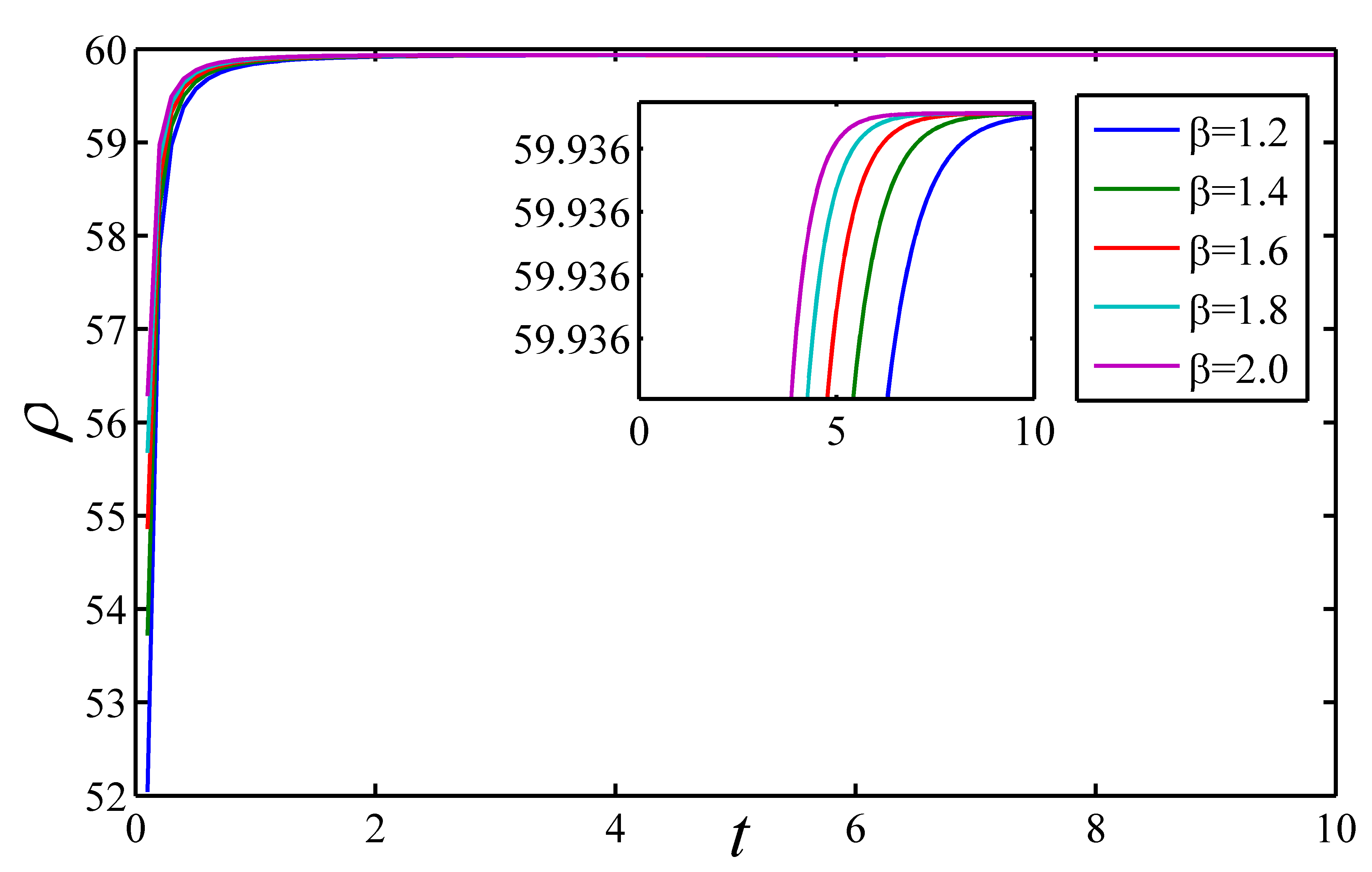}
\endminipage\hfill
\minipage{0.32\textwidth}
  \includegraphics[width=53mm]{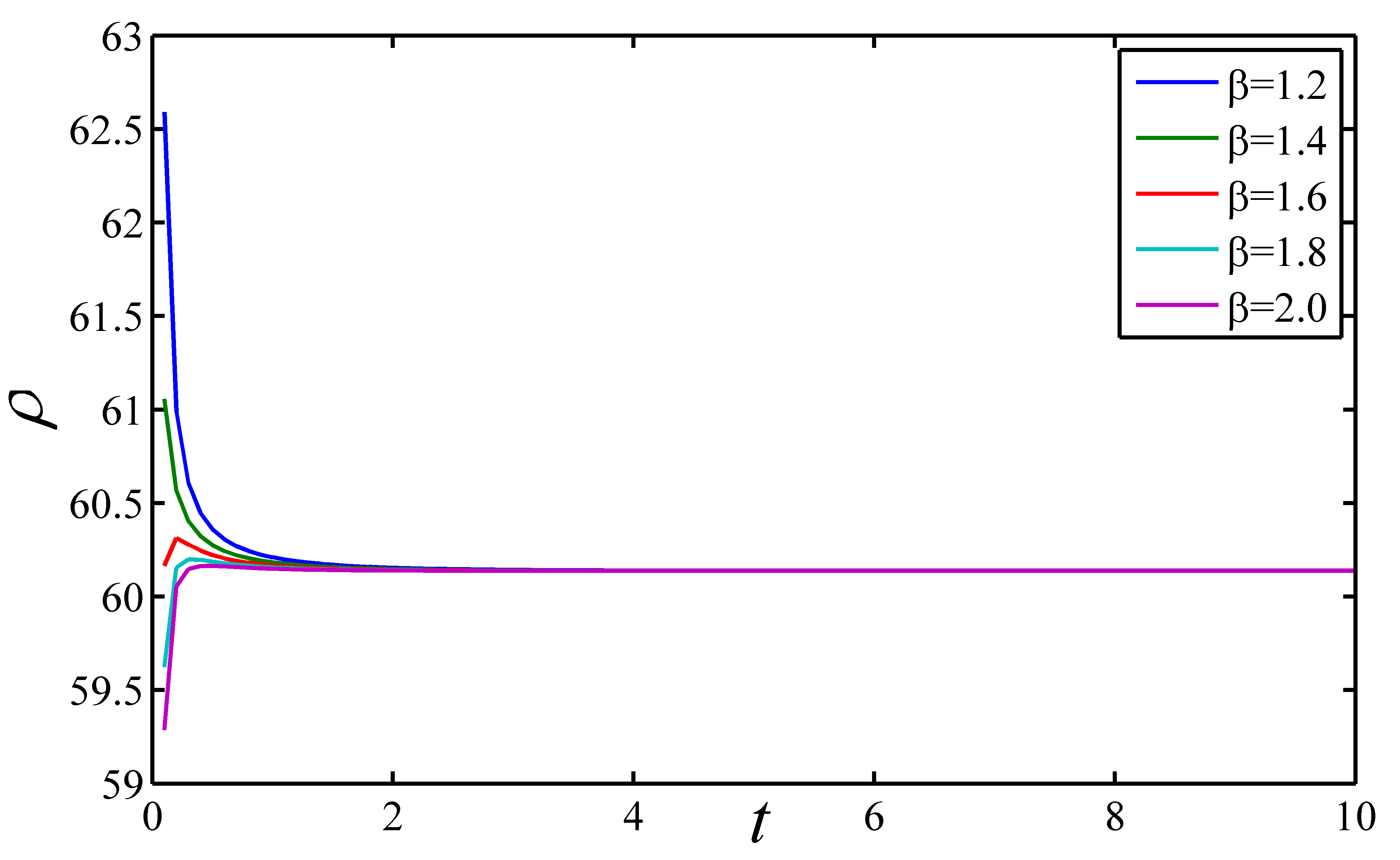}
\endminipage\hfill
\minipage{0.32\textwidth}%
  \includegraphics[width=53mm]{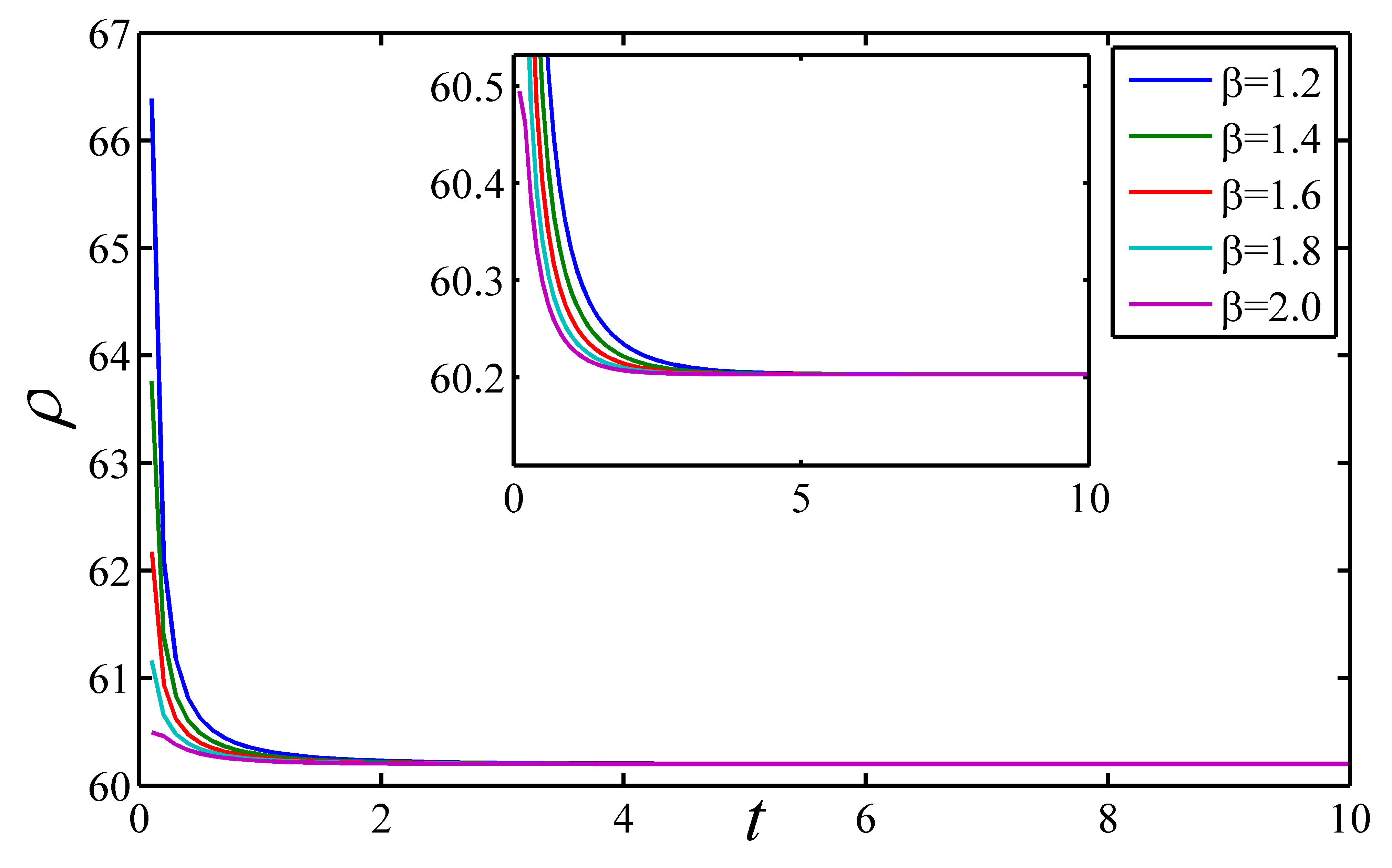}
\endminipage
\caption{Variation of $\rho$ against $t$ for $\lambda=0.1$, $B_\text{c}=60$ and diff. $\beta$ with $n=0.5$, $n=2.5$ and $n=3.6$ respectively}\label{ch4fig16}
\end{figure}

\begin{figure}[H]
\minipage{0.32\textwidth}
  \includegraphics[width=53mm]{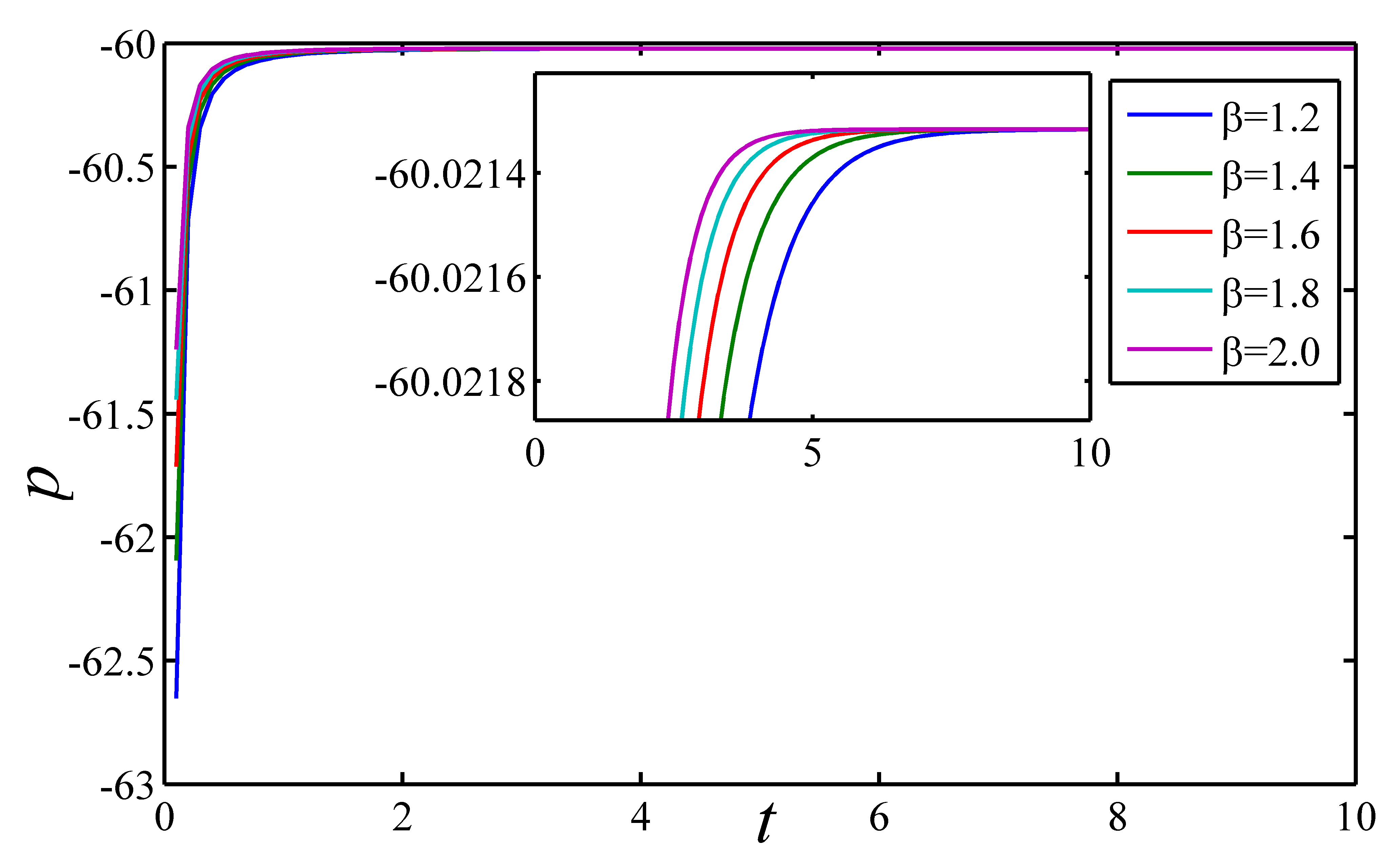}
\endminipage\hfill
\minipage{0.32\textwidth}
  \includegraphics[width=53mm]{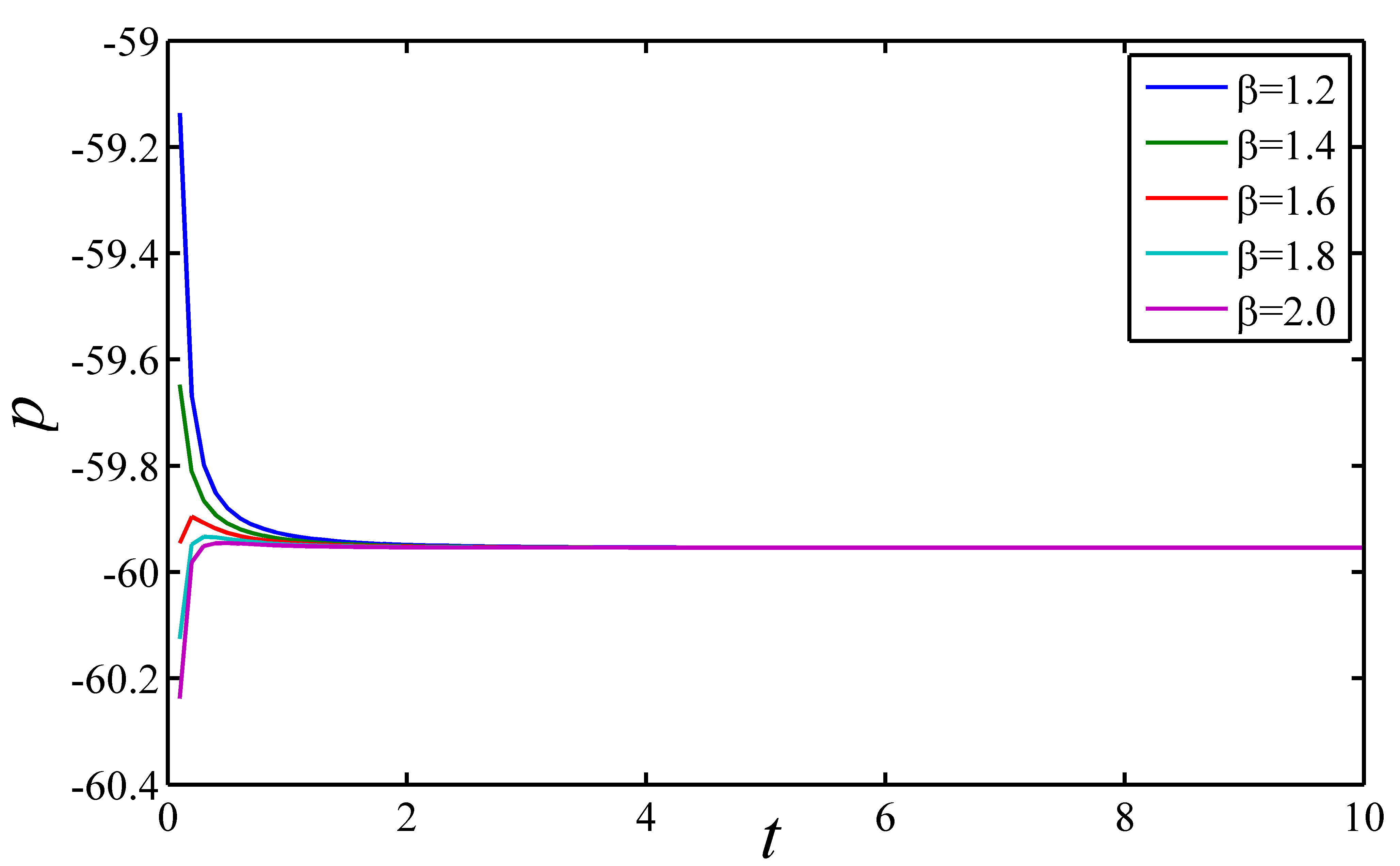}
\endminipage\hfill
\minipage{0.32\textwidth}%
  \includegraphics[width=53mm]{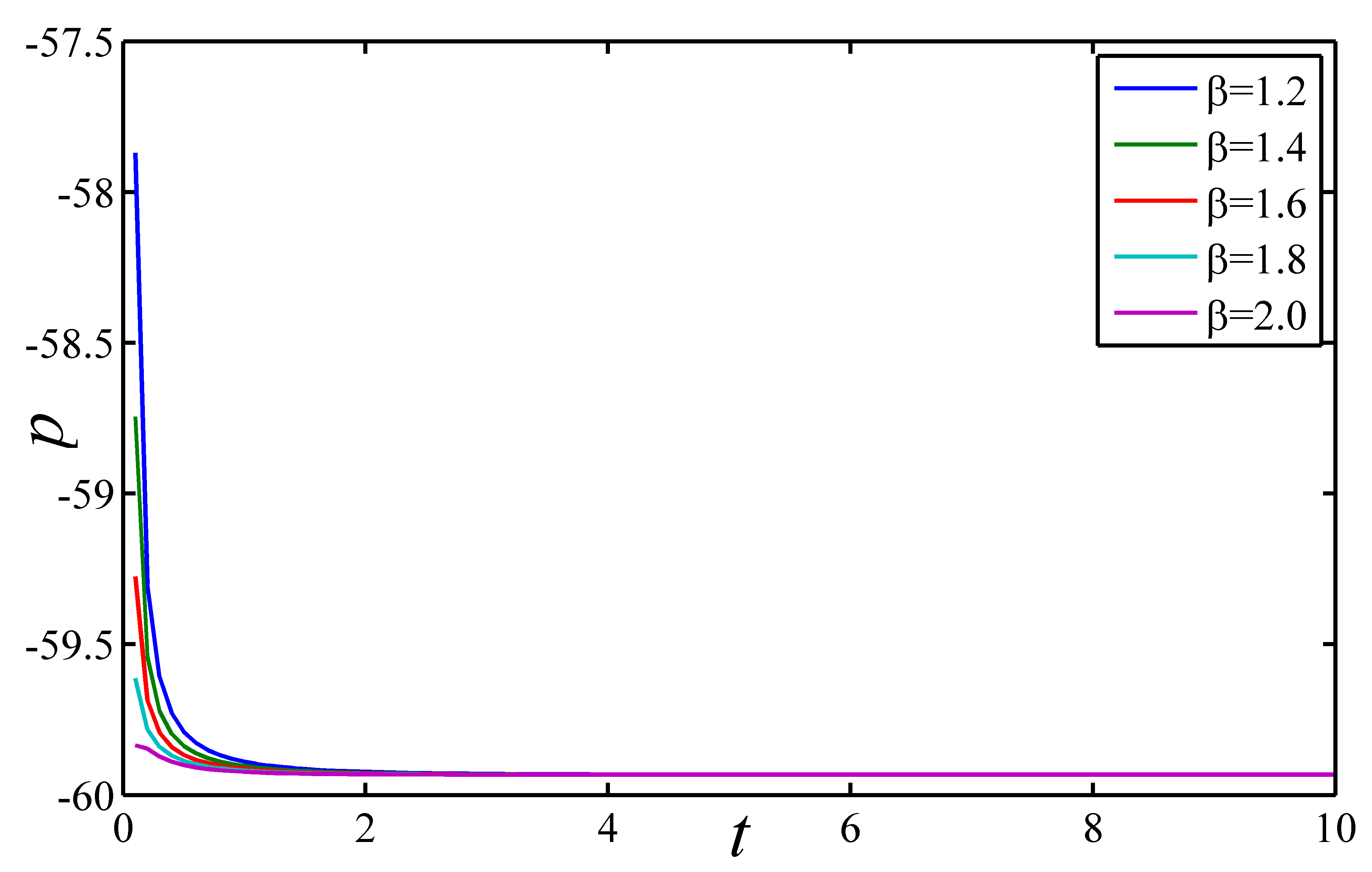}
\endminipage
\caption{Variation of $p$ against $t$ for $\lambda=0.1$, $B_\text{c}=60$ and diff. $\beta$ with $n=0.5$, $n=2.5$ and $n=3.6$ respectively}\label{ch4fig17}
\end{figure}

\begin{figure}[H]
\minipage{0.32\textwidth}
  \includegraphics[width=53mm]{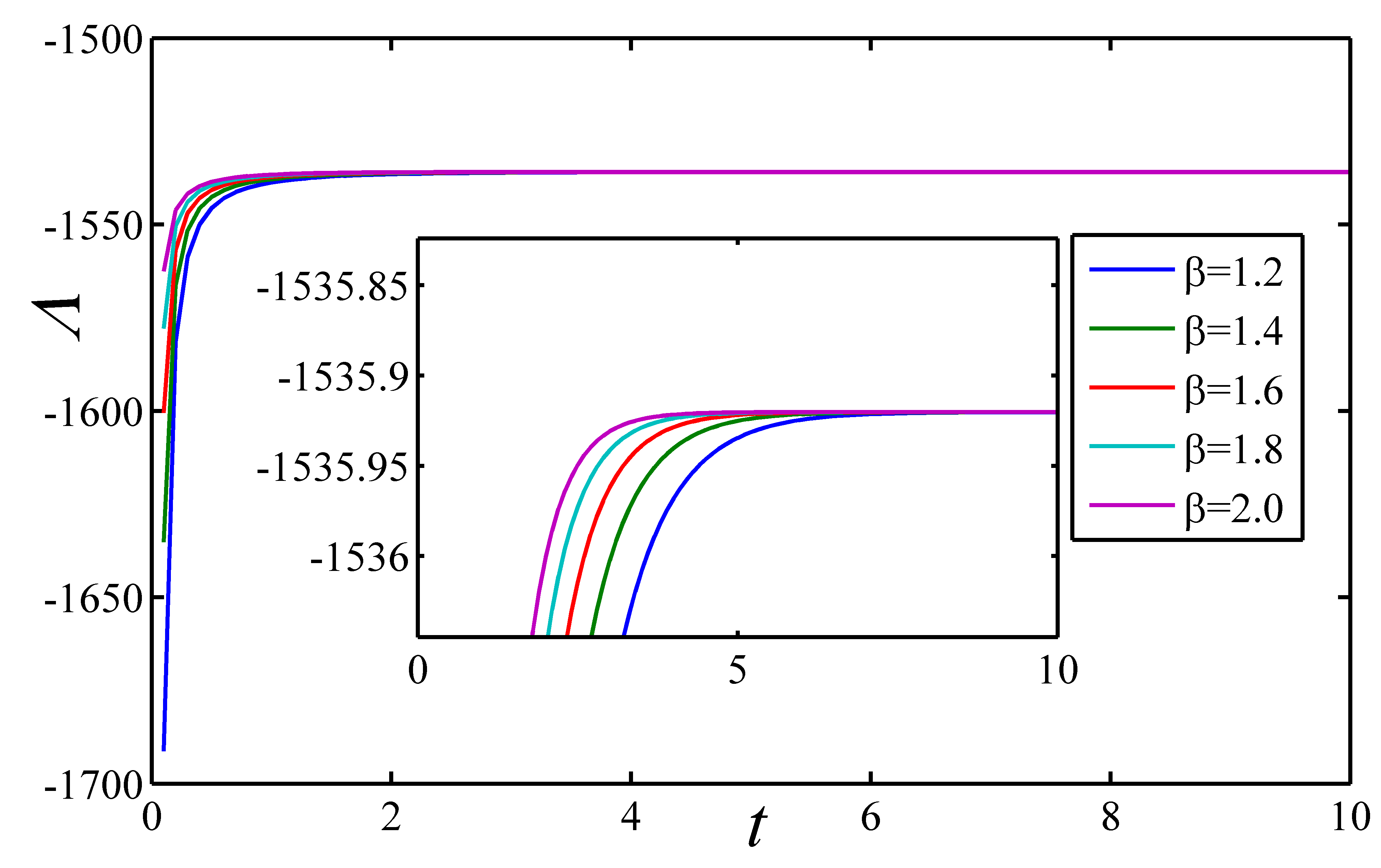}
\endminipage\hfill
\minipage{0.32\textwidth}
  \includegraphics[width=53mm]{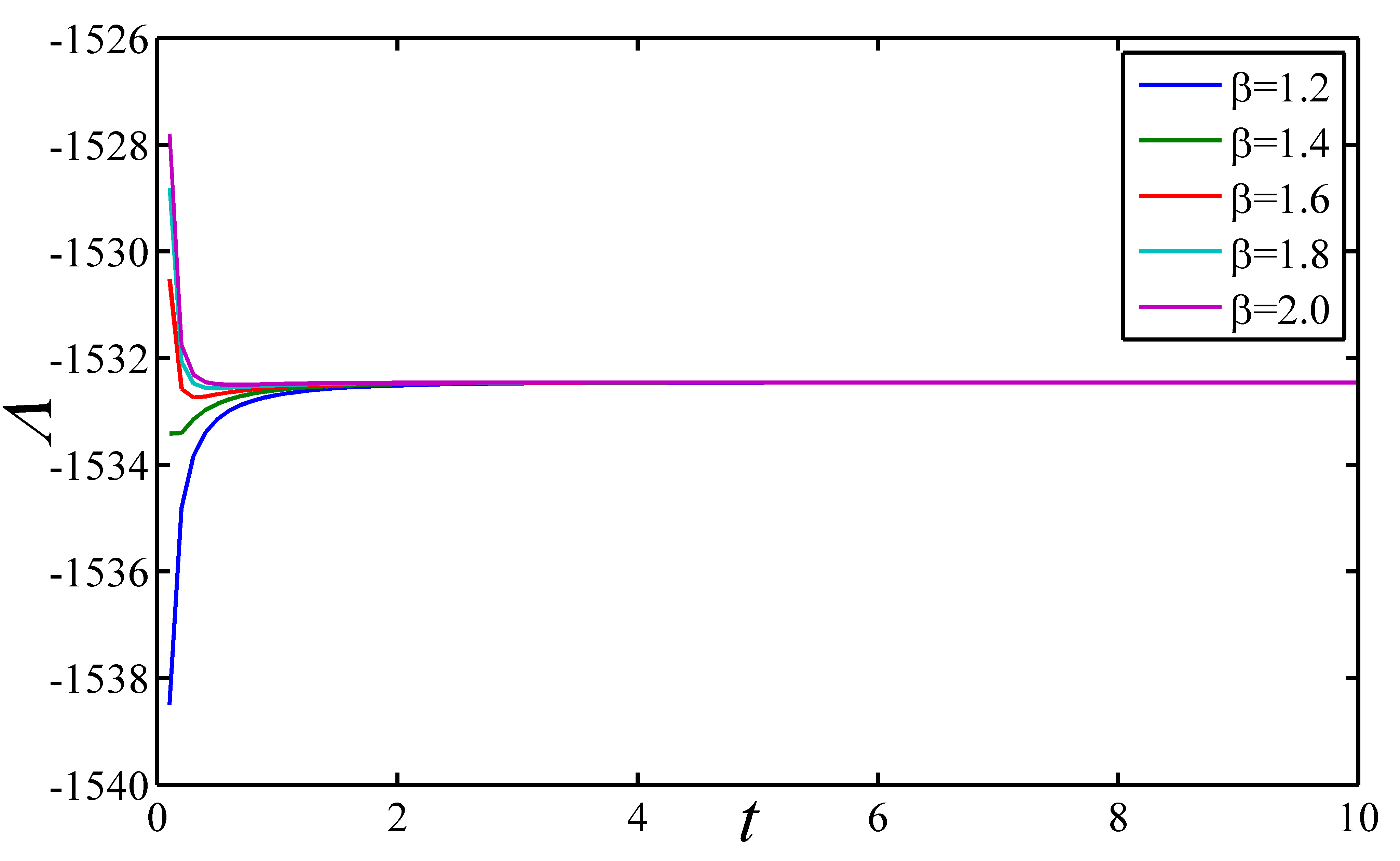}
\endminipage\hfill
\minipage{0.32\textwidth}%
  \includegraphics[width=53mm]{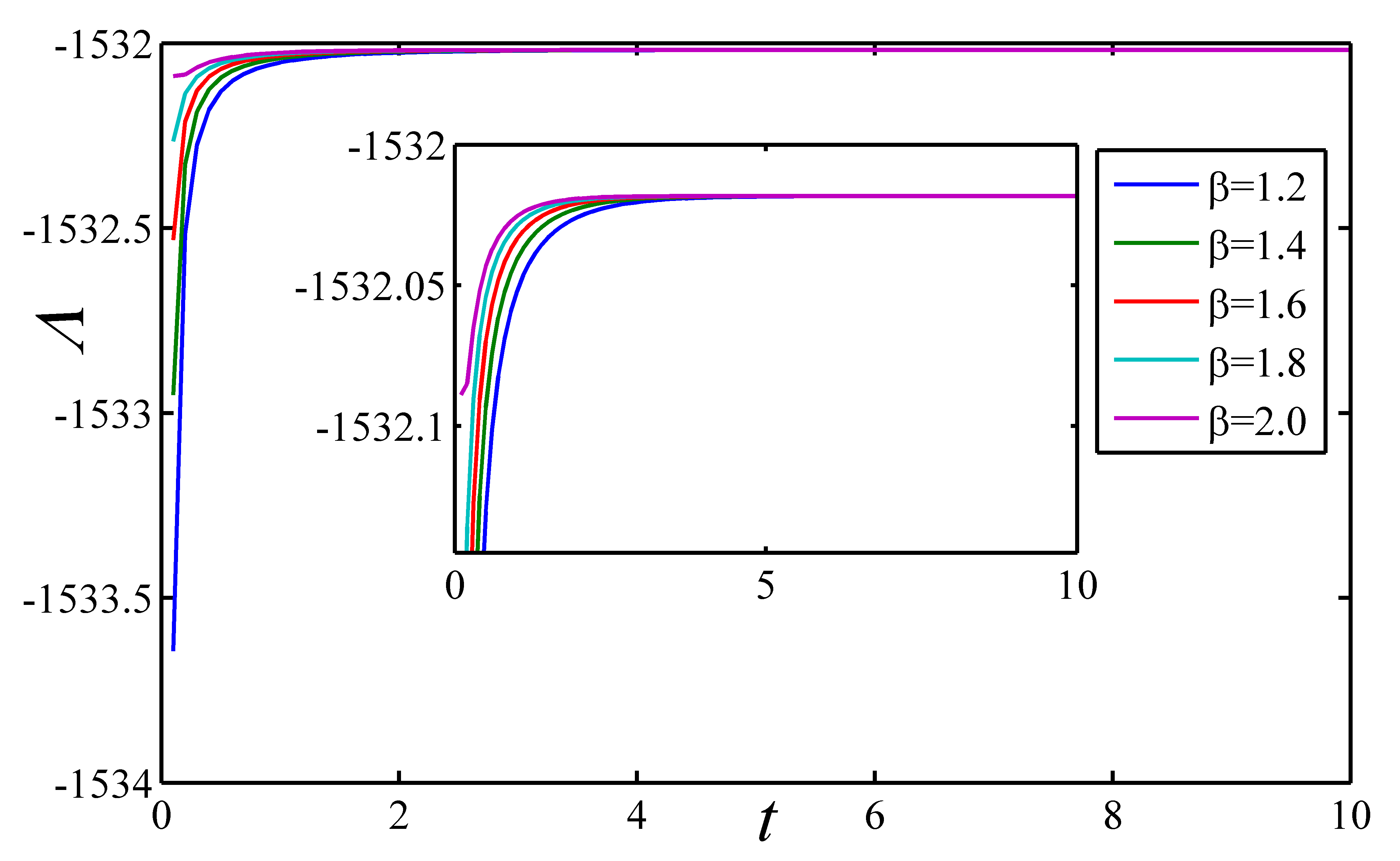}
\endminipage
\caption{Variation of $\Lambda$ against $t$ for $\lambda=0.1$, $B_\text{c}=60$ and diff. $\beta$ with $n=0.5$, $n=10$ and $n=240$ respectively}\label{ch4fig18}
\end{figure}

In this case, we need to notice the difference between the physical behaviors of model with special form of time varying DP \cite{Debnath2009}, which we already implemented in chapter  \ref{Chapter3}     
\begin{equation}\label{ch446}
q(t)=-1+\frac{\beta}{1+a^{\beta}},
\end{equation}
where $\beta >0$ is a constant. Consequently, the HP is
\begin{equation}\label{ch447}
H=A_1 (1+a^{-\beta}),
\end{equation}
where $A_1$ is an integrating constant. Again integrating the above equation, we have
\begin{equation}\label{ch448}
a=(e^{A_1 \beta t}-1)^{\frac{1}{\beta}}.
\end{equation}
The values of $\rho$, $h^2$ and $\Lambda$  are obtained as
\begin{equation}\label{ch4h3}
h^2=\frac{3(n-1)[-3+\beta e^{-A_1 \beta t}]}{2(4\pi+\lambda)(n+2)}H^2,
\end{equation}
\begin{equation}\label{ch4rho3}
\rho= \frac{-3}{4(4\pi+\lambda)}\biggl[\frac{9(n-1)}{(n+2)^2}+\frac{3[3+n\beta e^{-A_1 \beta t}-3n]}{(n+2)}\biggr]H^2+B_\text{c},
\end{equation}
\begin{equation}\label{ch4p3}
p= \frac{-1}{4(4\pi+\lambda)}\biggl[\frac{9(n-1)}{(n+2)^2}+\frac{3[3+n\beta e^{-A_1 \beta t}-3n]}{(n+2)}\biggr]H^2-B_\text{c},
\end{equation}
\begin{multline}\label{ch4Lambda3}
\Lambda=\biggl[\frac{3[(12n\pi+3n\lambda-n^2 \lambda+24\pi+10\lambda)(-1+\beta e^{-A_1 \beta t})]}{2(4\pi+\lambda)(n+2)^2}\\
+\frac{(-26\lambda+18n\lambda+6n^2 \lambda-76\pi)}{2(4\pi+\lambda)(n+2)^2}\biggr]H^2-(8\pi+4\lambda)B_\text{c}.
\end{multline}
The remaining physical parameters are as follows:
\begin{eqnarray}
V= (e^{A_1 \beta t}-1)^\frac{1}{\beta},\label{53}\\
\theta=3A_1 e^{A_1 \beta t}(e^{A_1 \beta t}-1)^{-1},\label{54}\\
\sigma^2=\frac{3(n-1)^2}{(n+2)^2}[A_1^2 e^{2A_1 \beta t}(e^{A_1 \beta t}-1)^{-2}], \label{55}\\
\Delta= \frac{2(n-1)^2}{(n+2)^2}.\label{56}
\end{eqnarray}
The variation of DP $q$, $H$ and $h^2$ with respect to time are presented in  Fig. \ref{ch4fig13},  Fig. \ref{ch4fig14}  and  Fig. \ref{ch4fig15} respectively. Here, we have noticed that, $q\in(0,-1)$ for $\beta \in (0,1]$ and $q\in(-1,1)$ for $\beta \in (1,2)$.
The DP has negative value within the interval of $\beta \in (0,1]$, whereas for $\beta \in (1,2)$ a phase transition takes place from positive to negative. We have found that the HP is a positive, decreasing valued function of time and approaches to zero with the increment of time.
The magnetic flux $h^2$ is also a positive and decreasing function of time for $n \in (0,1)$ and the given $\beta$. As we are interested in the case of phase transition, all the physical parameters are presented graphically with $\beta \in [1,2]$. The variation of $\rho$ with respect to time is depicted in the  Fig. \ref{ch4fig16}. One can observe in eqn. (\ref{ch4rho3}) that $\rho\rightarrow B_\text{c}$. Here, it is worth mentioning that, the approach of $\rho$ towards $B_\text{c}$ is different for different interval of $n$ and $\beta \in [1,2]$. That means for $n\in(0,2]$ and $n\geq 3.6$, $\rho$ approaches to $B_\text{c}$ from left and right of $B_\text{c}$ respectively. Then $\rho\rightarrow B_\text{c}$ from either side of $B_\text{c}$ for $\beta \in [1,2]$ and $n\in [2,3.6]$. As a representative case, we have chosen three different values of $n$ i.e.\ $n= 0.5, 2.5, 3.6$ and $\beta \in [1,2]$ (see  Fig. \ref{ch4fig16}). 
Pressure profile also has similar qualitative behavior as that of energy density but it approaches toward negative $B_\text{c}$ i.e. $p\rightarrow -B_\text{c}$ (see  Fig. \ref{ch4fig17}). The $\Lambda$ is also negative value function of time. Here, we have observed that, for $n\in(0,2)$, $n\in[2,235]$ and $n>235$, $\Lambda\rightarrow -(8\pi+4\lambda)B_\text{c}$ from left, either side and right of $-(8\pi+4\lambda)B_\text{c}$ respectively.  As a representative case, we have chosen three different values of $n$ i.e. $n= 0.5, 10, 240$ and $\beta \in [1,2]$ (see  Fig. \ref{ch4fig18}). The other physical quantities like volume, shear scalar, expansion scalar have the similar qualitative behavior as that of model \ref{ch4model1}.
\section{Conclusion}\label{ch4conclsn}
In this chapter we have investigated three cosmological models in the linear frame of $f(R,T)$ gravity (i.e $f(R,T)=R+2f(T)$) with $\Lambda$, where MSQM is considered as matter source. In this study, the exact solution of field equations. is obtained by using three different DPs.\\
The findings of this chapter are quite convincing, and thus the following conclusions can be drawn:
\begin{itemize}
\item The DP shows a phase transition for a universe which was decelerating in the past and is accelerating at present epoch. Furthermore, the first model of this chapter with a bilinear DP represents a transition of universe from early decelerating phase to a recent accelerating phase. In the second model, the universe lies at an accelerating phase. At last, the third model shows a transition of universe for $\beta> 1$ and again lies at an accelerating phase for $\beta\leq 1$ (see  Fig. \ref{ch4fig13}). The transitional behavior of this DP also can be observed from chapter  \ref{Chapter3}, which means that the given specific form of DP posses a phase transition of universe in the cosmological models. Summing up the results of this chapter, it can be concluded that, the DP plays a vital role in account of accelerated expansion of the universe that means the model with each time varying DP represents an expanding universe in accelerated phase.
\item  Each model represents an accelerated expansion of the universe with $V\rightarrow \infty $ at $t\rightarrow \infty$. In case of magnetic flux, it has more effect in the early universe and gradually reducing its effect in later.
\item The pressure and energy density of each model approaches to bag constant in negative and positive way at $t\rightarrow \infty$ i.e. $p\rightarrow -B_\text{c}$ and $\rho\rightarrow B_\text{c}$ at $t\rightarrow \infty$. The presence of negative pressure yields the existence of DE in the context of accelerated expansion of the universe. So the SQM within magnetic field epoch gives the same idea of existence of DE in the universe and supports the observations of the type Ia Supernovae \cite{Riess/1998}. Also these results agree with the study of Akta\c{s} and Ayg\"{u}n \cite{Aktas17}. They researched MSQM distribution in $f(R,T)$ gravity and found $p=-\rho$ DE model for $t\rightarrow \infty$. \item  The rate of expansion observed by scalar expansion is faster at the beginning then it slows down  later.
\item In each case, the shear scalar $\sigma^2 \neq 0$ and average anisotropy parameter gives a constant value i.e. $\Delta=\frac{2(n-1)^2}{(n+2)^2} \neq 0$. Henceforth, the obtained models with three different DP represent expanding, shearing and an anisotropic universe.
\end{itemize}
We have studied till here that the models show some initial singularity at $t=0$. But what would be the nature of the model in future time evolution of universe, whether it is accelerating forever or posses some certain jerks in future at some finite time? This would be a most interesting phenomena about the model behavior in the context of future evolution. Thus it is essential to develop a cosmological model in account of finite time future singularity, called Big Rip singularity, which is discussed in next chapter.

\chapter{Big Rip cosmological models in $f(R,T)$ gravity} 

\label{Chapter5} 

\lhead{Chapter 5. \emph{Big Rip Cosmological models in $f(R,T)$ gravity}} 
This chapter \blfootnote{The work presented in this chapter is covered by the following two publications: \\ 
\textit{Magnetized strange quark model with Big Rip singularity in $f(R,T)$ gravity}, Modern Physics Letters A, \textbf{32}, (2017) 1750105.\\
\textit{A periodic varying deceleration parameter in $f(R,T)$ gravity}, Modern Physics Letters A, \textbf{33}, (2018) 1850193.} is devoted to describe a phenomenon of accelerated expansion of the present as well as future universe and a cosmic transit aspect in the framework of $f(R,T)$ gravity. The first model of this chapter deals with the cosmic future singularity, which appeared due to the presence of linearly time varying DP (LVDP) in the process of obtaining the exact solution. In the second model, a periodic varying DP (PVDP) is discussed which yields a cosmic transit phenomenon of signature flipping behavior from early deceleration to late time acceleration.  The dynamical features of both the models are discussed in detail. Moreover, in second model the oscillatory behavior of the EoS parameter are studied and the violation of energy momentum conservation is also explored in $f(R,T)$ gravity theory. The stability of the model is investigated under linear homogeneous perturbation.
\section{Introduction}\label{ch5intro}
A billion years ago, the universe was  created through an inconceivable Big Bang as per the theorization of cosmological science. But what would be the ultimate fate of the universe after that? Whether it faces a Big crunch, a Big freeze or a Big bounce? In order to address these future possibilities of our Universe, there are several scientific literatures emerged in the 21st century \cite{Chimento/2004,Odintsov/2018}. Till 20th century, the possible fates of the universe were considered only through Big Bang models. Of which only two possible cases are apprehended, they are either an eternal expansion in an open/flat universe or an eventual recollapse known as Big crunch in a closed model. In other words, the universe would start from Big Bang, but would end in one the possible fates depending on the curvature of the universe. Moreover, the universe is accelerating as per the type Ia Supernovae, which is later supported by other observations such as the anisotropies in the CMB and BAO. In this case the destiny of the universe is not determined by geometry. This discovery insists the inclusion of extra component in the mass-energy content of the universe. It causes the speeding up of the expansion of the universe, a closed universe could expand indefinitely and an open universe could recollapse \cite{Krauss/1999}.\\
More specifically, the EoS parameter of DE will determine its ultimate fate in a fascinating way as we discussed in chapter \ref{Chapter1}. In a standard concordance cosmology the DE responsible for this behavior is the cosmological constant ($\Lambda$) with an EoS parameter $\omega=-1$. The universe will continue to expand exponentially into an empty de Sitter type universe, if it is dominated by $\Lambda$. Another leading candidate with $-1<\omega<\frac{-1}{3}$, which varies in time and space is known as quintessence. A special case of quintessence with $\omega<-1$ whose energy density increases with time is called phantom energy \cite{Caldwell/2002,Caldwell/1998}. Universe dominated by this phantom energy will expand towards Big Rip singularity \cite{Caldwell/2003}, where all the matter in the universe will also take part in the expansion and will hence be torn apart. Also, the Big Rip singularity violates all ECs, whereas the current cosmological data are consistent with a DE in terms of $\Lambda$. They are not yet able to rule out other more exotic candidates, including most intriguingly phantom energy which causes Big Rip singularity. This has encouraged us to the study the possibility of the occurrence of other
non-standard events in the future evolution of the universe. At the same time the inclusion of a negative pressure DE would be a solution to the acceleration problem. These cosmic singularities are classified into several categories as discussed in chapter \ref{Chapter1}.  Here, we have explored this ultimate fate (Big Rip) of the universe through a cosmological model within $f(R,T)$ gravity formalism.     
In the case of Big Rip, the very negative EoS parameter results in a super negative pressure of the dominating DE component, which together with its ever increasing energy density dissociate everything
in the universe. In which the matter energy starts from 
the largest structures of super clusters of galaxies and continuing to the smallest constituents of matter. This drastic end to the universe has not been ruled out by observations of type Ia supernovae, the CMB
and large scale structure. This discovery has been a great encouragement for searching other exotic possible fates for the universe. \\
In order to approach such type of fate of the universe, we have studied the first model of this chapter with a LVDP. While, the second model with a PVDP showing a new feature in the behavior of the universe, called as oscillatory behavior along with Big Rip singularity in matter content. 
The chapter is organized as follows: section \ref{ch5intro} contains the brief introduction and motivation regarding the present work. Here, two cosmological models are derived in $f(R,T)$ gravity in account of accelerated expansion and future singularity of the universe. In section \ref{ch5model1}, we have discussed a MSQM model in $f(R,T)$ gravity. The solution of field equations are determined by using LVDP. The discussion and graphical resolution of parameters for this model are presented in section \ref{ch5model1}. In section \ref{ch5model2}, a cosmological model with perfect fluid matter source is derived in details. The solution of field equations of this model are obtained by using PVDP. Finally, in  section \ref{ch5conclusion}, the conclusion and perspective of both the models are outlined.  
\section{Model I}\label{ch5model1}
This model is an extension of previous chapter that is adding a LVDP in the context of future evolution of universe. We can say it is the fourth cosmological model on basis of different DP for MSQM matter source in $f(R,T)$ gravity (see chapter \ref{Chapter4} including past three models). Therefore, the background details of this present model can be sourced from previous chapter. In particular, section \ref{ch4intro} consists of the basic idea about MSQM matter source and the importance of time varying DP, while the field equations within the linear frame of $f(R,T)$ gravity are seen in section \ref{ch4fldeqn}. The next section contains the solution models of the field equations in several aspects. Hence, by admitting a LVDP in the explicit expressions for $h^2$, $\rho$, $p$, and $\Lambda$ in eqns. (\ref{ch4h} - \ref{ch4Lambda}) available in section \ref{ch4sol}, we have obtained the model as bellow.\\      
Here, we have considered the LVDP in the form  \cite{Akarsu2012}
\begin{equation}\label{ch5dp1}
q(t)=-\frac{a\ddot{a}}{\dot{a}^2}=-kt+m-1,
\end{equation}
where $k \geq 0, m\geq 0$ are constants. The above DP leads to the following three different cases as:
\begin{itemize}
\item $q=-1,$ for $ k=0, m=0$,
\item $q=m-1,$ for $ k=0, m>0$,
\item $q=-kt+m-1,$ for $ k>0, m\geq 0$.
\end{itemize}
 Here, the first two cases i.e. for $k=0$ correspond to constant DP. Therefore, only the last case for $k>0$ renders a LVDP, which is compatible with the observational data of modern cosmology. 
\begin{figure}[H]
\centering
\includegraphics[width=78mm]{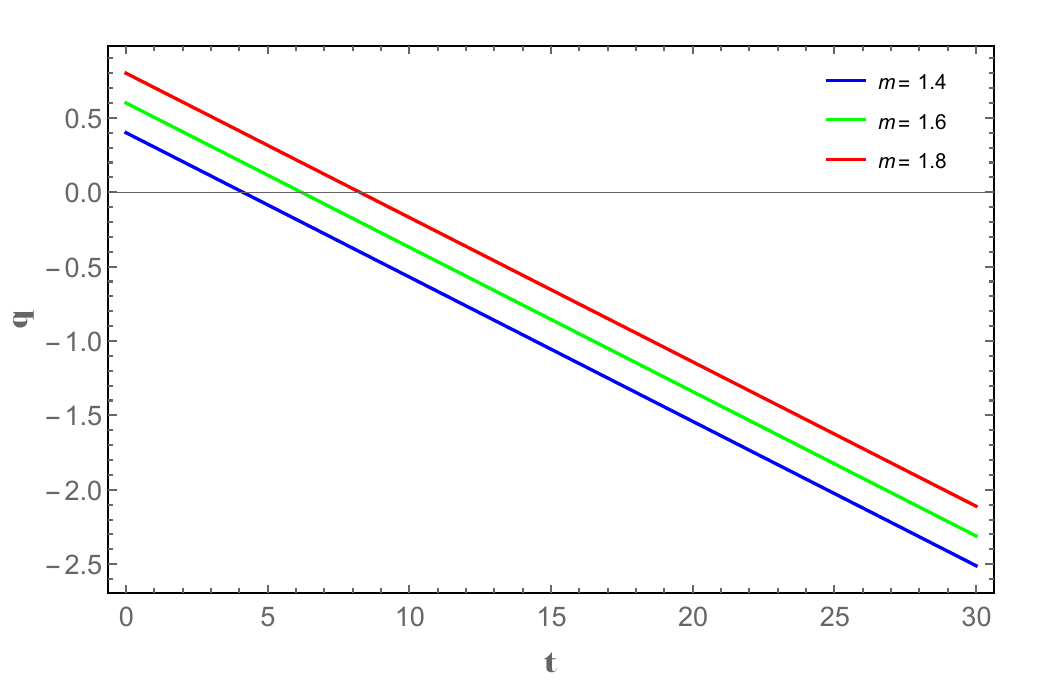}
\caption{Variation of DP against time with $k=0.097$ and different $m$}\label{ch5fig1}
\end{figure}
Fig. \ref{ch5fig1} shows a transitional behavior of universe from early deceleration to present acceleration. In which the universe starts with decelerating expansion for  $q=m-1>0$ and enters to accelerating phase at $t=\frac{m-1}{k}$. In particular, the universe enters to the acceleration phase at $t\approx4.1, 6.2, 8.2$ and present values of DP $q=-0.938, -0.738, -0.538$ at $t=13.798$ for $m=1.4, 1.6, 1.8$ respectively. Hence, these values are consistent with respect to the observational data. Since the universe experiences super exponential expansion for $q=-1$ at $t=\frac{m}{k}$ and ends with $q=-m-1$ at $t=\frac{2m}{k}$. One can get isotropic model at $t=\frac{2m}{k}$.\\
Considering the last case of eqn. (\ref{ch5dp1}) with $k>0$ and $m\geq0$ the scale factor $a$ is obtained as
\begin{equation}
a=c \exp\bigg[ \frac{2}{\sqrt{m^2-2l k}} \text{arctanh} \bigg(\frac{kt-m}{\sqrt{m^2-2l k}}\bigg)\bigg],
\end{equation}
where $c$, $l$ are integrating constants. Assuming the integrating constant $l=0$, the scale factor and corresponding mean HP, scalar expansion $\theta$, and the mean anisotropic parameter  are obtained as
\begin{eqnarray}
a=c\exp \biggl[\frac{2}{m}\text{arctanh}\left(\frac{kt}{m}-1\right)\biggr],\label{ch53}\\
H=-\frac{2}{kt(t-t_{BR})},\label{ch54}\\
\theta=-\frac{6}{kt(t-t_{BR})},\label{ch55}\\
\Delta=\frac{1}{3}\left(\frac{6n^2-16n+4}{(n+2)^2}\right).\label{ch56}
\end{eqnarray}
where $t_{BR}=\frac{2m}{k}$.
\begin{figure}[H]
\centering
\includegraphics[width=78mm]{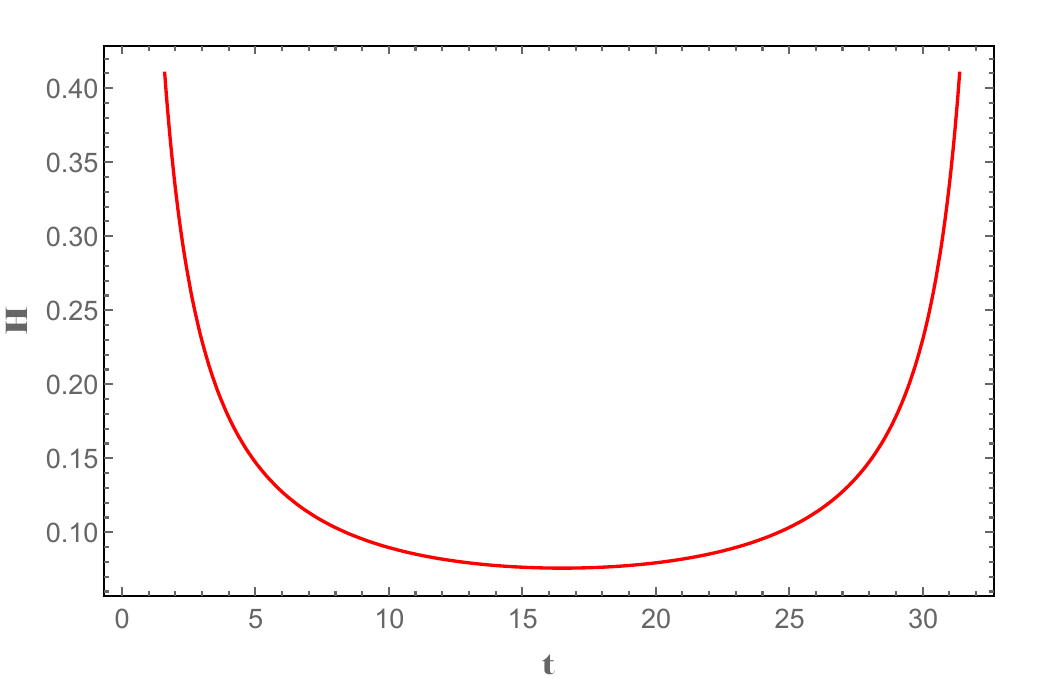}
\caption{Variation of $H$ against time with $k=0.097$ and $m=1.6$}\label{ch5fig2}
\end{figure}
The evolution of HP with respect to time $t$ is presented in Fig. \ref{ch5fig2}. In which $H$ approaches towards zero for large value of $t$ i.e. $H\rightarrow 0$ when $t\rightarrow \infty$. It can be noticed that the HP has singularities at $t=0 $ and $t=t_{BR}$. Hence, the HP and directional HPs both diverge at the beginning and at a finite time in future called the Big Rip.
The evolution of HP lies in  the intermediate phase between initial (the Big Bang) and  end of the universe (Big Rip). Hence, we can say that the model of the universe starts with Big Bang and ends with Big Rip and at transition phase the HP becomes $H=\frac{2k}{m^2-1}$ \cite{Nojiri2010}. Further we need the positive value of scalar expansion $\theta$ for an expanding universe.  So, the model of the universe is expanding for $t<t_{BR}$ from eqn. (\ref{ch55}).
\newline
The magnetic flux for the model becomes
\begin{equation}\label{ch5h}
h^2=\frac{3(n-1)(-kt+m-3)}{2(4\pi+\lambda)(n+2)}\left(\frac{-2}{kt(t-t_{BR})}\right)^2.
\end{equation}
\begin{figure}[H]
\centering
\includegraphics[width=78mm]{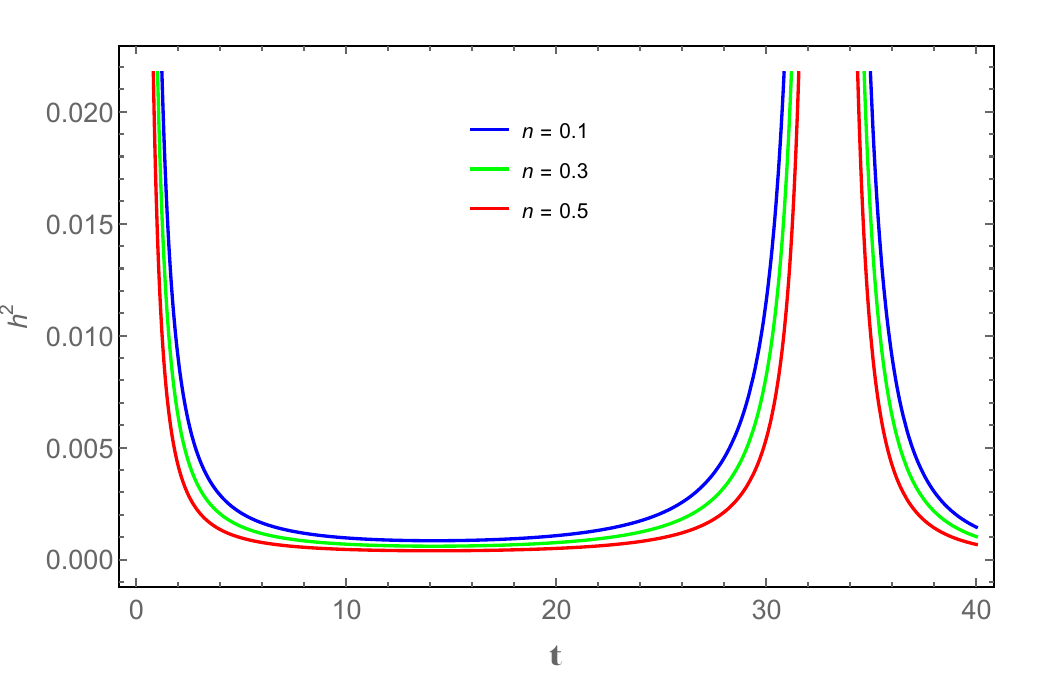}
\caption{Variation of $h^2$ against time with $k=0.097$, $m=1.6$, $\lambda=0.1$ and different $n$}\label{ch5fig3}
\end{figure}
The variation of magnetic flux $h^2$ with respect to time is described in Fig. \ref{ch5fig3}. Here, we have focused on the  positivity of $h^2$ which appears for $n\in (0,1)$  with $k=0.097$, $m=1.6$ and $\lambda=0.1$. Thus, we have neglected the case $h^2<0$, for $n>1$. Also we found that $h^2\rightarrow 0$ when $t\rightarrow \infty$. Moreover, it shows the the same singularity as that of HP. \\ 
Using the above values we obtain the energy density $\rho$ and pressure $p$ for the model as
\begin{equation}\label{ch5rho1}
\rho=\frac{-3}{4(4\pi+\lambda)}\biggl[\frac{9(n-1)}{(n+2)^2}+\frac{3[3-3n+(-kt+m)n]}{(n+2)}\biggr]H^2+B_c,
\end{equation}
\begin{equation}\label{ch5p1}
p=\frac{-1}{4(4\pi+\lambda)}\biggl[\frac{9(n-1)}{(n+2)^2}+\frac{3[3-3n+(-kt+m)n]}{(n+2)}\biggr]H^2-B_c.
\end{equation}
\begin{figure}[H]
\minipage{0.32\textwidth}
  \includegraphics[width=53mm]{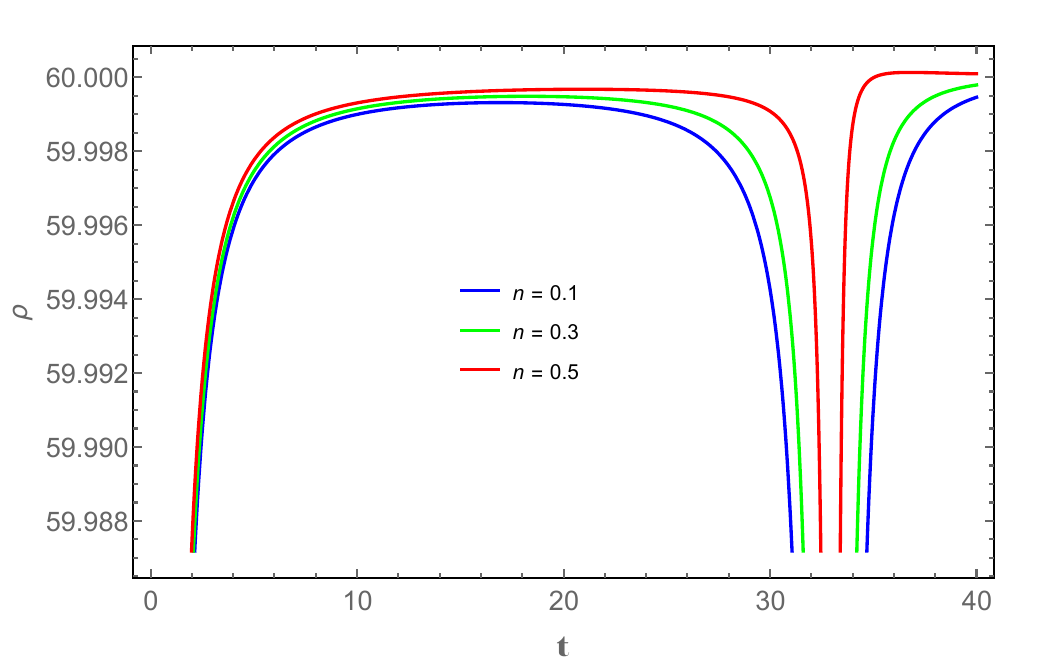}
\endminipage\hfill
\minipage{0.32\textwidth}
  \includegraphics[width=53mm]{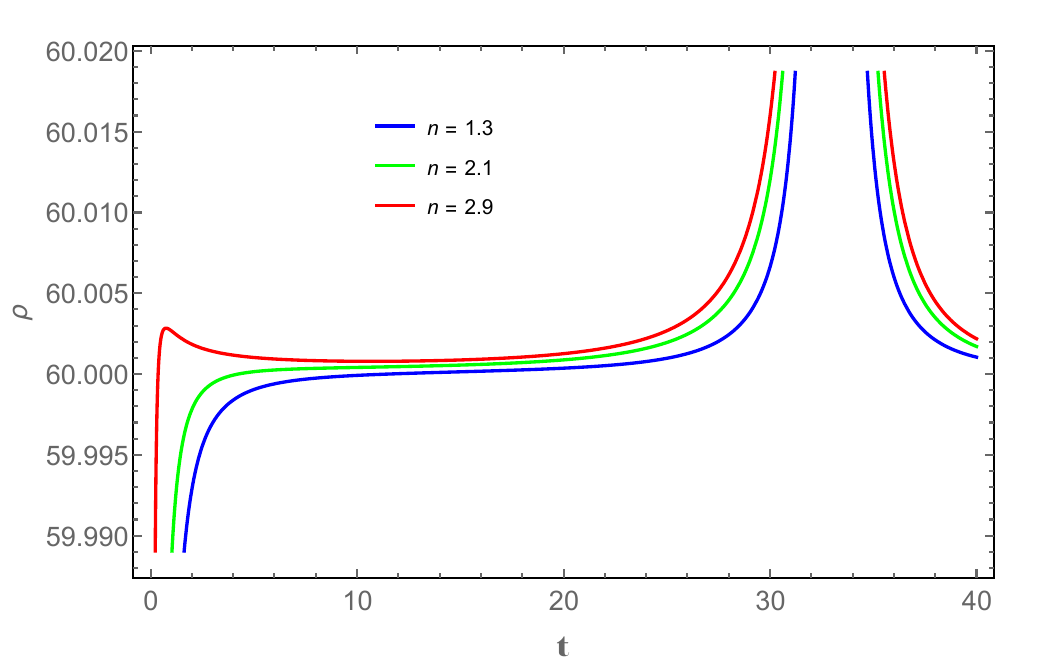}
\endminipage\hfill
\minipage{0.32\textwidth}%
  \includegraphics[width=53mm]{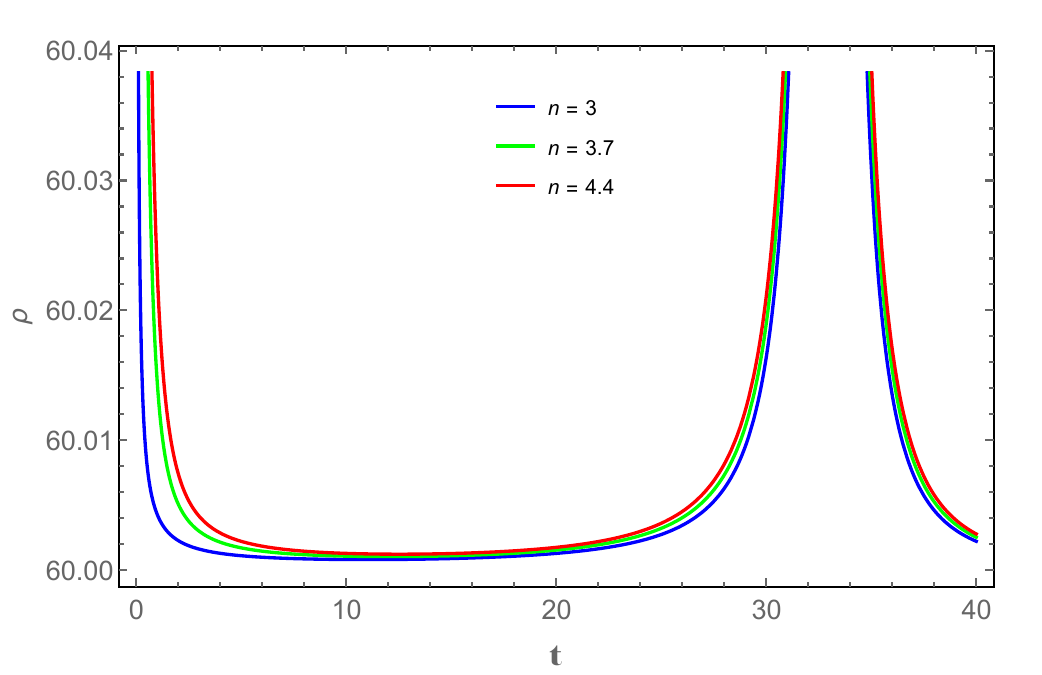}
\endminipage
\caption{Variation of $\rho$ against time with $k=0.097$, $m=1.6$, $\lambda=0.1$, $B_c=60$ and different $n$ i.e. $n\in (0,0.5]$, $n\in [0.6,3)$ and $n\in [3,\infty)$}\label{ch5fig4}
\end{figure}
Fig.  \ref{ch5fig4} describes the the evolution of energy density against time with suitable parameter values. In which one can observed that $\rho\rightarrow B_c$ when $t\rightarrow \infty$. The way of approach to $B_c$ are different for different interval of $n$ (see eqn. (\ref{ch5rho1})). Here, the energy density of the fluid diverges very fast as time increases and then leads a Big Rip singularity at finite time $t_{BR}=\frac{2m}{k}$.
\begin{figure}[H]
\minipage{0.32\textwidth}
  \includegraphics[width=53mm]{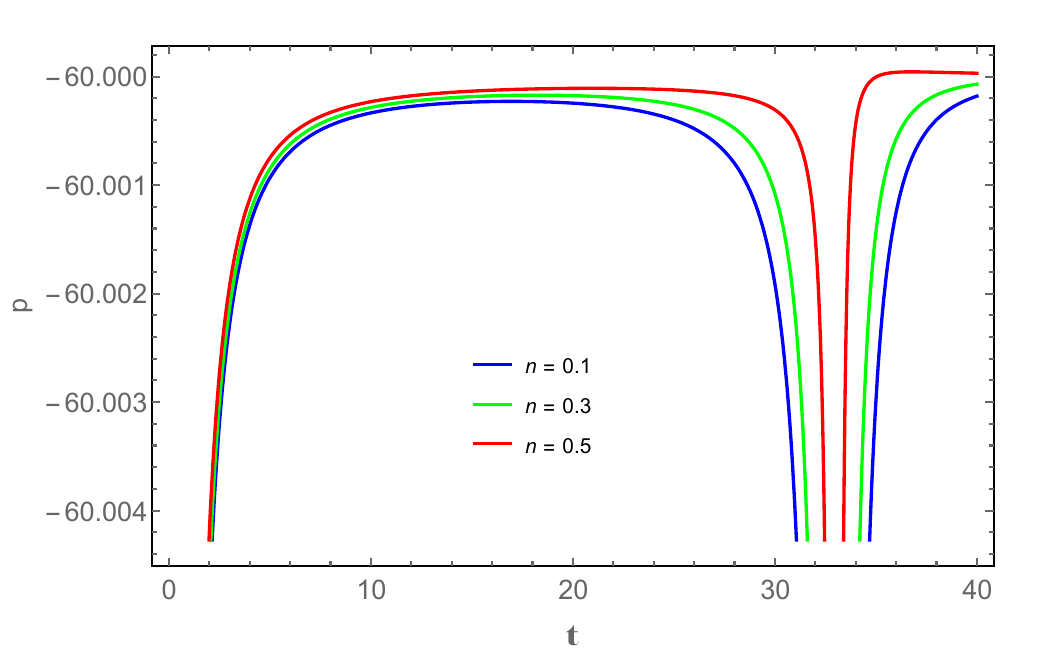}
\endminipage\hfill
\minipage{0.32\textwidth}
  \includegraphics[width=53mm]{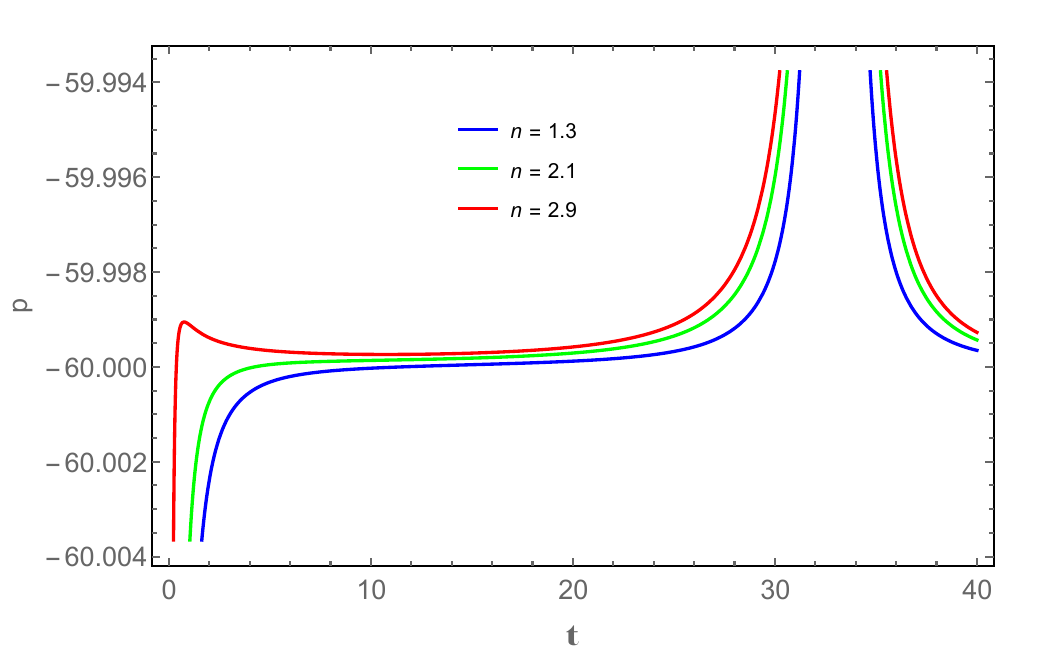}
\endminipage\hfill
\minipage{0.32\textwidth}%
  \includegraphics[width=53mm]{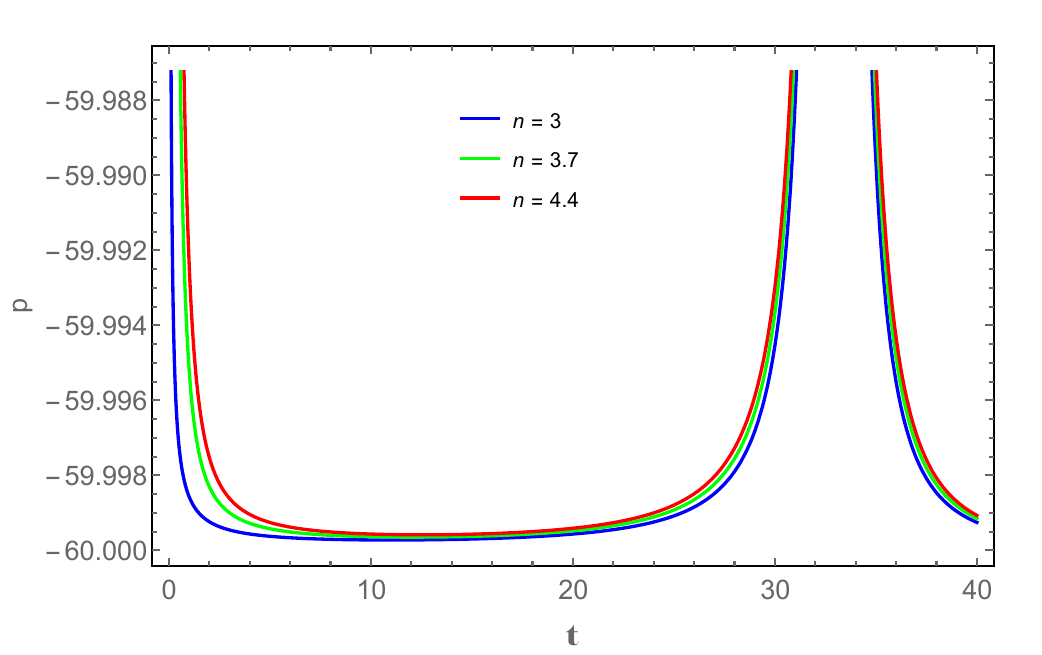}
\endminipage
\caption{Variation of $p$ against time with $k=0.097$, $m=1.6$, $\lambda=0.1$, $B_c=60$ and different $n$ i.e. $n\in (0,0.5]$, $n\in [0.6,3)$ and $n\in [3,\infty)$}\label{ch5fig5}
\end{figure}
Consequently, our EoS $\omega$ and cosmological constant $\Lambda$ are obtained as
\begin{equation}\label{ch5EoS1}
\omega=\frac{-{B_c}-\frac{\frac{3 (n (m-k t)-3 n+3)}{n+2}+\frac{9 (n-1)}{(n+2)^2}}{k^2 (\lambda +4 \pi ) t^2 \left(t-t_{BR} \right)^2}}{{B_c}-\frac{3 \left(\frac{3 (n (m-k t)-3 n+3)}{n+2}+\frac{9 (n-1)}{(n+2)^2}\right)}{k^2 (\lambda +4 \pi ) t^2 \left(t-t_{BR} \right)^2}},
\end{equation}
\begin{multline}\label{ch5Lambda1}
\Lambda=\biggl[\frac{3[(12n\pi+3n\lambda-n^2 \lambda+24\pi+10\lambda)(-kt+m-1)]}{2(4\pi+\lambda)(n+2)^2}\\+\frac{(-26\lambda+18n\lambda+6n^2 \lambda-76\pi)}{2(4\pi+\lambda)(n+2)^2}\biggr]H^2-(8\pi+4\lambda)B_c
\end{multline}
\begin{figure}[H]
\minipage{0.32\textwidth}
  \includegraphics[width=53mm]{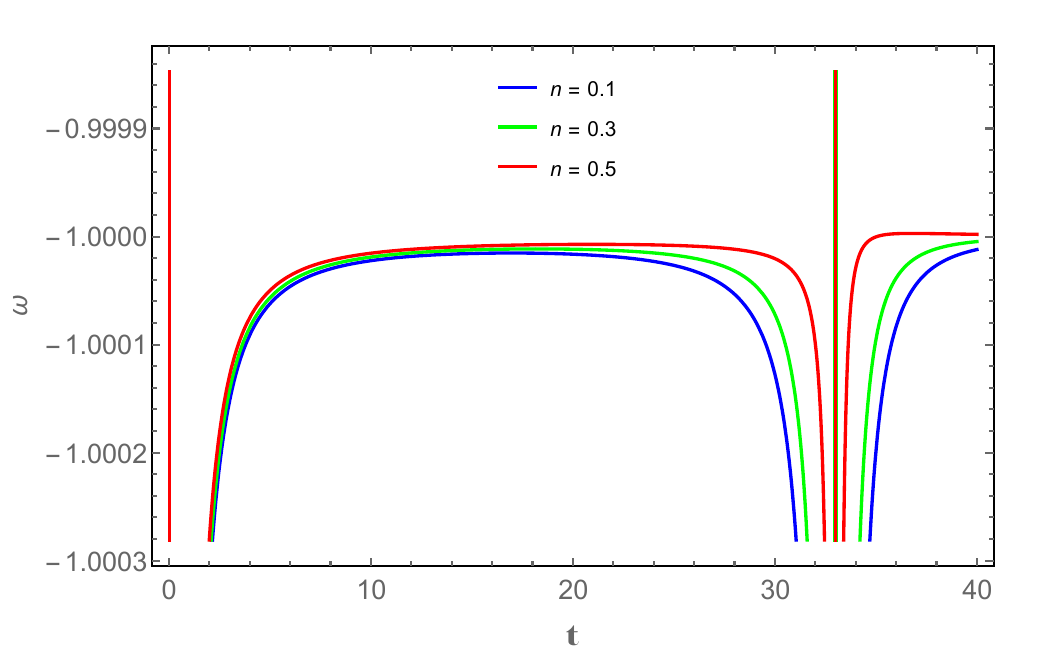}
\endminipage\hfill
\minipage{0.32\textwidth}
  \includegraphics[width=53mm]{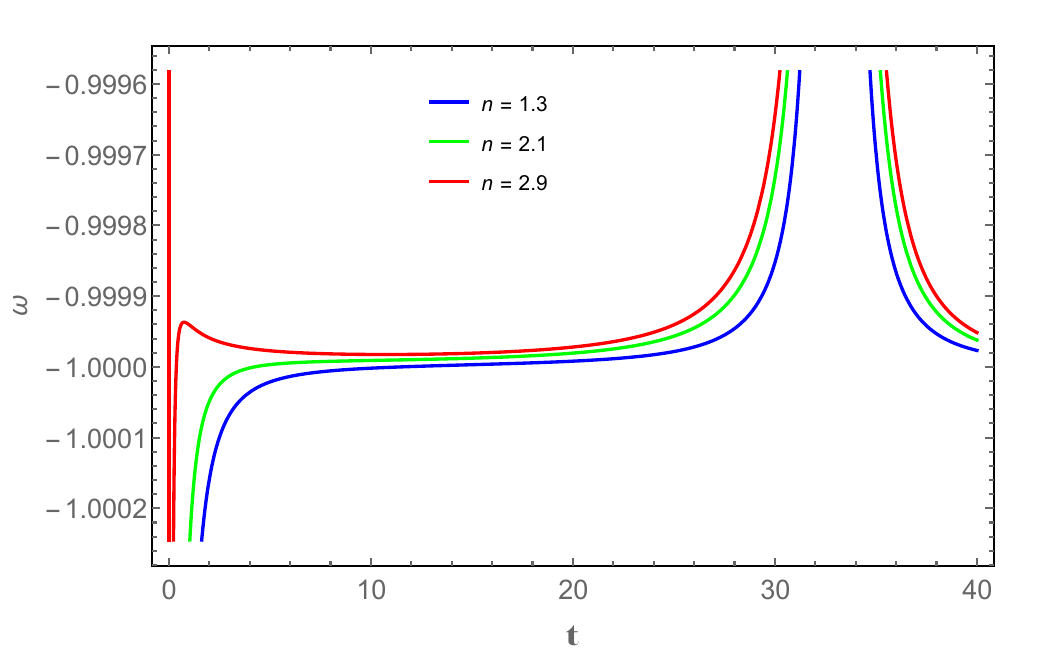}
\endminipage\hfill
\minipage{0.32\textwidth}%
  \includegraphics[width=53mm]{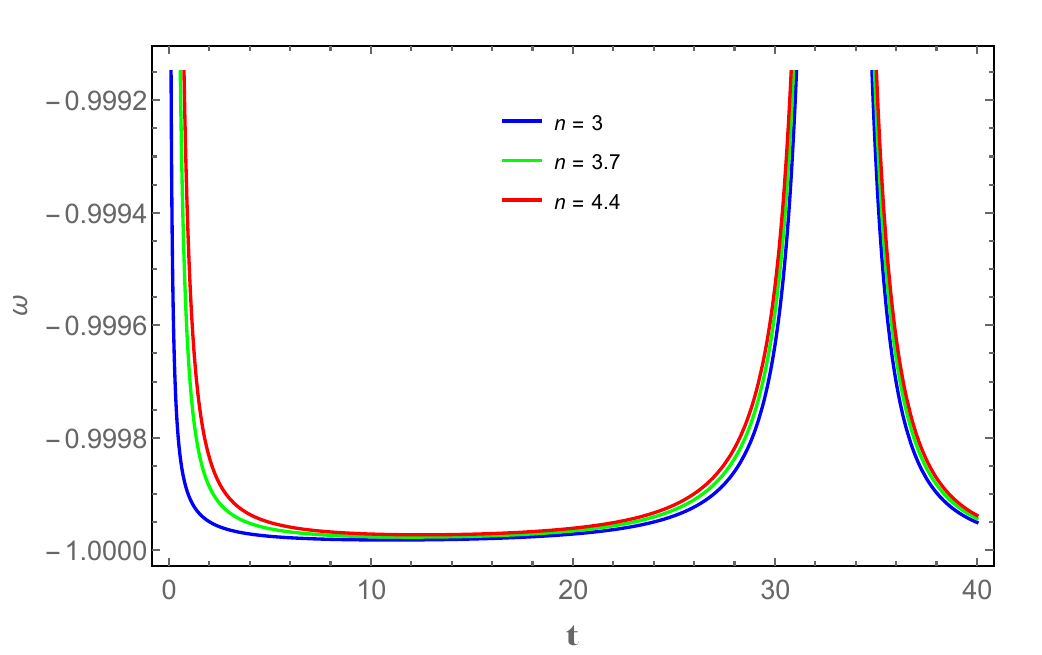}
\endminipage
\caption{Variation of $\omega$ against time with $k=0.097$, $m=1.6$, $\lambda=0.1$, $B_c=60$ and different $n$ i.e. $n\in (0,0.5]$, $n\in [0.6,3)$ and $n\in [3,\infty)$}\label{ch5fig6}
\end{figure}
\begin{figure}[H]
\minipage{0.32\textwidth}
  \includegraphics[width=53mm]{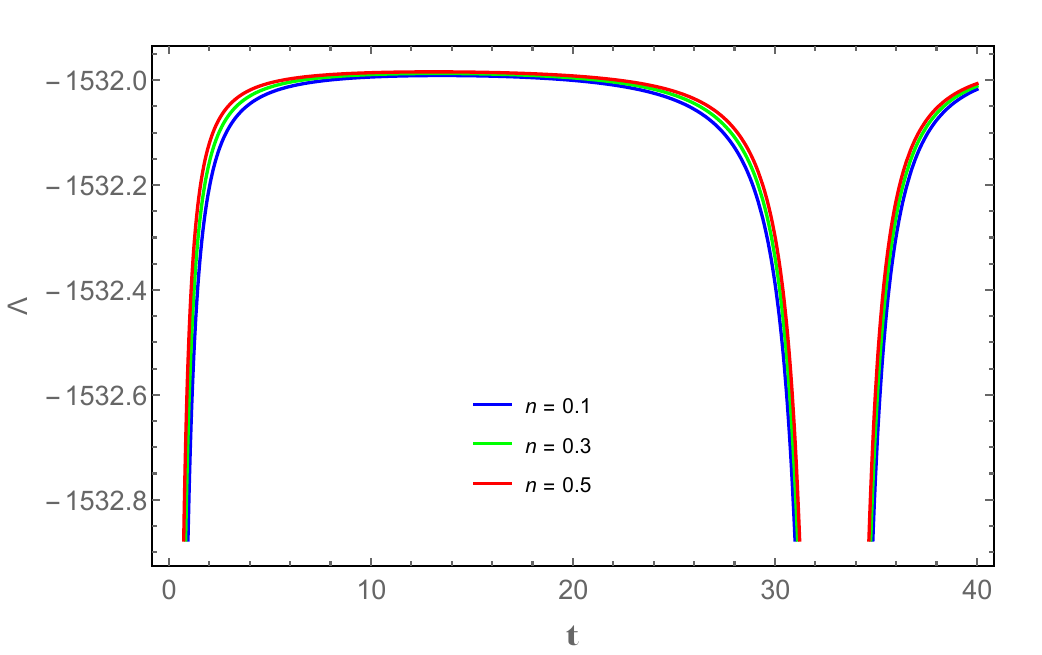}
\endminipage\hfill
\minipage{0.32\textwidth}
  \includegraphics[width=53mm]{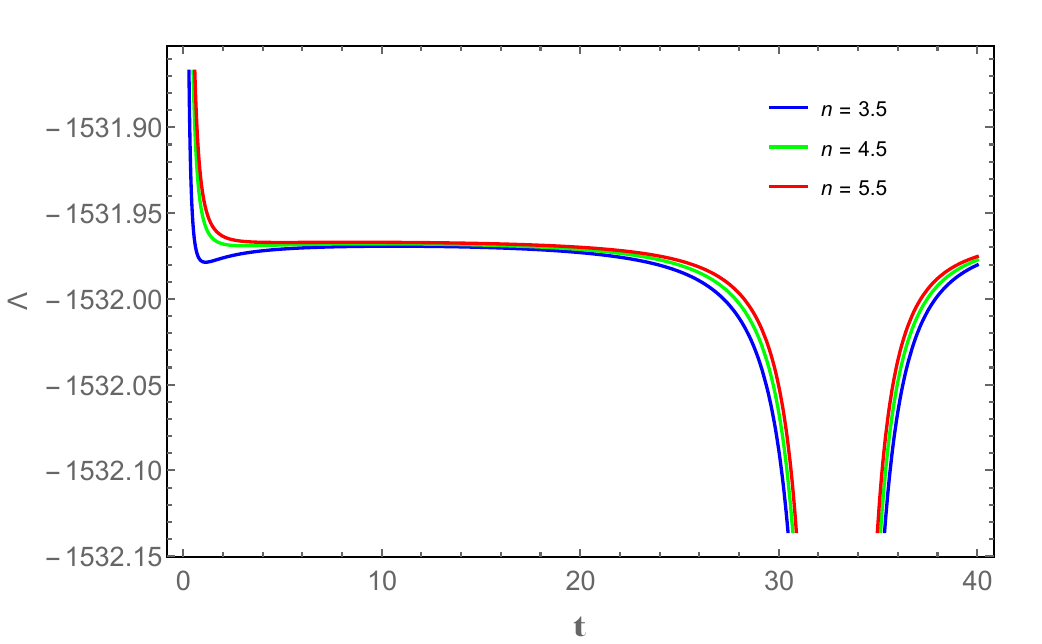}
\endminipage\hfill
\minipage{0.32\textwidth}%
  \includegraphics[width=53mm]{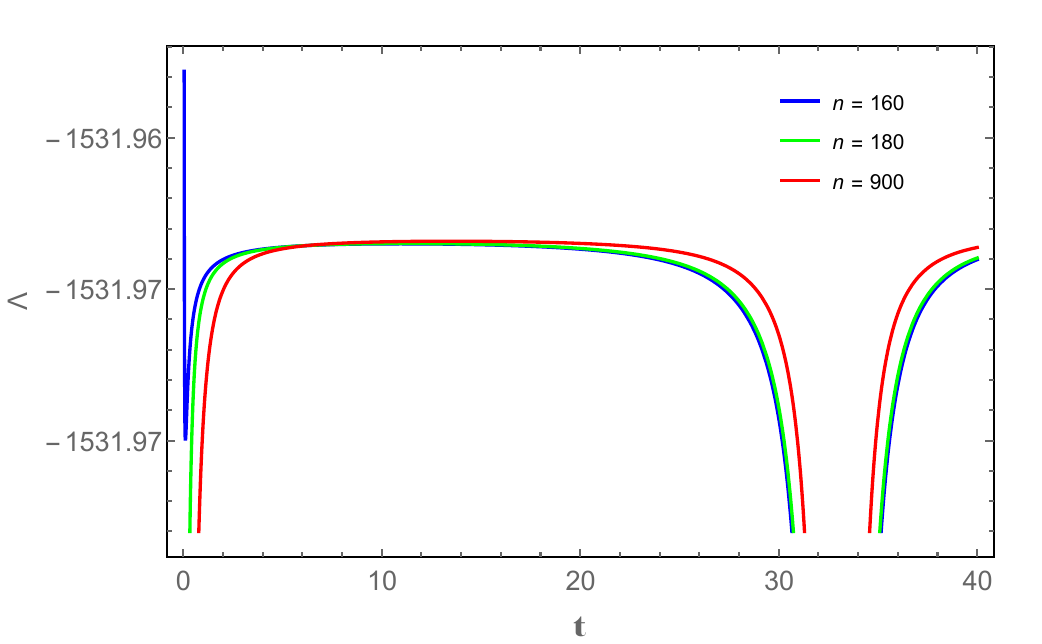}
\endminipage
\caption{Variation of $\Lambda$ against time with $k=0.097$, $m=1.6$, $\lambda=0.1$, $B_c=60$  and different $n$ i.e. $n\in (0,3]$, $n\in (3,160]$ and $n\in [160,\infty)$}\label{ch5fig7}
\end{figure}
It can be seen from Fig. \ref{ch5fig5} that 
the pressure profile shows the same singularity as that of HP. Then from eqn. (\ref{ch5p1}) one can observe that, $p\rightarrow -B_c$ when $t\rightarrow \infty$. Pressure is negative here and it approaches to $-B_c$ in different way for different interval of $n$ (see Fig. \ref{ch5fig5}).\\
The cosmological constant given in eqn. (\ref{ch5Lambda1}) is negative, which follow the observational data. Fig. \ref{ch5fig6} represents the variation of EoS parameter against time and it can be observed in eqn. (\ref{ch5EoS1}) that,  $\omega\rightarrow -1$ when $t\rightarrow \infty$. This approach  towards $-1$ is different for different interval of $n$ (see Fig. \ref{ch5fig6}). It follows the recent observational data and shows the same singularity as that of HP i.e. at the initial time and at the Big Rip $t_{BR}$.\newline
The profile of cosmological constant (see eqn. (\ref{ch5Lambda1})) against time is presented in Fig. \ref{ch5fig7}. Here,  $\Lambda\rightarrow -(8\pi+4\lambda)B_c$, when $t\rightarrow \infty$. Also, for different interval of $n$, $\Lambda$ approaches to $-(8\pi+4\lambda)B_c$ in different ways, which can be noticed from Fig. \ref{ch5fig7}. The parameter $\Lambda$ has also the same singularity as that of HP.\\
The Ricci scalar $R$ and the trace of energy momentum tensor $T$ are obtained as
\begin{equation}\label{ch5r1}
R = -\bigg[\frac{3n(2n+4)}{n+2}(-1-q)+\frac{9(2n^2+4n+6)}{(n+2)^2}\bigg]H^2,
\end{equation}
and
\begin{equation}\label{ch5t1}
T=\rho-3p+2h^2=4B_c +\frac{6(n-1)(q-2)}{2(4\pi+\lambda)(n+2)}H^2.
\end{equation}
Using eqns. (\ref{ch5r1}) and (\ref{ch5t1}), the function $f(R,T)$ can be obtained as
\begin{multline}\label{ch5frt1}
f(R,T)=8\lambda B_c+ \frac{[(4\pi+\lambda)(6n^2+12n)+6\lambda(n-1)](-kt+m-1)}{(n+2)(4\pi+\lambda)}\\+
\frac{6n^3+6n^2-12n-54}{(n+2)^2}-\frac{12\lambda(n-1)}{(n+2)(4\pi+\lambda)}.
\end{multline}
\begin{figure}[H]
\centering
\includegraphics[width=78mm]{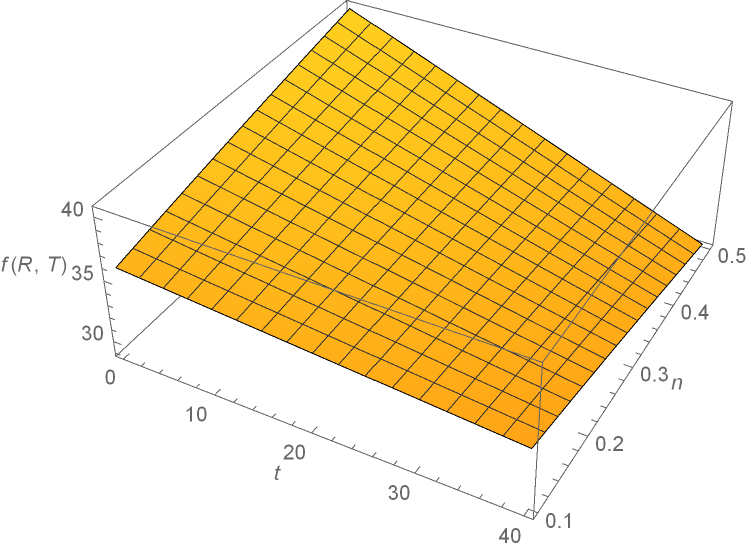}
\caption{Behaviour of $f(R,T)$ versus $t$ and $n$ with $\lambda=0.1$, $B_C=60$, $m=1.6$ and $k=0.097$ respectively.}\label{ch5fig8}
\end{figure}
Fig. \ref{ch5fig8} represents the behavior of $f(R,T)$ for this model.\\
If we take $\lambda=0$, we get $f(R,T)=R$ which agrees with GR results for LRS Bianchi type I universe with MSQM. Thus from eqn. (\ref{ch5h}), we can obtain the magnetic flux as follows
\begin{equation}\label{ch5h1}
h^2=\frac{3(n-1)(-kt+m-3)}{(8\pi)(n+2)}\left(\frac{-2}{kt(t-t_{BR})}\right)^2.
\end{equation}
Using eqns. (\ref{ch5rho1}) and (\ref{ch5p1}), we get energy density and pressure for this model as
\begin{equation}\label{ch5rho2}
\rho=-\frac{3}{(8\pi)}\biggl[\frac{9(n-1)}{(n+2)^2}+\frac{3[3-3n+(-kt+m)n]}{(n+2)}\biggr]H^2+B_c,
\end{equation}
\begin{equation}\label{ch5p2}
p=-\frac{1}{(8\pi)}\biggl[\frac{9(n-1)}{(n+2)^2}+\frac{3[3-3n+(-kt+m)n]}{(n+2)}\biggr]H^2-B_c,
\end{equation}
and from eqn. (\ref{ch5Lambda1}), we found cosmological constant value in GR as 
\begin{equation}\label{ch5Lambda2}
\Lambda=\biggl[\frac{3[(12n\pi+24\pi)(-kt+m-1)]}{8\pi(n+2)^2}+\frac{-19}{2(n+2)^2}\biggr]H^2-8\pi B_c.
\end{equation}
Here, we get same results with $f(R,T)$ gravitation theory for $t\rightarrow \infty$. In brief, we obtain magnetic flux value as $h^2\rightarrow 0$, from eqn. (\ref{ch5h1}), the cosmic density $\rho\rightarrow B_c$ and from eqn. (\ref{ch5rho2}), the cosmic pressure value as $p \rightarrow -B_c$, from eqn. (\ref{ch5p2}), and also we obtain different results for cosmological constant $\Lambda=-(8\pi)B_c$ from eqn. (\ref{ch5Lambda2}) in GR.
\subsection{The dynamics of the model}
The scale factor $a$ and the HP in terms of redshift parameter $z$ are written as
\begin{eqnarray}
a=\frac{a_0}{1+z},\\
H=H_0(1+z)^m\left(\frac{t_0}{t}\right)^2,
\end{eqnarray}
where, $a_0$ is the present scale factor and $H_0$ is the value of present HP.\\
The distance modulus $\mu_{dm}(z)$ is defined as
\begin{equation}
\mu_{dm}(z)=5 \log d_L+25
\end{equation}
where $d_\text{L}$ is the luminosity distance and defined as
\begin{equation}
d_\text{L}=r_1(1+z)a_0
\end{equation}
and
\begin{multline}
r_1=\int_t^{t_0}\frac{dt}{a}= \int_t^{t_0}\frac{dt}{e^{\frac{2}{m}\text{arctanh}\left(\frac{kt}{m}-1\right)}}\\=\frac{1}{c_1(9m-1)}\left\{mt\left(\frac{2m}{kt}\right)^\frac{1}{m}\times_2F_1\left[1-\frac{1}{m},-\frac{1}{m},2-\frac{1}{m},\frac{kt}{2m}\right]\right\}_t^{t_0}
\end{multline}
here, $r_1$ is a function of time $t$ at which the light we see at present time $t_0$ is emitted by the object.
The DP $q$ in terms of $z$ is
\begin{equation}
q=2m-1-m\ \text{tanh}\left[\frac{m}{2}\text{ln}(z+1)-\text{arctanh}\left(\frac{1+q_0}{m}-2\right)\right].
\end{equation}
where, $q_0=q_{z=0}$ is the present DP.
\begin{figure}[H]
\centering
\begin{minipage}[b]{.45\textwidth}
\includegraphics[width=72mm]{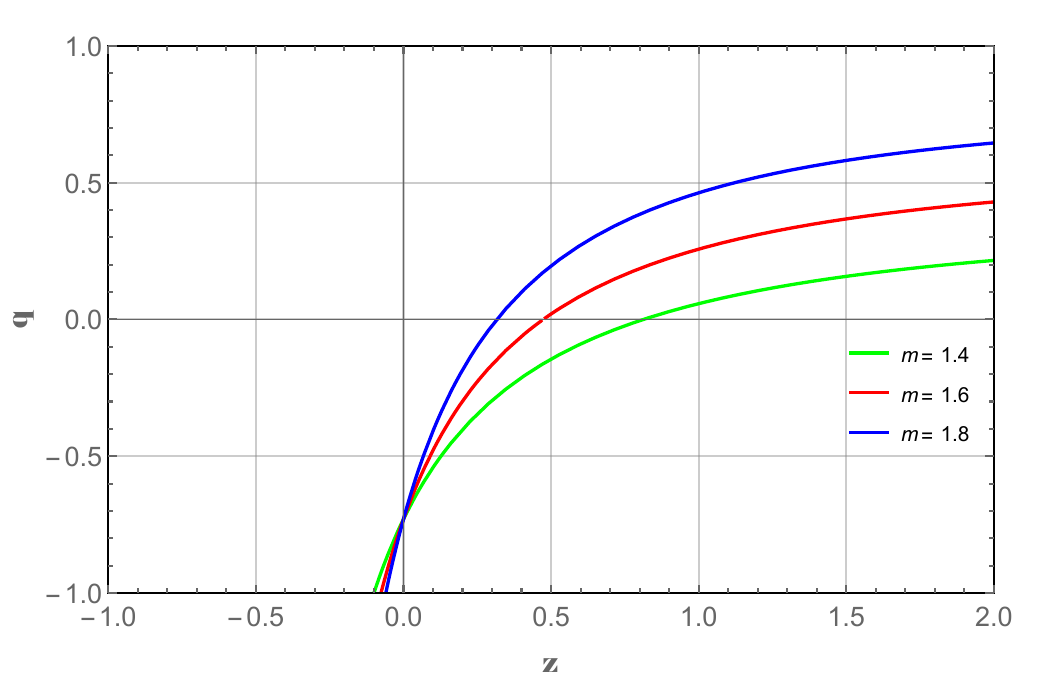}
\caption{Variation of $q$ versus $z$ with $q_0=-0.73$ and different $m$.}\label{ch5fig9}
\end{minipage}\qquad
\begin{minipage}[b]{.45\textwidth}
\includegraphics[width=72mm]{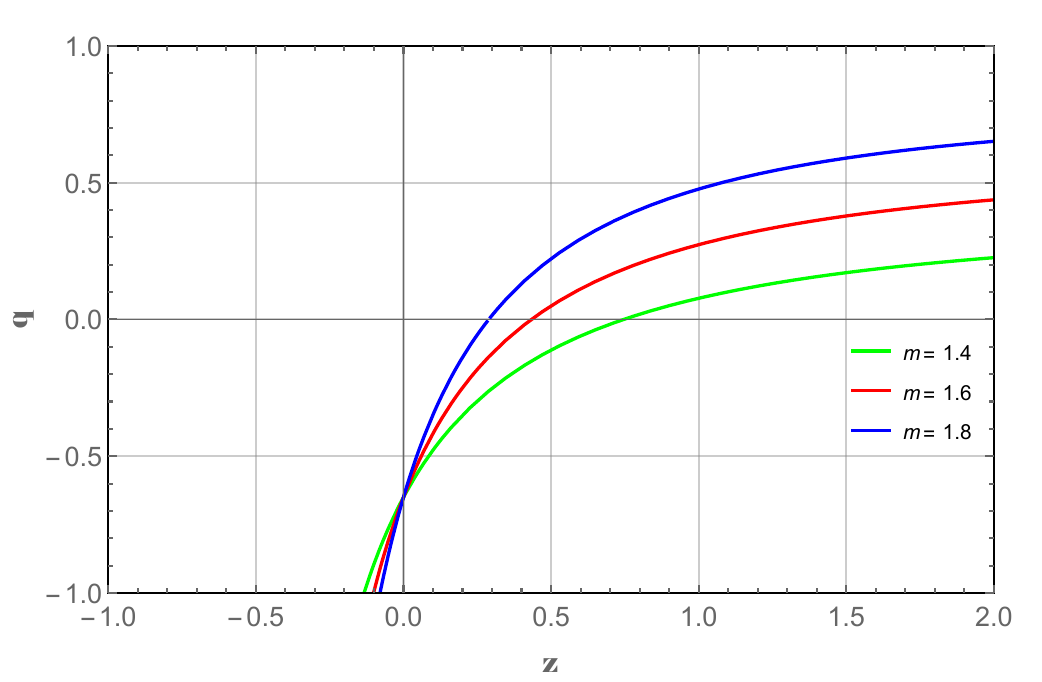}
\caption{Variation of $q$ versus $z$ with $q_0=-0.65$ and different $m$.}\label{ch5fig10}
\end{minipage}
\end{figure}
Here, we have considered two different values of $q_{z=0}$, $q_{z=0}=-0.73$ and $q_{z=0}=-0.65$ as per the kinematic data analysis of Cuhna \cite{Cunha09} and Li et al. \cite{Li11} in the Fig. \ref{ch5fig9} and Fig. \ref{ch5fig10} respectively. Again, for $q_{z=0}=-0.73$ the transition from deceleration to acceleration occurs at redshift $z_{tr}=0.82,\ \ 0.48,\ \ 0.327$ corresponding to $m=1.4,\ \ m=1.6,\ \ m=1.8$ respectively. Similarly, in the right figure for $q_{z=0}=-0.65$ the transition redshift values are $z_{tr}=0.75,\ \ 0.44,\ \ 0.29$ corresponding to $m=1.4,\ \ m=1.6,\ \ m=1.8$ respectively. Our $z_{tr}$ values of transition redshift fit with the observational data  \cite{Capozziello14,Capozziello15,Farooq17}.\\
In addition,  we intend to compare the model with $\Lambda$CDM model by plotting the evolution trajectories of the $\{q,j\}$ and $\{j,s\}$. The jerk parameter $j$ has the value
\begin{equation}
j=\frac{3 k^2 t^2}{2}-3 m (k t+1)+3 k t+2 m^2+1
\end{equation}
The $s$ parameter is defined as \cite{Akarsu14}
\begin{equation}
s=\frac{-6 m (k t+1)+3 k t (k t+2)+4 m^2}{6 (-k t+m-2)}
\end{equation}
\begin{figure}[H]
\centering
\begin{minipage}[b]{.45\textwidth}
\includegraphics[width=72mm]{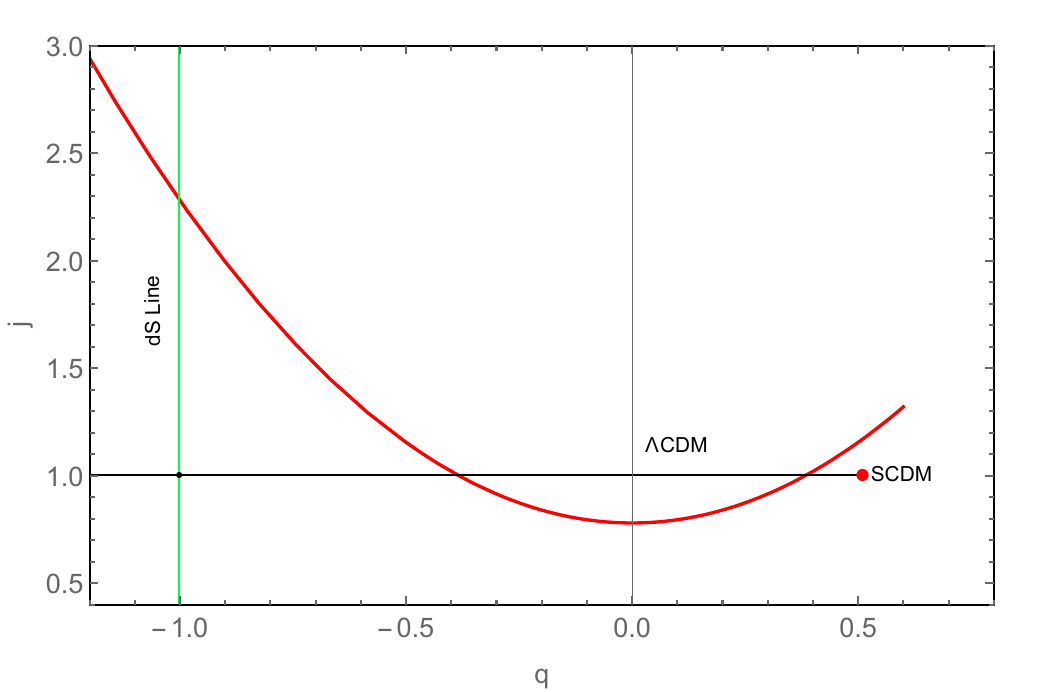}
\caption{Variation of $q$ versus $j$ with $m=1.6$ and $k=0.097$.}\label{ch5fig11}
\end{minipage}\qquad
\begin{minipage}[b]{.45\textwidth}
\includegraphics[width=72mm]{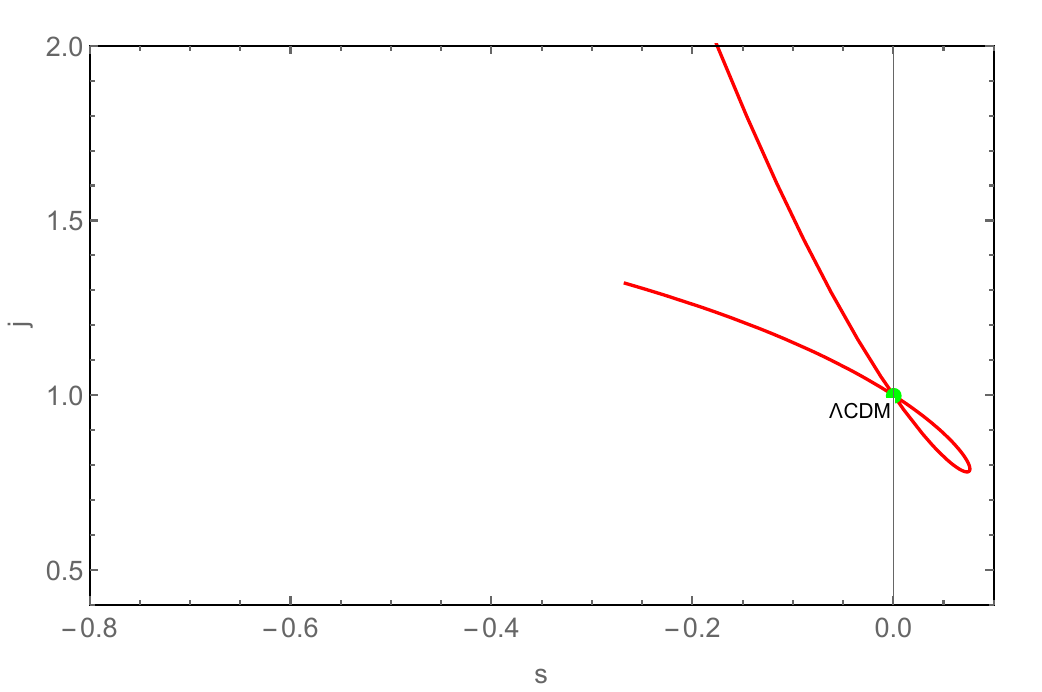}
\caption{Variation of $s$ versus $j$ with $m=1.6$ and $k=0.097$.}\label{ch5fig12}
\end{minipage}
\end{figure}
In Fig. \ref{ch5fig11} the vertical line is the de Sitter state at $q=-1$. The LVDP $q-j$ curve as shown in the Fig. \ref{ch5fig11} crosses the de Sitter line and going up due to Big Rip. Similarly, Fig. \ref{ch5fig12} shows the LVDP $s-j$ curve crosses the $\Lambda$CDM statefinder pair $(0,1)$ two times. Hence, we observe from both the figures that the LVDP model evolves and crosses the de Sitter line and reaches to the super-exponential expansion.
\section{Model II}\label{ch5model2}
In this model, we have considered the energy-momentum tensor for a perfect fluid distribution of the universe,
\begin{equation}\label{ch5ent2}
T_{\mu \nu}=(\rho+p)u_\mu u_\nu -pg_{\mu \nu}.
\end{equation}
Here, by considering $f(R,T)=R+2f(T)$, the corresponding field equations become
\begin{equation}
R_{\mu \nu}-\frac{1}{2}Rg_{\mu \nu}=\kappa T_{\mu \nu}+2f_T T_{\mu \nu}+\left[f(T)+2pf_T\right]g_{\mu \nu}, \label{eq:3}
\end{equation}
where $f_T$ denotes the partial derivative of $f$ with respect to $T$. Assuming $f(T)=\lambda T$, $\lambda$ being a constant, the field equations for a flat  FLRW metric (\ref{ch5met2}) with perfect fluid matter source (\ref{ch5ent2})
\begin{equation}\label{ch5met2}
ds^2=dt^2-a^2(t)\left(dx^2+dy^2+dz^2\right), 
\end{equation} 
are obtained as 
\begin{equation}\label{ch5eqn1}
3H^2=(1+3\lambda)\rho-\lambda p, 
\end{equation}
\begin{equation}\label{ch5eqn2}
2\dot{H}+3H^2=\lambda \rho-(1+3\lambda)p.
\end{equation}
In the above equations, we have chosen the unit system such that $\kappa=1$. It contains two equations with three unknowns, $\rho$, $p$, and $a$. In order to get an exact solution, we have employed a PVDP as below. \\
\textbf{Periodically varying deceleration parameter}\\
In the context of the late time cosmic acceleration phenomena with a cosmic transit from a phase of deceleration to acceleration at some redshift $z_{tr} \sim 1$, one can speculate a signature flipping of the DP. Obviously, at a decelerated phase, $q$ is positive and at the accelerating phase, it becomes negative. Geometrical parameters such as the DP and jerk parameter are usually extracted from observations of high redshift Supernova. However, the exact time dependence of these parameters are not known to a satisfactory extent. In the absence of any explicit form of these parameters, many authors have used parametrized forms especially that of the DP to address different cosmological issues. One of them is linear parametrization of the DP, which shows quite natural phenomena towards the future evolution of the universe, either it expands forever or ends up with Big Rip in finite future (see model \ref{ch5model1}). Such a parametrization has been used frequently in literature \cite{Akarsu2012, Sahoo15}. It is worth noting that the general dynamical behavior can be assessed through the values of the DP in the negative domain. While de Sitter expansion occurs for $q=-1$, for $-1<q<0$, power-law expansion is achieved, and for $q<-1$, a super-exponential expansion of the universe occurs. Even though, there is uncertainty in the determination of the DP from observational data, most of the studies in recent times constrain this parameter in the range $-0.8 \leq q \leq -0.4$. Keeping in view the signature flipping nature of $q$, in the present model, we assume a periodic time varying DP \cite{Shen2014} 
\begin{equation}\label{ch5dp2}
q = m\cos kt-1, 
\end{equation}
where $m$ and $k$ are positive constants. Here, $k$ decides the periodicity of the PVDP and can be considered as a cosmic frequency parameter. $m$ is an enhancement factor that enhances the peak of the PVDP. This model simulates a positive DP $q=m-1$  (for $m>1$) at an initial epoch and evolves into a negative peak of $q=-m-1$. After the negative peak, it again increases and comes back to the initial states. The evolutionary behavior of $q$ is periodically repeated. In other words, the universe in the model, starts with a decelerating phase and evolves into a phase of super-exponential expansion in a cyclic history.

Integration of eqn. (\ref{ch5dp2}) yields
\begin{equation}\label{ch5HP2}
H=\frac{k}{m\sin kt+k_1},
\end{equation}
where $k_1$ is a constant of integration. Using the definitions $q= -1-\frac{\dot{H}}{H^2}$ and $\dot{a}=aH$ we found $\dot{H}=-mH^2\cos kt$. Without loss of generality, we may consider $k_1=0$ and then Hubble function becomes 
\begin{equation}\label{ch5HP21}
H=\frac{k}{m\sin kt}.
\end{equation}
The scale factor $a$ is obtained by integrating the Hubble function in eqn. (\ref{ch5HP21}) as
\begin{equation}\label{ch5sp2}
a=a_0\left[\tan \left(\frac{1}{2}kt\right)\right]^{\frac{1}{m}},
\end{equation}
where $a_0$ is the scale factor at the present epoch  and can be taken as 1. Inverting eqn. (\ref{ch5sp2}), we obtain
\begin{equation}\label{ch5time2}
t=\frac{2 \tan ^{-1}\left[\frac{1}{(z+1)^m}\right]}{k}.
\end{equation}
\begin{figure}[H]
\centering
\begin{minipage}[b]{.45\textwidth}
\includegraphics[width=74mm]{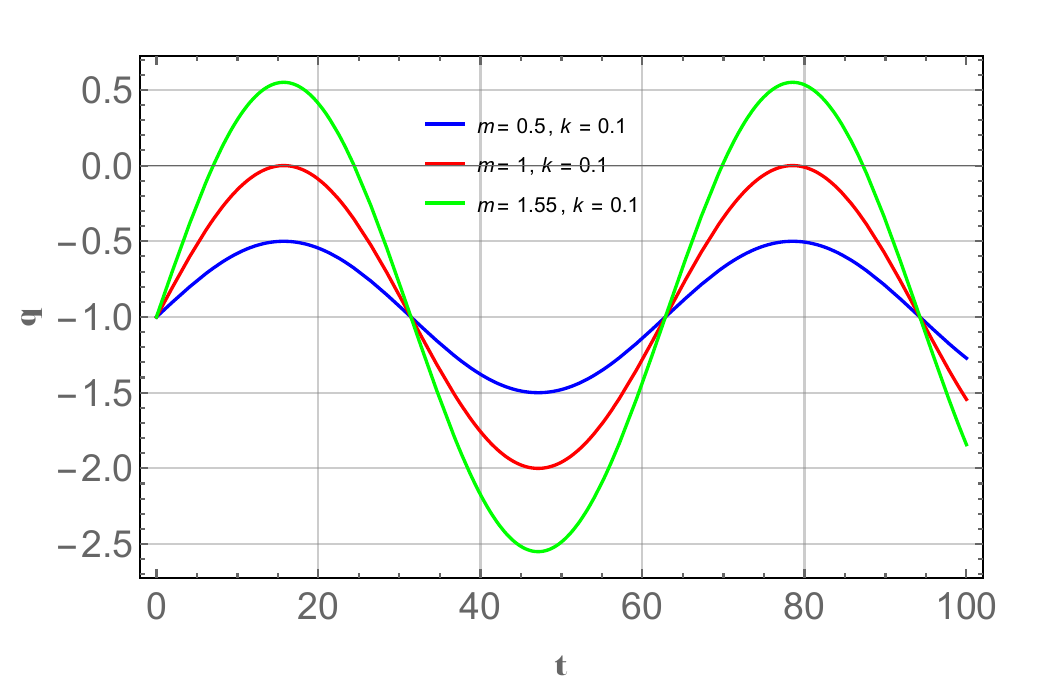}
  \caption{Evolution of DP for three representative values of $m$ and $k=0.1$.}\label{ch5fig13}
\end{minipage}\qquad
\begin{minipage}[b]{.45\textwidth}
\includegraphics[width=74mm]{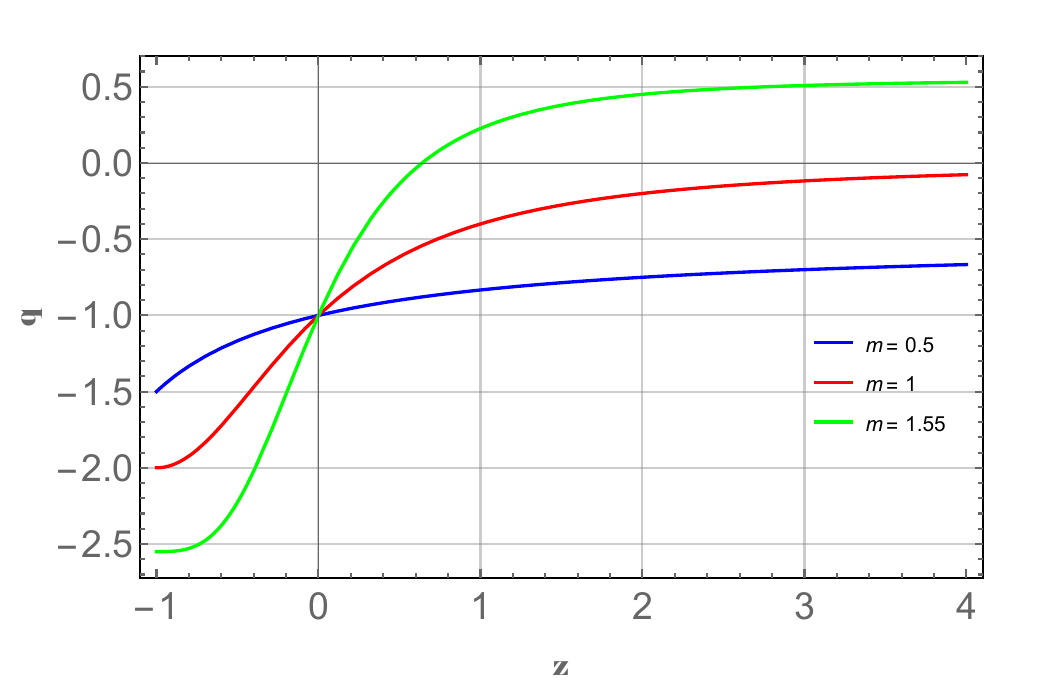}
  \caption{Evolution of DP as a function of redshift. The cosmic transit behavior is obtained for $m>1$.}\label{ch5fig14}
\end{minipage}
\end{figure}
Here, the redshift parameter $z$ is defined through the relation $z=\frac{1}{a}-1$. Hence for present epoch, we can derive from eqn. (\ref{ch5time2}) as $t=\left(\frac{8n+1}{k}\right)\frac{\pi}{2}$ corresponding to $a_0=1$, where $n=0,1,2,3,\cdots$ is a positive integer including zero. Then the evolutionary aspect of the DP as a function of cosmic time is shown in Fig. \ref{ch5fig13} for three different domain of the parameter $m$ namely $m<1, m=1$ and $m>1$. In which the periodic nature of the PVDP is clearly visible in the figure. Similarly the evolutionary aspect of the DP as a function of redshift is shown in Fig. \ref{ch5fig14}. It can be observed that the total evolutionary behavior of the PVDP is greatly affected by the choice of the parameter $m$.  In particular, the DP oscillates in between $m-1$ and $-m-1$, and for $m=0$, it becomes a constant quantity with a value of $-1$ which leads to a de-Sitter kind of expansion. It varies periodically in the negative domain and provides accelerated models for $0<m\leq 1$. However, for $m>1$, $q$ evolves from a positive region to a negative region showing a signature flipping at some redshift $z_{tr}$. Hence, it is worth mentioning here that, the parameter $m$ describes the transition redshift as per it's choice, and can be constrained from the cosmic transit behavior and transit redshift $z_{tr}$. We have adjusted the values of $m$ so as to get a $z_{tr}$ compatible with that extracted from observations \cite{Capozziello14, Capozziello15, Farooq17, Moraes16}. After that the signature flipping of DP is shown in Fig. \ref{ch5fig14}  for $m=1.55$ at $z_{tr}=0.64$. In the event of non availability of any observational data regarding cosmic oscillation and corresponding frequency, we consider $k$ as a free parameter. In the present model, we are interested for a time varying DP that oscillates in between the decelerating and accelerating phase to simulate the cosmic transit phenomenon. In order to assess the dynamical features of the model through numerical plots, we assume a small value for $k$, say $k=0.1$.
\begin{figure}[H]
\centering
\begin{minipage}[b]{.45\textwidth}
\includegraphics[width=74mm]{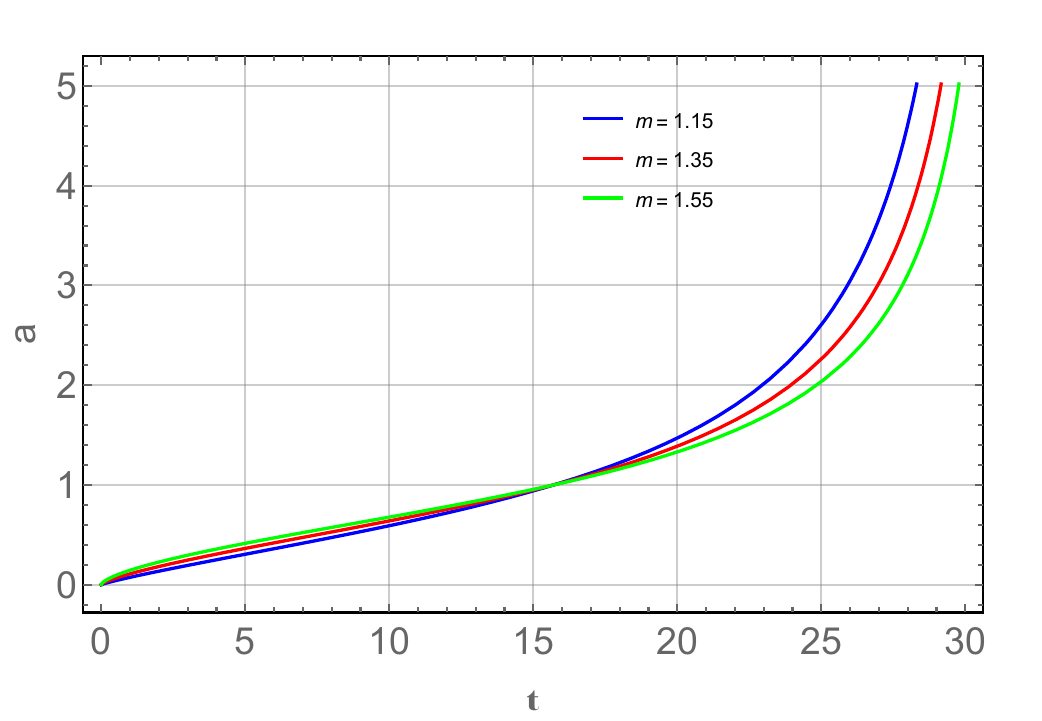}
  \caption{Scale factor as a function of cosmic time for $k=0.1$ and three representative values of $m$.}\label{ch5fig15}
\end{minipage}\qquad
\begin{minipage}[b]{.45\textwidth}
\includegraphics[width=74mm]{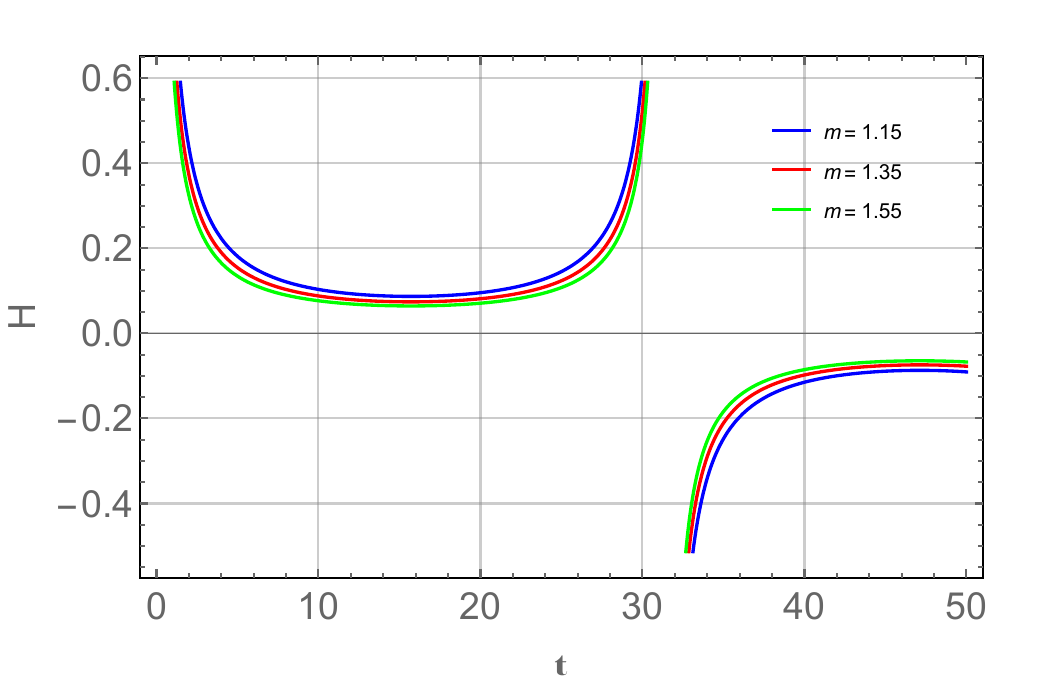}
  \caption{HP as a function of cosmic time for $k=0.1$ and three representative values of $m$.}\label{ch5fig16}
\end{minipage}
\end{figure}
In Fig. \ref{ch5fig15} and Fig. \ref{ch5fig16}, the scale factor and the HP are represented in some specific time frame. Within the specified time frame, the scale factor increases with cosmic time whereas the HP decreases with time. However, the evolutionary behavior of the scale factor is governed by a $tan$ function and that of the HP is governed by a $sine$ function and therefore both can either be positive or negative at some epoch. 
\subsection{Dynamical properties of the model}\label{IV}
The energy density and pressure are obtained from eqns. (\ref{ch5eqn1})- (\ref{ch5eqn2}) as
\begin{eqnarray}
\rho &=&\frac{(3+6\lambda)H^2-2\lambda \dot{H}}{(1+3\lambda)^2-\lambda^2},\label{ch5rho3}\\ 
p &=& \frac{-(3+6\lambda)H^2-2(1+3\lambda) \dot{H}}{(1+3\lambda)^2-\lambda^2}.\label{ch5p3}
\end{eqnarray}
For a PVDP as defined eqn. (\ref{ch5dp2}), the above expressions reduce to 
\begin{eqnarray}
\rho &=&\left[\frac{2\lambda m\cos kt+3(2\lambda+1)}{(3\lambda+1)^2-\lambda^2}\right]\frac{k^2}{m^2\sin^2 kt},\label{ch5rho4}\\
p &=& \left[\frac{2(3\lambda+1) m\cos kt-3(2\lambda+1)}{(3\lambda+1)^2-\lambda^2}\right]\frac{k^2}{m^2\sin^2 kt}.\label{ch5p4}
\end{eqnarray}
\begin{figure}[H]
\centering
\begin{minipage}[b]{.45\textwidth}
\includegraphics[width=74mm]{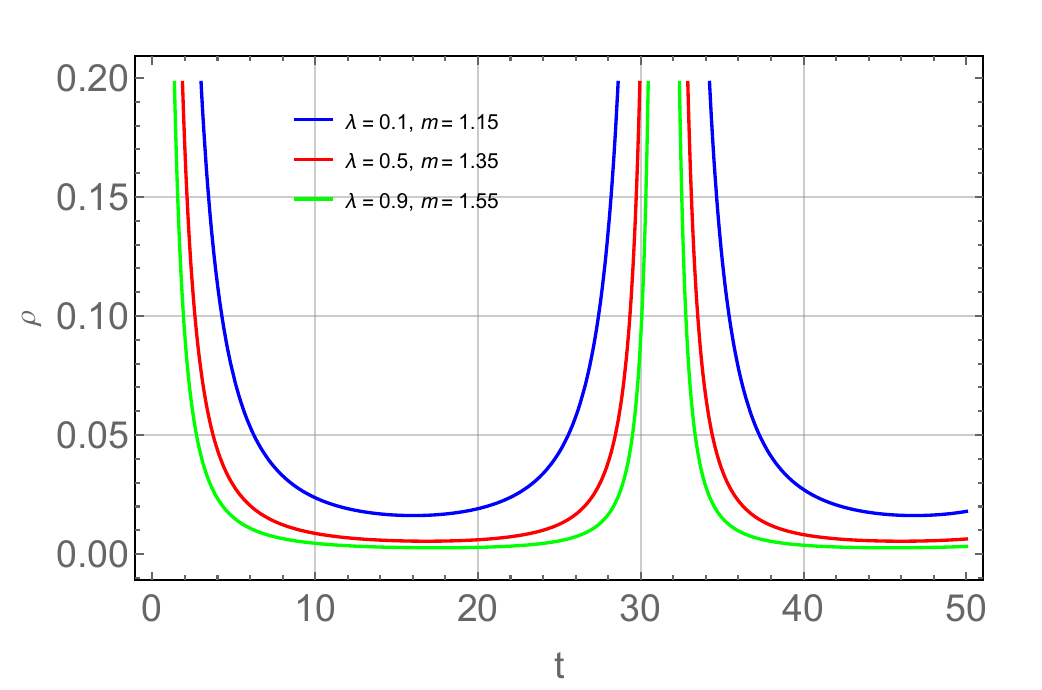}
  \caption{Time variation of energy density.}\label{ch5fig17}
\end{minipage}\qquad
\begin{minipage}[b]{.45\textwidth}
\includegraphics[width=74mm]{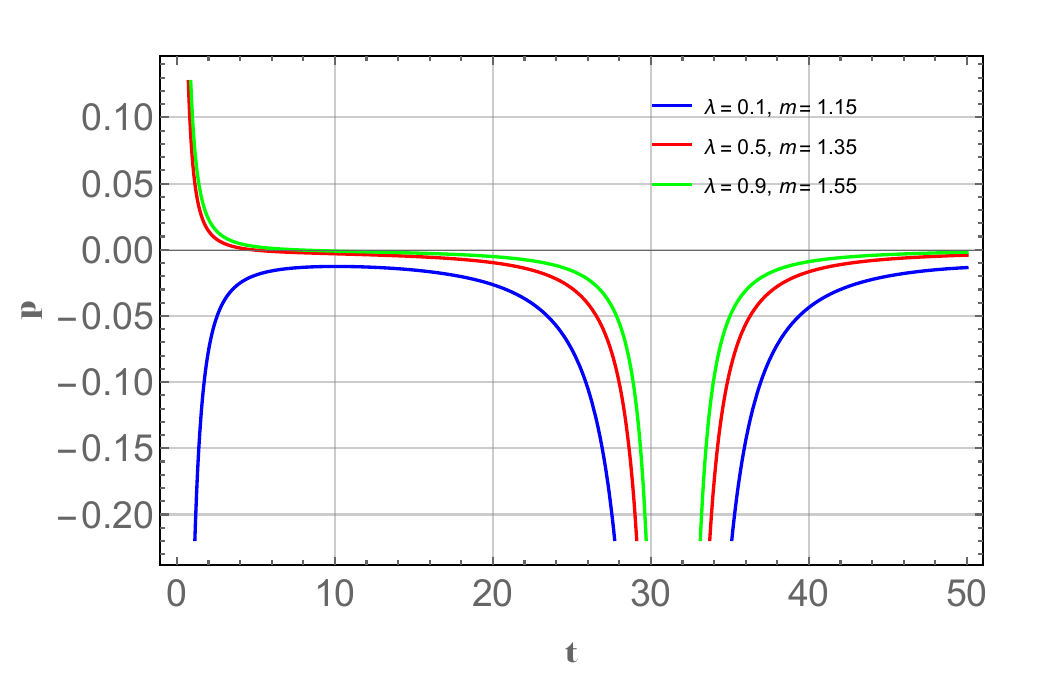}
  \caption{Time variation of pressure.}\label{ch5fig18}
\end{minipage}
\end{figure}
In order to get a positive energy density, eqn. (\ref{ch5rho4}) sets up a condition for the parameters $\lambda$ and $m$ as
\begin{equation}\label{ch5cond1}
6\lambda+3 > 2\lambda m.
\end{equation}
For a signature flipping behavior of DP, we need to fix 
$m$ to be greater than 1 (refer Fig. \ref{ch5fig14}). 
So we have constrained $m$ from the cosmic transit redshift $z_{tr}$ to be $1.55$. From eqn.  (\ref{ch5cond1}), it is certain that, this value of $m$ allows any positive values for $\lambda$. In view of this, one may take $\lambda$ as a free parameter with positive values only. For this purpose we have considered here three moderate values, $\lambda = 0.1, 0.5$ and $0.9$ for numerical calculation of the dynamical parameters. The periodic evolution of energy density with respect to time is shown in Fig. \ref{ch5fig17} for the considered values of $\lambda$. Here, it can be observed that the energy density has periodic singularities at the cosmic times $t=\frac{n\pi}{k}$, $n=0,1,2,3,\cdots$ is an integer, and this periodic variation clearly depends on the choice of $k$.  Since we have taken $k=0.1$, the cosmic singularity occurs corresponding to the time period, $t=0, 31.4, 62.8, \cdots$. The interesting feature here is that, in a given cosmic cycle, it starts from a very large value at an initial time ($t\rightarrow 0$) and decreases to a minimum, $\rho_{min}$, and again increases with the growth of time. The minimum in energy density occurs at a time given by $t=\frac{(n+1)\pi}{2k}$. The evolutionary trend of the energy density is not changed by a variation of $\lambda$,  rather an increase in $\lambda$ simply decreases the value of $\rho$ at a given time. In other words, with an increase in $\lambda$, there occurs a decrements in $\rho_{min}$.

At the same time, the evolutionary behavior of pressure is depicted in Fig. \ref{ch5fig18} and shows same periodic variation with singularities at $t=\frac{n\pi}{k}$ or at  $t=0,31.4, 62.8, \cdots$. Within a given cycle, pressure decreases from large positive values at the beginning to large negative values and then reverses the trend. However, pressure is a negative quantity at the present epoch in a given cycle. The choice of the parameter $\lambda$ has some effects on the evolutionary trend. In a given cycle, in general, lower value of $\lambda$ results in a pressure curve that lies to the left side in the figure. The crossing over time from positive domain to negative domain is decided by the value of $\lambda$. 

The EoS parameter, $\omega=\frac{p}{\rho}$ can be obtained from eqns. (\ref{ch5rho4}) and (\ref{ch5p4}) as
\begin{equation}\label{ch5EoS2}
\omega=\frac{2 (3 \lambda +1) m \cos k t-3(2\lambda +1)}{2 \lambda m \cos k t+3(2 \lambda +1)}.
\end{equation}

\begin{figure}[H]
\centering
\begin{minipage}[b]{.45\textwidth}
\includegraphics[width=74mm]{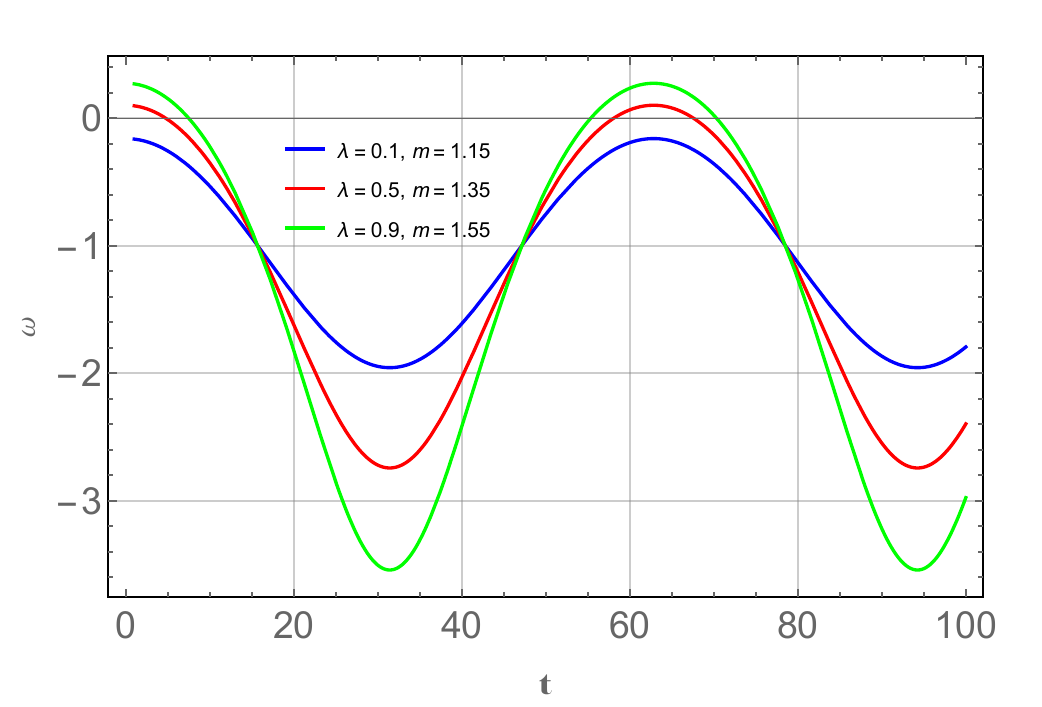}
  \caption{EoS parameter as function of cosmic time.}\label{ch5fig19}
\end{minipage}\qquad
\begin{minipage}[b]{.45\textwidth}
\includegraphics[width=74mm]{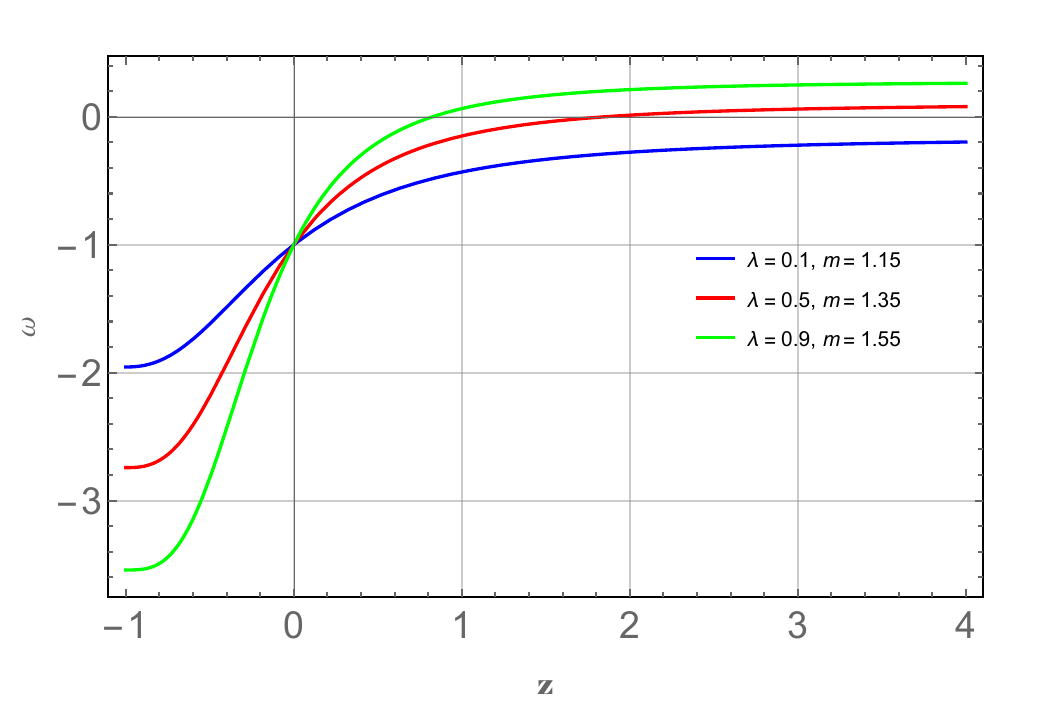}
  \caption{EoS parameter as function of redshift.}\label{ch5fig20}
\end{minipage}
\end{figure}
\begin{figure}[H]
\centering
\includegraphics[width=78mm]{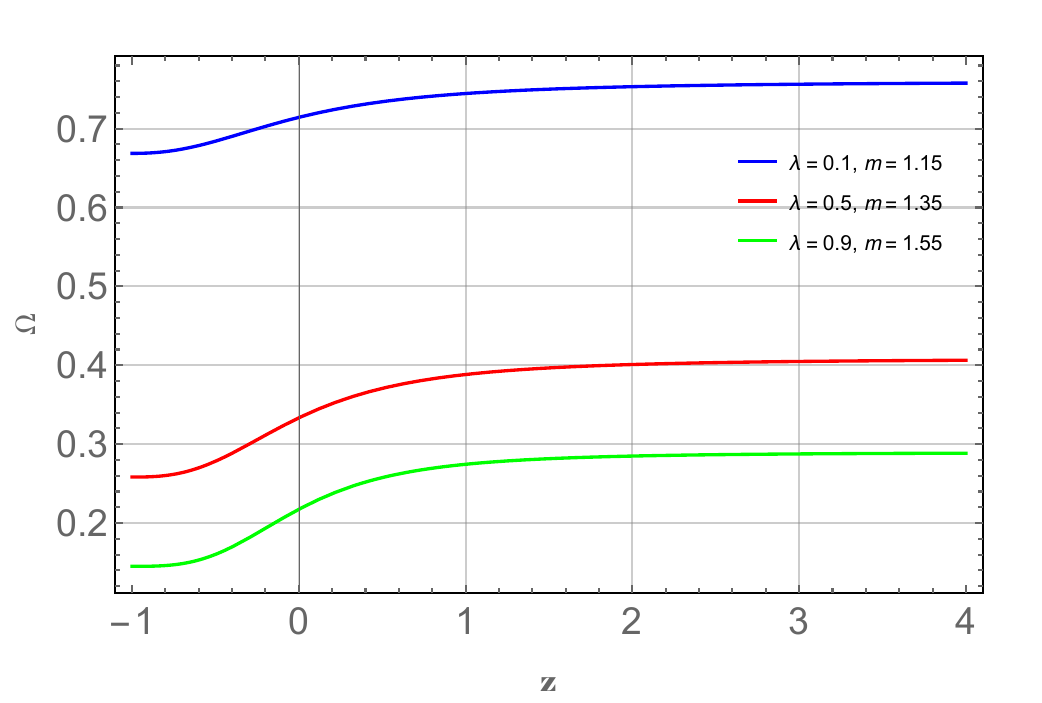}
\caption{Evolution of Density parameter.}\label{ch5fig21}
\end{figure}

In Fig. \ref{ch5fig19}, the plot of EoS parameter $\omega$ with respect to time, exhibits an oscillatory behavior. In the first half of the cosmic cycle, $\omega$ decreases from a positive value close to $\frac{1}{3}$ to negative values after crossing the phantom divide at $\omega =-1$. After attending a minimum it again increases to positive value at the end of the cycle. One interesting feature of the EoS parameter is that, unlike the energy density and pressure, it does not acquire any singular values during the cosmic cycle. This fact is due to the cancellation of the $\frac{1}{\sin^2 kt}$ factor from the pressure by the same factor of energy density and depends only the value of the PVDP. Since the PVDP does not have singularity, the same thing occurs in the EoS parameter. The oscillatory behavior comes only from the $\cos kt$ factor appearing both in the numerator and denominator of $\omega$. The evolutionary trend is affected by the choice of $\lambda$. Curves of $\omega$ with low values of $\lambda$ remain on top before the phantom divide whereas after the phantom divide, they remain in the bottom of all the curves. At an equivalent present epoch $\left(t=\frac{8n+1}{k}\frac{\pi}{2}\right)$ in any given cycle, the EoS parameter remains within the quintessence region with a value close to $-1$. One can decipher the detail evolution of the EoS parameter from chapter \ref{Chapter1}. At the present epoch, the model predicts an EoS that behaves more like a cosmological constant and the model is somewhat close to that of $\Lambda$CDM model. This aspect of the EoS parameter is clearly visible in Fig. \ref{ch5fig20}. It is clear from Fig. \ref{ch5fig20} that, at $z=0$, $\omega =-1$. It is interesting to note that, the overlapping of the present model with $\Lambda$CDM at equivalent present epochs is independent of the choice of the parameter $\lambda$. In this model from Fig. \ref{ch5fig20}, it can be observed that the transitional evolution of $\omega$ against redshift from  $\omega <-1$ at low redshift to $\omega>-1$ at higher redshift. This transitional behavior of $\omega$ in this model fits with SNL3 data \cite{Feng/2005}. In particular, the use of SNL3 data suggests that BAO data is also partly responsible for this. Some recent reconstruction of the EoS from different observational data sets including the high redshift Lyman-$\alpha$ forest (Ly$\alpha$FB) measurement favor a non-constant dynamical DE. In these reconstructed models, the EoS evolves with time and crosses the phantom divide \cite{Zhao12, Zhao17}. The behavior of EoS $\omega$ in this model is consistent with quintom model which allows  $\omega$ to cross $-1$. As inferred in ref. \cite{Zhao17}, the departure from $\omega=-1$ is more evident in the reconstruction history of the dynamical DE with more recent data sets including the Ly$\alpha$FB measurement.\\
For the present model with a PVDP, we obtain, the density parameter $\Omega = \frac{\rho}{3H^2}$ as 
\begin{equation}\label{ch5Omega1}
\Omega =\frac{1}{3}\left[\frac{2\lambda m\cos kt+3(2\lambda+1)}{(3\lambda+1)^2-\lambda^2}\right],
\end{equation}
which can be expressed in terms of redshift as
\begin{equation}\label{ch5Omega2}
\Omega =\frac{1}{3}\left[\frac{2\lambda m\cos \left(2 \tan ^{-1}\frac{1}{(z+1)^m}\right)+3(2\lambda+1)}{(3\lambda+1)^2-\lambda^2}\right].
\end{equation}
In Fig. \ref{ch5fig21}, we have plotted the density parameter as a function of redshift for three representative values of $\lambda$. $\Omega$ remains almost unaltered in the range of redshift greater than 1 for all the three values of $\lambda$ considered in this model. However, with the cosmic evolution,  $\Omega$ decreases with cosmic time after $z=1$. The density parameter, at a given redshift, is observed to have lower value for higher values of $\lambda$.
\subsection{Violation of energy-momentum conservation}\label{V}
Friedman models in GR ensure the energy conservation through the continuity equation
\begin{equation}
\dot{\rho}+3H(\rho+p)=0, 
\end{equation}
which implies $\text{d}(\rho V)=-p\text{d}V$. Here, $V=a^3$, the volume scale factor of universe and the quantity $\rho V$ give an account of the total energy. As the universe expands the amount of DE in an expanding volume increases in proportion to the volume. If that space-time is standing completely still, the total energy is constant; if it is evolving, the energy changes in a completely unambiguous way. It either decreases or increases in time. However, in modified gravity theories, one may get a different picture. Taking a covariant derivative of eqn. (\ref{eqnfld1}) (see chapter \ref{Chapter1}), one can obtain \cite{Harko14,alva13,bar14, das17}
\begin{equation} \label{ch5eq:21}
\nabla^{\mu}T_{\mu \nu}=\frac{F_T(R,T)}{\kappa-F_T(R,T)}\biggl((T_{\mu \nu}+\Theta_{\mu \nu})\nabla^{\mu} ln F_T(R,T)+ \nabla^{\mu}\Theta_{\mu \nu} -\frac{1}{2}g_{\mu \nu}\nabla^{\mu}T\biggr).
\end{equation}
With the substitution of $f(R,T)=R+2\lambda T$ and $\kappa =1$, eqn. (\ref{ch5eq:21}) reduces to
\begin{equation}
\nabla^{\mu}T_{\mu \nu}=-\frac{2\lambda}{1+2\lambda} \left(\nabla^{\mu}(pg_{\mu \nu})+\frac{1}{2}g_{\mu \nu} \nabla^{\mu}T\right).\label{eq:22}
\end{equation}
It is worth noting here, for $\lambda =0$, one would get $\nabla^{\mu}T_{\mu \nu} =0$. However for $\lambda \neq 0$, the conservation of energy-momentum is violated. Recently some researchers have investigated the consequence of the violation of energy-momentum conservation (i.e. $\dot{\rho}+3H(\rho+p) \neq0 $) in modified gravity theories. The non-conservation of energy-momentum may arise due to non unitary modifications of quantum mechanics and in phenomenological models motivated by quantum gravity theories with space-time discreteness at the Planck scale \cite{Josset17}. In the context of unimodular gravity, Josset et al. \cite{Josset17} have shown that a non-conservation of energy-momentum leads to an effective cosmological constant which increases or decreases with the creation or annihilation of energy during the cosmic expansion and can be reduced to a constant when matter density diminishes. Shabani and Ziaie \cite{Shabani17} have studied the same in some classes of $f(R,T)$ gravity with pressure-less cosmic fluid and showed a violation of energy-momentum conservation in modified theories of gravity can provide accelerated expansion. Also, the non-conservation of the energy-momentum tensor implies in non-geodesic motions for test particles in gravitational fields as it is deeply investigated in \cite{Baffou17}. In \cite{Shabani18,Moraes18} the authors have constructed a formalism in which an effective fluid is conserved in $f(R,T)$ gravity, rather than the usual energy-momentum tensor non-conservation. We have investigated the non-conservation of energy-momentum for the class of models in $f(R,T)$ gravity with the suggested PVDP. We quantify the violation of energy-momentum conservation through a deviation factor $S$, defined as
\begin{equation}\label{ch5S}
S=\dot{\rho}+3H(\rho+p).
\end{equation}
\begin{figure}[H]
\centering
\includegraphics[width=78mm]{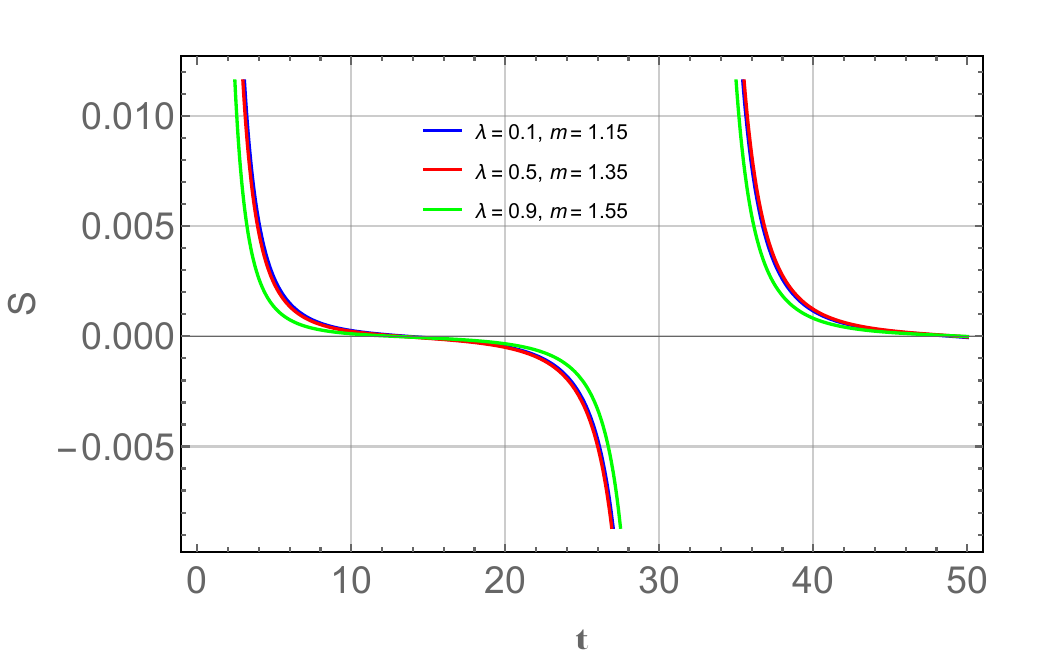}
\caption{Energy-momentum non conservation.}\label{ch5fig22}
\end{figure}
In case of the model, satisfying energy-momentum conservation, we have $S=0$, otherwise, we get a non zero value for this quantity. $S$ can be positive or negative depending on whether the energy flows away from or into the matter field. In Fig. \ref{ch5fig22}, the non-conservation of energy-momentum is plotted for a periodic cosmic cycle. It is clear that, except for a very limited period, the conservation is violated along with the cosmic evolution. However, the nature of energy flow changes periodically. This behavior is obtained for all the values of $\lambda$ taken in the model. At an equivalent present epoch in a given cosmic cycle, at least within the purview of the present model, there is a signal of non-conservation.
\subsection{Energy conditions}\label{VI}
The ECs of GR are a variety of different ways to show positive energy density more precisely. The ECs take the form of various linear combinations of the stress-energy tensor components (at any specified point in space-time) should be positive, or at least non-negative (see section \ref{ch1ECs} for details expressions). The ECs are the fundamental tools for the study of wormholes (WHs) and black holes in different physical scenario. The study of singularities in the space-time is based on ECs. 

\begin{figure}[H]
\centering
\begin{minipage}[b]{.45\textwidth}
\includegraphics[width=74mm]{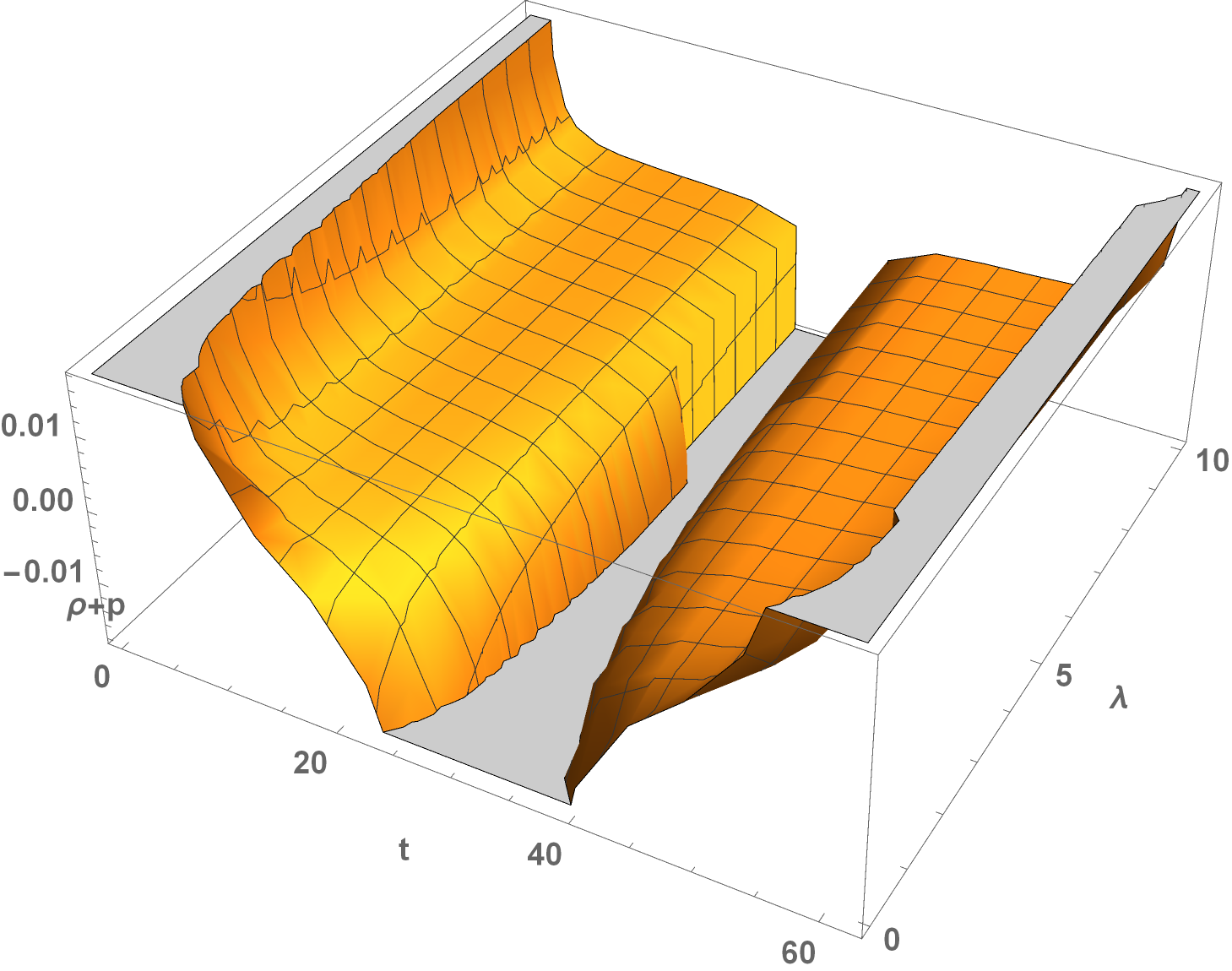}
  \caption{ Violation of NEC ($\rho+p \geq 0$) versus $\lambda$ and $t$.}\label{ch5fig23}
\end{minipage}\qquad
\begin{minipage}[b]{.45\textwidth}
\includegraphics[width=74mm]{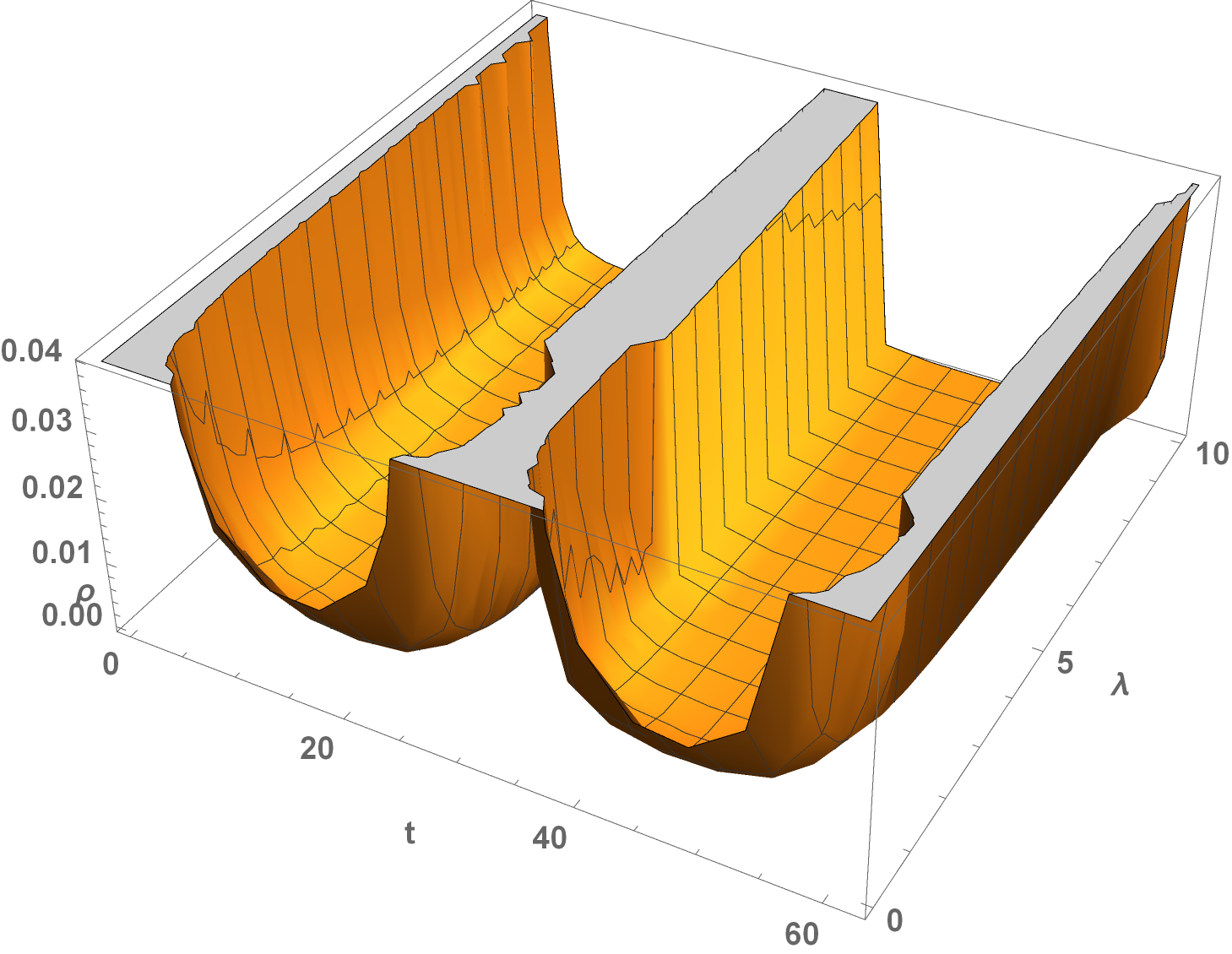}
  \caption{$\rho \geq 0$ versus $\lambda$ and $t$.}\label{ch5fig24}
\end{minipage}
\end{figure}

\begin{figure}[H]
\centering
\begin{minipage}[b]{.45\textwidth}
\includegraphics[width=74mm]{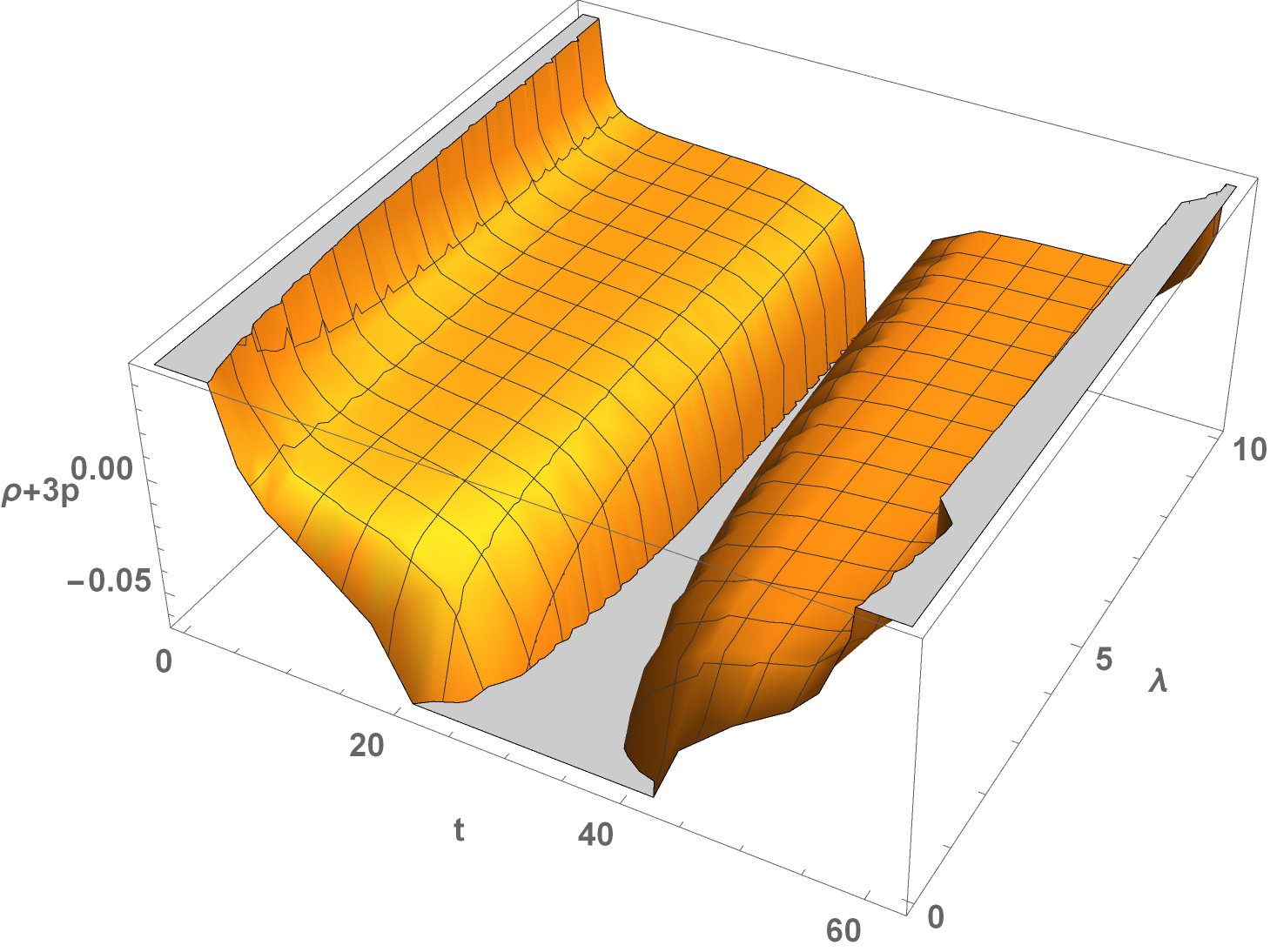}
  \caption{Violation of SEC ($\rho+3p \geq 0$) versus $\lambda$ and $t$.}\label{ch5fig25}
\end{minipage}\qquad
\begin{minipage}[b]{.45\textwidth}
\includegraphics[width=74mm]{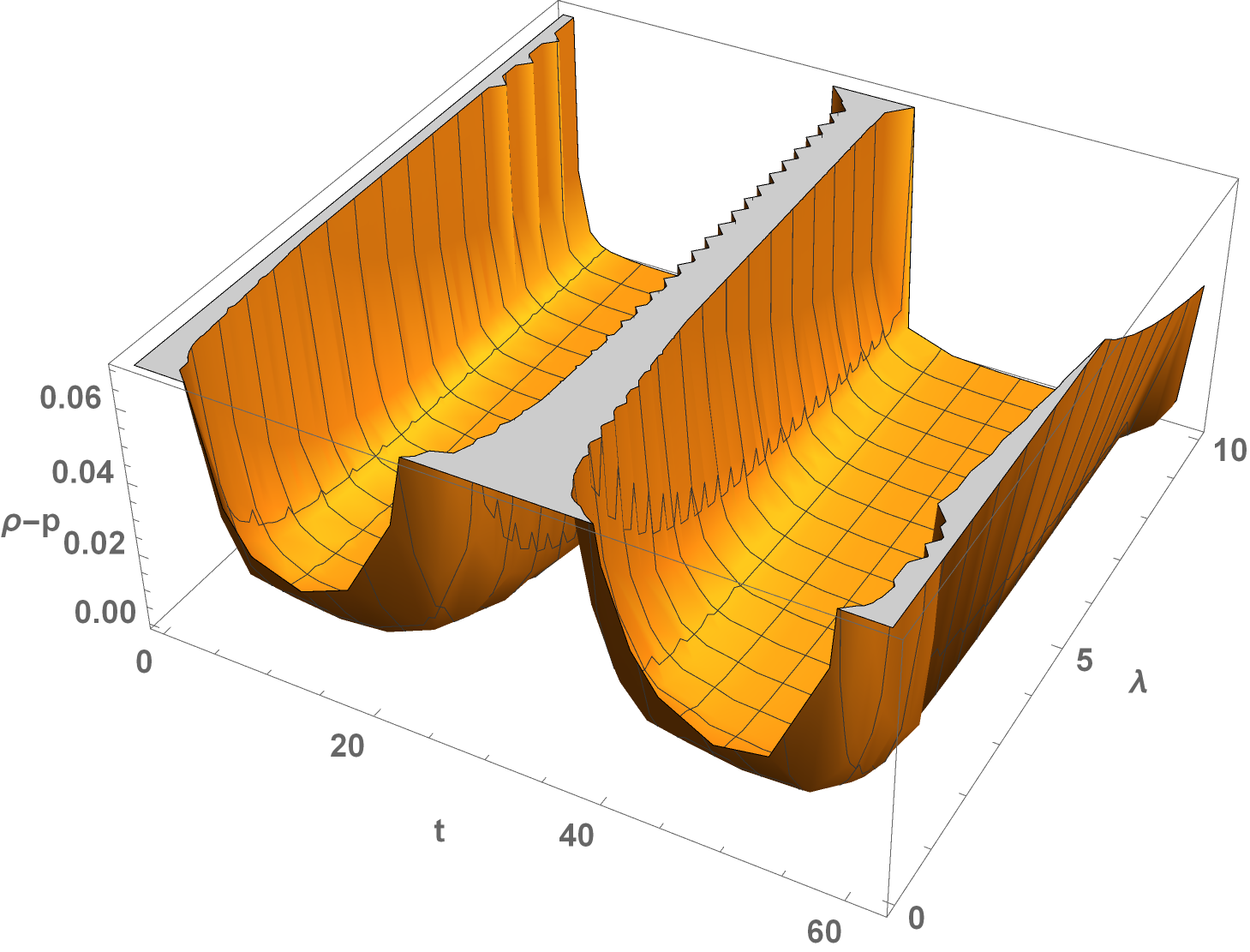}
  \caption{DEC, $\rho \geq \vert p\vert $ versus $\lambda$ and $t$.}\label{ch5fig26}
\end{minipage}
\end{figure}

In fact, Alvarenga et al. \cite{Alvarenga/2012} and Sharif et al. \cite{Sharif13} have analyzed the ECs in $f(R,T)$ gravity. The ECs for this model are plotted in Fig. \ref{ch5fig23} to Fig. \ref{ch5fig26} with $m=1.55$, $k=0.1$ and varying $\lambda$.
It can be observed from the above Fig. \ref{ch5fig23} to Fig. \ref{ch5fig26} that all the ECs are behaving periodically for fixed values of $m$ and $k$ with accepted range of $\lambda$ in this model. The free parameters are considered, based on the positivity of energy density as shown in Fig. \ref{ch5fig24}. In the present model DEC is satisfied and all other ECs are violated.
\subsection{Stability analysis}\label{VIIa}
In this section we wish to analyze the stability of the model under linear homogeneous perturbations in the FLRW background. We consider linear perturbations for the HP and the energy density as \cite{Sharif 013}
\begin{eqnarray}
H(t)&=& H_\text{b}(t)\left(1+\delta(t)\right),\label{ch5eq:28}\\
\rho(t)&=& \rho_b\left(1+\delta_\text{m}(t)\right),\label{ch5eq:29}
\end{eqnarray}
where $\delta (t)$ and $\delta_\text{m}(t)$ are the perturbation parameters. Here, we have assumed a general solution $H(t)=H_\text{b}(t)$ which satisfies the background FLRW equations. The matter energy density can be expressed in terms of $H_b$ as 
\begin{equation}
\rho_\text{b} = \frac{(3+6\lambda)H_\text{b}^2-2\lambda \dot{H}_\text{b}}{(1+3\lambda)^2-\lambda^2}.
\end{equation}
The Friedman equation and the trace equation for the modified gravity model with a functional $f(R,T)=R+2\lambda T$ can be obtained as
\begin{eqnarray}
\theta^2 &=& 3[\rho+2\lambda (\rho+p)+f(R,T)],\label{ch5eq:30}\\
R &=& -(\rho-3p)-2\lambda (\rho+p)-4f(R,T).\label{ch5eq:31}
\end{eqnarray}
Here, $\theta=3H$ is the expansion scalar. For a standard matter field, we can have the first order perturbation equation
\begin{equation}\label{ch5eq:32}
\dot{\delta}_\text{m}(t)+3H_\text{b}(t)\delta(t)=0.
\end{equation}
Using eqns. (\ref{ch5eq:28})- (\ref{ch5eq:30}), one can obtain
\begin{equation}\label{ch5eq:33}
(1+3\lambda)T\delta_\text{m}(t)=6H_\text{b}^2\delta(t).
\end{equation}
The first order matter perturbation equation can be obtained by the elimination of $\delta(t)$ from eqns. (\ref{ch5eq:32}) and (\ref{ch5eq:33}) as
\begin{equation}\label{ch5eq:34}
\dot{\delta}_\text{m}(t)+\frac{T}{2H_b}\left(1+3\lambda\right)\delta_\text{m}(t)=0.
\end{equation}
Integration of eqn. \ref{ch5eq:34} leads to 
\begin{equation}
\delta_\text{m}(t)= C \exp \left[-\left(\frac{1+3\lambda}{2}\right)\int \frac{T}{H_\text{b}}~\text{d}t \right],
\end{equation}
where $C$ is a non zero positive constant. Consequently, the evolution of the perturbation $\delta(t)$ becomes
\begin{equation}
\delta(t)=\frac{(1+3\lambda)CT}{6H_\text{b}^2} \exp \left[-\left(\frac{1+3\lambda}{2}\right)\int \frac{T}{H_\text{b}}~\text{d}t \right].
\end{equation}
Since 
\begin{equation}
\frac{T}{H_\text{b}}=\frac{-k}{[(1+3\lambda)^2-\lambda^2]}\left[\frac{(16\lambda+6)\cos kt}{\sin kt}+\frac{6(2\lambda+1)}{m~\sin kt}\right],
\end{equation}
the factor $\int \frac{T}{H_\text{b}}~\text{d}t$ is evaluated as
\begin{equation}
\int \frac{T}{H_\text{b}}\text{d}t= -\frac{\splitfrac{(6 \lambda +(8 \lambda +3) m+3) \log (1-\cos (k t))}{+(-6 \lambda +(8 \lambda +3) m-3) \log (\cos (k t)+1)}}{\left(8 \lambda ^2+6 \lambda +1\right) m}.
\end{equation}
The growth and decay of the perturbation depend on the factors $k$ and $\lambda$ periodically. We have found that the considered values for $k$ and $\lambda$ in the physical parameters $\rho$, $p$ and $\omega$ are compatible with the decay of perturbation.
\section{Conclusion}\label{ch5conclusion}
The conclusion of this chapter is based on two cosmological models in the linear frame of $f(R,T)$ gravity. In the first model, we have investigated LRS Bianchi type I model with MSQM. While the second model consists of the background cosmology of an isotropic flat universe with a perfect fluid matter source. The exact solution of the first model is obtained by using HP, EoS for SQM and a LVDP. Similarly, in the second model we have employed a PVDP to obtain the exact solution. Since the present universe undergoes an accelerated expansion, the universe might have transitioned from an early decelerated phase to an accelerated phase. This behavior clearly hints towards a time varying DP which should evolve from a positive value in past to negative values at present time. That means, evolving DP displays a signature flipping behavior. Therefore, keeping in view the signature flipping nature of the DP, model II deals with a periodically varying DP to reconstruct the cosmic history. The model I shows the transitional behavior of universe with future singularity at finite time. The results of model I are summarized in the following paragraph.\\
In Fig. \ref{ch5fig3}, it is clearly seen that magnetic flux is effective and non-vanish for LRS Bianchi type I universe model and changes with cosmic time as $h^2\rightarrow 0$ when $t\rightarrow \infty$. It can be interpreted that at the end of the universe the magnetic field may lose its effect. In addition, the Bag constant $B_{c} $ is effective on pressure, density and cosmological constant. The Bag constant $B_{c}$ increases the energy density value towards positive, it decreases towards negative for cosmic pressure value. Also,  we get constant density for $t\rightarrow 0$, and when $t$ increases, we obtain $\rho \rightarrow B_c$. In this model, we obtain small, constant and negative cosmological constant value as $\Lambda=-(8\pi+4\lambda)B_c$. When $t$ increases we get negative pressure value, i.e., $p \rightarrow -B_c$.\\ 
On the basis of these results, we can claim that SQM may be a source of DE also we may agree with strange quark stars because of the obtained constant pressure and density in this model. However, these results are compatible with the previous study of \cite{can} in $f(R,T)$ gravitation theory.\\
In the second model assumed in this chapter, of the two adjustable parameters of PVDP, one of them can be constrained by the cosmic transit behavior. The assumed DP oscillates in between two limits usually set by the transit redshift. Consequently, the universe in this model, starts with a decelerating phase and evolves into a phase of super-exponential expansion with a periodic repetition of the phenomenon. Further, the energy density and pressure vary cyclically within a given cosmic period decided by the cosmic frequency parameter of the model. At some finite time, the magnitude of these physical parameters become infinitely large. This behavior leads to a future type I singularity as classified by Nojiri et. al. \cite{Nojiri/2005}. There appears to be a Big Rip at certain finite time during the cosmic repetition of the phenomenon, because $a\rightarrow \infty, \rho \rightarrow \infty$ and $|p|\rightarrow \infty$. Since the parameters repeats their behavior after a time period $t=\frac{n\pi}{k}$, the Big Rip also occurs  periodically after a time gap of $t=\frac{n\pi}{k}$.\\
The EoS parameter for the PVDP has a cyclic behavior that repeat with time.  It may cross the phantom divide $\omega=-1$ for some cosmic time range. How far the well of the EoS parameter will go beyond the phantom divide is decided by the coupling constant $\lambda$. At present, the model predicts an EoS that behaves more like a cosmological constant.\\
In addition, the violation of energy-momentum conservation is also investigated in this chapter, which is an important aspect of $f(R,T)$ gravity theory. As pointed out by Josset and Perez \cite{Josset17}, modified gravity models can explain the accelerated expansion at the cost of energy-momentum non conservation. We have shown that, a PVDP leads to a kind of universe model, within a given cycle, the energy momentum conservation is continuously violated except for a small period of cosmic time. Also, we have shown the violation/validity of the ECs both analytically as well as graphically in section \ref{VI}. Finally, we have discussed the stability of the solutions under linear homogeneous perturbations. The stability depends on the values of the parameters $k$ and $\lambda$ as they behaves periodically. Hence, we can say that the model is appropriate to investigate some ultimate fate of the universe.

\chapter{$f(R,T)$ gravity models with higher order curvature scalar} 

\label{Chapter6} 

\lhead{Chapter 6. \emph{$f(R,T)$ gravity models with higher order curvature scalar}} 

This chapter \blfootnote{The work, in this chapter, is covered by the following two publications: \\
 \textit{$f(R,T)=f(R)+\lambda T$ gravity models as alternatives to cosmic acceleration}, European Physical Journal C, \textbf{78} (2018) 736.\\
\textit{Wormholes in $R^2$-gravity within the $f(R, T)$ formalism}, European Physical Journal C, \textbf{78} (2018) 46.} presents cosmological models that arise in a subclass of $f(R,T)=f(R)+f(T)$ gravity models, with different $f(R)$ functions and fixed $T$-dependence. That is, the gravitational Lagrangian is considered as $f(R,T)=f(R)+\lambda T$, with constant $\lambda$.  The modified gravitational field equations are obtained through the metric formalism for the FLRW metric and WH metric with signature $(+,-,-,-)$. We work with $f(R)=R+\alpha R^2-\frac{\epsilon^4}{R}$, $f(R)=R+k\ln(\gamma R)$, $f(R)=R+me^{(-nR)}$ and $f(R)=R+\alpha R^2$ with $\alpha, \epsilon, k, \gamma, m$ and $n$ as free parameters, which lead to three different cosmological models for FLRW universe and one for WH solution. For the choice of $\lambda=0$, this reduces to widely discussed $f(R)$ gravity models. This chapter clearly describes the effects of adding the trace of the energy-momentum tensor in the $f(R)$ Lagrangian. The exact solution of the modified field equations are obtained under the hybrid expansion law for first three models. In fourth model, a specific form of shape function is used to obtain exact solutions. Also the quadratic form of geometry and linear material corrections of this model make the matter content of WH to obey the ECs. Moreover, this model concerns about a new approach of WH analysis in such functional form within $f(R,T)$ gravity theory.
\section{Introduction}\label{ch6intro}
One of the first and simplest modifications to EH actions is called $f(R)$ gravity, with $f(R)$ being a function of the Ricci scalar. The unification of early-time inflation and late-time acceleration can be studied through $f(R)$ gravity models \cite{Nojiri/2008,Appleby/2010}. In the literature, it has been found that the higher order curvature terms in $f(R)$ gravity model play a vital role to avoid cosmological singularities \cite{Kanti/1999,Odintsov/2008,Bamba08}.
The $f(R)=R^{(1+\delta)}$ and $f(R)=R-\frac{\beta}{R^n}$ type models suffer in passing the solar system tests \cite{Chiba/2003} and from gravitational instabilities \cite{Dolgov03}. Also, these theories are incapable of producing standard matter dominated era followed by accelerated expansion \cite{Amendola/2007a,Amendola/2007}. The $f(R)=R+\frac{\alpha}{R^m}-\frac{\beta}{R^n}$ type models have difficulties in satisfying the set of constraints coming from early and late-time acceleration, Big Bang nucleosynthesis and fifth-force experiments \cite{Brookfield/2006}. 
In order to resolve all these issues corresponding to most of the $f(R)$ models, we will consider here some $f(R,T)$ models. The $f(R,T)$ gravity is the recent generalization of $f(R)$ gravity, with addition of trace of stress energy tensor $T$. Thereafter, a wide literature was developed in the context of $f(R,T)$ gravity (refer previous chapters). There are still so many cosmological questions to investigate in $f(R,T)$ gravity. In this chapter we have chosen the following form for the $f(R,T)$ gravity function: $f(R,T)=f(R)+\lambda T$, with constant $\lambda$, i.e. , we fix the $T$-dependence of the theory on its simplest case while investigating different cases for the $R$-dependence. We shall investigate if the $T$-term is capable of evading the shortcomings one faced in $f(R)$ cosmological models. The accelerated expansion of the universe can indeed be described through modified gravity, but sometimes it faces a number of instabilities \cite{Chiba/2003,Dolgov03} which yields further modifications in cosmological models. Nojiri and Odintsov \cite{Nojiri/2004} have discussed a modified gravity with terms proportional to $\ln(R)$ or $R^{-n}(\ln R)^m$, which grow at small curvature. The presence of $\ln(R)$ or $R^{-n}(\ln R)^m$ terms in $f(R)$ gravity may be responsible for the acceleration of the universe. Again, Nojiri and Odinstov \cite{Nojiri/2007} have discussed the $f(R)$ gravity cosmology by considering $f(R)=R+\gamma R^{-n}\left(\ln \frac{R}{\mu^2}\right)^m.$ These forms for the $f(R)$ function are also used in \cite{Nojiri/2004,Nojiri/2007,Nozari/2009} to study different aspects of the theory. In \cite{Paul/2009}, the authors have shown that all these models exhibit current accelerating phase of the universe and the duration of the accelerating phase depends on the coupling constants of the gravitational action. In this chapter, we have considered three different choices for the $f(R)$ function as given in Ref.\cite{Paul/2009} for first three models. In these models, the mixed form for $f(R)$ is considered, namely a positive and a negative power of the curvature $R$, which is normally assumed to study the inflationary scenario of the early universe and the accelerating phase of the present universe. Such a functional form reads (A) $f(R,T)=R+\alpha R^2-\frac{\epsilon^4}{R}+\lambda T$, where the constants $\alpha$ and $\epsilon$ have dimension of $R^{-1}$ (i.e., $(time)^2$) and $R^{\frac{1}{2}}$ (i.e., $ (time)^{-1}$) \cite{Barrow/1983, Capozziello/2003}. The other two models will be followed as (B) $f(R,T)=R+k\ln(\gamma R)+\lambda T$ and (C) $f(R,T)=R+me^{(-nR)}+\lambda T$ where $k, \gamma, m$ and $n$ are constants. In fourth model the choice of $f(R)$ is considered as given by the Starobinsky model ($f(R)=R+\alpha R^2$) \cite{starobinsky/1980} (check also \cite{Starobinsky/2007}). This Starobinsky model has gained importance in recent years for its role in the analysis of matter density perturbations, inflation and many other applications \cite{Fu/2010}-\cite{Kaneda/2016}. Here, we have focused on the analysis of WH solutions from the Starobinsky model within the $f(R,T)$ gravity. \\
In the last few years, there has been a growing interest in developing some exact WH models that account for the minimization of the violation or even the validation of NEC, which in fact can be attained in MGTs (check, for instance, \cite{Garcia/2011,jawad/2016}). WH solutions can be seen in the literature in the framework of $f(R,T)$ gravity in different aspects. In ref. \cite{Azizi/2013}, WHs are firstly analysed within the linear frame of $f(R,T)$ gravity, i.e., $f(R,T)=R+2\lambda T$, with the matter Lagrangian $\mathcal{L}_m=-\rho$. This concept is also analysed in the next chapter \ref{Chapter7}, but with the matter Lagrangian $\mathcal{L}_m=-\mathcal{P}$, where $\mathcal{P}$ is the total pressure. In the same context, Moraes et al. \cite{mcl/2017} have investigated some theoretical predictions for static WHs obtained from $f(R,T)$ gravity. Static numerical solutions have been obtained for different WH matter contents and stable solutions were attained in \cite{Zubair16,Yousaf/2017}. In ref. \cite{ms/2017b}, a static WH model has been constructed from various shape functions. In ref. \cite{Yousaf/2017a}, the authors have studied spherical WH models with different fluid configurations. Motivated by the above references, we are going to check here if it is possible to obtain stable WH solutions in $R^{2}$-gravity within the $f(R,T)$ formalism. 
It is worth remarking that the functional form to be used here, named $f(R,T)=R+\alpha R^2+\lambda T$, has already been applied to the analysis of compact astrophysical objects in references \cite{noureen/2015,noureen/2015b,zubair/2015}, but no WH analysis has been made so far for such a functional form. In these references, it is shown that such a model is well motivated since it is consistent with stable stellar configurations because the second order derivative with respect to $R$ remains positive for the assumed choice of the parameters and this prior choice agrees with the validation of ECs in the present model. Moreover, Starobinsky has shown that his model (free from the $T$-dependence) predicts an overproduction of scalarons in the very early universe \cite{Starobinsky/2007}. This issue was also addressed, for instance, in \cite{koshelev/2016b,gorbunov/2015}. On the other hand, it has been shown that the presence of the trace of the energy-momentum tensor of a scalar field in a gravity formalism can well address the inflationary era \cite{ms/2016}. So, the $T$-dependence inserted in the Starobinsky formalism  indeed raises as a promising model of gravity to be deeply and widely investigated. 
The chapter is organized as follows: section \ref{ch6intro} contains the brief introduction and motivation regarding the present work and the general field equations. Here, we have discussed four cosmological models in $f(R,T)$ gravity with different aspects in the following ways; In section \ref{ch6model1}, we have discussed the details of first model i.e. $f(R,T)=R+\alpha R^2-\frac{\epsilon^4}{R}+\lambda T$ model. The details of other three models are presented in section \ref{ch6model2}, section \ref{ch6model3} and section \ref{ch6model4} respectively. Finally, the conclusion and perspective of all the models are outlined in section \ref{ch6conclusion}. \\
\textbf{\textit{Field equations and Solutions}}\\
From chapter \ref{Chapter1}, eqn. (\ref{eqnfld1}) can be rewritten as:
\begin{equation}\label{ch6e5}
G_{\mu \nu}=T_{\mu \nu}^{eff},
\end{equation}
where
\begin{multline}\label{ch6e6}
T_{\mu \nu}^{eff}=\frac{1}{f_{R}(R,T)}\biggl[(8\pi+f_{T}(R,T))T_{\mu \nu}+pf_{T}(R,T)g_{\mu \nu}+\frac{f(R,T)-Rf_R(R,T)}{2}g_{\mu \nu}\\
-(g_{\mu \nu}\Box -\nabla _{\mu}\nabla
_{\nu})f_{R}(R,T)\biggr].
\end{multline}
Here, we will concentrate for first three models on a spatially flat FLRW metric with a time-dependent scale factor $a(t)$ such that the metric reads,
\begin{equation}\label{ch6met1}
ds^{2}=dt^{2}-a^{2}(t)\left[dr^{2}+r^2(d\theta^{2}+\sin^2 \theta d\phi^{2})\right].
\end{equation}
The energy-momentum tensor for a perfect fluid is considered as:
\begin{equation}\label{ch6ent1}
T_{\mu \nu}=(\rho+p)u_\mu u_\nu-pg_{\mu \nu}.
\end{equation}
The general $f(R,T)$ gravity field equations for $f(R,T)=f(R)+\lambda T$ and the above metric (\ref{ch6met1}) is given as:
\begin{equation}\label{ch6eqn1}
3H^2=\frac{1}{f_{R}}\left[\left(8\pi +\frac{3\lambda}{2}\right)\rho-\frac{\lambda}{2} p\right]+\frac{1}{f_{R}}\left(\frac{f(R)-Rf_R}{2}-3H\dot{R}f_{RR}\right),
\end{equation}
\begin{multline}\label{ch6eqn2}
2\dot{H}+3H^2=\frac{1}{f_{R}}\left[-\left(8\pi +\frac{3\lambda}{2}\right)p+\frac{\lambda}{2} \rho\right]-\frac{1}{f_{R}}\biggl(-\frac{f(R)-Rf_R}{2} \\+\dot{R}^2f_{RRR}+2H\dot{R}f_{RR}+\ddot{R}f_{RR}\biggr).
\end{multline}
The Ricci scalar $R$ for metric (\ref{ch6met1}) is obtained as:
\begin{equation}\label{ch6e11}
R=-6(\dot{H}+2H^2).
\end{equation}
From eqns. (\ref{ch6eqn1}) and (\ref{ch6eqn2}), the pressure $p$, the energy density $\rho$ and the EoS parameter $\omega=p/\rho$ can be explicitly expressed as
 \begin{multline}\label{ch6rho}
\rho=\frac{f_R}{2}\left(\frac{-2\dot{H}}{8\pi+\lambda}+\frac{2\dot{H}+6H^2}{8\pi+2\lambda} \right)
+\left(\frac{H\dot{R}-\ddot{R}}{8\pi+\lambda}+\frac{5H\dot{R}+\ddot{R}}{8\pi+2\lambda}\right)\frac{f_{RR}}{2}\\+\left(\frac{\dot{R}^2}{8\pi+2\lambda}-\frac{\dot{R}^2}{8\pi+\lambda}\right)\frac{f_{RRR}}{2}
-\frac{f(R)-Rf_R}{2(8\pi+2\lambda)},
\end{multline}
 \begin{multline}\label{ch6p}
p=\frac{f_R}{2}\left(\frac{-2\dot{H}}{8\pi+\lambda}-\frac{2\dot{H}+6H^2}{8\pi+2\lambda} \right)
+\left(\frac{H\dot{R}-\ddot{R}}{8\pi+\lambda}-\frac{5H\dot{R}+\ddot{R}}{8\pi+2\lambda}\right)\frac{f_{RR}}{2}\\+\left(\frac{-\dot{R}^2}{8\pi+2\lambda}-\frac{\dot{R}^2}{8\pi+\lambda}\right)\frac{f_{RRR}}{2}+\frac{f(R)-Rf_R}{2(8\pi+2\lambda)},
\end{multline}
\begin{equation}\label{ch6omega}
\omega=\dfrac{\frac{f_R}{2}\left(\frac{-2\dot{H}}{8\pi+\lambda}-\frac{2\dot{H}+6H^2}{8\pi+2\lambda} \right)
+\left(\frac{H\dot{R}-\ddot{R}}{8\pi+\lambda}-\frac{5H\dot{R}+\ddot{R}}{8\pi+2\lambda}\right)\frac{f_{RR}}{2}+\left(\frac{-\dot{R}^2}{8\pi+2\lambda}-\frac{\dot{R}^2}{8\pi+\lambda}\right)\frac{f_{RRR}}{2}+\frac{f(R)-Rf_R}{2(8\pi+2\lambda)}}{\frac{f_R}{2}\left(\frac{-2\dot{H}}{8\pi+\lambda}+\frac{2\dot{H}+6H^2}{8\pi+2\lambda} \right)
+\left(\frac{H\dot{R}-\ddot{R}}{8\pi+\lambda}+\frac{5H\dot{R}+\ddot{R}}{8\pi+2\lambda}\right)\frac{f_{RR}}{2}+\left(\frac{\dot{R}^2}{8\pi+2\lambda}-\frac{\dot{R}^2}{8\pi+\lambda}\right)\frac{f_{RRR}}{2}-\frac{f(R)-Rf_R}{2(8\pi+2\lambda)}}.
\end{equation}
Here, the  exact solutions are derived by using the hybrid expansion law for the scale factor as following \cite{Akarsu/2014}
\begin{equation}\label{ch6sp1}
a=t^{\eta}e^{\beta t},
\end{equation}
where $\eta$ and $\beta$ are positive constants. Such a scale factor yields the DP and HP as
\begin{eqnarray}
q=-1+\frac{\eta}{(\beta t+\eta)^2},\label{ch6DP}\\
H=\frac{\eta +\beta  t}{t}.\label{ch6HP}
\end{eqnarray}
The time-redshift relation is obtained from the relation $a(t)=\frac{1}{1+z}$, where $z$ and $a_0 = 1$ are the redshift and the present scale factor respectively, such that
\begin{equation}\label{ch6time}
t=\frac{\eta}{\beta} W\left[\frac{\beta\left(\frac{1}{z+1}\right)^{1/\eta}}{\eta }\right],
\end{equation}
where $W$ denotes the Lambert function (also known as ``product logarithm'').
\begin{figure}[H]
\centering
  \includegraphics[width=78mm]{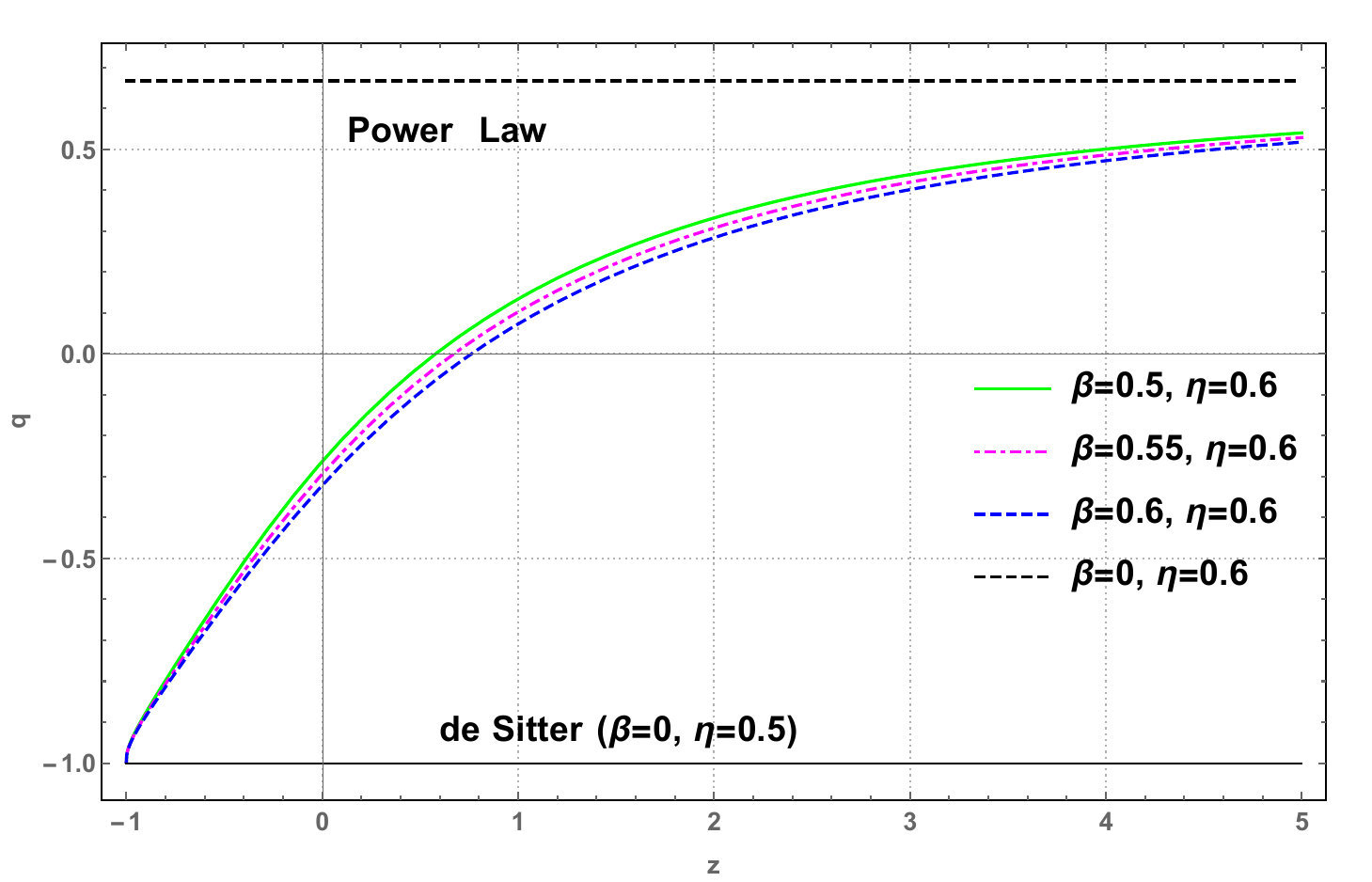}
   \caption{Variation of DP $q$ against redshift $z$.} \label{ch6fig2}
\end{figure}
In order to check the reliability of the model, the plot of DP with respect to redshift is required, which provides a transition from the deceleration stage to the present acceleration era of the universe. It can be observed from Fig. \ref{ch6fig2} that the transition occurs at transition redshift $z_{tr} =0.5662,0.6691,0.7574$, corresponding to a fixed value for $\eta=0.6$,  and various values for $\beta$, as $\beta=0.5,0.55,0.6$. The transition values for this model are in accordance with the observational data, as one can check in \cite{Capozziello14,Capozziello15,Farooq17}.
\section{ The $f(R,T)=R+\alpha R^2-\frac{\epsilon^4}{R}+\lambda T$ model}\label{ch6model1}
\begin{figure}[H]
\minipage{0.48\textwidth}
\includegraphics[width=75mm]{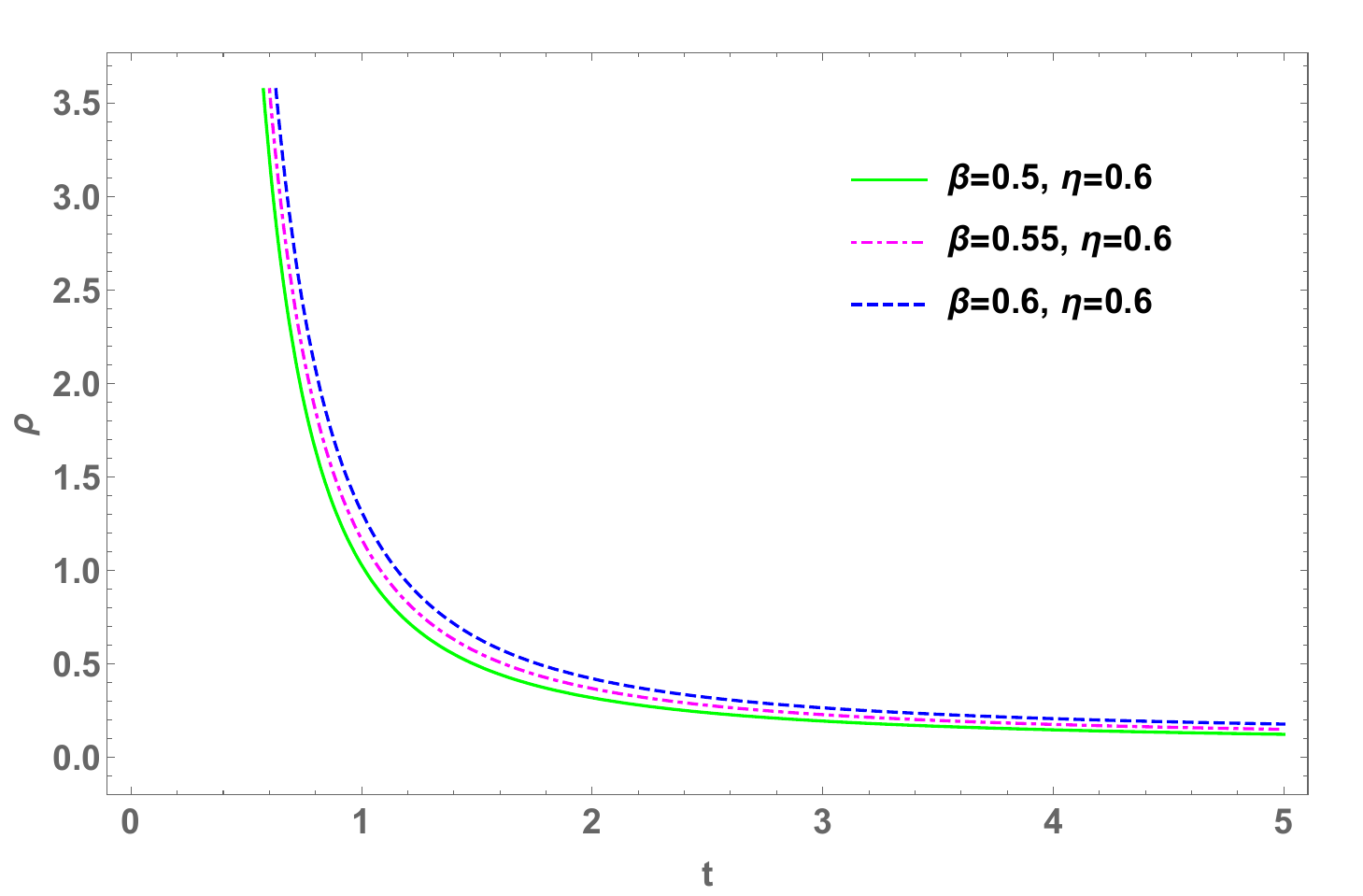}
  \caption{Plot of $\rho$ against time with $\alpha=0.2$, $\epsilon=-1$, $\lambda=-8$.}\label{ch6fig3}
\endminipage\hfill
\minipage{0.50\textwidth}
\includegraphics[width=75mm]{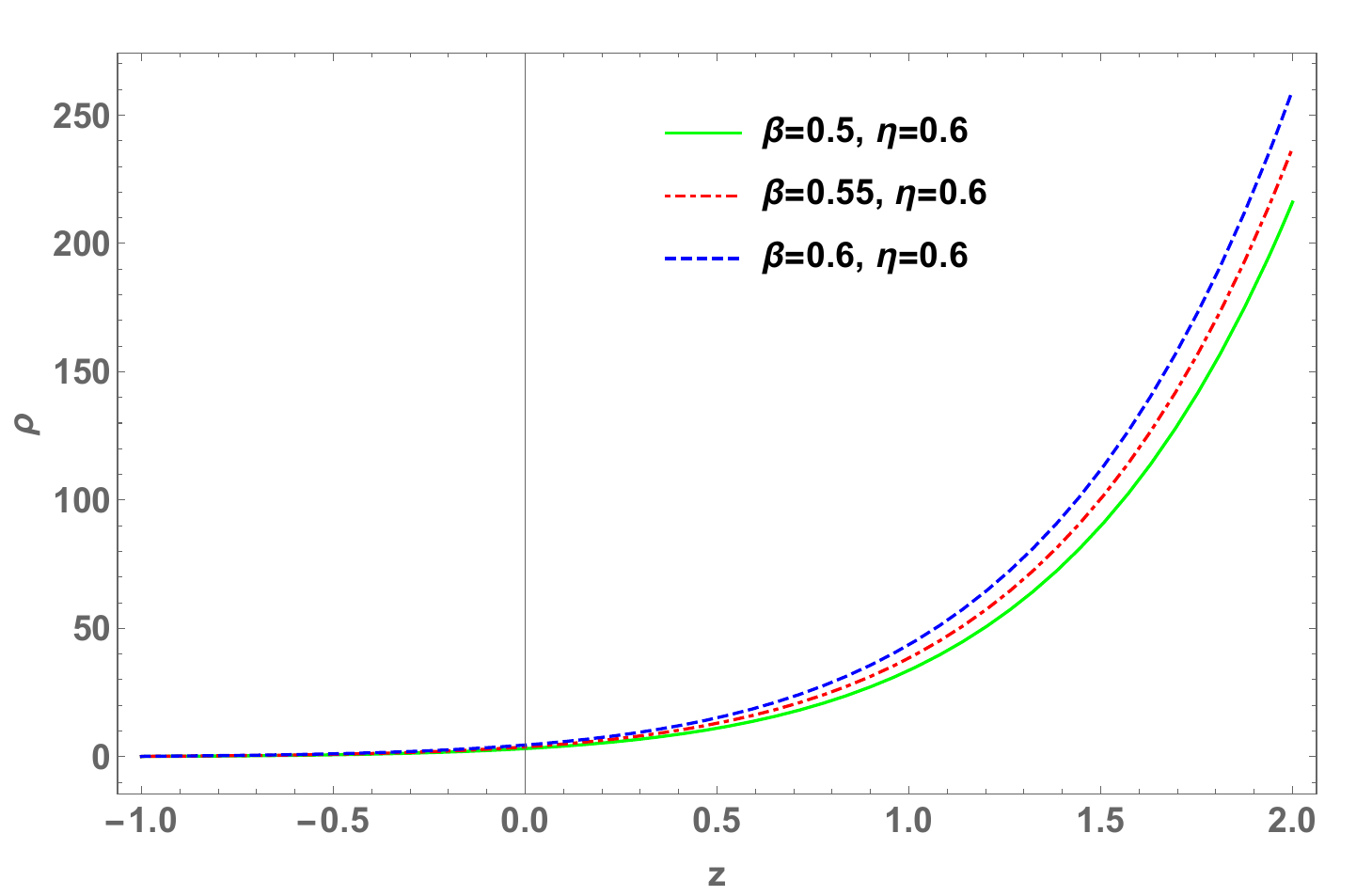}
  \caption{Plot of $\rho$ against $z$ with $\alpha=0.2$, $\epsilon=-1$, $\lambda=-8$.}\label{ch6fig3z}
\endminipage
\end{figure}

\begin{figure}[H]
\minipage{0.48\textwidth}
\includegraphics[width=75mm]{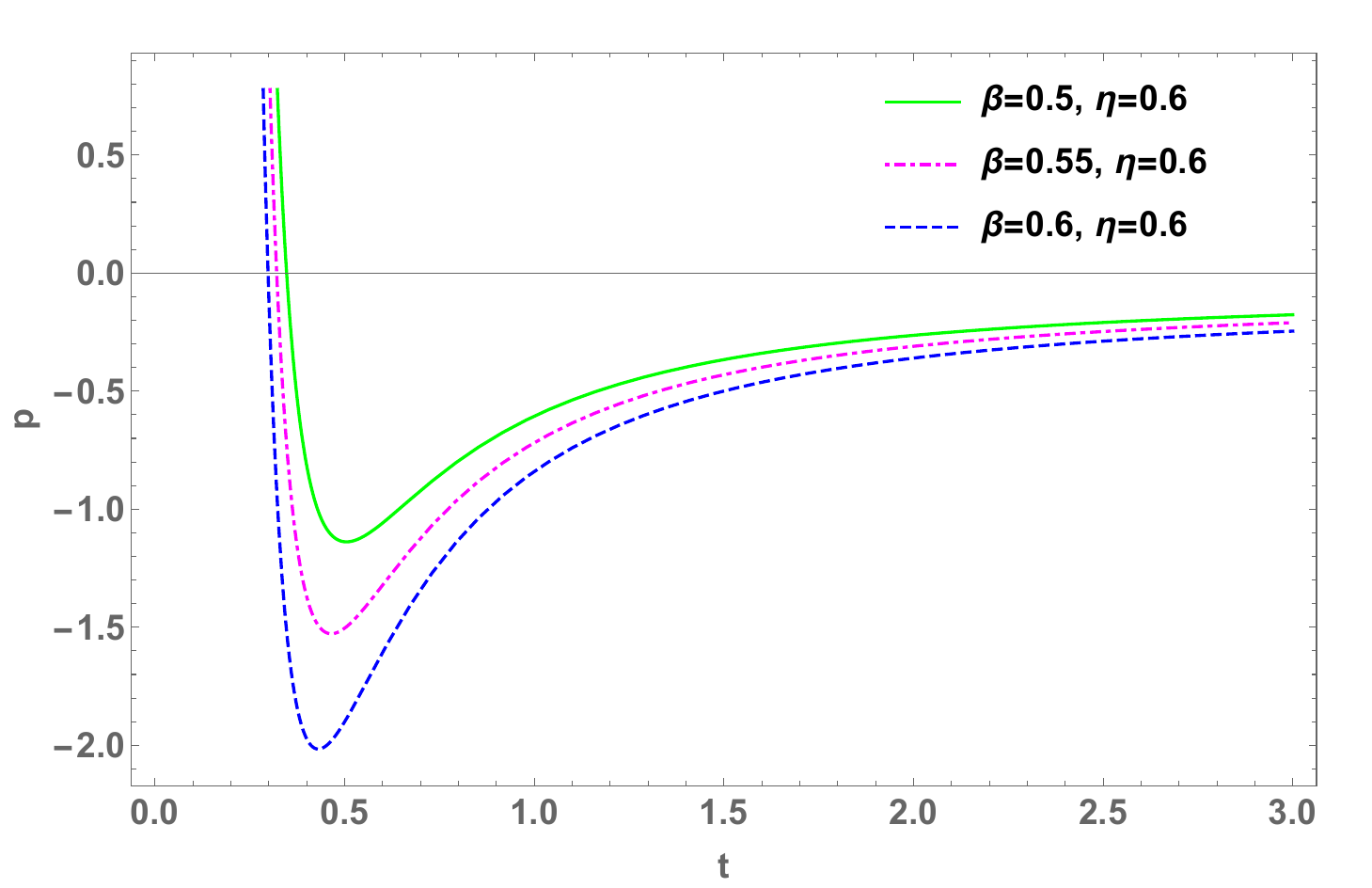}
  \caption{Plot of $p$ against time with $\alpha=0.2$, $\epsilon=-1$, $\lambda=-8$.}\label{ch6fig4}
\endminipage\hfill
\minipage{0.50\textwidth}
\includegraphics[width=75mm]{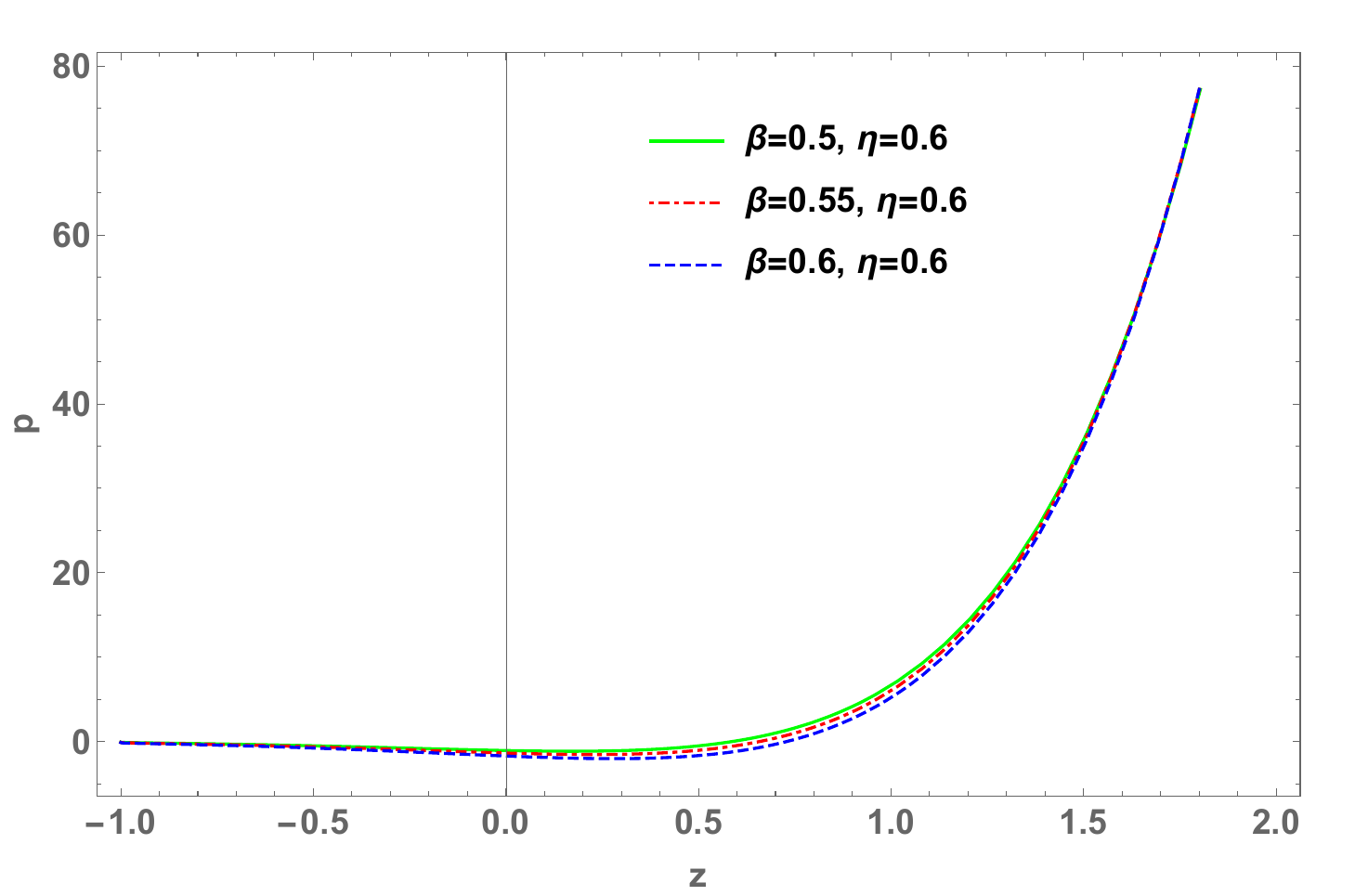}
  \caption{Plot of $p$ against $z$ with $\alpha=0.2$, $\epsilon=-1$, $\lambda=-8$.}\label{ch6fig4z}
\endminipage
\end{figure}

\begin{figure}[H]
\minipage{0.48\textwidth}
\includegraphics[width=75mm]{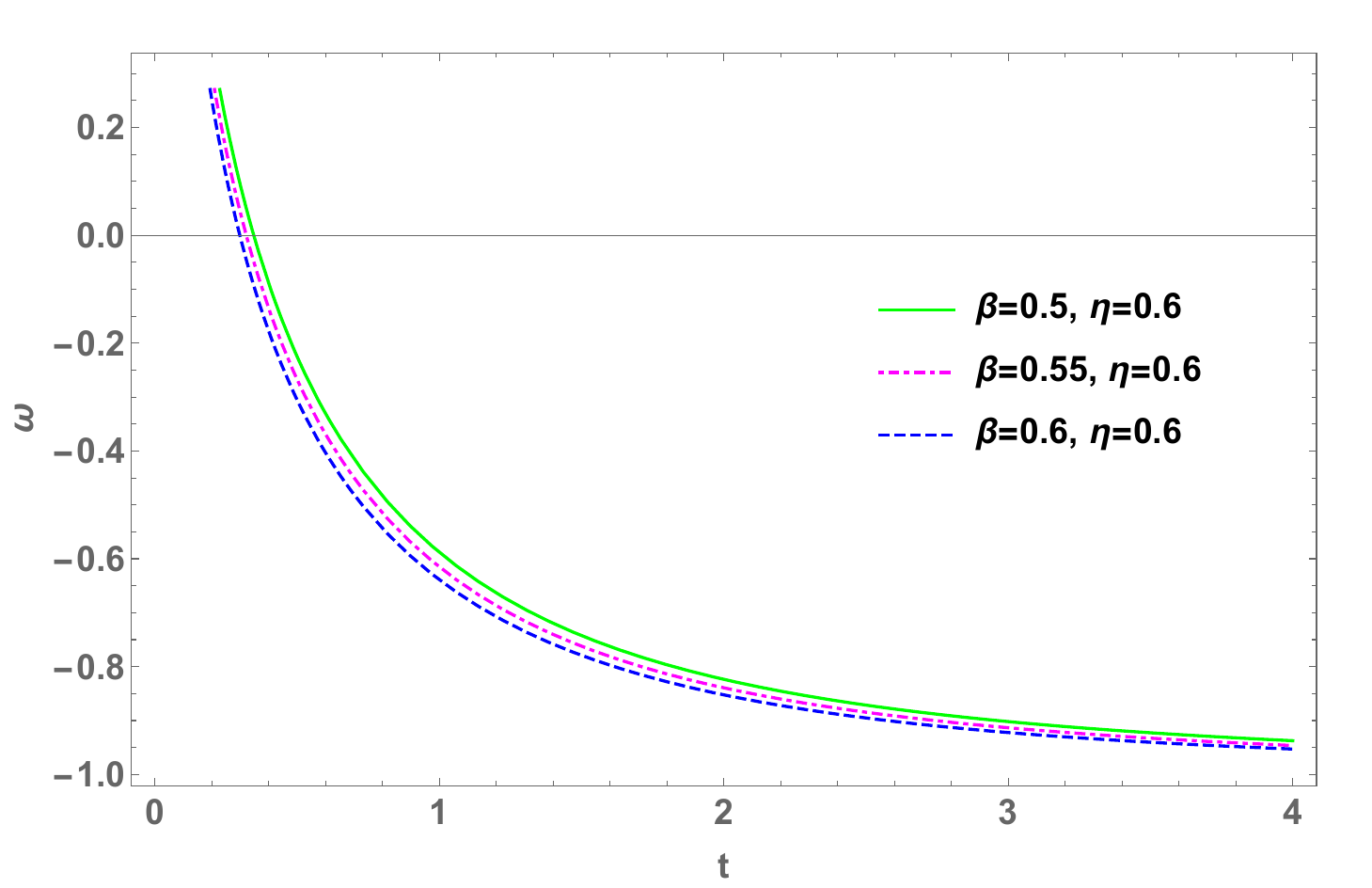}
  \caption{Plot of EoS Parameter against time with $\alpha=0.2$, $\epsilon=-1$, $\lambda=-8$.}\label{ch6fig5}
\endminipage\hfill
\minipage{0.50\textwidth}
\includegraphics[width=75mm]{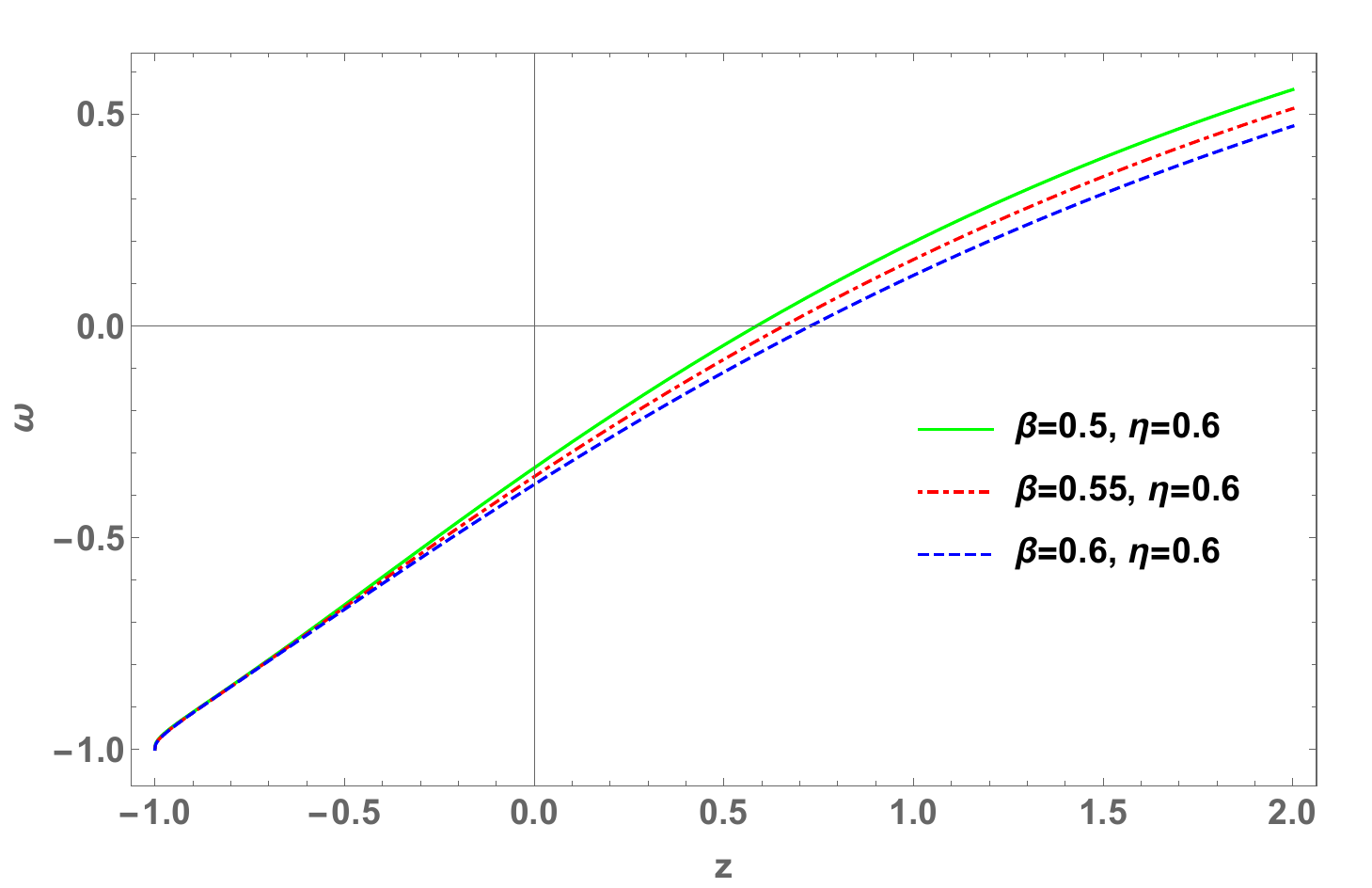}
  \caption{Plot of EoS Parameter against $z$ with $\alpha=0.2$, $\epsilon=-1$, $\lambda=-8$.}\label{ch6fig5z}
\endminipage
\end{figure}

In this case, by using eqn. (\ref{ch6sp1}) for $f(R,T)=R+\alpha R^2-\frac{\epsilon^4}{R}+\lambda T$,  in eqns. (\ref{ch6rho} - \ref{ch6omega}), the analytical forms for $\rho$  $p$, and $\omega$ are expressed as follows 
\begin{multline}\label{ch6rho1}
\rho=\frac{1}{2t^2} \left[\frac{2 \eta }{\lambda +8 \pi }+\frac{3 (\eta +\beta  t)^2-\eta }{\lambda +4 \pi }\right] \left \{ \frac{\epsilon ^4 t^4}{36 \left[\eta -2 (\eta +\beta  t)^2\right]^2}-\frac{12 \alpha  \left[2 (\eta +\beta  t)^2-\eta \right]}{t^2}+1\right\}\\+\frac{3 \alpha G_{11}(t)}{t^4} +\frac{\epsilon ^4 t^2 G_{21}(t)}{36 \left[\eta -2 (\eta +\beta  t)^2\right]^4},
\end{multline}
\begin{multline}\label{ch6p1}
p= \frac{1}{2t^2} \left[\frac{2 \eta }{\lambda +8 \pi }-\frac{3 (\eta +\beta  t)^2-\eta }{\lambda +4 \pi }\right] \left\{\frac{\epsilon ^4 t^4}{36 \left[\eta -2 (\eta +\beta  t)^2\right]^2}-\frac{12 \alpha  \left[2 (\eta +\beta  t)^2-\eta \right]}{t^2}+1\right\}\\
 -\frac{36 \alpha F_{11}(t) }{(\lambda +4\pi) (\lambda +8\pi) t^4} +\frac{\epsilon ^4 t^2 F_{21}(t) }{36 \left[\eta -2 (\eta +\beta  t)^2\right]^2},
\end{multline}
\begin{equation}\label{ch6omega1}
\omega=\dfrac{\splitfrac{\frac{1}{2t^2} \left[\frac{2 \eta }{\lambda +8 \pi }-\frac{3 (\eta +\beta  t)^2-\eta }{\lambda +4 \pi }\right] \left\{\frac{\epsilon ^4 t^4}{36 \left[\eta -2 (\eta +\beta  t)^2\right]^2}-\frac{12 \alpha  \left(2 (\eta +\beta  t)^2-\eta \right)}{t^2}+1\right\}\\
 -\frac{36 \alpha F_{11}(t) }{(\lambda +4\pi) (\lambda +8\pi) t^4} }{+\frac{\epsilon ^4 t^2 F_{21}(t) }{36 \left[\eta -2 (\eta +\beta  t)^2\right]^2}}}
{\splitfrac{\frac{1}{2t^2} \left[\frac{2 \eta }{\lambda +8 \pi }+\frac{3 (\eta +\beta  t)^2-\eta }{\lambda +4 \pi }\right] \left \{ \frac{\epsilon ^4 t^4}{36 \left[\eta -2 (\eta +\beta  t)^2\right]^2}-\frac{12 \alpha  \left[2 (\eta +\beta  t)^2-\eta \right]}{t^2}+1\right\}+\frac{3 \alpha G_{11}(t)}{t^4} }{+\frac{\epsilon ^4 t^2 G_{21}(t)}{36 \left[\eta -2 (\eta +\beta  t)^2\right]^4}}},
\end{equation}
where
\begin{equation}
G_{11}(t)= \dfrac{4 \eta  \left[2 \eta ^2+5 \eta +4 \beta  \eta  t+\beta  t (2 \beta  t+3)-3\right]}{\lambda +8 \pi}+\dfrac{\splitfrac{3 \left[\eta -2 (\eta +\beta  t)^2\right]^2-2 \eta  (6 \eta +4 \beta  t-3)}{+10 \eta  (\eta +\beta  t) (2 \eta +2 \beta  t-1)}}{\lambda +4 \pi},
\end{equation}
\begin{multline}
G_{21}(t)=\frac{\splitfrac{ 2 \eta[\eta(\eta -3)(1-2 \eta )^2+4 \beta ^4 t^4+2 \beta ^3 (8 \eta +3) t^3+2 \beta ^2 [2 \eta  (6 \eta -1)-3] t^2}{+\beta  \eta  (2 \eta -1) (8 \eta -9) t]}}{\lambda +8 \pi }\\
+\dfrac{\splitfrac{6 \eta ^2 (2 \eta +2 \beta  t-1)^2-3 \left[2 (\eta +\beta  t)^2-\eta \right]^3+\eta  t (2 \eta +2 \beta  t-1) \left[2 (\eta +\beta  t)^2-\eta \right]}{+5 \eta  (\eta +\beta  t) (2 \eta +2 \beta  t-1) \left[2 (\eta +\beta  t)^2-\eta \right]}}{\lambda +4 \pi },
\end{multline}
\begin{multline}
F_{11}(t)=\eta  \left[\eta ^4 (\lambda +25.1327)+8.37758 \eta ^3+\eta^2  (-3.25 \lambda -60.7375)+\eta(1.5 \lambda +25.1327)\right]\\
 +\beta ^4 (1. \lambda +25.1327) t^4+\beta ^3 \eta  (4 \lambda +100.531) t^3+\beta ^2 \eta  t^2 [\eta  (6 \lambda +150.796)+8.37758]+\\
 \beta  \eta  t \left[\eta ^2 (4 \lambda +100.531)+16.7552 \eta -2.5 \lambda -50.2655\right],
\end{multline}
\begin{multline}
F_{21}(t)=\frac{\splitfrac{8\eta \beta ^4 t^4+4 \eta \beta ^3 (8 \eta +3) t^3+4\eta \beta ^2 \left(12 \eta ^2-2 \eta -3\right) t^2+2 \eta^2\beta \left(16 \eta ^2-26 \eta +9\right)t}{+2 \eta(\eta^2 -3\eta) (1-2 \eta )^2}}{(\lambda +8 \pi ) \left[\eta  (2 \eta -1)+2 \beta ^2 t^2+4 \beta  \eta  t\right]^2}\\
-\dfrac{\splitfrac{6 \eta ^2 (2 \eta +2 \beta  t-1)^2-3 \left[2 (\eta +\beta  t)^2-\eta \right]^3+\eta  t (2 \eta +2 \beta  t-1) \left[2 (\eta +\beta  t)^2-\eta \right]}{+5 \eta  (\eta +\beta  t) (2 \eta +2 \beta  t-1) \left[2 (\eta +\beta  t)^2-\eta \right]}}{(\lambda +4 \pi ) \left[\eta -2 (\eta +\beta  t)^2\right]^2}.
\end{multline}
The plot for the above obtained physical parameters with respect to time $t$ and redshift $z$ are presented in Fig. \ref{ch6fig3} to Fig. \ref{ch6fig5z} respectively.
\section{The $f(R,T)=R+k\ln(\gamma R)+\lambda T$ model}\label{ch6model2}
\begin{figure}[H]
\minipage{0.48\textwidth}
\includegraphics[width=75mm]{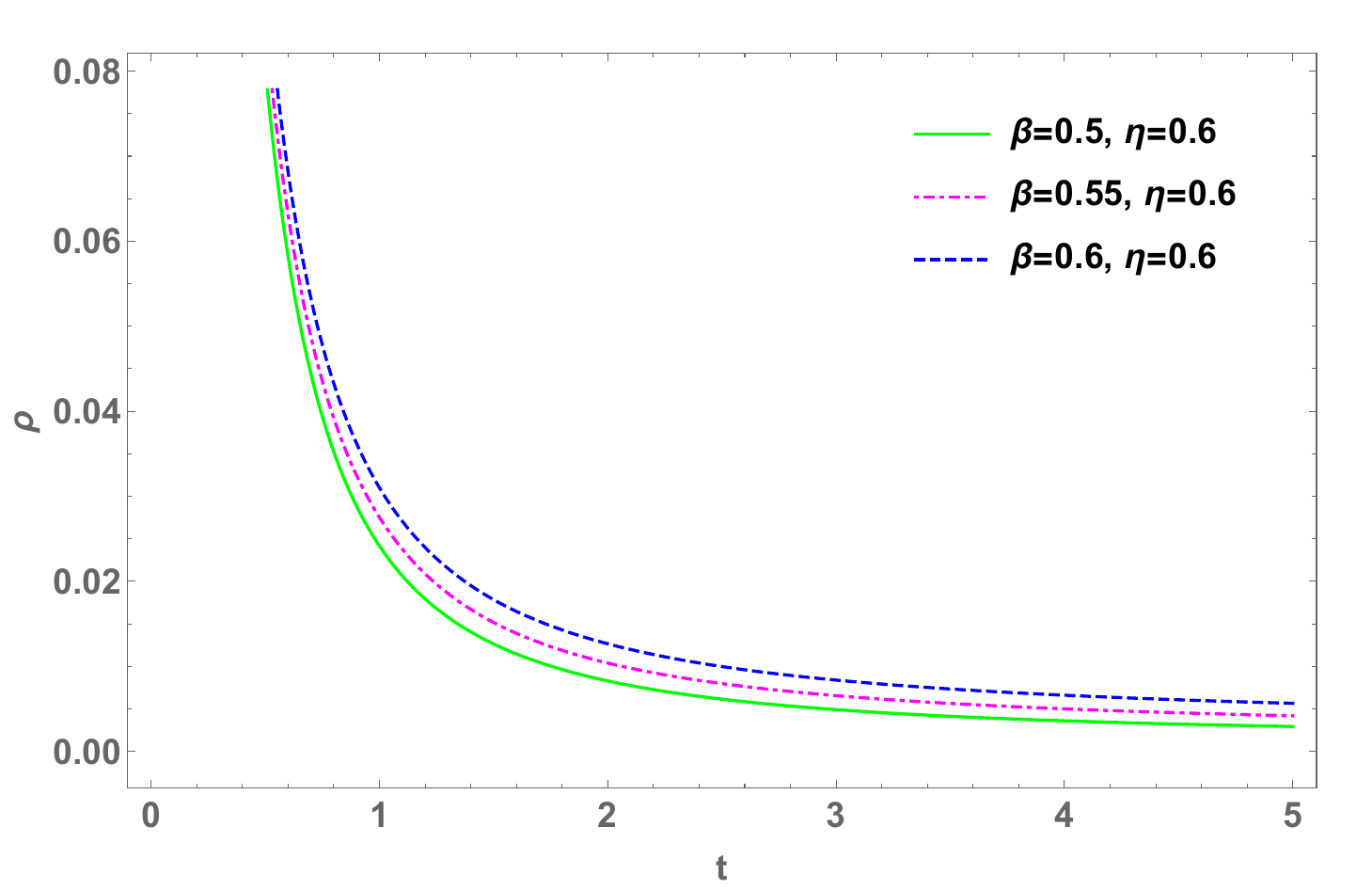}
  \caption{Plot of $\rho$ against time with $k=1$, $\gamma=-2$, $\lambda=35$.}\label{ch6fig6}
\endminipage\hfill
\minipage{0.50\textwidth}
  \includegraphics[width=75mm]{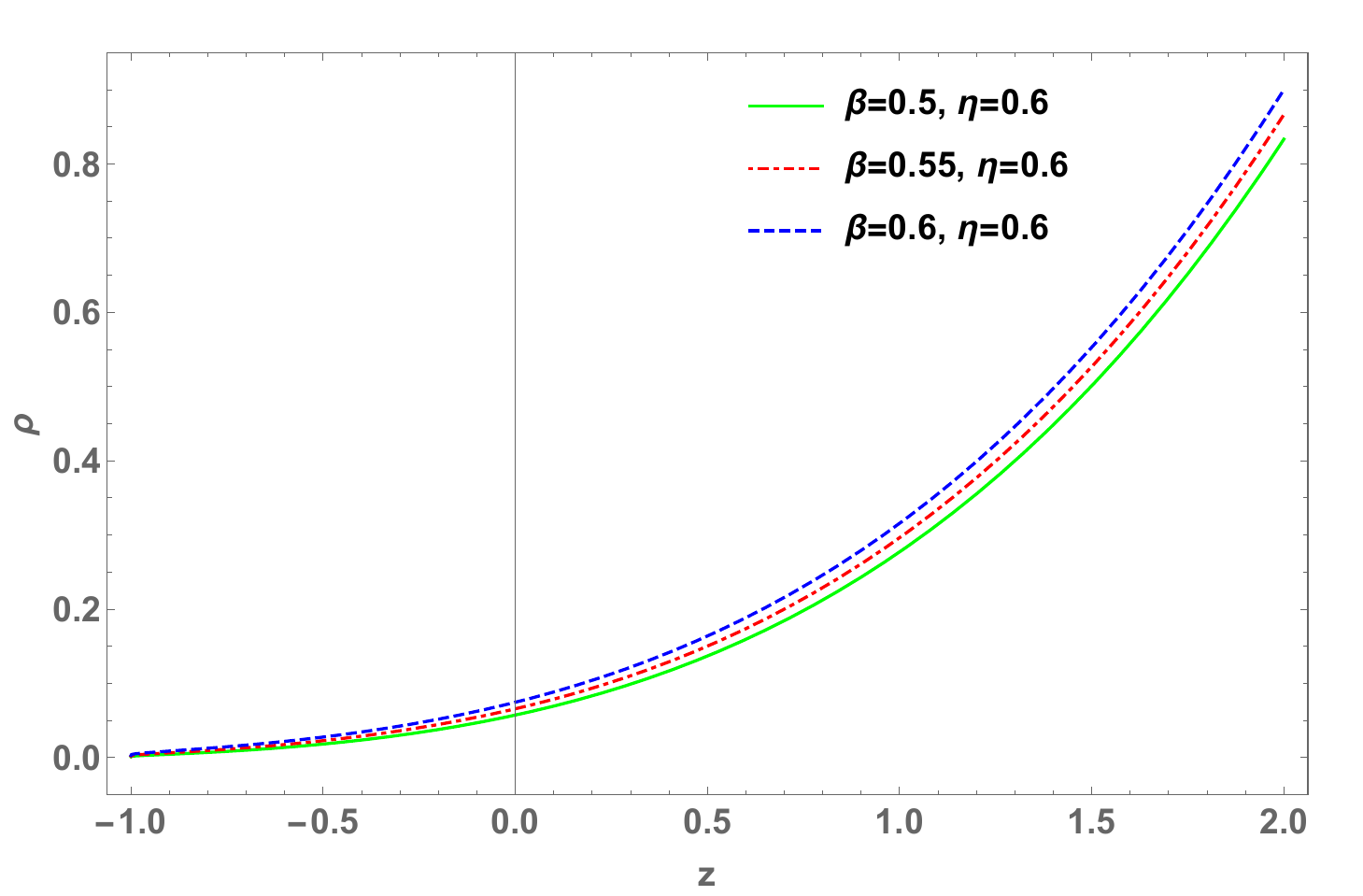}
  \caption{Plot of $\rho$ against $z$ with $k=1$, $\gamma=-2$, $\lambda=35$.}\label{ch6fig6z}
\endminipage
\end{figure}

\begin{figure}[H]
\minipage{0.48\textwidth}
\includegraphics[width=75mm]{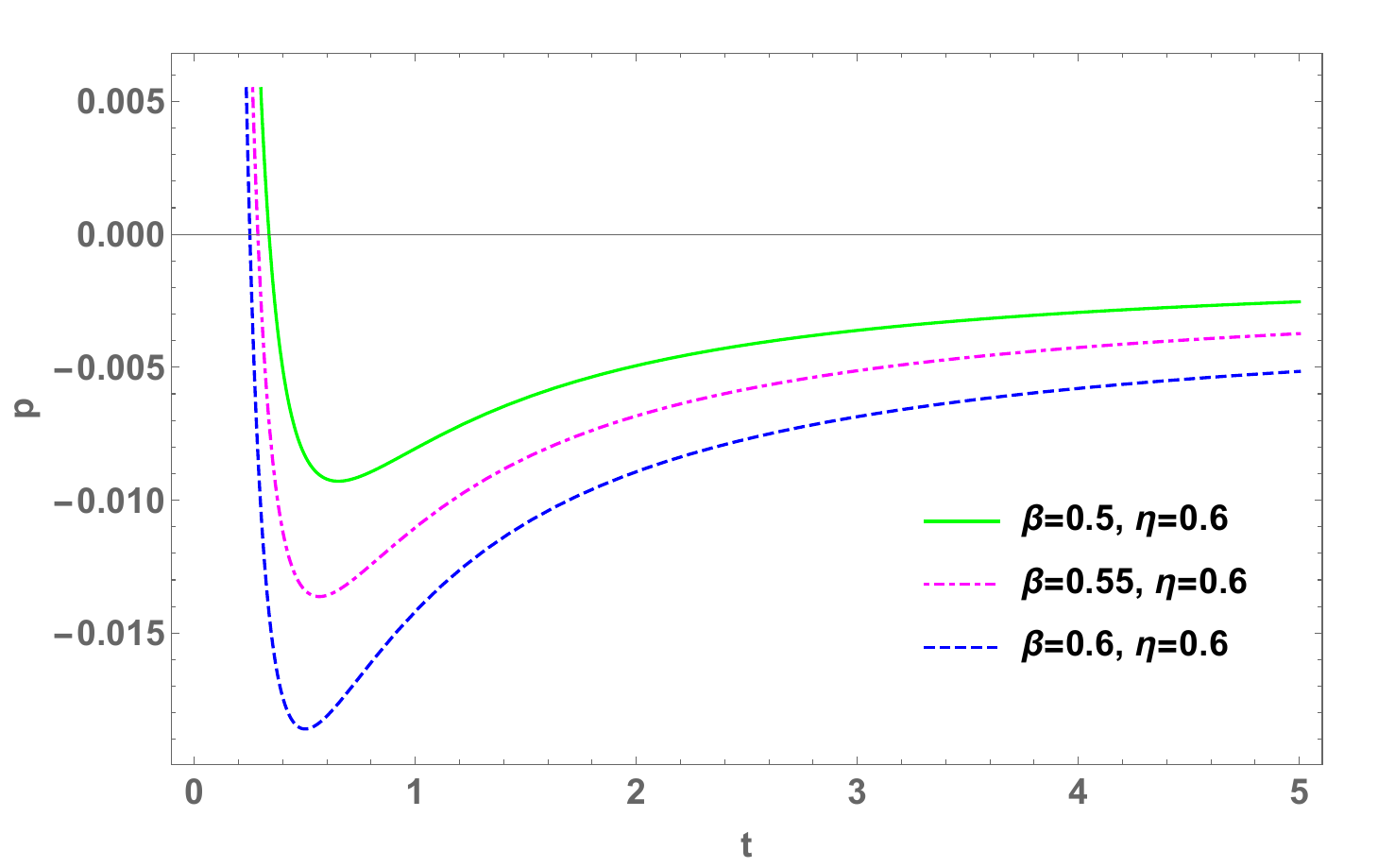}
  \caption{Plot of $p$ against time with $k=1$, $\gamma=-2$, $\lambda=35$.}\label{ch6fig7}
\endminipage\hfill
\minipage{0.50\textwidth}
  \includegraphics[width=75mm]{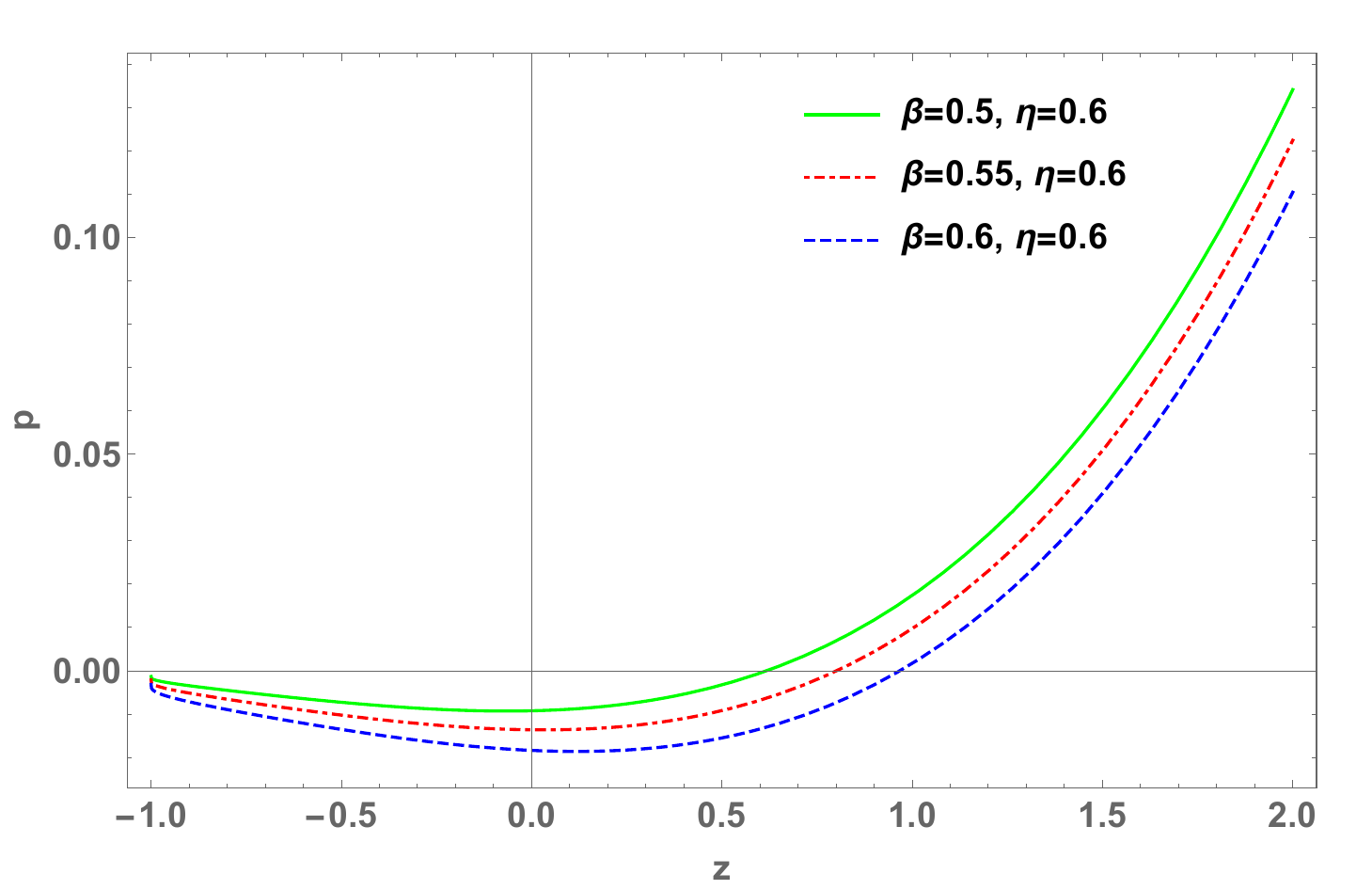}
  \caption{Plot of $p$ against $z$ with $k=1$, $\gamma=-2$, $\lambda=35$.}\label{ch6fig7z}
\endminipage
\end{figure}

\begin{figure}[H]
\minipage{0.48\textwidth}
\includegraphics[width=75mm]{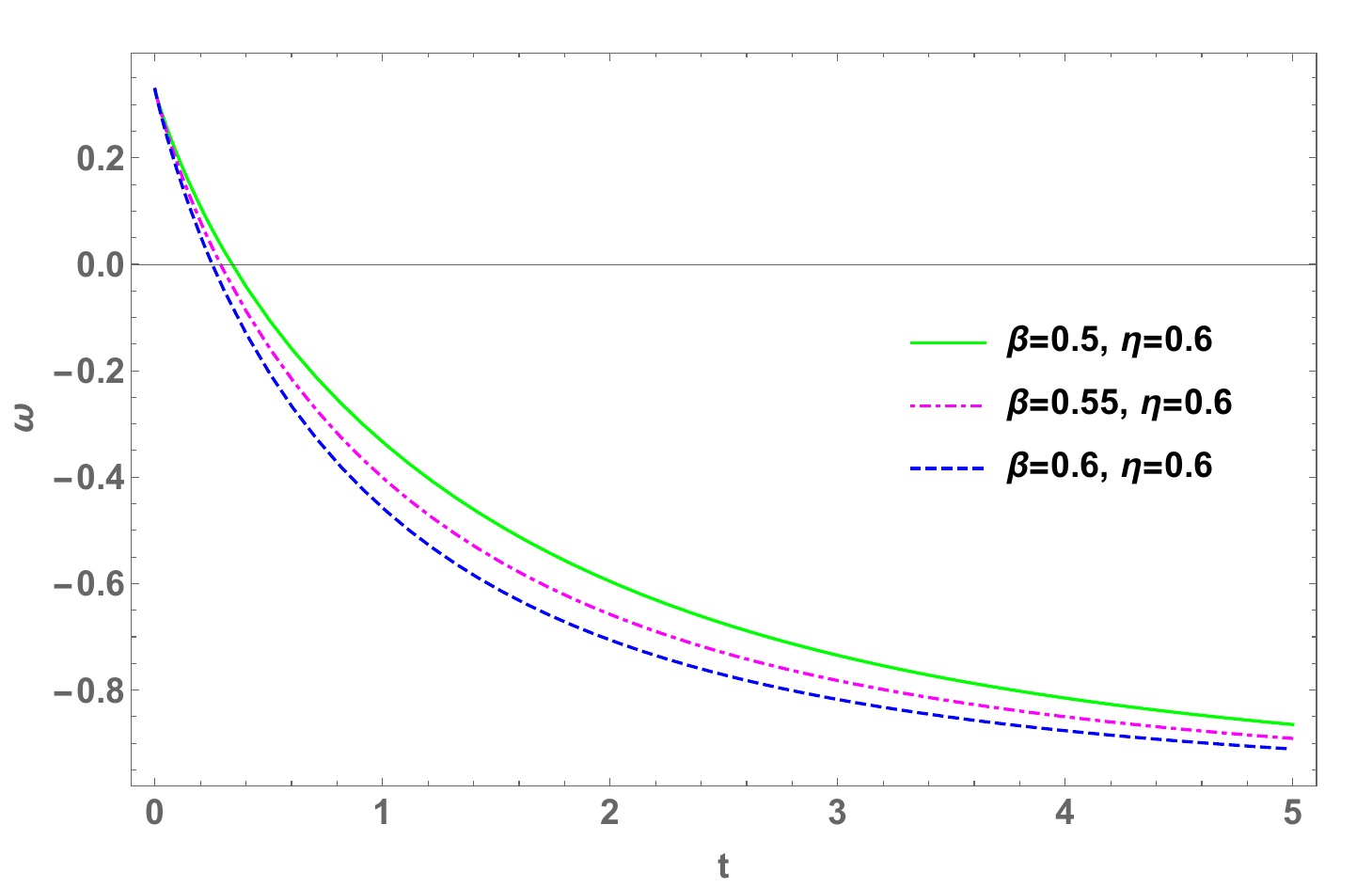}
  \caption{Plot of EoS Parameter against time with $k=1$, $\gamma=-2$, $\lambda=35$.}\label{ch6fig8}
\endminipage\hfill
\minipage{0.50\textwidth}
  \includegraphics[width=75mm]{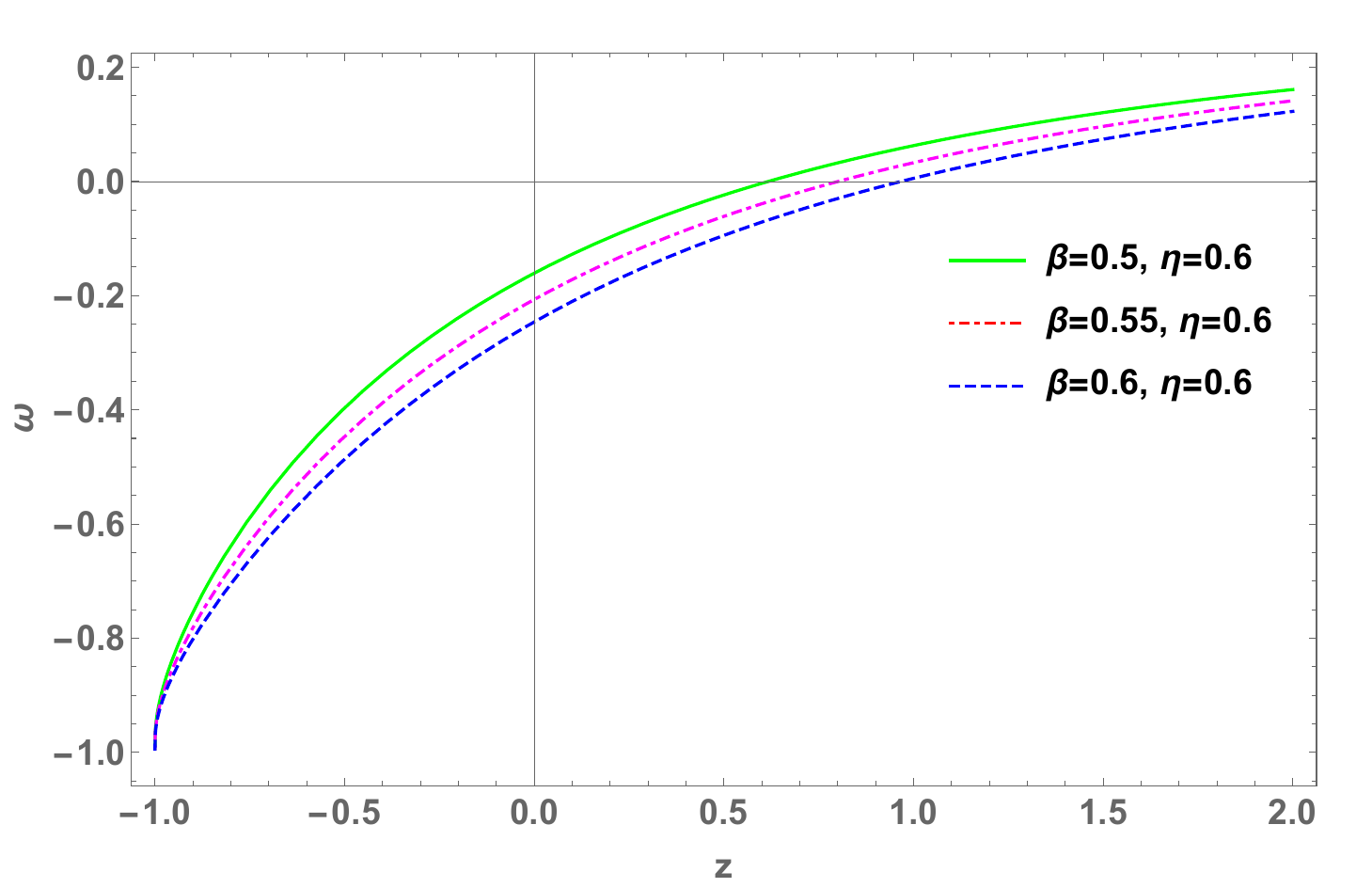}
  \caption{Plot of EoS Parameter against $z$ with $k=1$, $\gamma=-2$, $\lambda=35$.}\label{ch6fig8z}
\endminipage
\end{figure}

By using $f(R,T)=R+k\ln(\gamma R)+\lambda T$ with eqn. (\ref{ch6sp1}) in eqns. (\ref{ch6rho}-\ref{ch6omega}), the analytical forms for $p$, $\rho$ and $\omega$ are written as 
\begin{multline}\label{ch6rho2}
\rho=\frac{-0.0833333}{2 (\eta +\beta  t)^2-\eta } \left[k-\frac{ 12 (\eta +\beta  t)^2-6\eta}{t^2}\right] \left[\frac{2 \eta }{\lambda +8 \pi }+\frac{3 (\eta +\beta  t)^2-\eta }{\lambda +4 \pi }\right]+\\
\frac{\eta  k G_{12}(t)}{12 (\lambda +4 \pi ) (\lambda +8 \pi ) \left[2 (\eta +\beta  t)^2-\eta \right]^3} -\frac{k}{4\lambda +16 \pi} \left\{\log \left(-\frac{6 \gamma  \left[2 (\eta +\beta  t)^2-\eta \right]}{t^2}\right)-1\right\},
\end{multline}
\begin{multline}\label{ch6p2}
p=\frac{-0.0833333 t^2}{2 (\eta +\beta  t)^2-\eta } \left[k-\frac{6 \left[2 (\eta +\beta  t)^2-\eta \right]}{t^2}\right] \left[\frac{2 \eta }{(\lambda +8 \pi ) t^2}-\frac{6 (\eta +\beta  t)^2-2 \eta}{(2 \lambda +8 \pi)t^2}\right]\\+
\frac{\eta  k F_{12}(t)}{12 (\lambda +4 \pi) (\lambda +8 \pi) \left[2 (\eta +\beta  t)^2-\eta \right]^3}
+\frac{k}{4\lambda +16 \pi} \left\{\log \left(-\frac{6 \gamma  \left[2 (\eta +\beta  t)^2-\eta \right]}{t^2}\right)-1\right\},
\end{multline}
\begin{equation}\label{ch6omega2}
\omega=\dfrac{\splitfrac{\frac{-0.0833333 t^2}{2 (\eta +\beta  t)^2-\eta } \left[k-\frac{6 \left[2 (\eta +\beta  t)^2-\eta \right]}{t^2}\right] \left[\frac{2 \eta }{(\lambda +8 \pi ) t^2}-\frac{6 (\eta +\beta  t)^2-2 \eta}{(2 \lambda +8 \pi)t^2}\right]\\+
\frac{\eta  k F_{12}(t)}{12 (\lambda +4 \pi) (\lambda +8 \pi) \left[2 (\eta +\beta  t)^2-\eta \right]^3}
}{+\frac{k}{4\lambda +16 \pi} \left\{\log \left(-\frac{6 \gamma  \left[2 (\eta +\beta  t)^2-\eta \right]}{t^2}\right)-1\right\}}}
{\splitfrac{\frac{-0.0833333}{2 (\eta +\beta  t)^2-\eta } \left[k-\frac{ 12 (\eta +\beta  t)^2-6\eta}{t^2}\right] \left[\frac{2 \eta }{\lambda +8 \pi }+\frac{3 (\eta +\beta  t)^2-\eta }{\lambda +4 \pi }\right]+\\
\frac{\eta  k G_{12}(t)}{12 (\lambda +4 \pi ) (\lambda +8 \pi ) \left[2 (\eta +\beta  t)^2-\eta \right]^3} }{-\frac{k}{4\lambda +16 \pi} \left\{\log \left(-\frac{6 \gamma  \left[2 (\eta +\beta  t)^2-\eta \right]}{t^2}\right)-1\right\}}},
\end{equation}
where
\begin{multline}
G_{12}(t)=\lambda [-\eta  (7 \eta -1)(1-2 \eta )^2 -28 \beta ^4 t^4+2 \beta ^3 (3-56 \eta ) t^3+2 \beta ^2 \left(-84 \eta ^2+22 \eta +3\right) t^2
\\-7 \beta  \eta  (2 \eta -1) (8 \eta -1) t]-48 \pi  (\eta +\beta  t) (2 \eta +2 \beta  t-1) \left[2 (\eta +\beta  t)^2-\eta \right],
\end{multline}
\begin{multline}
F_{12}(t)=4 \beta ^4 t^4 (3 \lambda +32 \pi)+2 \beta ^3 t^3 [3 \lambda (8 \eta -5) +32\pi (8 \eta -3)]+2 \beta ^2 t^2 \{6 \eta\lambda (6 \eta -5) +9 \lambda\\
 +16\pi[2 \eta (12 \eta -7)+3]\}+\beta  \eta  (2 \eta -1) (8 \eta -1)(3 \lambda +32 \pi) t+(1-2 \eta )^2 [3 \eta \lambda (\eta +1) +16 \eta \pi (2 \eta +1)].
\end{multline}
In Fig. \ref{ch6fig6} to Fig. \ref{ch6fig8z}, we have plotted  
the above values with respect to time $t$ and redshift $z$ as follows.
\section{The $f(R,T)=R+m e^{(-nR)}+\lambda T$ model}\label{ch6model3}
\begin{figure}[H]
\minipage{0.48\textwidth}
 \includegraphics[width=75mm]{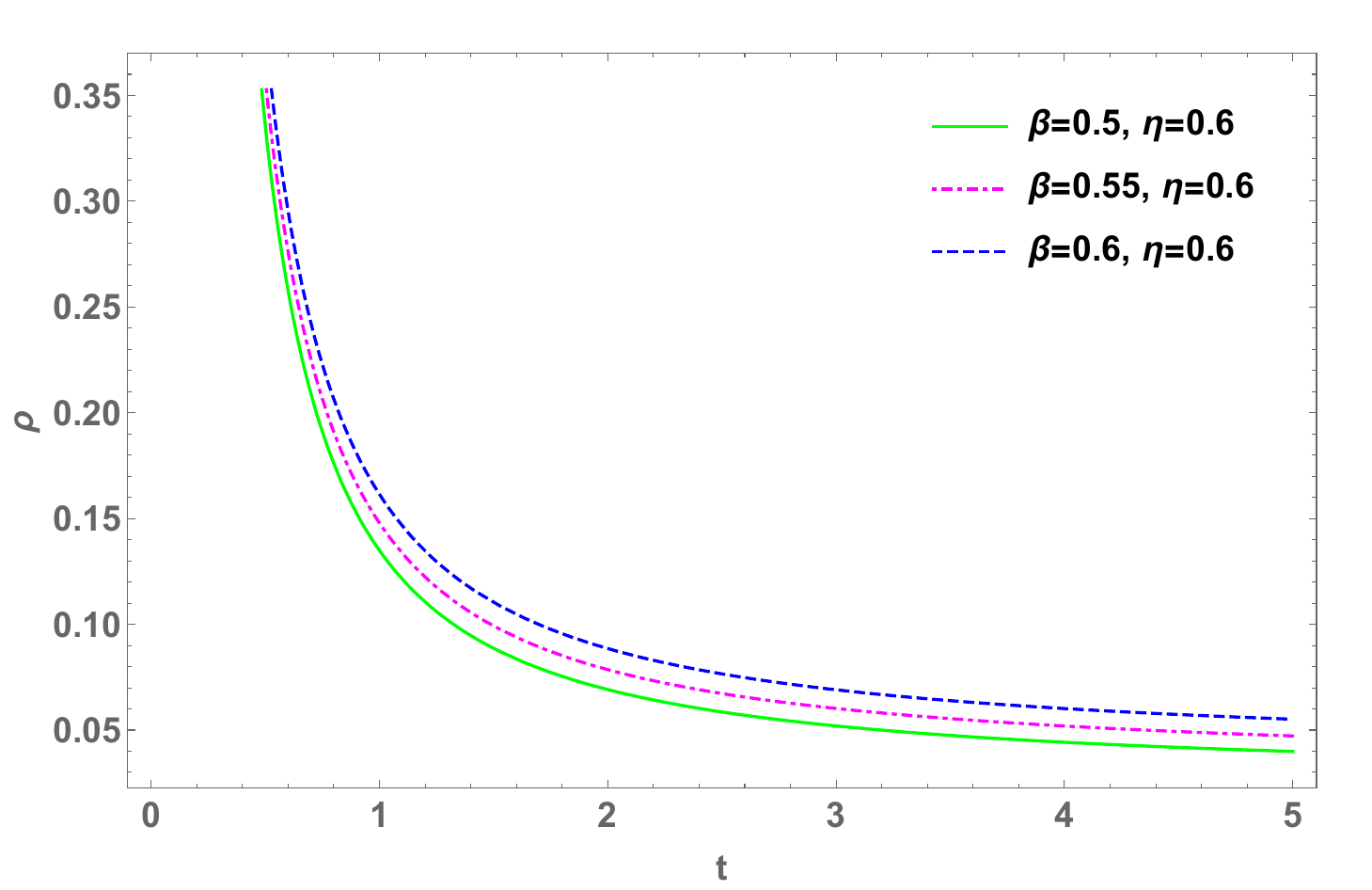}
  \caption{Plot of $\rho$ against time with $m=0.2$, $n=0.05$, $\lambda=0.5$.}\label{ch6fig9}
\endminipage\hfill
\minipage{0.50\textwidth}
 \includegraphics[width=75mm]{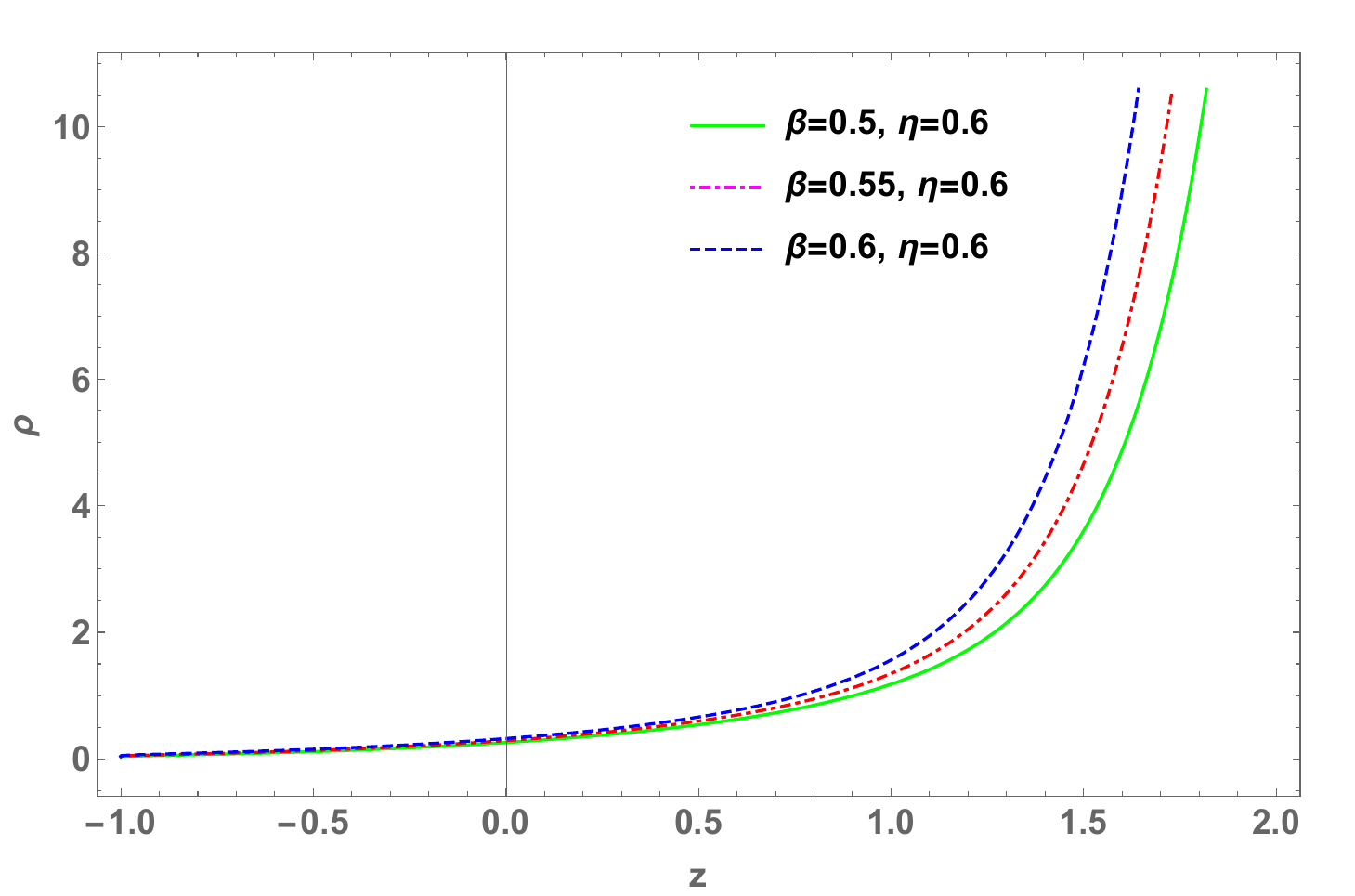}
  \caption{Plot of $\rho$ against $z$ with $m=0.2$, $n=0.05$, $\lambda=0.5$.}\label{ch6fig9z}
\endminipage
\end{figure}

\begin{figure}[H]
\minipage{0.48\textwidth}
 \includegraphics[width=75mm]{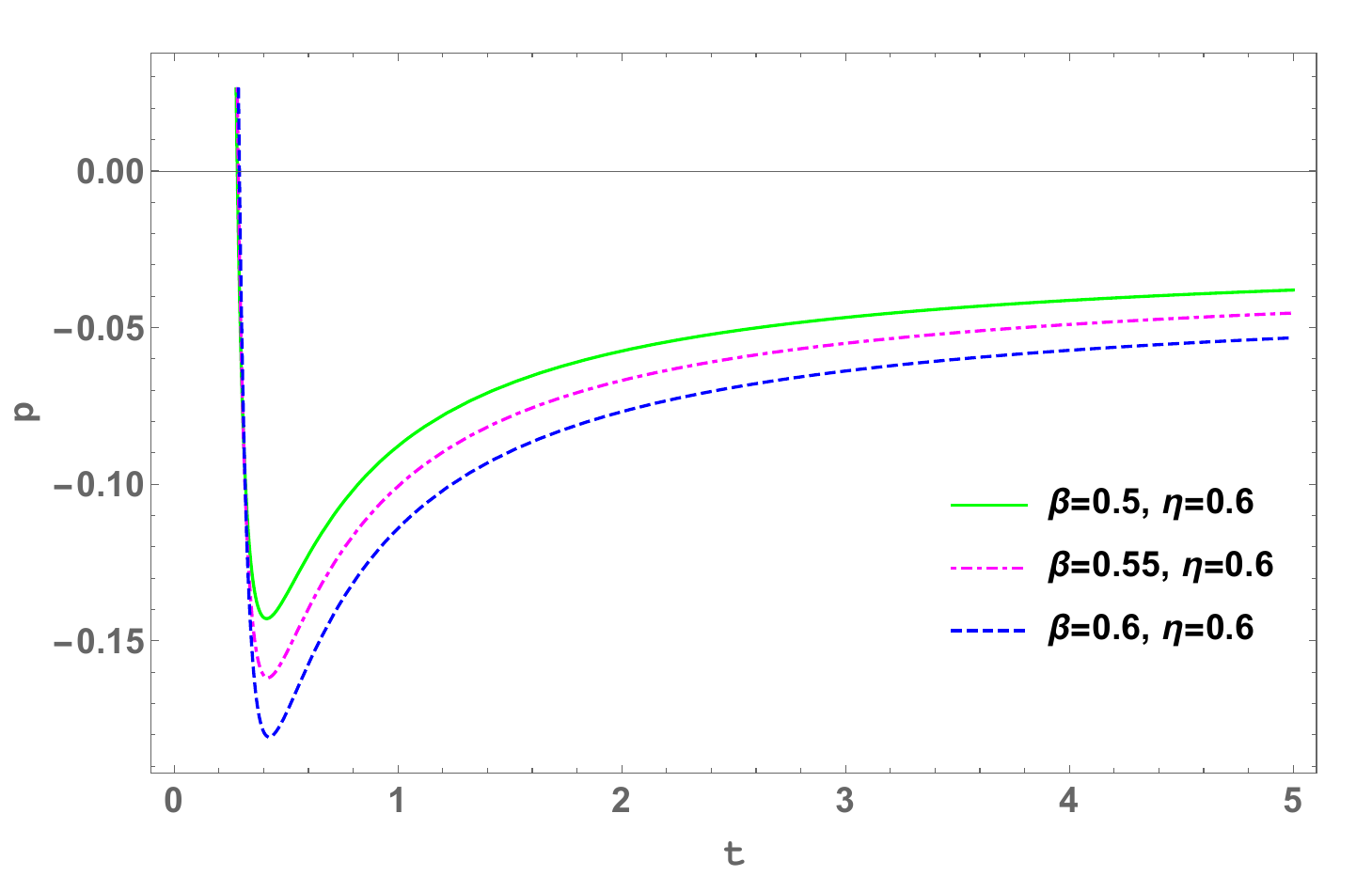}
  \caption{Plot of $p$ against time with $m=0.2$, $n=0.05$, $\lambda=0.5$.}\label{ch6fig10}
\endminipage\hfill
\minipage{0.50\textwidth}
 \includegraphics[width=75mm]{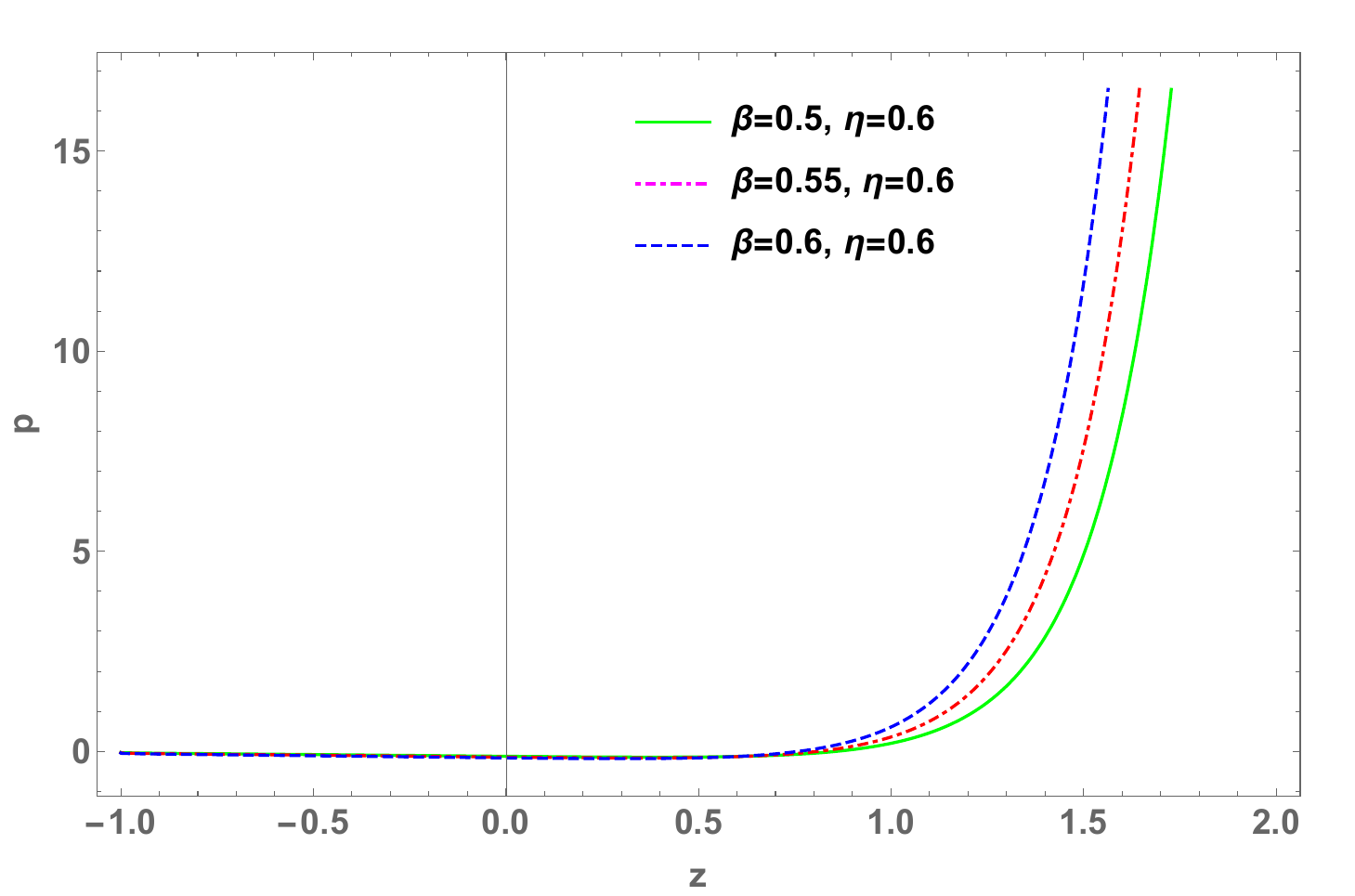}
  \caption{Plot of $p$ against $z$ with $m=0.2$, $n=0.05$, $\lambda=0.5$.}\label{ch6fig10z}
\endminipage
\end{figure}

\begin{figure}[H]
\minipage{0.48\textwidth}
 \includegraphics[width=75mm]{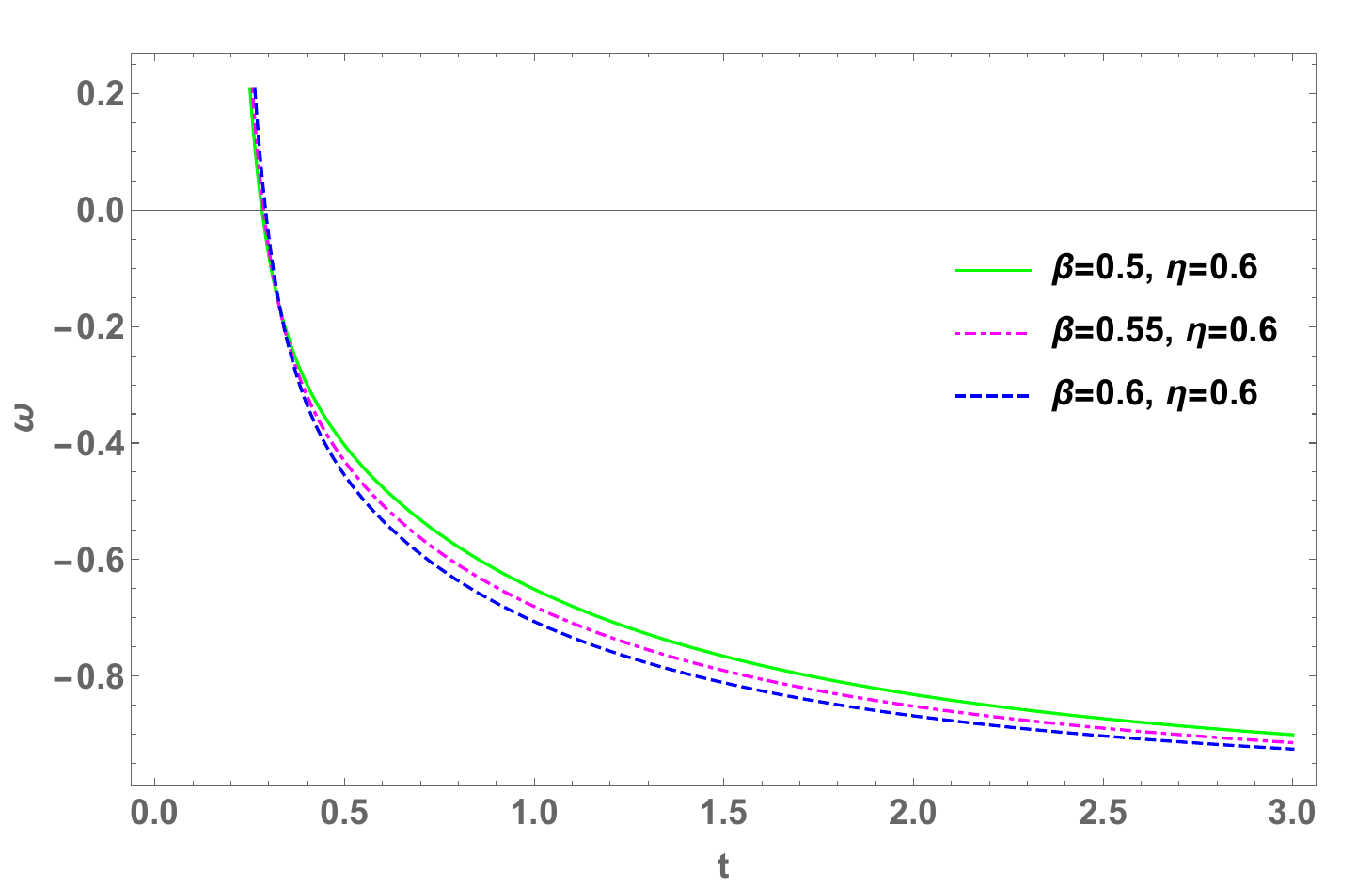}
  \caption{Plot of EoS Parameter against time with $m=0.2$, $n=0.05$, $\lambda=0.5$.}\label{ch6fig11}
\endminipage\hfill
\minipage{0.50\textwidth}
 \includegraphics[width=75mm]{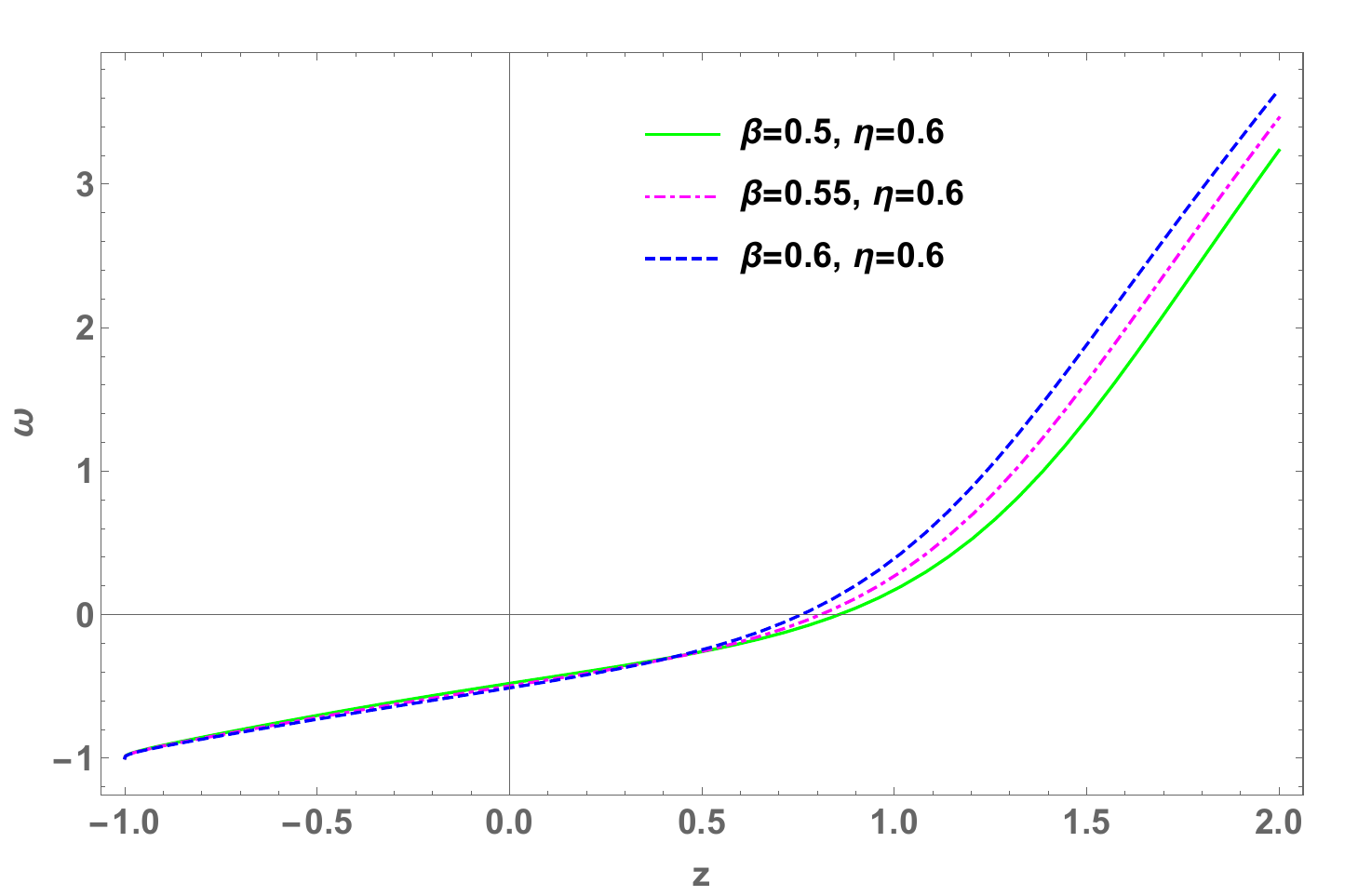}
  \caption{Plot of EoS Parameter against $z$ with $m=0.2$, $n=0.05$, $\lambda=0.5$.}\label{ch6fig11z}
\endminipage
\end{figure}

By taking $f(R,T)=R+m e^{(-nR)}+\lambda T$ and eqn. (\ref{ch6sp1}) in eqns. (\ref{ch6rho}-\ref{ch6omega}), the analytical forms for $p$, $\rho$ and $\omega$ are expressed as
\begin{multline}\label{ch6rho3}
\rho=\left\{1-m n e^{\frac{6 n \left[2 (\eta +\beta  t)^2-\eta \right]}{t^2}}\right\}\left[\frac{ \eta }{t^2 (\lambda +8 \pi)}-\frac{3 (\eta +\beta  t)^2- \eta}{(2 \lambda +8 \pi)t^2}\right]
+\frac{3 \eta  m n^2 e^{\frac{6 n \left[2 (\eta +\beta  t)^2-\eta \right]}{t^2}}G_{13}(t)}{t^6} \\
-\frac{m e^{\frac{6 n \left[2 (\eta +\beta  t)^2-\eta \right]}{t^2}}}{4 (\lambda +4 \pi )} \left\{1-\frac{6 n \left[2 (\eta +\beta  t)^2-\eta \right]}{t^2}\right\},
\end{multline}
\begin{multline}\label{ch6p3}
p=\left\{1-m n e^{\frac{6 n \left[2 (\eta +\beta  t)^2-\eta \right]}{t^2}}\right\}\left[\frac{ \eta }{(\lambda +8 \pi ) t^2}-\frac{3 (\eta +\beta  t)^2- \eta}{(2 \lambda +8 \pi)t^2 }\right] +\frac{3 \eta  m n^2 e^{\frac{6 n \left[2 (\eta +\beta  t)^2-\eta \right]}{t^2}}F_{13}(t) }{t^6}\\
+\frac{m e^{\frac{6 n \left[2 (\eta +\beta  t)^2-\eta \right]}{t^2}}}{4 (\lambda +4 \pi)} \left\{1-\frac{6 n \left[2 (\eta +\beta  t)^2-\eta \right]}{t^2}\right\},
\end{multline}
\begin{equation}\label{ch6omega3}
\omega=\dfrac{\splitfrac{\left\{1-m n e^{\frac{6 n \left[2 (\eta +\beta  t)^2-\eta \right]}{t^2}}\right\}\left[\frac{ \eta }{(\lambda +8 \pi ) t^2}-\frac{3 (\eta +\beta  t)^2- \eta}{(2 \lambda +8 \pi)t^2 }\right] +\frac{3 \eta  m n^2 e^{\frac{6 n \left[2 (\eta +\beta  t)^2-\eta \right]}{t^2}}F_{13}(t) }{t^6}}{
+\frac{m e^{\frac{6 n \left[2 (\eta +\beta  t)^2-\eta \right]}{t^2}}}{4 (\lambda +4 \pi)} \left\{1-\frac{6 n \left[2 (\eta +\beta  t)^2-\eta \right]}{t^2}\right\}}}
{\splitfrac{\left\{1-m n e^{\frac{6 n \left[2 (\eta +\beta  t)^2-\eta \right]}{t^2}}\right\}\left[\frac{ \eta }{t^2 (\lambda +8 \pi)}-\frac{3 (\eta +\beta  t)^2- \eta}{(2 \lambda +8 \pi)t^2}\right]
+\frac{3 \eta  m n^2 e^{\frac{6 n \left[2 (\eta +\beta  t)^2-\eta \right]}{t^2}}G_{13}(t)}{t^6}}{
-\frac{m e^{\frac{6 n \left[2 (\eta +\beta  t)^2-\eta \right]}{t^2}}}{4 (\lambda +4 \pi )} \left\{1-\frac{6 n \left[2 (\eta +\beta  t)^2-\eta \right]}{t^2}\right\}}},
\end{equation}
where
\begin{multline}
G_{13}(t)=\frac{12 \eta  n (2 \eta +2 \beta  t-1)^2-5 t^2 (\eta +\beta  t) (2 \eta +2 \beta  t-1)+t^2 (6 \eta +4 \beta  t-3)}{\lambda +4 \pi}\\
+\frac{2 \left[12 \eta  n (2 \eta +2 \beta  t-1)^2+t^2 (\eta +\beta  t) (2 \eta +2 \beta  t-1)+t^2 (6 \eta +4 \beta  t-3)\right]}{\lambda +8 \pi},
\end{multline}
\begin{multline}
F_{13}(t)=\frac{12 \eta  n (2 \eta +2 \beta  t-1)^2-5 t^2 (\eta +\beta  t) (2 \eta +2 \beta  t-1)+t^2 (6 \eta +4 \beta  t-3)}{\lambda +4 \pi}\\
+\frac{2 \left[12 \eta  n (2 \eta +2 \beta  t-1)^2+t^2 (\eta +\beta  t) (2 \eta +2 \beta  t-1)+t^2 (6 \eta +4 \beta  t-3)\right]}{\lambda +8 \pi}.
\end{multline}
The plot of this model values with respect to time $t$ and redshift $z$ are represented in Fig. \ref{ch6fig9} to Fig. \ref{ch6fig11z} as follows.
\textbf{\textit{Om Diagnostic Analysis}}\\
In order to differentiate DE models, the state finder parameters and Om diagnostic analysis are widely used in the literature \cite{Sahni/2008}. In cosmological model understanding, the HP, DP and EoS parameters play an important role. It is known from the literature that DE models produce a positive HP and a negative DP. So $H$ and $q$ cannot be used to differentiate effectively between different DE models. Thus Om diagnostic analysis plays a crucial role in such analysis. The Om diagnosis has also been applied to Galileon models \cite{Jamil/2013,Fromont/2013}. The Om$(z)$ parameter for spatially flat universe is given by \cite{Sahni/2008,Zunckel/2008}
\begin{equation}
Om(z)=\frac{\left(\frac{H(z)}{H_0}\right)^2-1}{(1+z)^3-1}.
\end{equation}
Here, $H_0$ is the present value of the HP. One can observe that the Om$(z)$ parameter involves first derivatives of the scale factor, so Om diagnosis is a simpler diagnostic than the state finder diagnosis. The positive, negative and zero values of Om$(z)$ represent the phantom ($\omega<-1$),  quintessence ($\omega>-1$) and $\Lambda$CDM DE models, respectively \cite{Shahalam/2015}.

\begin{figure}[H]
\centering
\includegraphics[width=75mm]{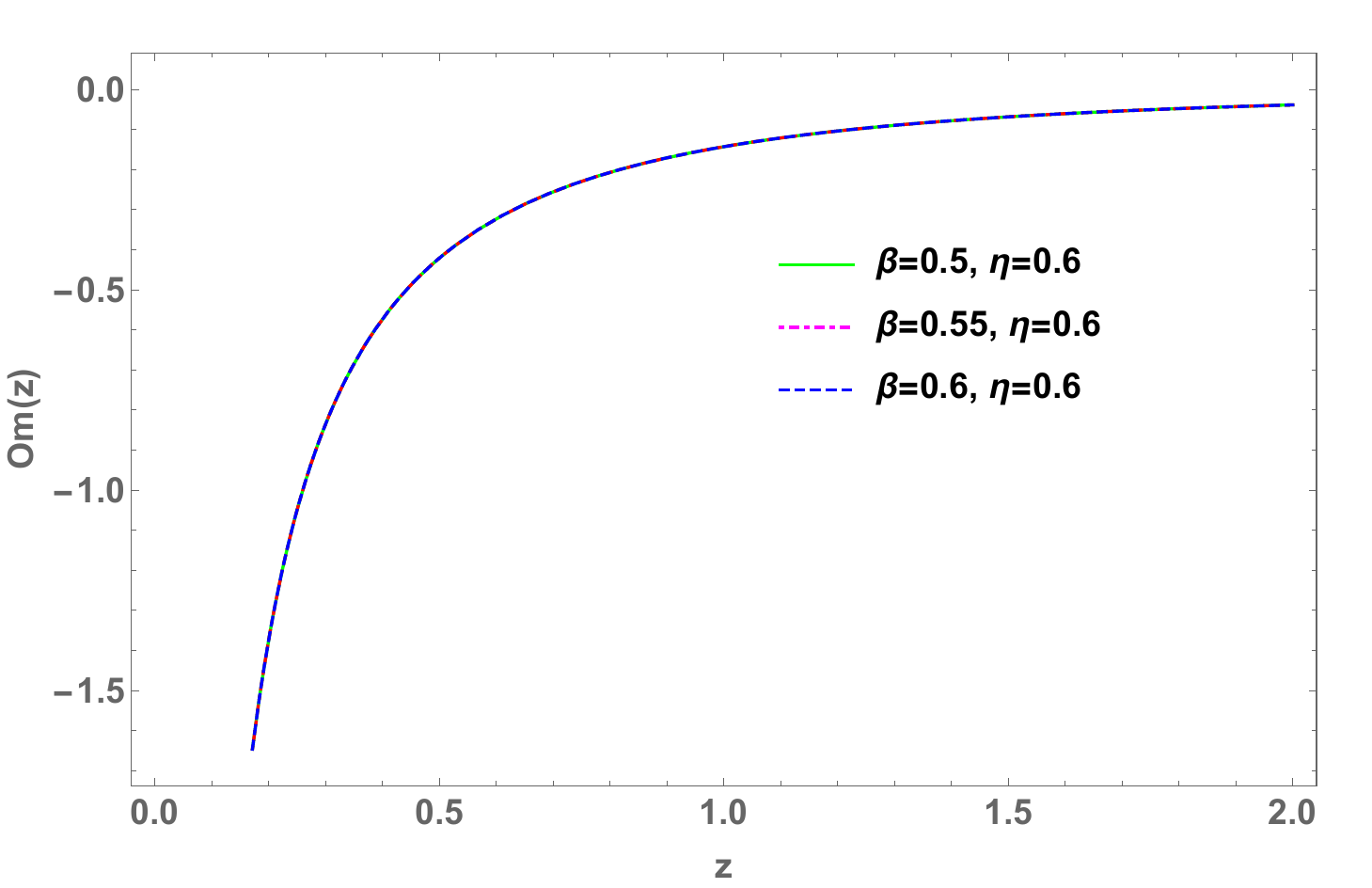}
  \caption{Variation of Om(z) against $z$ with $H_0=67.77$ km s$^{-1}$M pc$^{-1}$.}\label{ch6fig20}
\end{figure}

In the discussed models, the Om$(z)$ parameter takes the form
\begin{equation}
Om(z)=\frac{(\beta^2-H_0^2)W^2\left[\frac{\beta  \left(\frac{1}{z+1}\right)^{1/\eta }}{\eta }\right]+2\beta^2W\left[\frac{\beta  \left(\frac{1}{z+1}\right)^{1/\eta }}{\eta }\right]+\beta^2}{W^2\left[\frac{\beta  \left(\frac{1}{z+1}\right)^{1/\eta }}{\eta }\right]H_0^2z(3+3z+z^2)},
\end{equation}
and its behavior can be observed in the Fig. \ref{ch6fig20}. Here, we have plotted $Om(z)$ for the redshift range $0\leq z\leq 2$. We observe that when the redshift $z$ is increasing within the interval $0\leq z\leq 2$, the $Om(z)$ is monotonically increasing, which also indicates the accelerated expansion of the universe.
\section{The $f(R,T)=R+\alpha R^2+\lambda T$ model}\label{ch6model4}
In this model, we consider the energy-momentum tensor for the relativistic source of anisotropic fluid as
\begin{equation}\label{ch6ent2}
T_{\mu \nu}=(\rho+p_t)u_\mu u_\nu-p_t g_{\mu \nu}+(p_r-p_t)X_\mu X_\nu,
\end{equation}
where $\rho, p_r,$ and $p_t$ are the energy density, radial pressure and tangential pressure, respectively. Moreover $u_\mu$ and $X_\mu$ are the four-velocity vector and radial unit four-vector, respectively, which satisfy the relations $u^{\mu} u_\mu=1$ and $X^\mu X_\nu=-1$. 

By considering the matter Lagrangian of this model as $\mathcal{L}_\text{m}=-\mathcal{P}$, $\theta_{\mu \nu}$ can be rewritten as
\begin{equation}\label{ch6m45}
\theta_{\mu \nu}=-2T_{\mu \nu}-\mathcal{P} g_{\mu \nu},
\end{equation}
where $\mathcal{P}=\frac{p_r+2p_t}{3}$ is the total pressure, Hence, the $f(R,T)$ gravity field eqns. (\ref{ch6e5}) with (\ref{ch6m45}) take the form
\begin{equation}\label{ch6m46}
R_{\mu \nu}-\frac{1}{2}Rg_{\mu \nu}= T_{\mu \nu}^{eff},
\end{equation}
where $T_{\mu \nu}^{eff}$ redefined as 
\begin{multline}\label{ch6m47}
 T_{\mu \nu}^{eff}=\frac{1}{f_R(R,T)}\left\{[8\pi+f_T(R,T)]T_{\mu \nu}+\mathcal{P}g_{\mu \nu}f_R(R,T)\right\}\\
 + \frac{1}{f_R(R,T)}\left\{\frac{1}{2}[f(R,T)-Rf_R(R,T)]g_{\mu \nu}\right\}\\
-\frac{1}{f_R(R,T)} \left[(g_{\mu \nu}\square-\nabla_{\mu}\nabla_{\nu})f_R(R,T)\right].
\end{multline}
\subsection{WH solution in $R^2$-gravity model} 
The static spherically symmetric WH metric is considered here in Schwarzschild coordinates $(t,r,\theta, \phi)$ as described in eqn. (\ref{eqnwh}) of WH section \ref{ch1WH} in chapter \ref{Chapter1}.
The effective field eqns. (\ref{ch6m46}) for the metric in eqn. (\ref{eqnwh}) with constant redshift function $a(r)$ and the specific functional form of $f(R,T)$ gravity, i.e. $f(R,T)=R+\alpha R^2+\lambda T$, read as
\begin{equation}
\frac{b'}{r^2}=\frac{\left[\left(8\pi+\frac{3\lambda}{2}\right)\rho-\frac{\lambda(p_r+2p_t)}{6}-\frac{2\alpha b'^2}{r^4}\right]}{2\alpha R+1},\label{ch6eqn3}
\end{equation}
\begin{equation}
\frac{b}{r^3}=\frac{\left[-\left(8\pi+\frac{7\lambda}{6}\right)p_r+\frac{\lambda}{2}\left(\rho-\frac{2p_t}{3}\right)-\frac{2\alpha b'^2}{r^4}\right]}{2\alpha R+1},\label{ch6eqn4}
\end{equation}
\begin{equation}
\frac{b'r-b}{2r^3}=\frac{\left[-\left(8\pi+\frac{4\lambda}{3}\right)p_t+\frac{\lambda}{2} \left(\rho-\frac{p_r}{3}\right)-\frac{2\alpha b'^2}{r^4}\right]}{2\alpha R+1}.\label{ch6eqn5}
\end{equation}
From the above eqns. (\ref{ch6eqn3}-\ref{ch6eqn5}), the explicit form of $\rho$, $p_r$ and $p_t$, can be written as 
\begin{equation}\label{ch6m4rho}
\rho=\frac{ b' \left[\lambda  \left(2 r^2-5 \alpha   b'\right)+12 \pi  \left(r^2-2 \alpha   b'\right)\right]}{3 (\lambda +4 \pi ) (\lambda +8 \pi ) r^4},
\end{equation}
\begin{equation}\label{ch6m4pr}
p_r=-\frac{-12 \alpha  b \lambda  b'-48 \pi  \alpha  b b'-\lambda  r^3 b'+7 \alpha  \lambda  r b'^2+24 \pi  \alpha  r b'^2+3 b \lambda  r^2+12 \pi  b r^2}{3 (\lambda +4 \pi ) (\lambda +8 \pi ) r^5},
\end{equation}
\begin{equation}\label{ch6m4pt}
p_t=-\frac{12 \alpha  b \lambda  b'+48 \pi  \alpha  b b'+\lambda  r^3 b'+12 \pi  r^3 b'+2 \alpha  \lambda  r b'^2-3 b \lambda  r^2-12 \pi  b r^2}{6 (\lambda +4 \pi ) (\lambda +8 \pi ) r^5}.
\end{equation}
In order to get an exact solution of this model, a specific form of shape function is considered as:  \cite{Heydarzade/2015}
\begin{equation}\label{ch6m416}
b(r)=r_0+\beta_1 r_0\left[\left(\frac{r}{r_0}\right)^{\beta_2}-1\right] ,\,\,\,\,\ 0<\beta_2<1,
\end{equation}
where $\beta_1$ is an arbitrary constant only for this model. The flaring out condition $\frac{b-b'r}{b^2}>0$ implies that $1-\beta_1+\beta_1\left(\frac{r}{r_0}\right)^{\beta_2}(1-\beta_2)>0$.

Similarly, at the throat, for $b'(r_0)<1$, we have $\beta_1 \beta_2 <1$. Also, $\frac{b(r)}{r}=(1-\beta_1)\frac{r_0}{r}+\beta_1\left(\frac{r_0}{r}\right)^{1-\beta_2}\rightarrow 0$ when $r\rightarrow \infty$ for $\beta_2 <1$. 

\begin{figure}[H]
\centering
  \includegraphics[width=75mm]{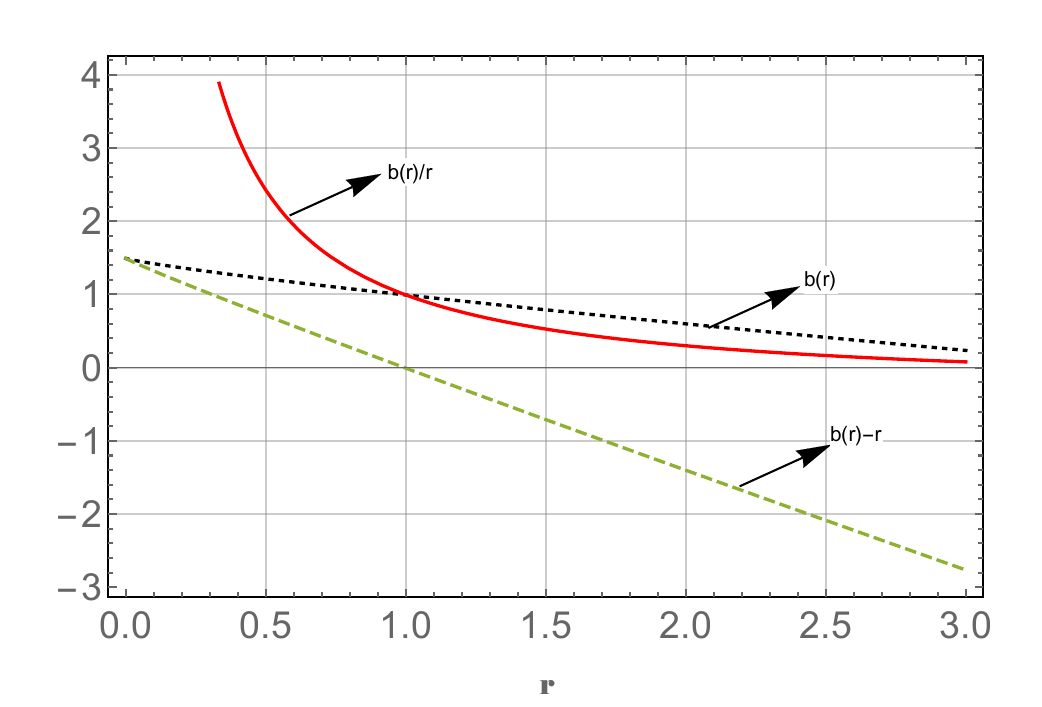}
  \caption{$b(r)$, $b(r)/r$ and $b(r)-r$ as functions of $r$ for $\beta_1=-0.5$, $\beta_2=0.84$ and $r_0=0.995$.}\label{ch6m4fig0}
\end{figure}

\begin{figure}[H]
\minipage{0.48\textwidth}
\includegraphics[width=75mm]{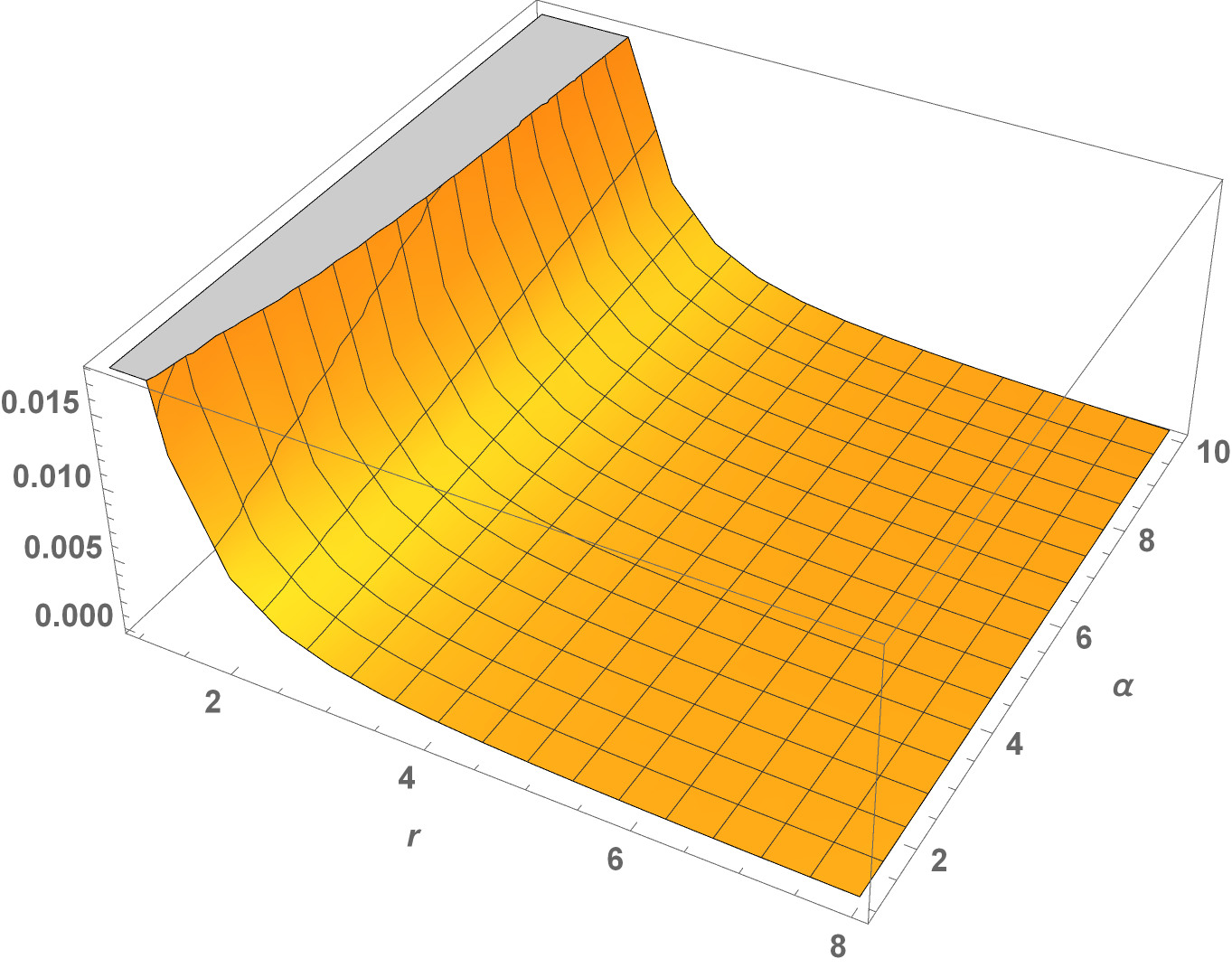}
  \caption{Variation of energy density for $\lambda=-35$ and different $\alpha $.}\label{ch6m4fig1}
\endminipage\hfill
\minipage{0.50\textwidth}
\includegraphics[width=75mm]{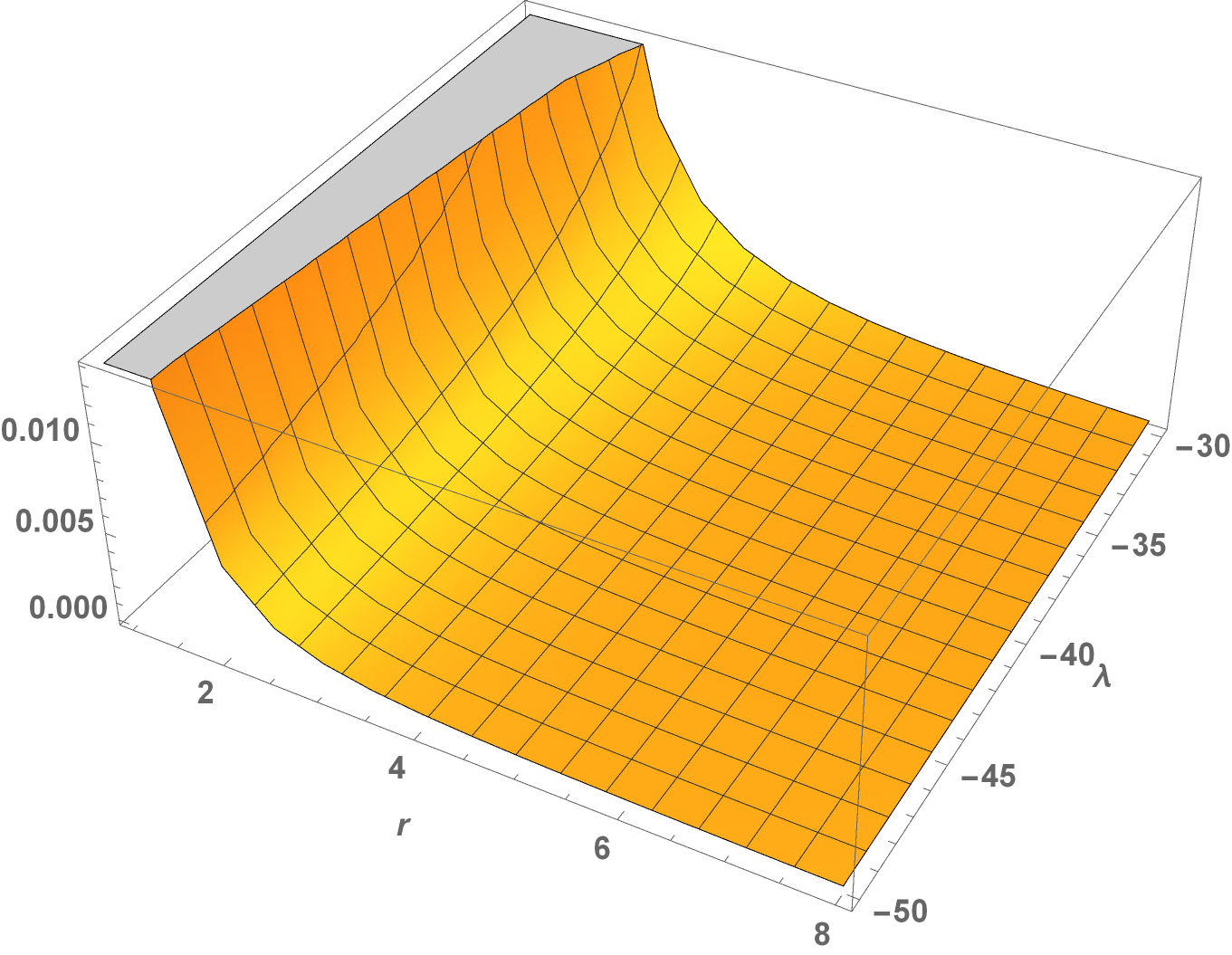}
  \caption{Variation of energy density for $\alpha=5$ and different $\lambda $.}
\label{ch6m4fig2} 
\endminipage
\end{figure}

In order to clarify the metric conditions mentioned in section \ref{ch1WH}, we have plotted here the graph of
 $b(r)$, $b(r)/r$ and $b(r)-r$, for $\beta_1=-0.5$, $\beta_2=0.84$ and $r_0=0.995$, as shown in the Fig. \ref{ch6m4fig0}.
Using the shape function given by eqn. (\ref{ch6m416}) in eqns. (\ref{ch6m4rho} - \ref{ch6m4pt}), we obtain the values of $\rho$, $p_r$ and $p_t$ as
\begin{equation}\label{ch6m4rho1}
\rho=-\frac{\beta_1 \beta_2 r_0 \left(\frac{r}{r_0}\right)^{\beta_2} \left\{-12 \pi  \left[r^3-2 \beta_1 \alpha  \beta_2  r_0 \left(\frac{r}{r_0}\right)^{\beta_2}\right]+5 \beta_1 \alpha  \beta_2  \lambda  r_0 \left(\frac{r}{r_0}\right)^{\beta_2}-2 \lambda  r^3\right\}}{3 (\lambda +4 \pi ) (\lambda +8 \pi ) r^6},
\end{equation}
\begin{multline}\label{ch6m4pr1}
p_r=\frac{\splitfrac{r_0 \biggl[-\beta_1^2 \alpha  \beta_2  r_0 [(7 \beta_2 -12) \lambda +24 \pi  (\beta_2 -2)] \left(\frac{r}{r_0}\right)^{2 \beta_2 }-\beta_1 \left(\frac{r}{r_0}\right)^{\beta_2 } \{\lambda (12 (\beta_1-1) \alpha  \beta_2  r_0-}{(\beta_2 -3) r^3)+12 \pi  \left((4\beta_1-4) \alpha  \beta_2 r_0+r^3\right)\}\biggr]}}{3 (\lambda +4 \pi ) (\lambda +8 \pi ) r^6}\\
+\frac{r_0 \left[3 (\beta_1-1) (\lambda +4 \pi ) r^3\right]}{3 (\lambda +4 \pi ) (\lambda +8 \pi ) r^6},
\end{multline}
\begin{multline}\label{ch6m4pt1}
p_t=\frac{\splitfrac{r_0 \biggl[-2 \beta_1^2 \alpha  \beta_2 r_0 [(\beta_2 +6) \lambda +24 \pi ] \left(\frac{r}{r_0}\right)^{2 \beta_2 }+\beta_1 \left(\frac{r}{r_0}\right)^{\beta_2 } \{\lambda  \left(12 (\beta_1-1) \alpha  \beta_2 r_0-(\beta_2 -3) r^3\right)}{+12 \pi  \left(4 (\beta_1-1) \alpha  \beta_2 r_0-(\beta_2 -1) r^3\right)\}\biggr]}}{6 (\lambda +4 \pi ) (\lambda +8 \pi ) r^6}
\\-\frac{r_0 \left[3 (\beta_1-1) (\lambda +4 \pi ) r^3\right]}{6 (\lambda +4 \pi ) (\lambda +8 \pi ) r^6}.
\end{multline}
From the quantities above, we have plotted the energy density for  $\beta_1=-0.5$, $\beta_2=0.84$, and $r_0=0.995$ as shown in Fig. \ref{ch6m4fig1} and Fig. \ref{ch6m4fig2}.
\subsection{Energy conditions}
\begin{figure}[H]
\minipage{0.48\textwidth}
\includegraphics[width=75mm]{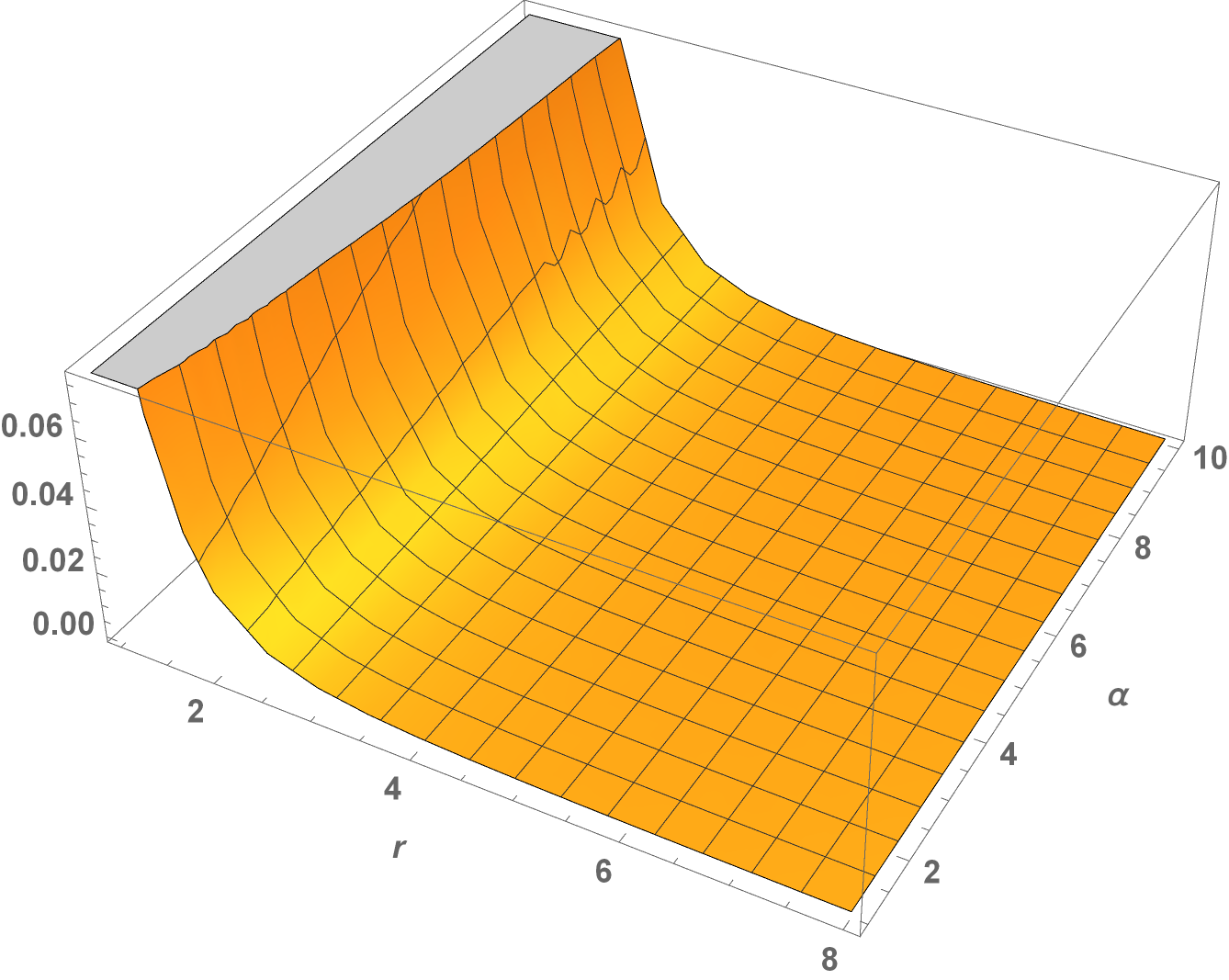}
  \caption{Validity of NEC, $\rho+p_r\geq 0$, for $\lambda=-35$ and different $\alpha $.}\label{ch6m4fig3}
\endminipage\hfill
\minipage{0.50\textwidth}
\includegraphics[width=75mm]{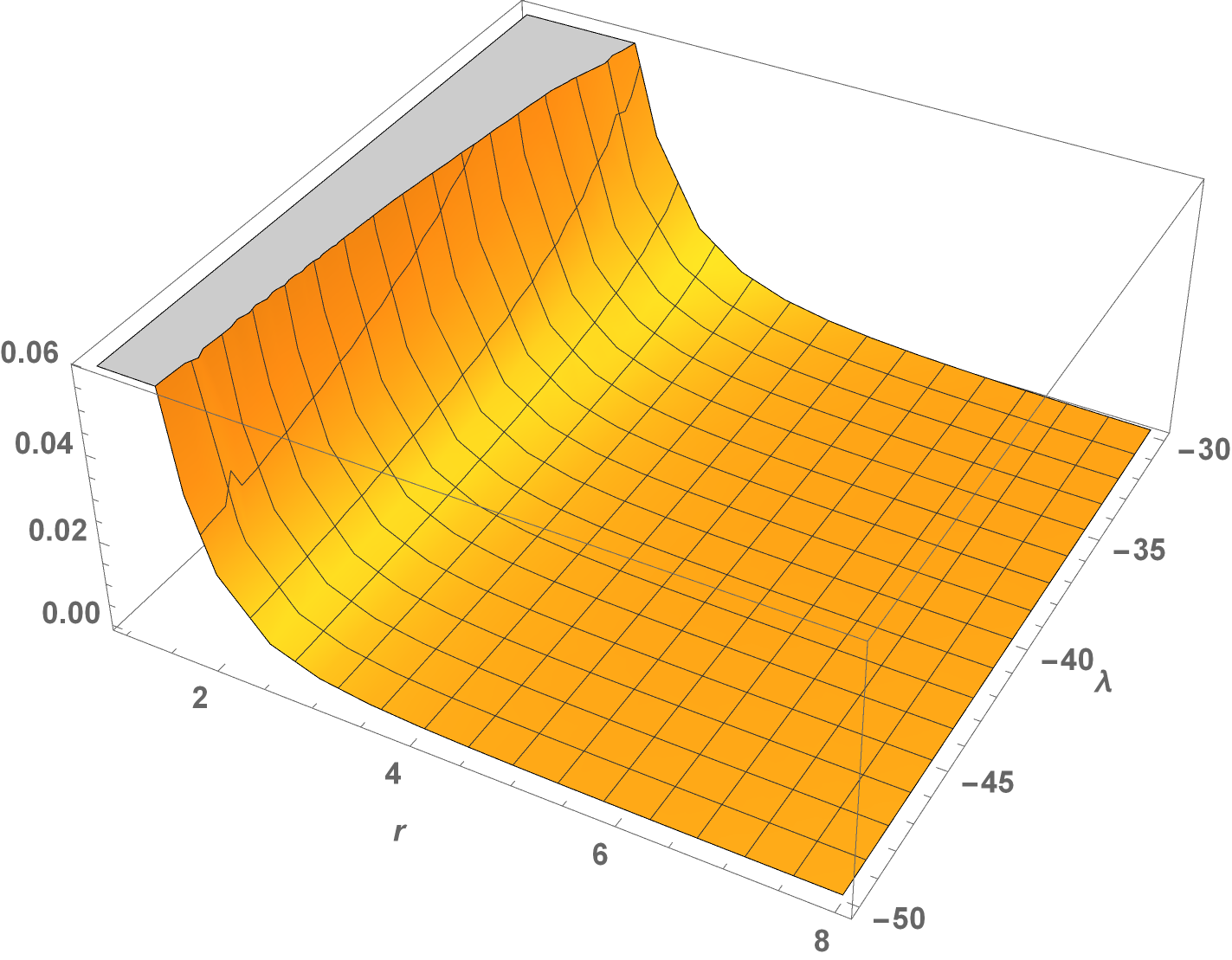}
  \caption{Validity of NEC, $\rho+p_r\geq 0$, for $\alpha=5$ and different $\lambda $.}\label{ch6m4fig4}
\endminipage
\end{figure}

\begin{figure}[H]
\minipage{0.48\textwidth}
\includegraphics[width=75mm]{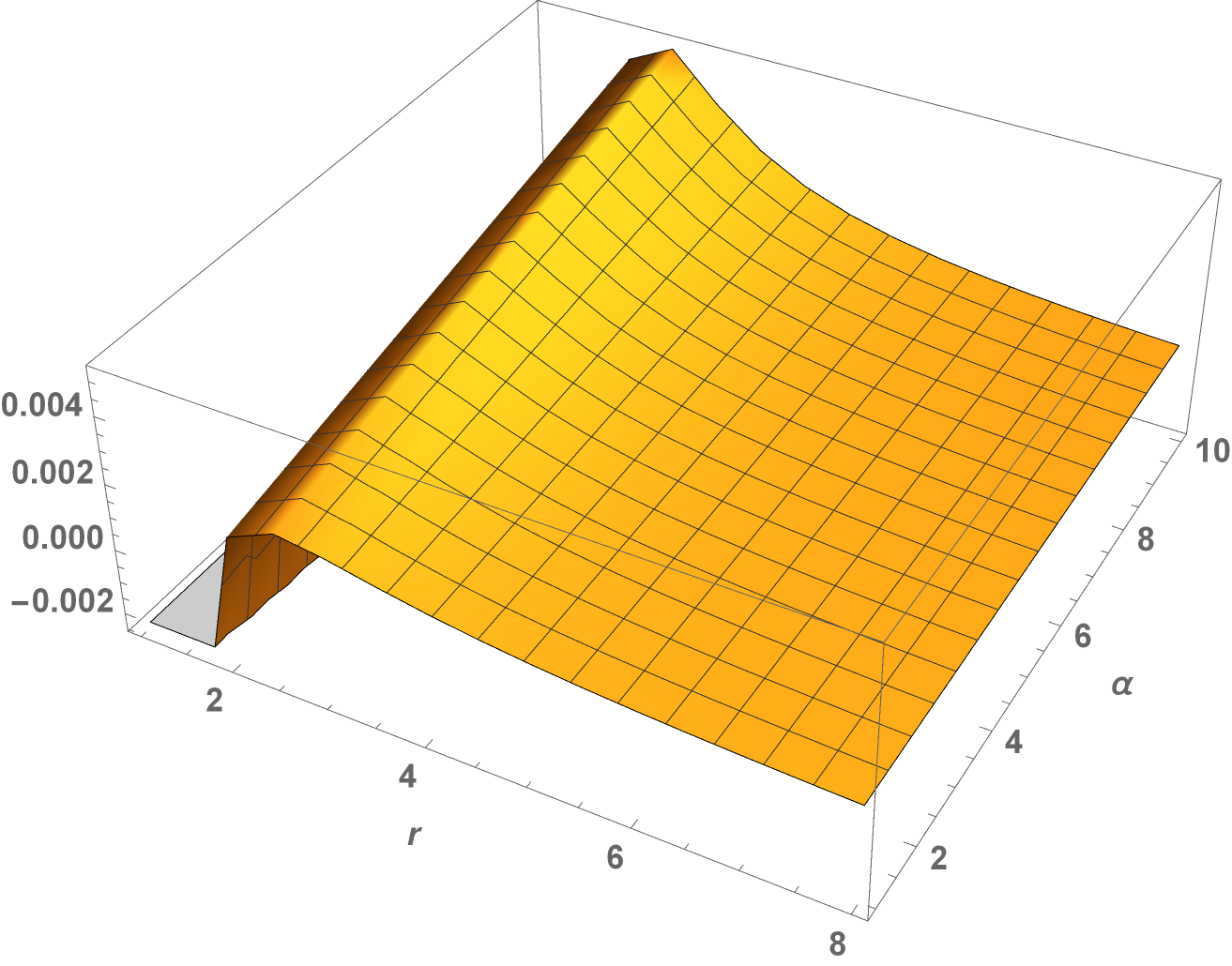}
  \caption{NEC, $\rho+p_t\geq 0$, for $\lambda=-35$ and different $\alpha $.}\label{ch6m4fig5}
\endminipage\hfill
\minipage{0.50\textwidth}
\includegraphics[width=75mm]{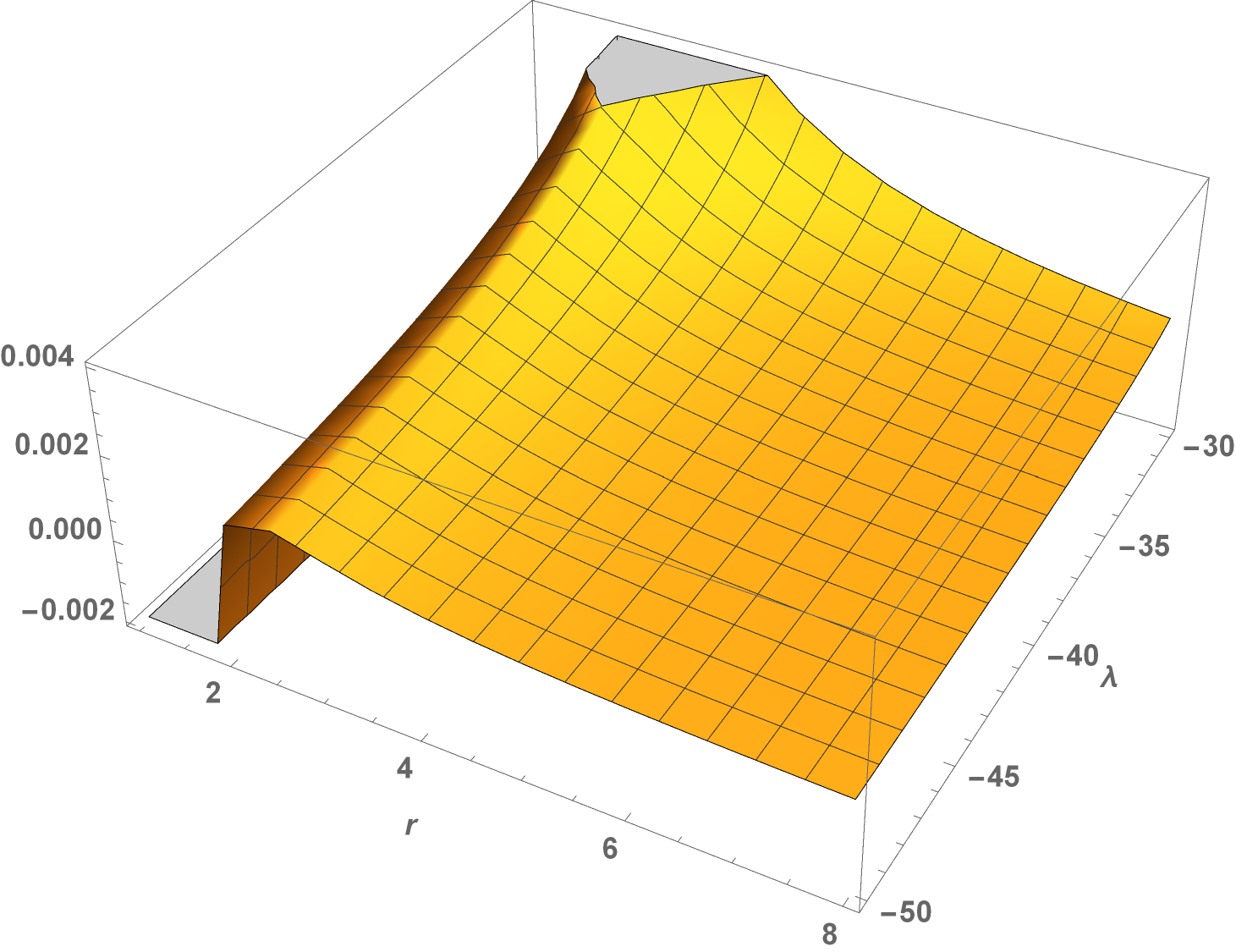}
  \caption{NEC, $\rho+p_t\geq 0$, for $\alpha=5$ and different $\lambda $.}\label{ch6m4fig6}
\endminipage
\end{figure}

\begin{figure}[H]
\minipage{0.48\textwidth}
\includegraphics[width=75mm]{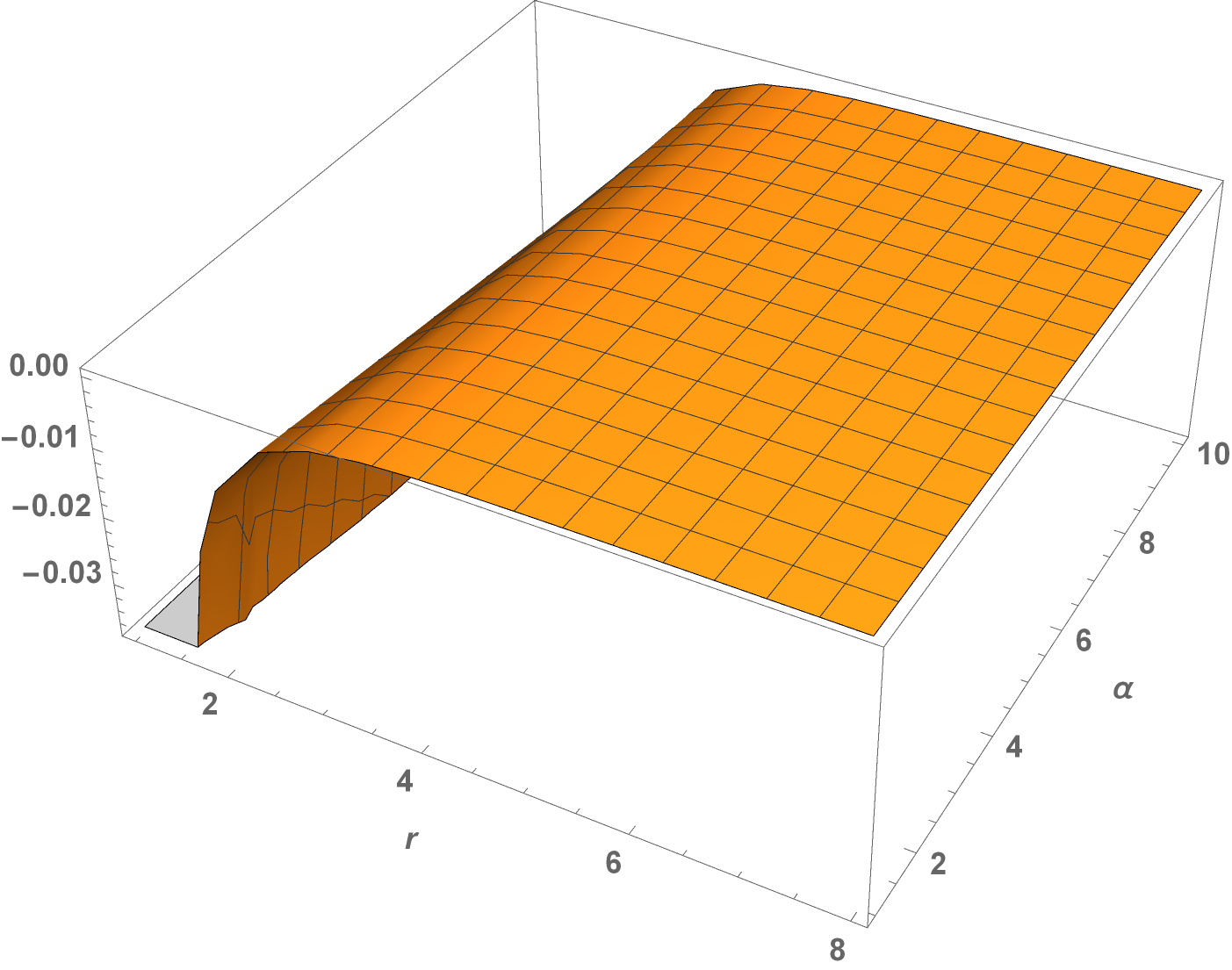}
  \caption{Violation of DEC, $\rho\geq \vert p_r\vert$, for $\lambda=-35$ and different $\alpha $.}\label{ch6m4fig7}
\endminipage\hfill
\minipage{0.50\textwidth}
\includegraphics[width=75mm]{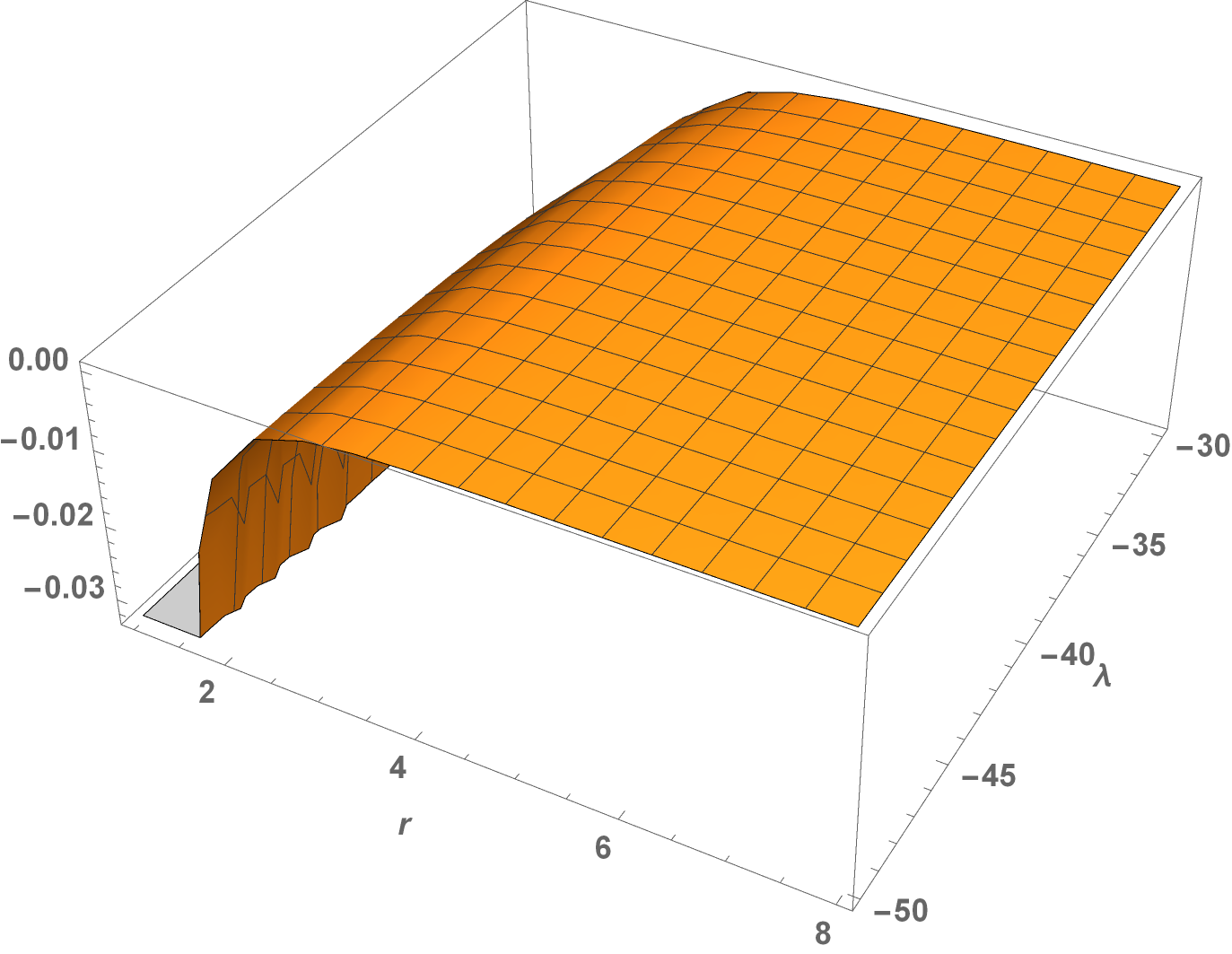}
  \caption{Violation of DEC, $\rho\geq \vert p_r\vert$, for $\alpha=5$ and different $\lambda $.}\label{ch6m4fig8}
\endminipage
\end{figure}

\begin{figure}[H]
\minipage{0.48\textwidth}
\includegraphics[width=75mm]{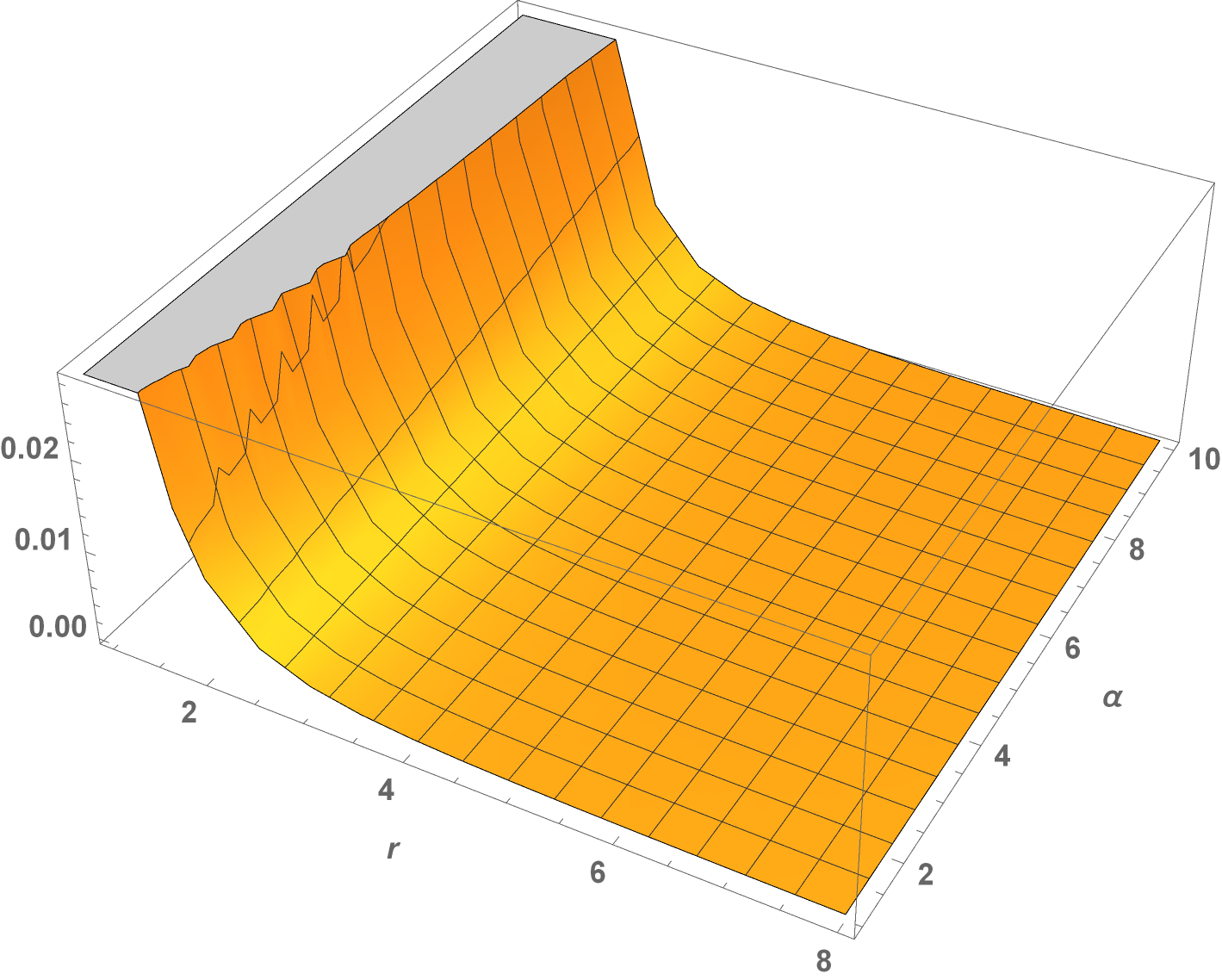}
  \caption{Validation of DEC, $\rho\geq \vert p_t\vert$, for $\lambda=-35$ and different $\alpha $.}\label{ch6m4fig9}
\endminipage\hfill
\minipage{0.50\textwidth}
\includegraphics[width=75mm]{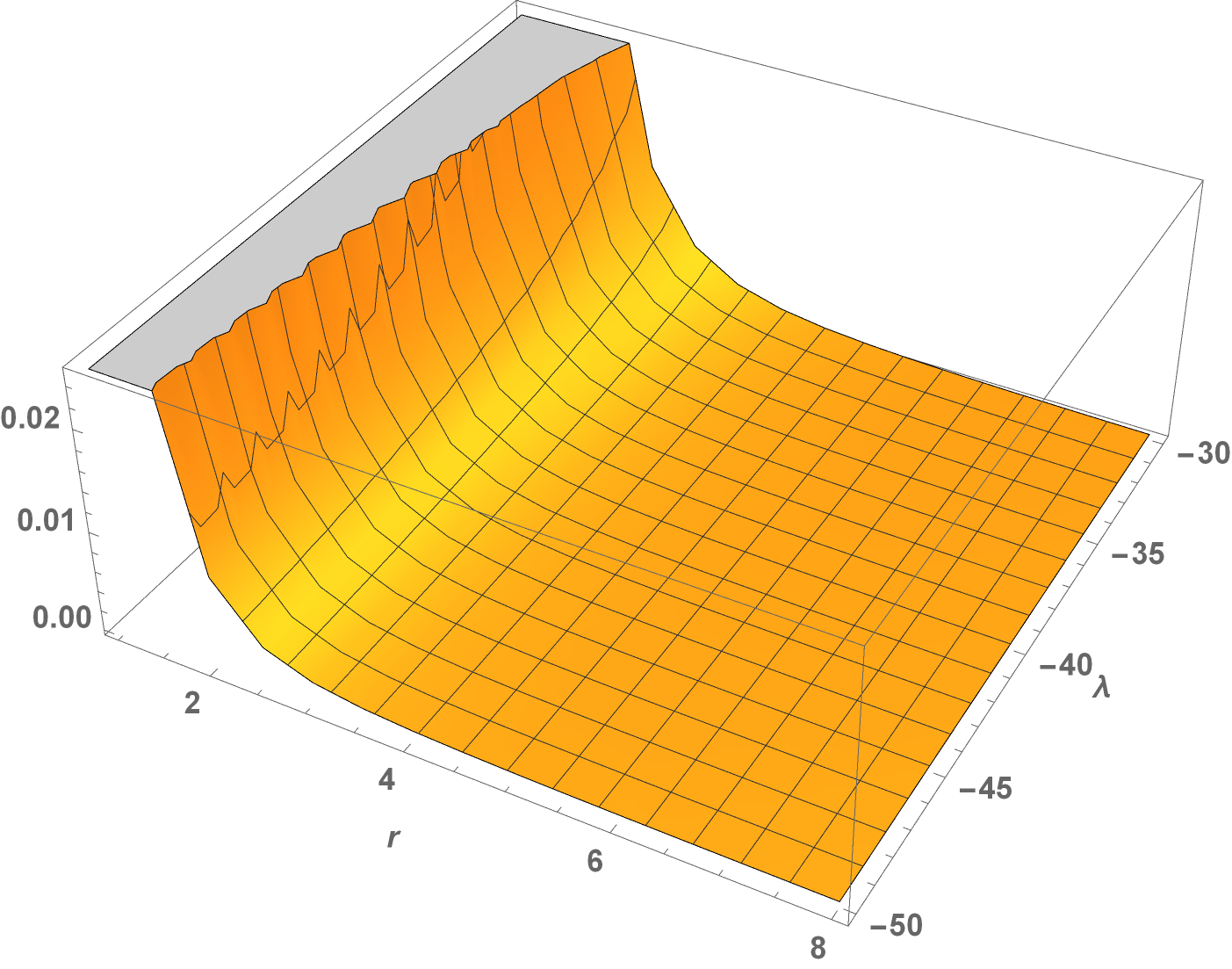}
  \caption{Validation of DEC, $\rho\geq \vert p_t\vert$, for $\alpha=5$ and different $\lambda $.}\label{ch6m4fig10}
\endminipage
\end{figure}

\begin{figure}[H]
\minipage{0.48\textwidth}
\includegraphics[width=75mm]{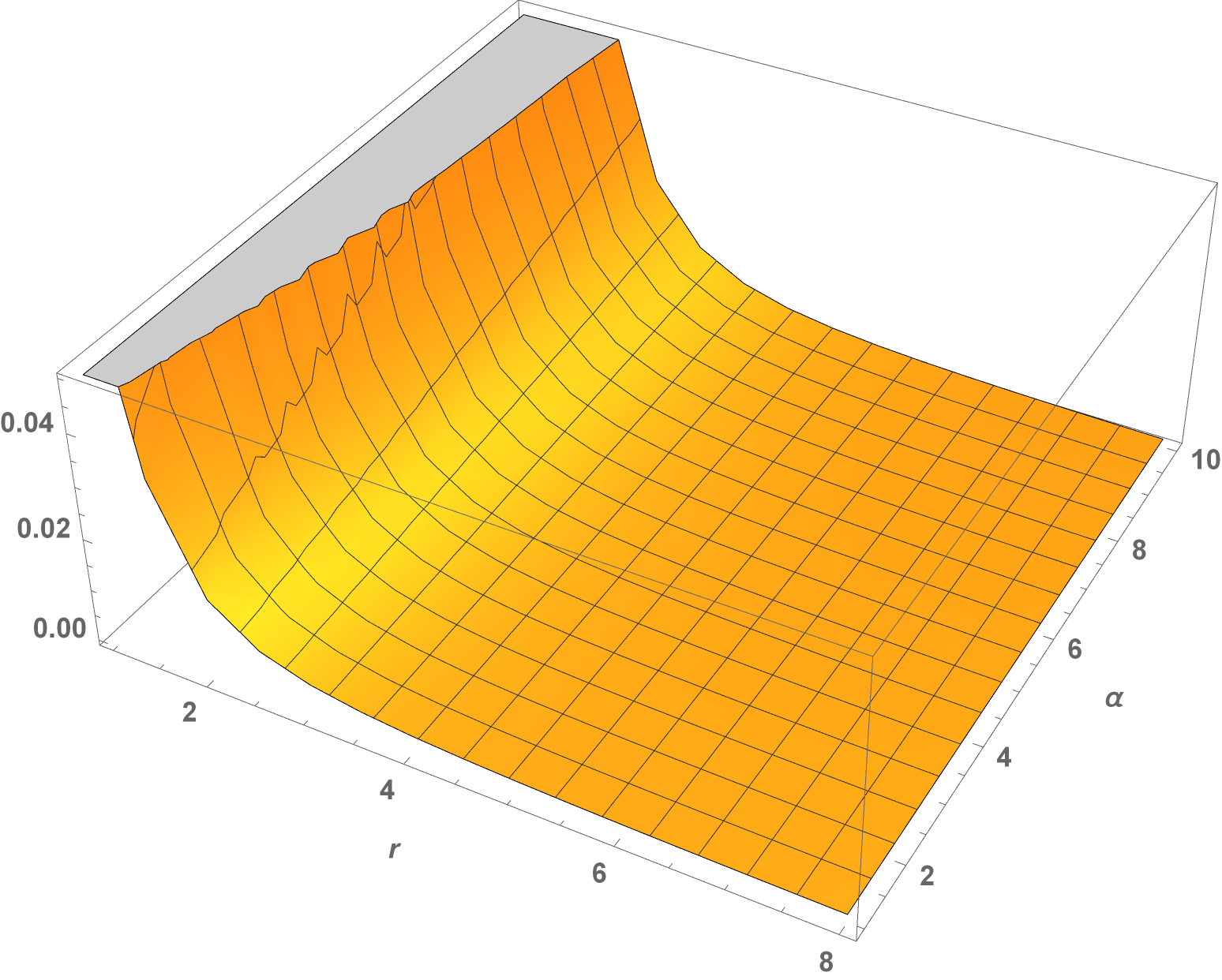}
  \caption{Validation of SEC, $\rho+p_r+2p_t\geq0$, for $\lambda=-35$ and different $\alpha $.}\label{ch6m4fig11}
\endminipage\hfill
\minipage{0.50\textwidth}
\includegraphics[width=75mm]{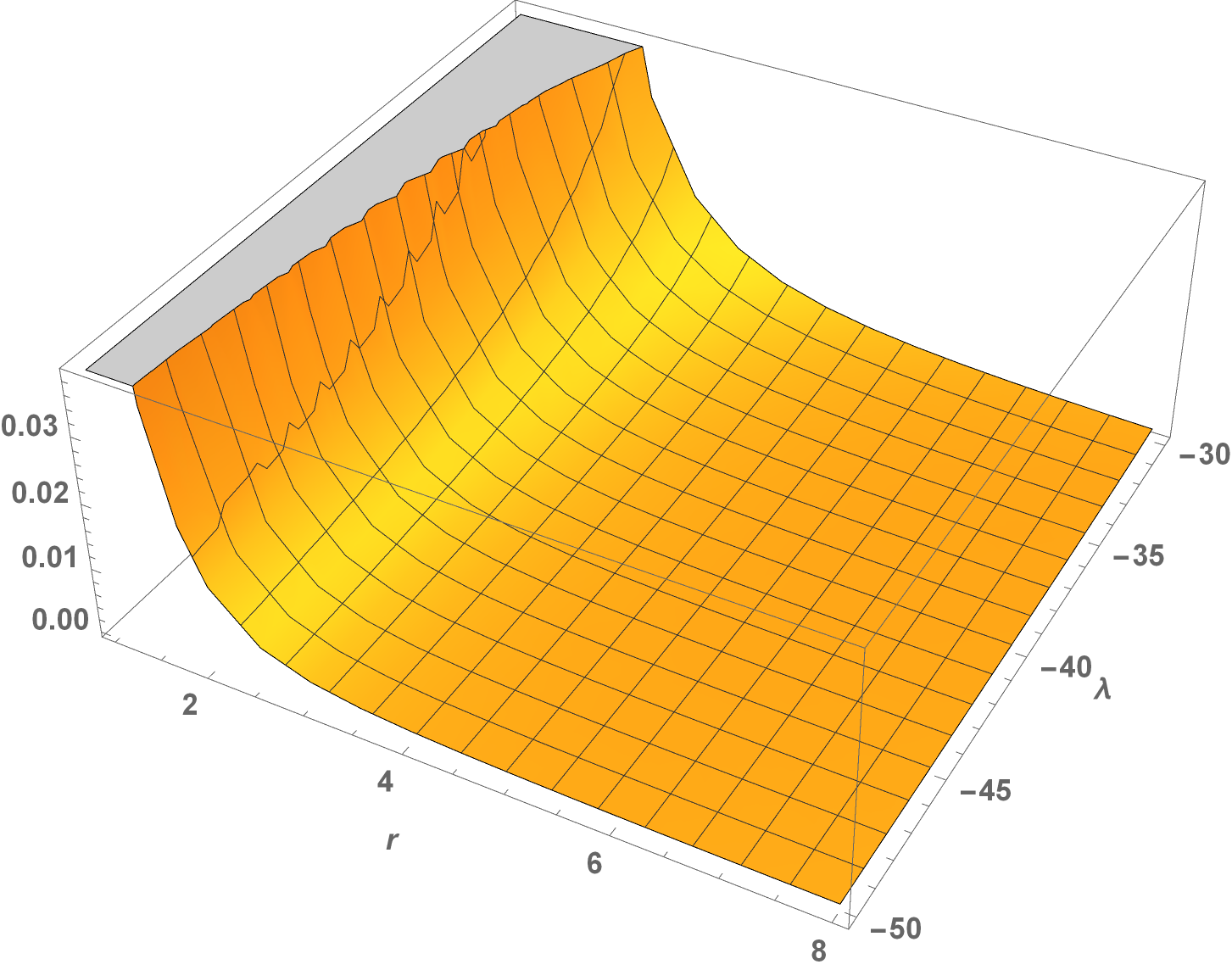}
  \caption{Validation of SEC, $\rho+p_r+2p_t\geq0$, for $\alpha=5$ and different $\lambda $.}\label{ch6m4fig12}
\endminipage
\end{figure}

In possession of eqns. (\ref{ch6m4rho1} - \ref{ch6m4pt1}) we plotted the WH ECs, namely NEC, DEC and SEC in Fig.   \ref{ch6m4fig3} to Fig. \ref{ch6m4fig12}. We mention that the WEC $\rho\geq0$ obedience can also be appreciated in Fig. \ref{ch6m4fig1} and Fig. \ref{ch6m4fig2}.

Fig. \ref{ch6m4fig3} and Fig. \ref{ch6m4fig4} shows the validity of NEC for radial pressure, $\rho+p_r\geq0$, for different $\alpha$ and different $\lambda$, respectively. While Fig. \ref{ch6m4fig5} and Fig. \ref{ch6m4fig6} also represents NEC in terms of $p_t$ showing the violation of NEC only for small values of $r$. For DEC, Fig. \ref{ch6m4fig7} and Fig.  \ref{ch6m4fig8}, in terms of $p_r$, shows its violation for most values of $r$ and Fig. \ref{ch6m4fig9} and Fig. \ref{ch6m4fig10}, in terms of $p_t$, indicate its respectability for all $r$. Similarly, Fig. \ref{ch6m4fig11} and Fig. \ref{ch6m4fig12} show that SEC is respected for the present WH matter content, for different values of $\alpha$ and $\lambda$, respectively.
The respectability of the ECs in a wide region or even in the whole WH, which yields non-exotic matter to fill in this object, is attained as a consequence of the extra (or correction) terms of the theory, namely $\alpha R^{2}$ and $\lambda T$. As remarked above, non-exotic matter WHs are non-trivial to be attained and we believe that once the material correction terms, predicted by the $f(R,T)$ theory, are related with the possible existence of imperfect fluids in the universe. This results may provide a breakthrough for the understanding of WHs in this modified gravity. 

Particularly, the physical reasons for the validity of the ECs as a consequence of the extra terms of the theory are worth a deeper discussion. Apparently, the cosmological and astrophysical observational issues we face nowadays may be overcome by either an alternative gravity theory or a non-standard EoS to describe the matter content concerned. Let us take the DE problem, for example. It is well known that the present universe undergoes an accelerated phase of expansion \cite{Perlmutter/1999,Riess/1998}. The counter-intuitive effect of acceleration may be described either by an exotic EoS for the matter filling the universe, namely $\mathcal{P}\sim-\rho$ \cite{hinshaw/2013}, or by alternative gravity models, as it can be checked in \cite{Starobinsky/2007}. The recently detected massive pulsars \cite{demorest/2010,antoniadis/2013} can also be attained by particular EoS \cite{ozel/2016,fortin/2015} or alternative gravity \cite{mam/2016}.

The same picture can be visualized for the WH case, i.e., the extra degrees of freedom of an extended theory of gravity may also allow WHs to be filled by non-exotic EoS matter, departing from the GR case. The $T$-dependence of the $f(R,T)$ theory may characterize the first steps in describing quantum effects in a gravity theory \cite{Harko11,ms/2017} and such a description, which is missing in GR, can explain the ECs obedience. 
\section{Conclusion}\label{ch6conclusion}
This section deals with the concluding remarks derived from the  described models. The given cosmological solutions admit the observations and theoretical prospective of the present scenario.
Particularly, in case of the EoS parameter evolution Fig.  \ref{ch6fig5}, Fig. \ref{ch6fig8} and Fig. \ref{ch6fig11} show a remarkable feature. The present scenario of evolution of EoS parameter is consistent with three different stages of the universe evolution, namely radiation, matter and DE eras, as we described below.
One can see that for small values of time, $\omega\sim1/3$, which is the EoS parameter value for the primordial stage of the universe in its dynamics was dominated by radiation \cite{ryden/2003}, whose high temperature did not allow, for a period of time, the formation of the first atoms. As the universe cooled down, it allowed the formation of the atoms and {\it a posteriori} the formation of stars, galaxies, clusters of galaxies etc. These objects, namely matter or pressure less matter, dominate the dynamics of the universe as a fluid with EoS $\omega=0$ \cite{ryden/2003}. From Fig. \ref{ch6fig5}, Fig. \ref{ch6fig8} and Fig. \ref{ch6fig11}, one can see that after describing a radiation-dominated period, $\omega$ indeed passes through $0$, indicating the matter-dominated phase of the universe expansion. Finally, for high values of time, $\omega\rightarrow-1$, in accordance with recent observational data on fluctuations of temperature in the cosmic microwave radiation \cite{hinshaw/2013}. In standard model, the cosmological constant is the ``mechanism'' responsible for taking the universe to a DE-dominated phase, in which a negative pressure fluid accelerates its expansion. In the present approach, rather, the extra terms in $f(R)$ and $f(T)$ are  responsible for such an important feature, which remarkably evades the cosmological constant problem \cite{weinberg/1989,Peebles/2003,padmanabhan/2003}. It is important to highlight that the description of three different stages of the evolution of the universe in a continuous and analytical form is not only a novelty in $f(R,T)$ gravity but also in the broad literature.\\
Moreover, we have obtained fourth model for the first time in the literature, WH solutions in the $f(R,T)=R+\alpha R^{2}+\lambda T$ gravity. Such a gravity theory can be seen as the simplest and more natural theory which presents corrections in both the geometrical and material sectors. Although it has already been applied to the study of compact astrophysical objects, yielding valuable results \cite{noureen/2015}-\cite{zubair/2015}, this is the first time WHs are analysed from such a functional form for $f(R,T)$.
The $R^2$-correction only is well motivated by its applications in cosmology and astrophysics \cite{starobinsky/1980}, \cite{ellis/2013}-\cite{koshelev/2016}. The $T$-correction only is inserted with the purpose of considering quantum effects in a gravity theory, or even the existence of imperfect fluids in the universe \cite{Harko11}. WHs material content is described by an anisotropic imperfect fluid thus the WH analysis in theories with material corrections is well motivated.\\
In the present model, the WH analysis is made by considering eqn. (\ref{ch6m416}) as the shape function. Such an assumption is also taken in other references, such as \cite{Heydarzade/2015,lobo/2013}. From eqn. (\ref{ch6m416}), we are able to obtain the material solutions of the WH, i.e., $\rho$, $p_r$ and $p_t$, as eqns. (\ref{ch6m4rho1} - \ref{ch6m4pt1}), respectively. The considered matter content of the WHs is different as given in literature \cite{Azizi/2013}-\cite{Yousaf/2017a}. Since the WH EoS is still poorly known, this is a considerable advantage of the present solutions.\\ 
By comparing these results with present literature we are inspired to conclude that the particular forms of $f(R)$ together with the linear term on $T$ are responsible for the remarkable features of the present model.


\chapter{Phantom fluid supporting traversable WHs in $f(R,T)$ gravity with extra material terms} 

\label{Chapter7} 

\lhead{Chapter 7. \emph{Phantom fluid supporting traversable WHs in $f(R,T)$ gravity with extra material terms}} 

This chapter \blfootnote{The work, in this chapter, is covered by the following publication: \\
\textit{Phantom fluid supporting traversable wormholes in alternative gravity with extra material terms}, International Journal of Modern Physics D, \textbf{27} (2018) 1950004.} contains the modelling of WHs within an alternative gravity theory (namely $f(R,T)$ gravity) that proposes an extra material (rather than geometrical) term in its gravitational action. The solutions of this chapter are obtained from well-known particular cases of the WH metric potentials, named redshift and shape functions, and yield the WHs to be filled by a phantom fluid, i.e., a fluid with EoS parameter $\omega<-1$. WHs are obtained originally as a solution for Einstein's GR. According to GR they are needed to be filled by an exotic kind of anisotropic matter, which leads to violation of the ECs. In possession of the solutions for the WH material content, one can study  all the features of the model by admitting ECs to them.
\section{Introduction}\label{ch7intro}
A traversable WH can be visualized as a tunnel in space-time with two ends (or mouths), through which observers may safely traverse. The whole concept of traversable WHs is quite exciting because they admit the superluminal travel as a global effect of space-time topology. This was demonstrated by Morris and Thorne in a metric representing a static traversable WH \cite{Morris/1988}. According to GR, traversable WHs are only possible if exotic matter exists at their throat, which involves an energy-momentum tensor violating the NEC \cite{Morris/1988}, which is in turn a part of the WEC, whose physical meaning is that the energy density is non-negative in any reference frame.  

On the other hand, modified $f(R,T)$ gravity has been deeply analysed due to some GR apparent incompleteness in some regimes  \cite{alavirad/2013,capozziello/2008}. Also, the same can be observed from previous chapters. The concept in these theories is basically to use an arbitrary but appropriate function in the gravitational action which generates extra terms in the field equations of GR. We have focused here mainly on the importance of the redshift and shape functions in different cases, analysing the possibility of generating a traversable WH. It is important to quote that $f(R,T)$ gravity has been applied to different areas of astrophysics and cosmology, yielding interesting and observationally testable results. One can check the recent $f(R,T)$ gravity applications as: some analysis about compact stellar structures in $f(R,T)$ gravity are made in \cite{sharif/2018,yousaf/2018}. Particularly, the hydrostatic equilibrium configurations of strange stars are obtained \cite{deb/2018,deb/2018b}, with the latter reference regarding an anisotropic distribution of matter inside such stars. Anisotropic stellar filaments evolving under expansion-free condition are analysed \cite{zubair/2018} and the dynamical stability of shearing viscous anisotropic fluid with cylindrical symmetry is investigated  \cite{azmat/2018}.

From some particular well addressed cases for the WHs redshift and shape functions, in this chapter we have obtained the material content solutions for WHs. In possession of those we applied the ECs to them. Those disfavour a constant redshift function, which is broadly assumed in the literature. We also have derived the anisotropic dimensionless parameter (recall that WHs material content is described by an anisotropic energy-momentum tensor) and have showed that these WHs are filled and supported by a phantom fluid. The latter conclusion may yield some new thoughts on a cosmological perspective and those are presented and discussed. 
The chapter is organized in such way that, section \ref{ch7intro} contains the brief introduction and motivation regarding the present work and the general field equations for WH metric are derived in section \ref{ch7sec2}. In section \ref{ch7sec:whm} we have discussed WH models in $f(R,T)$ gravity with two type of redshift functions. Finally, the model conclusions are covered in section \ref{ch7conclusion}.
\section{WH metric and field equations of an alternative gravity with extra material terms}\label{ch7sec2}
Nowadays the most popular alternative gravity theory is the $f(R)$ gravity (refer section \ref{ch1f(R)}), which takes general terms of $R$ in its gravitational action. Instead of taking general terms of geometrical aspect in the action, one can also take material extra terms. Let us consider a gravity theory that takes terms proportional to the trace of the energy-momentum $T$ in its action, which yields $f(R,T)$ gravity \cite{Harko11} (For more details refer section \ref{ch1f(R,T)}).
The $f(R,T)$ gravity field eqns. (\ref{eqnfld3}) for $f(R,T)=R+2f(T)$ and with stress energy-momentum tensor given in eqn.  (\ref{ch6ent2}) take the form
\begin{equation}\label{ch76}
R_{\mu \nu}-\frac{1}{2}Rg_{\mu \nu}=(8 \pi+2 \lambda) T_{\mu \nu}+\lambda g_{\mu \nu}(\rho-\mathcal{P}),
\end{equation}
where $f(T)=\lambda T$ and $\lambda$ is an arbitrary constant.

An interesting and intriguing property of the $f(R,T)$ theory of gravity, that can be extracted from eqn. (\ref{ch76}), is the non-conservation of the energy-momentum tensor. From eqn. \eqref{ch76},
\begin{equation}\label{ch76.1}
\nabla^{\mu}T_{\mu \nu}=\left(\frac{\lambda}{8\pi+\lambda}\right)\nabla^{\mu}g_{\mu \nu}(\mathcal{P}-\rho).
\end{equation} 
The particular consequences of the non-conservation of the energy-momentum tensor in $f(R,T)$ gravity have been explored in chapter- \ref{Chapter5}. Also, inspired by \cite{Josset17}, the cosmological consequences of the energy-momentum tensor non-conservation of $f(R,T)$ gravity are deeply investigated in \cite{Shabani17}. These non-conservation of the energy-momentum tensor implies in non-geodesic motions for test particles in gravitational fields as it is studied in  \cite{Baffou17}. In \cite{Shabani18,Moraes18} a different approach is considered. The authors have independently constructed a formalism in which an effective fluid is conserved in $f(R,T)$ gravity, rather than the usual energy-momentum tensor non-conservation.\\
The static spherically symmetric WH metric is considered here in Schwarzschild coordinates $(t,r,\theta, \phi)$ as described in eqn. (\ref{eqnwh}).
Further, the main conditions of shape function $b(r)$ are related to the shape of the WH, which is determined by the mathematics of embedding in eqn. (\ref{ch1flareout}) 
at or near the throat. The function $z=z(r)$ determines the profile of the embedding diagram of the WH,
\begin{equation}\label{ch7e2}
z(r)=\pm \int_{r_0}^{r} \frac{\text{d}r}{\sqrt{\frac{r}{b(r)}-1}},
\end{equation}
which is obtained by rotating the graph of the function $z(r)$ around the vertical $z$-axis.\\
The general field eqns. (\ref{ch76}) for the metric (\ref{eqnwh}) are given as 
\begin{equation}\label{ch7eqn1}
 \frac{b(r)+a'r-a'r^2}{r^3}= -(8\pi + 2\lambda)p_r + \lambda\left(\rho - \frac{p_r+2p_t}{3}\right),
\end{equation}
\begin{equation}\label{ch7eqn2}
\frac{(2 r^2 a''+r^2 a'^2) (b-r)+r a' \left(r b'+b-2 r\right)+2 \left(r b'-b\right)}{4 r^3}=
-(8 \pi + 2\lambda)p_t + \lambda\left(\rho - \frac{p_r+2p_t}{3}\right),
\end{equation}
\begin{equation}\label{ch7eqn3}
\frac{b'(r)}{r^2}= (8\pi + 2\lambda)\rho + \lambda\left(\rho - \frac{p_r+2p_t}{3}\right).
\end{equation}

Also, by developing eqn. \eqref{ch76.1}, we obtain
\begin{equation}\label{ch710.1}
(8\pi+3\lambda)\rho'+\left(8\pi+\frac{5}{3}\lambda\right)(p_r'+2p_t')=0.
\end{equation}
From the above eqns. (\ref{ch7eqn1} - \ref{ch7eqn3}), the explicit form of the WH matter content, namely $\rho$, $p_r$ and $p_t$, are obtained as
 \begin{equation}\label{ch7rho}
 \rho=rF_1(r)\left(48\pi b'-\lambda\left(F_2(r)(b-r)+a'(F_3(r)+2)-16b'\right)\right),
 \end{equation}
 \begin{equation}\label{ch7pr}
 p_r=F_1(r) \left(48(r-1)ra'+\lambda r\left(-rF_2(r)+a'F_4(r)+8b'\right)+b\lambda(r(F_2(r)+a')-24)-48\pi b\right),
 \end{equation}
 \begin{equation}\label{ch7pt}
 p_t=\splitdfrac{F_1(r) (\lambda(2ra'+r(r(5F_2(r)+8a')-F_6(r))+b(12-5r(F_2(r)+a'))}{-12\pi\left(r(F_5(r)-F_2(r)-2a')+b(r(F_2(r)+a'')-2))\right)}.
 \end{equation}
Here, for mathematical simplifications, the functions $F_i(r)$, where $i$ runs from 1 to 6, are written as
\begin{eqnarray}
F_1(r)\equiv\left(48(\lambda +2\pi)(\lambda +4\pi)r^3\right)^{-1}, \\
F_2(r)\equiv2ra''+ra'^2, \\
F_3(r)\equiv r(b'-4)+b, \\
F_4(r)\equiv r(b'+20)-22,\\
F_5(r)\equiv b(ra'+2), \\
F_6(r)\equiv b'(5ra'+4).
\end{eqnarray}
Furthermore, the dimensionless anisotropy parameter for anisotropic pressures, as the present case, is defined as \cite{Lobo/2013} 
\begin{equation}\label{ch7an}
\Delta=\frac{p_t-p_r}{\rho}.
\end{equation}
Since $\rho >0$, the relation $\frac{\rho \Delta}{r}$ represents a force due to the anisotropic nature of the WH model. Geometry is attractive if $p_t<p_r$, i.e. $\Delta<0$, and repulsive if $p_t>p_r$, i.e. $\Delta>0$. The fluid is isotropic for $\Delta=0$, i.e. $p_r=p_t$.
\section{WH models with hyperbolic shape function}\label{ch7sec:whm}
\begin{figure}[H]
 \centering
  \includegraphics[width=75mm]{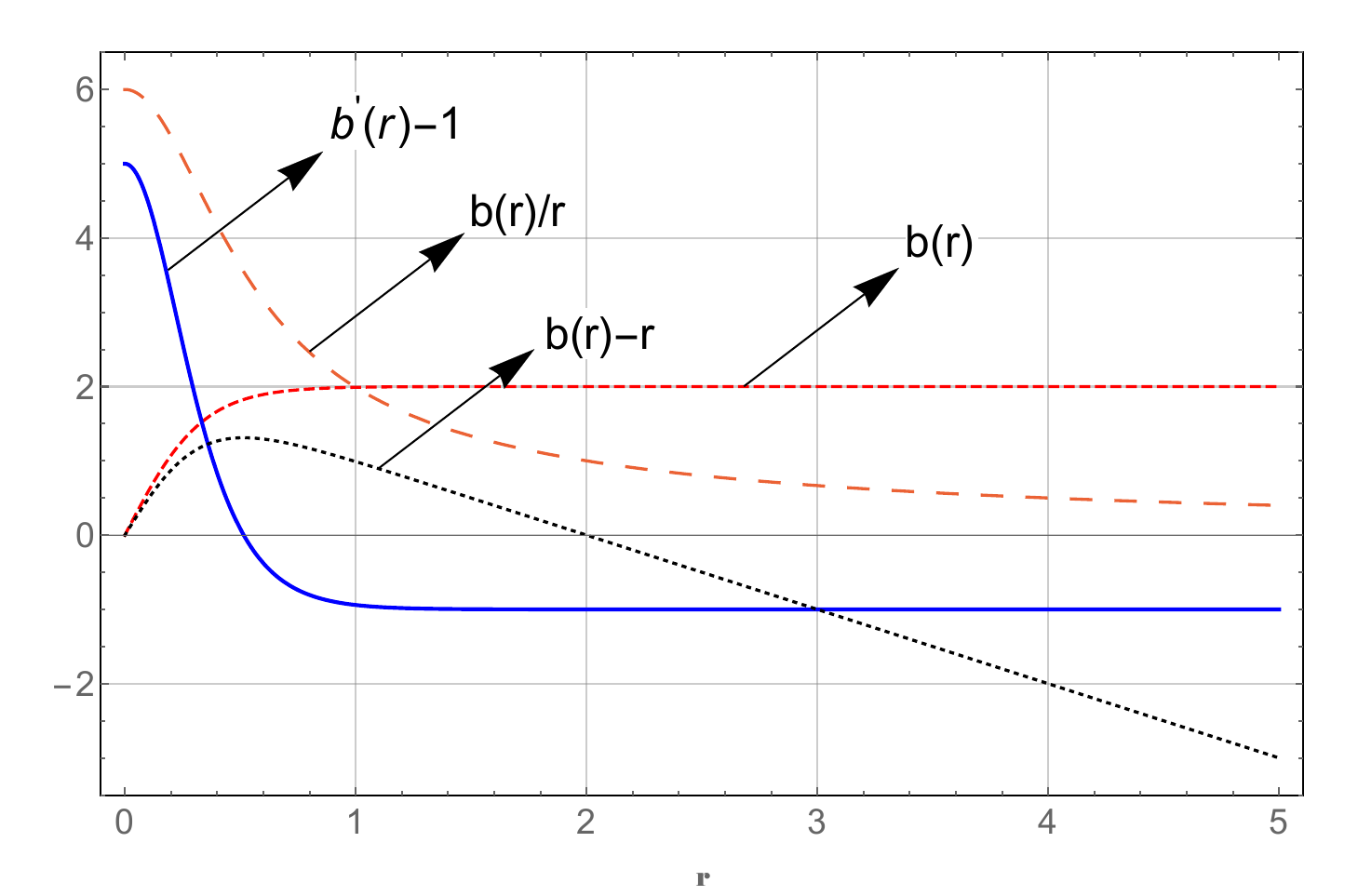}
  \caption{Variation of shape function with $m=2$ and $n=3$.}\label{ch7fig1}
  \end{figure}

We have considered the following specific form for the shape function \cite{Farook/2008}
 \begin{equation}\label{ch7b(r)}
 b(r)=m\tanh (nr),
 \end{equation}
where $m$ and $n>0$ are constants. As we have quoted above, to admit the necessary metric conditions of WHs we have $b(r_0)=r_0$ and the flaring out condition $b'(r_0)-1<0$ as it is represented in the Fig. \ref{ch7fig1}. 
From such a figure, we have $b(r)<r$ for $r>r_0$ and $\frac{b(r)}{r}\rightarrow 0$ as $r\rightarrow \infty$ showing an asymptotically flat behavior. In the present chapter we have considered the WH throat where $b(r)-r$ cuts $r-$axis, i.e. $r_0=2$ as $b(2)\approx 1.99998$ with $m=2$ and $ n=3$.  
\subsection{Logarithmic redshift function}\label{ch7ss:lrf}
In order to obtain traversable WHs, we consider a logarithmic form for the redshift function as \cite{Pavlovic/2015}
\begin{equation}\label{ch7a(r)1}
a(r)=\ln\left(\frac{r_0}{r}+1\right),
\end{equation} 
By substituting eqns. (\ref{ch7b(r)}) and (\ref{ch7a(r)1}) in eqns. (\ref{ch7rho}-\ref{ch7pt}), we obtain $\rho$, $p_r$, $p_t$ and the radial EoS parameter $\omega_r=\frac{p_r}{\rho}$ as
\begin{equation}\label{ch7rho1}
\rho=\frac{F_1(r)}{(r+r_0)^2}\left[G_1(r) G_4(r)\sec h^2(n r)-\lambda r_0(m \tan h(nr)(3r+2r_0)+G_7(r))\right],
\end{equation}
\begin{multline}\label{ch7pr1}
p_r=\frac{F_1(r)}{(r+r_0)^2}\biggl[-m\tan h(nr)G_3(r)+\lambda G_1(r)(8r+7r_0)\sec h^2(nr)\\
+\lambda G_2(r)-48\pi r_0(r-1)(r+r_0)\biggr],
\end{multline}
\begin{equation}\label{ch7pt1}
p_t=\frac{F_1(r)}{(r+r_0)^2}\left[m \tan h(nr)G_5(r)-G_1(r) \sec h^2(nr)(\lambda(4r-r_0)+12\pi(2r+r_0))+G_6(r)\right],
\end{equation}
\begin{equation}\label{ch7omega1}
\omega_r(r)=\frac{-m \tan h(nr) G_3(r)+\lambda (G_1(r) (8r+7r_0)\sec h^2(nr)+G_2(r))-48\pi r_0(r-1)(r+r_0)}{G_1(r) G_4(r)\sec h^2(nr)-\lambda r_0(m \tan h(nr)(3r+2r_0)+G_7(r))},
\end{equation}
where $G_j(r)$, with $j$ running from 1 to 7, are expressed by the following equations
\begin{eqnarray}
G_1(r) & = & mnr (r+r_0),\\
G_2(r) & = & r_0(r(-24r-23r_0+22)+22r_0),\\
G_3(r) & = & \lambda(24r^2+4r r_0+22r_0^2)+48(r+r_0)^2,\\
G_4(r) & = & 48\pi (r+r_0)+\lambda(16r+17r_0),\\
G_5(r) & = & \lambda(12r^2+9r r_0+r_0^2)+12\pi r(2r+r_0),\\
G_6(r) & = & \lambda r_0(r(12r+7r_0-2)-2r_0)+12\pi r r_0(2r+r_0),\\
G_7(r) & = & r_0(r-2)-2r.
\end{eqnarray}  
\subsubsection{Energy conditions}\label{ch7sss:ec1}
The main ECss, such as the NEC, WEC, SEC and DEC can be expressed directly in terms of $\rho, p_r, p_t$ as represented in section \ref{ch1ECs}.
%
The role of these ECs are defined as; NEC represents the attractive nature of gravity; DEC states that the velocity of energy transfer cannot be higher than the speed of light and SEC stems from the attractive nature of gravity and its form is the direct result of considering a spherically symmetric metric in the GR framework. 
Hence, one can find that the ECs may be obtained to traversable WHs in modified gravity.

\begin{figure}[H]
\minipage{0.48\textwidth}
\includegraphics[width=75mm]{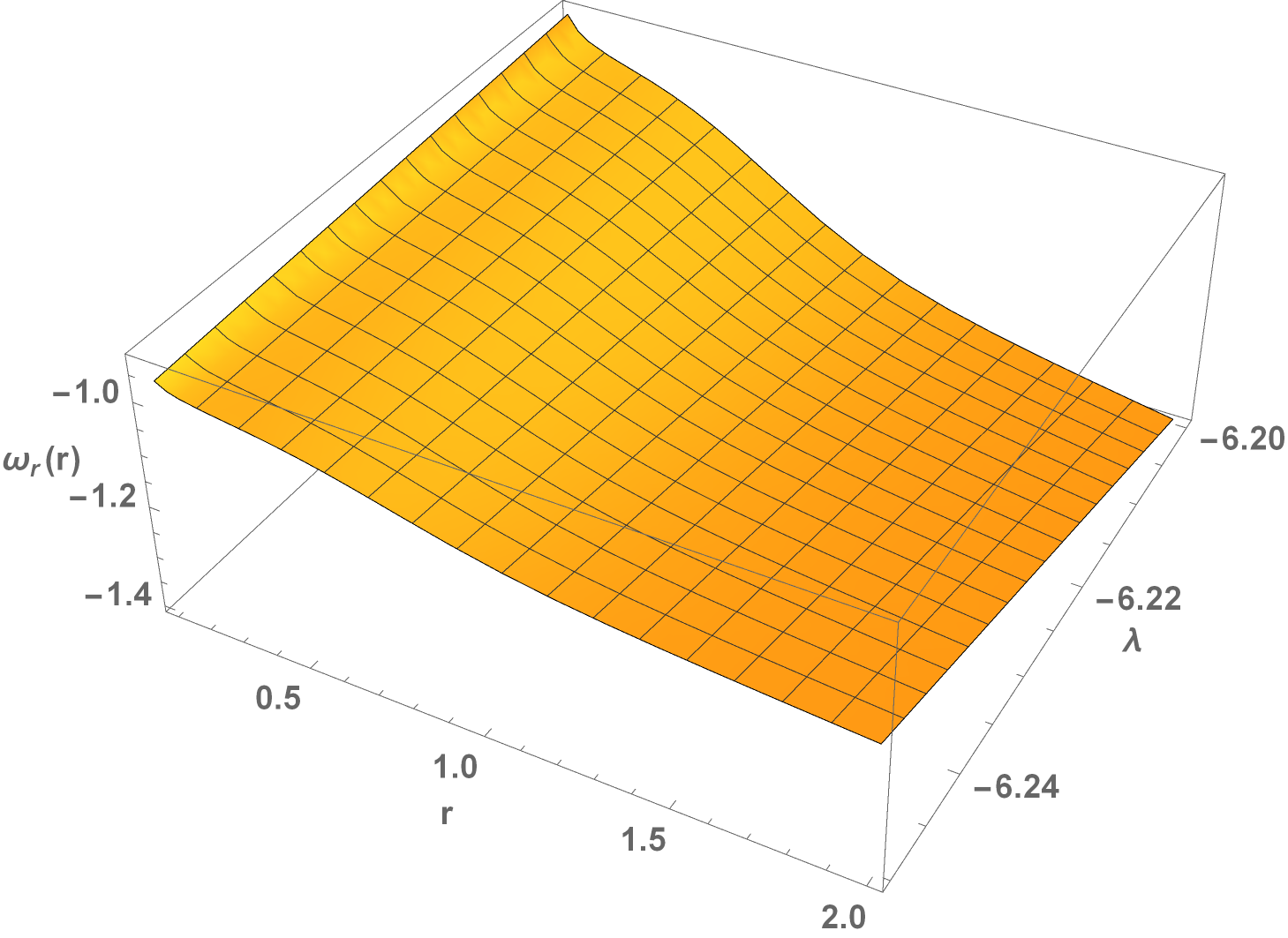}
  \caption{Radial EoS parameter $\omega_r$ as a function of $r$ and $\lambda$.}\label{ch7fig9}
\endminipage\hfill
\minipage{0.50\textwidth}
\includegraphics[width=75mm]{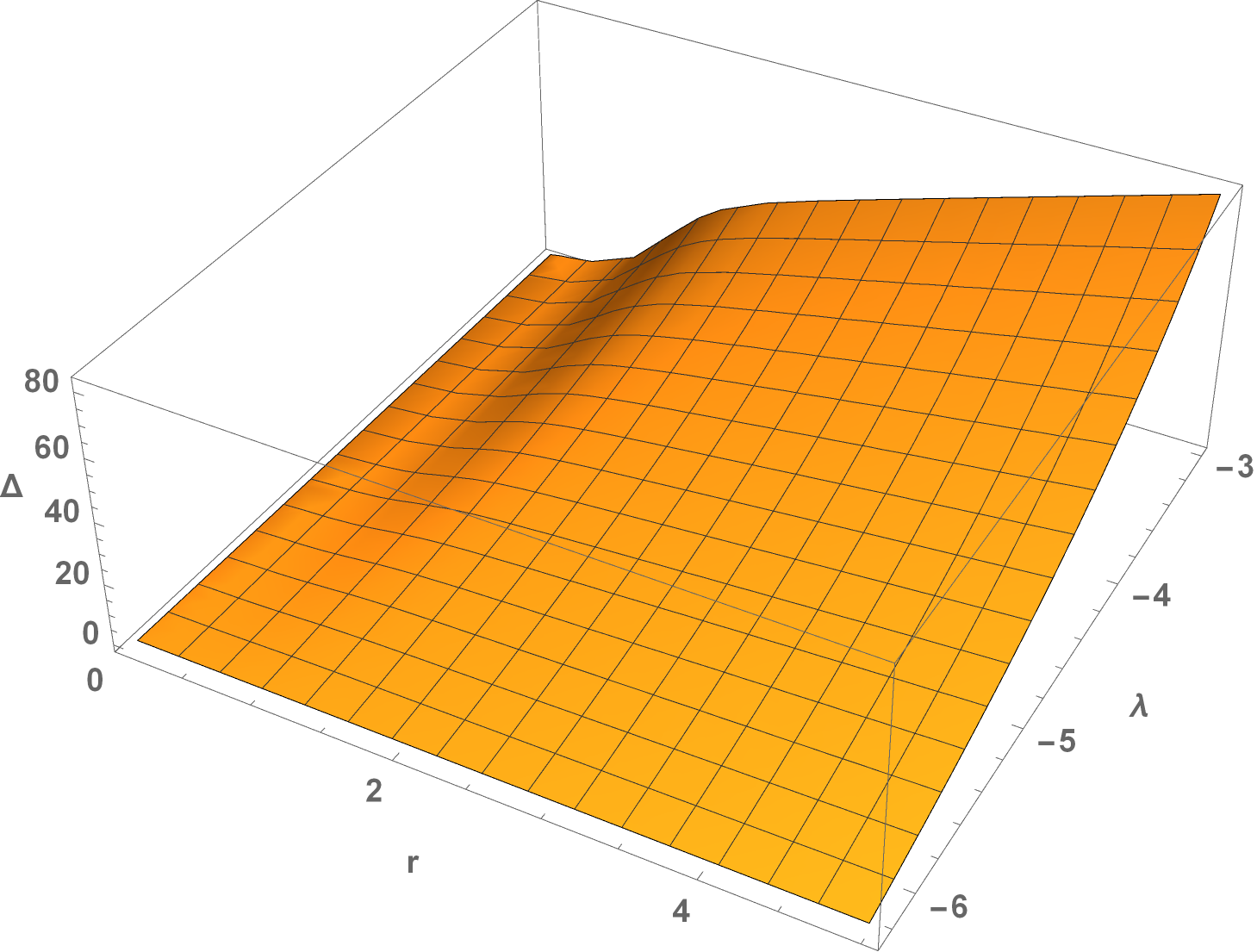}
  \caption{Dimensionless anisotropic parameter $\Delta$ as a function of $r$ and $\lambda$.}\label{ch7fig9a}
\endminipage
\end{figure}

From the above quantities, we have plotted here the energy density as well as the ECs in Fig.  \ref{ch7fig3} to Fig. \ref{ch7fig8} below. In all figures, we consider the free parameters $m=2$ and $n=3$.
\begin{figure}[H]
\minipage{0.48\textwidth}
\includegraphics[width=75mm]{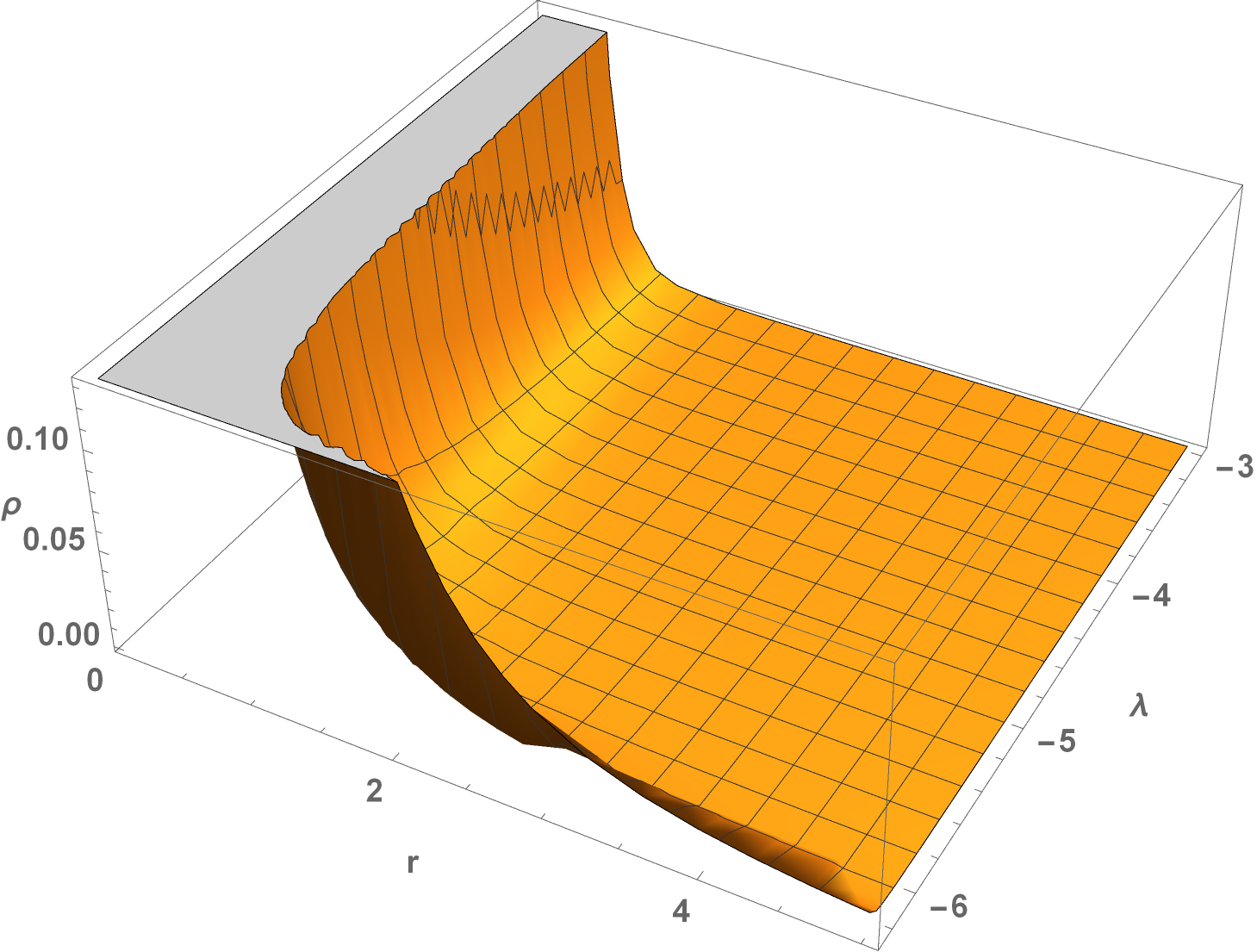}
  \caption{Energy density as a function of $r$ for different $\lambda$.}\label{ch7fig3}
\endminipage\hfill
\minipage{0.50\textwidth}
\includegraphics[width=75mm]{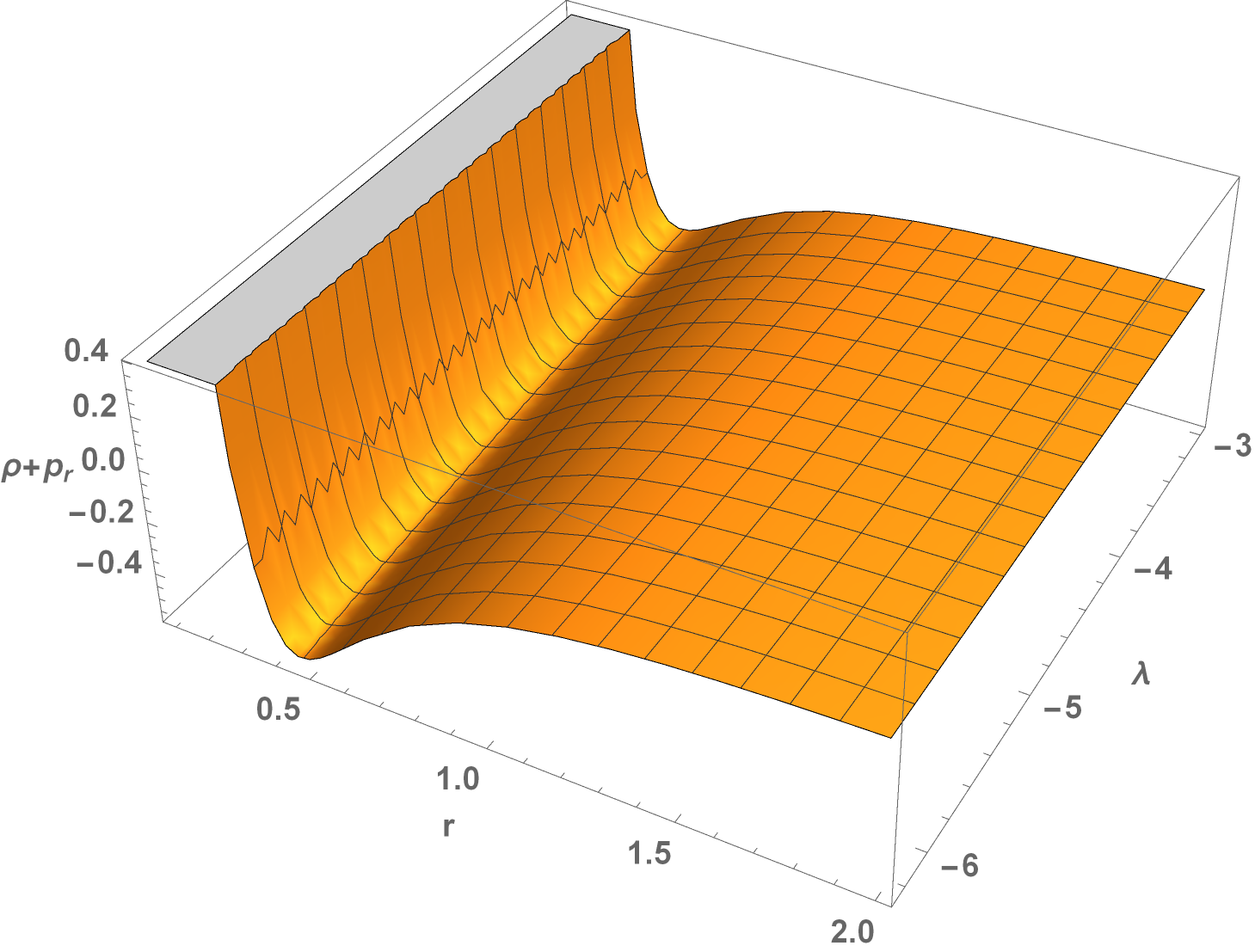}
  \caption{NEC, $\rho+p_r\geq 0$.}\label{ch7fig4}
\endminipage
\end{figure}
From Fig.  \ref{ch7fig4} one can observe that NEC for $\rho+p_r$ validates for small $r$. It is to be noted here that the violation of NEC implies that WEC will be also violated; while  WEC is valid, it does not imply that the NEC is satisfied. 
Fig. \ref{ch7fig6} shows the validation of DEC for radial case. 
Moreover, in Fig. \ref{ch7fig8} to Fig. \ref{ch7fig9a} we have plotted the SEC, radial EoS parameter and the dimensionless anisotropic parameter.
\begin{figure}[H]
\minipage{0.48\textwidth}
\includegraphics[width=75mm]{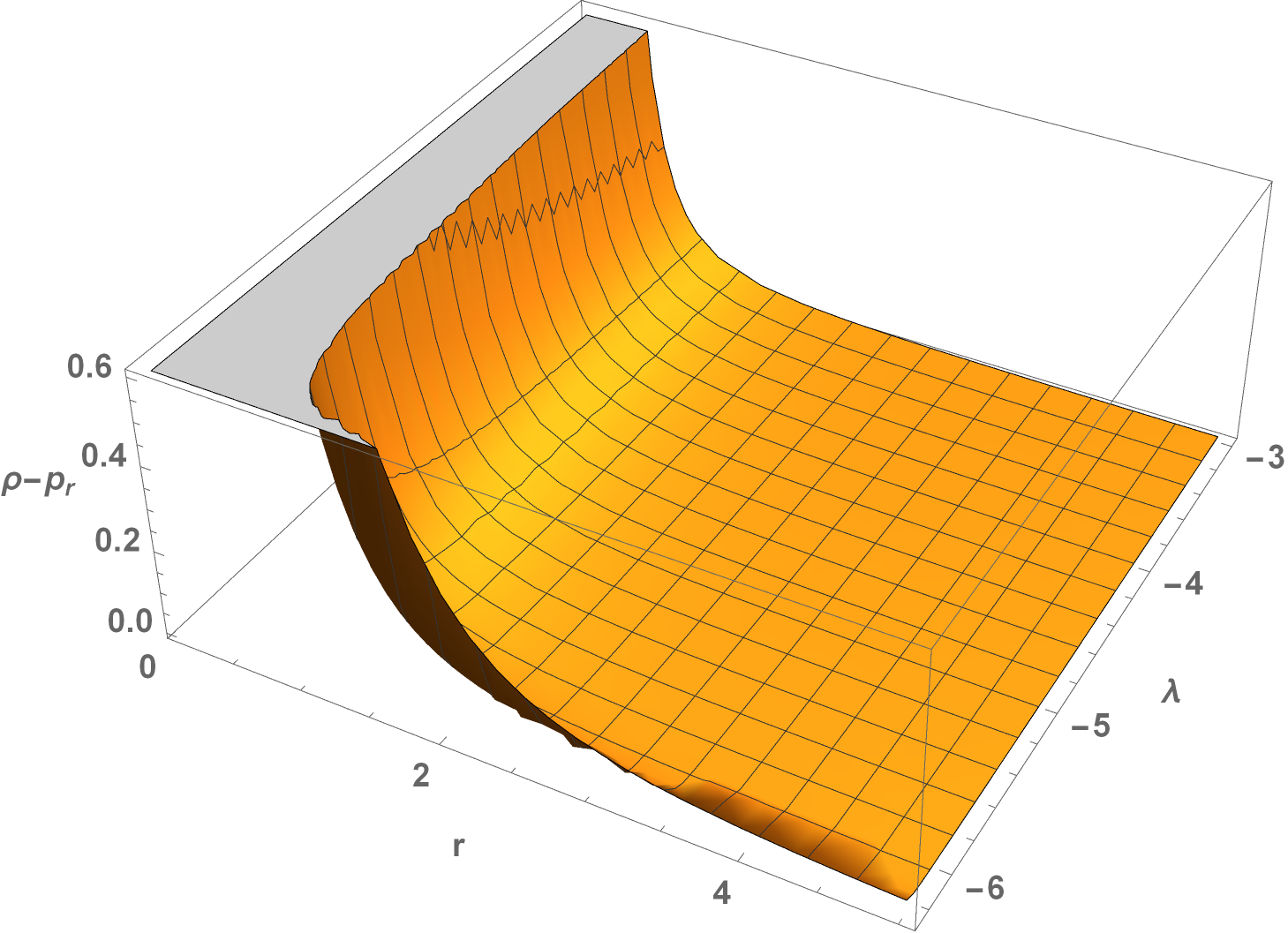}
  \caption{DEC, $\rho\geq \vert p_r\vert$.}\label{ch7fig6}
\endminipage\hfill
\minipage{0.50\textwidth}
\includegraphics[width=75mm]{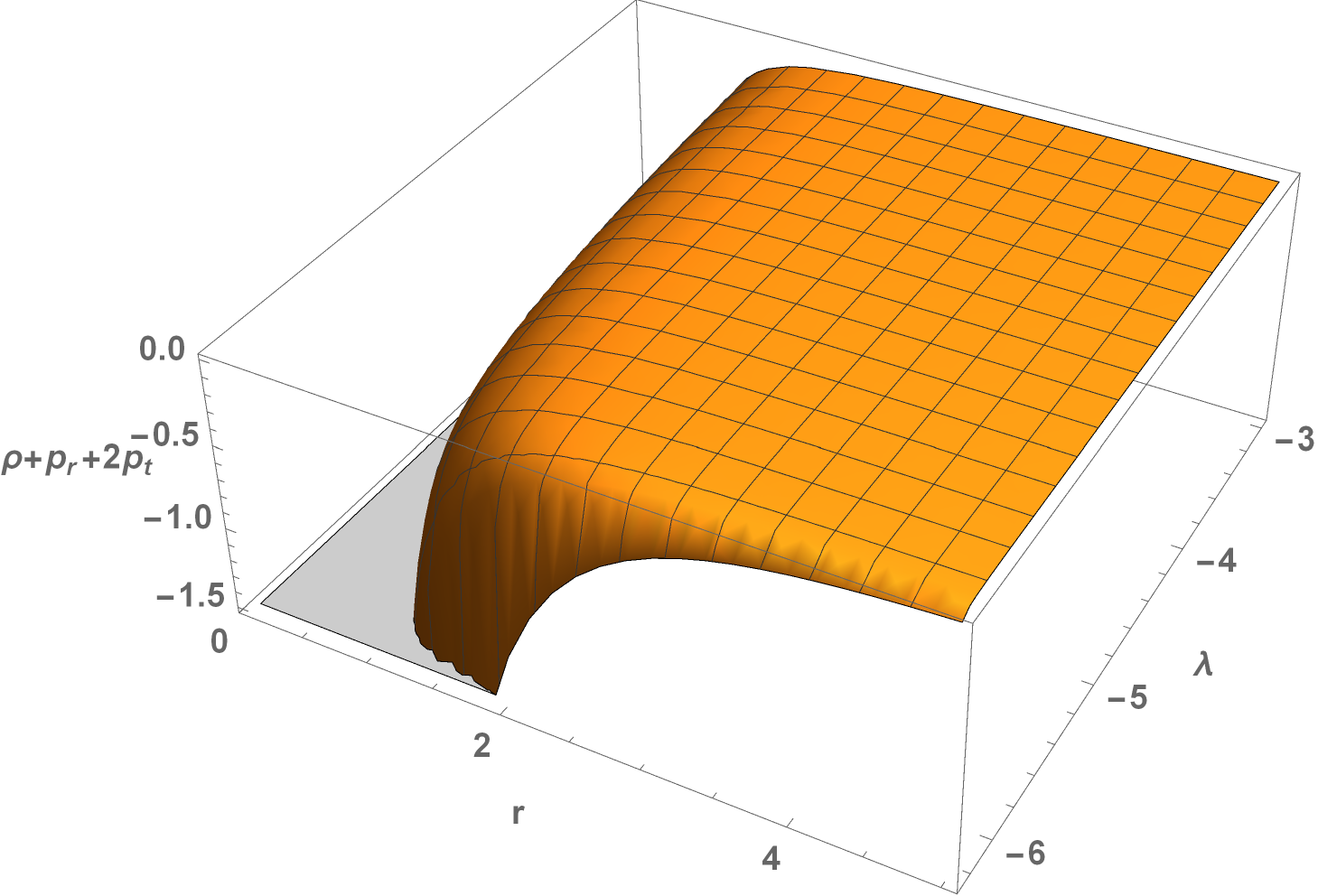}
  \caption{SEC, $\rho+p_r+2p_t\geq 0 $.}\label{ch7fig8}
\endminipage
\end{figure}
\subsection{Constant redshift function}\label{ch7ss:crf}
In this case, we set the redshift function $a(r)=1$, and by using this redshift function in eqns. (\ref{ch7rho}) - (\ref{ch7pt}), the explicit form of matter quantities reduces to
\begin{equation}\label{ch7rho2}
\rho=\frac{rF_1(r)}{16} \left(3\pi+\lambda \right)b',
\end{equation}
\begin{equation}\label{ch7pr2}
p_r=-\frac{F_1(r)}{8} \left[-\lambda r b'+(6\pi +3\lambda)b\right],
\end{equation}
\begin{equation}\label{ch7pt2}
p_t=-\frac{F_1(r)}{4} \left[(6\pi +\lambda)rb'-(6\pi +3\lambda)b\right].
\end{equation}
The explicit expressions for the matter quantities and radial EoS parameter are obtained by substituting eqn. (\ref{ch7b(r)}) in the above equations, leading to
\begin{equation}\label{ch7rho3}
\rho=\frac{rF_1(r)}{16}mn (\lambda +3 \pi ) \text{sech}^2(n r),
\end{equation}
\begin{equation}\label{ch7pr3}
p_r=\frac{F_1(r)}{8} m \left[n \lambda  r \text{sech}^2(n r)-3 (\lambda +2 \pi ) \tanh (n r)\right],
\end{equation}
\begin{equation}\label{ch7pt3}
p_t=\frac{F_1(r)}{2} m \text{sech}^2(n r) \left[3 (\lambda +2 \pi ) \sinh (2 n r)-2 n (\lambda +6 \pi ) r\right],
\end{equation}
\begin{equation}\label{ch7omega2}
\omega_r(r)=\frac{2 \lambda  n r-3 (\lambda +2 \pi ) \sinh (2 n r)}{4 (\lambda +3 \pi ) n r}.
\end{equation}
\subsubsection{Energy conditions}\label{ch7sss:ec2}
\begin{figure}[H]
\minipage{0.48\textwidth}
\includegraphics[width=75mm]{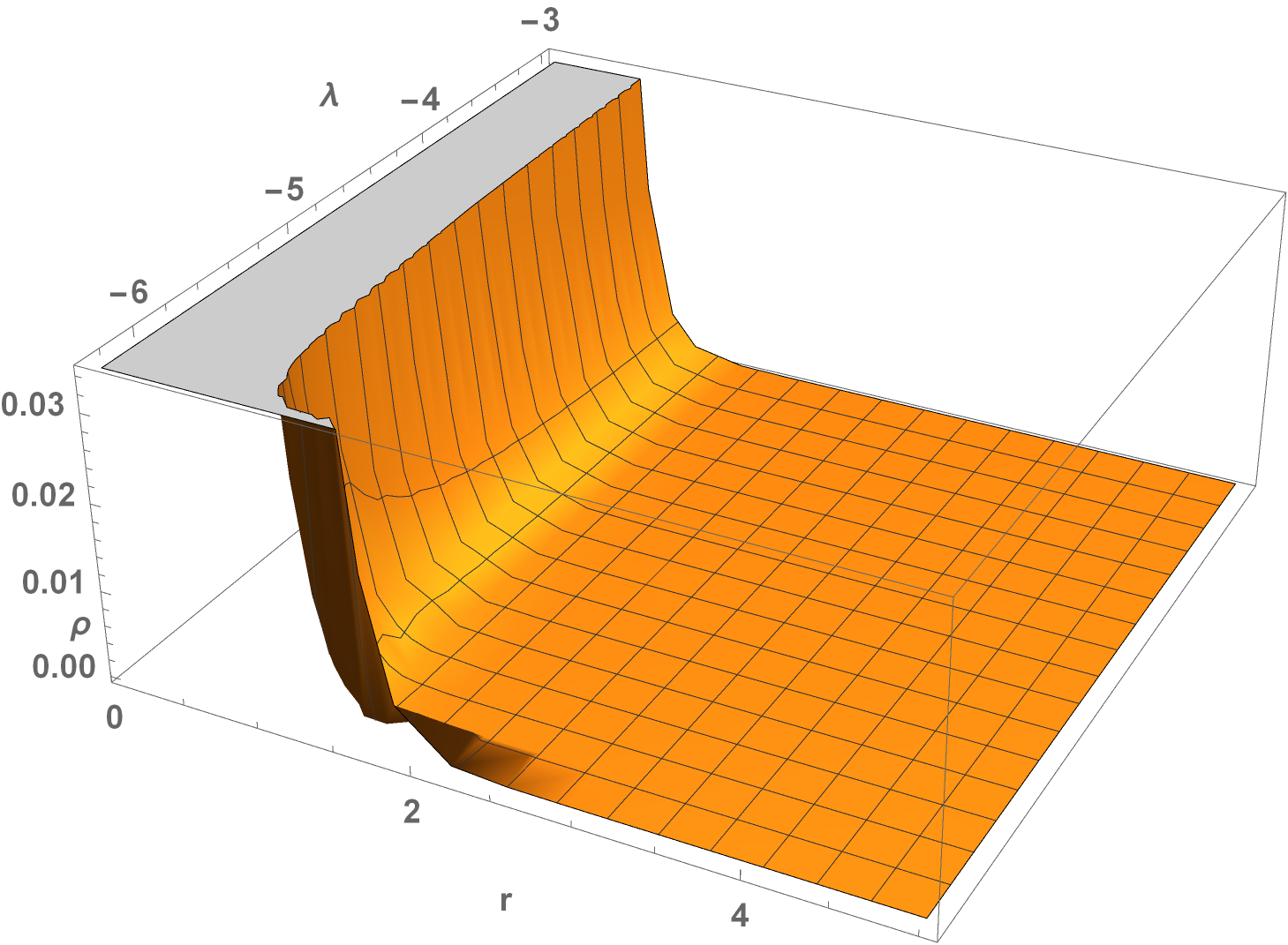}
  \caption{Variation of the energy density.}\label{ch7fig10}
\endminipage\hfill
\minipage{0.50\textwidth}
\includegraphics[width=75mm]{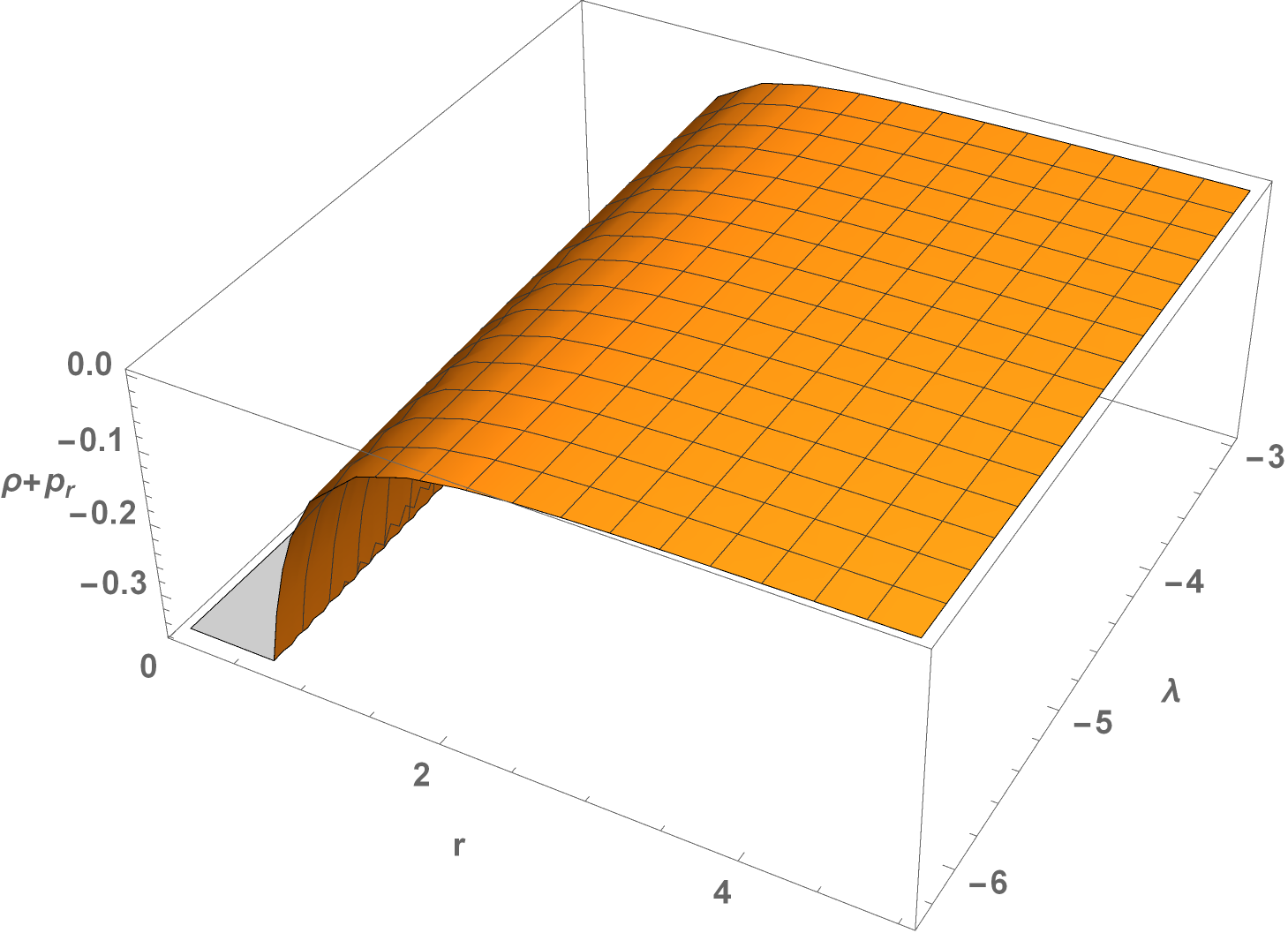}
  \caption{NEC, $\rho+p_r\geq 0$.}\label{ch7fig11}
\endminipage
\end{figure}

\begin{figure}[H]
\minipage{0.48\textwidth}
\includegraphics[width=75mm]{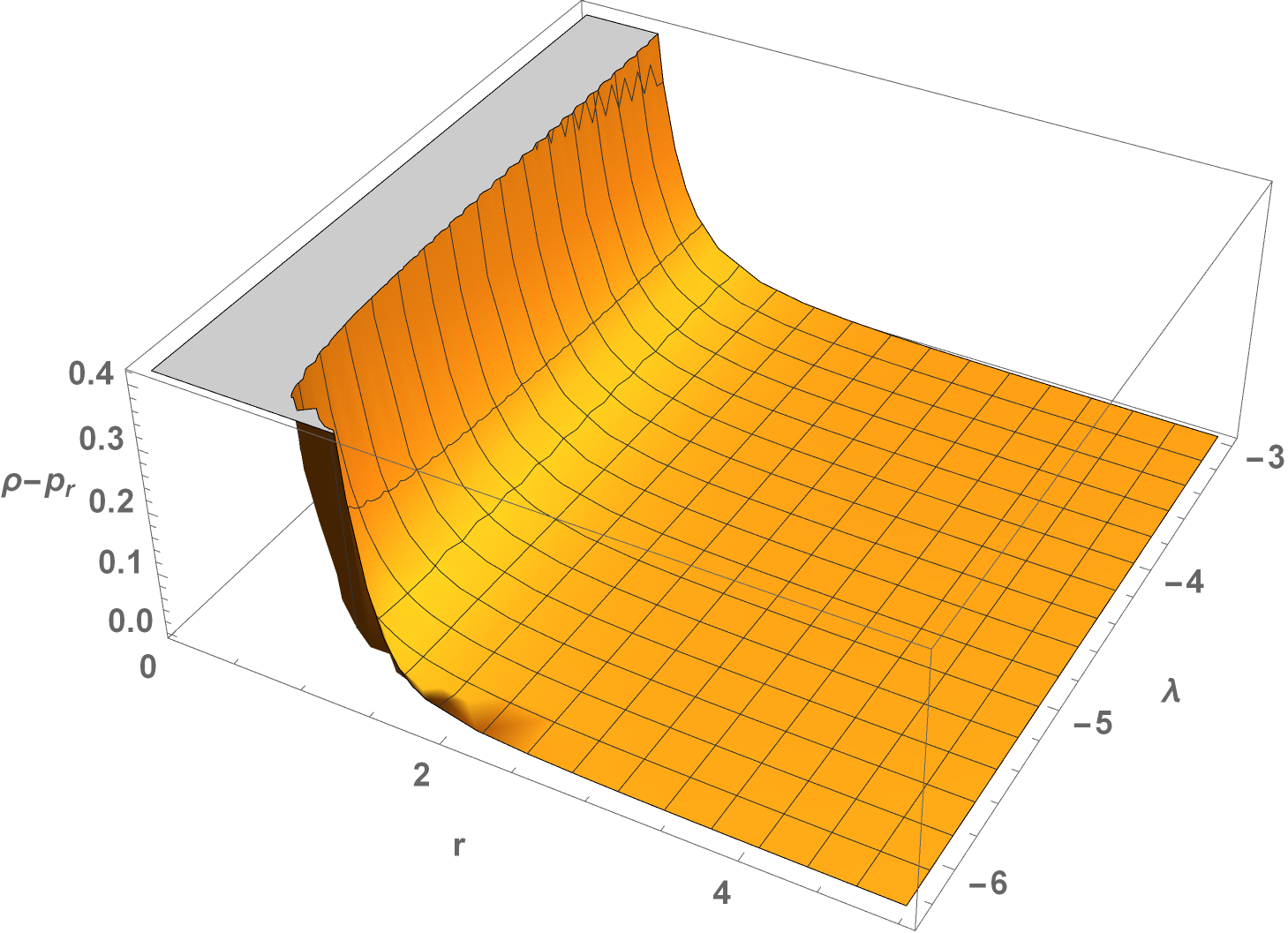}
  \caption{DEC, $\rho\geq \vert p_r\vert$.}\label{ch7fig13}
\endminipage\hfill
\minipage{0.50\textwidth}
\includegraphics[width=75mm]{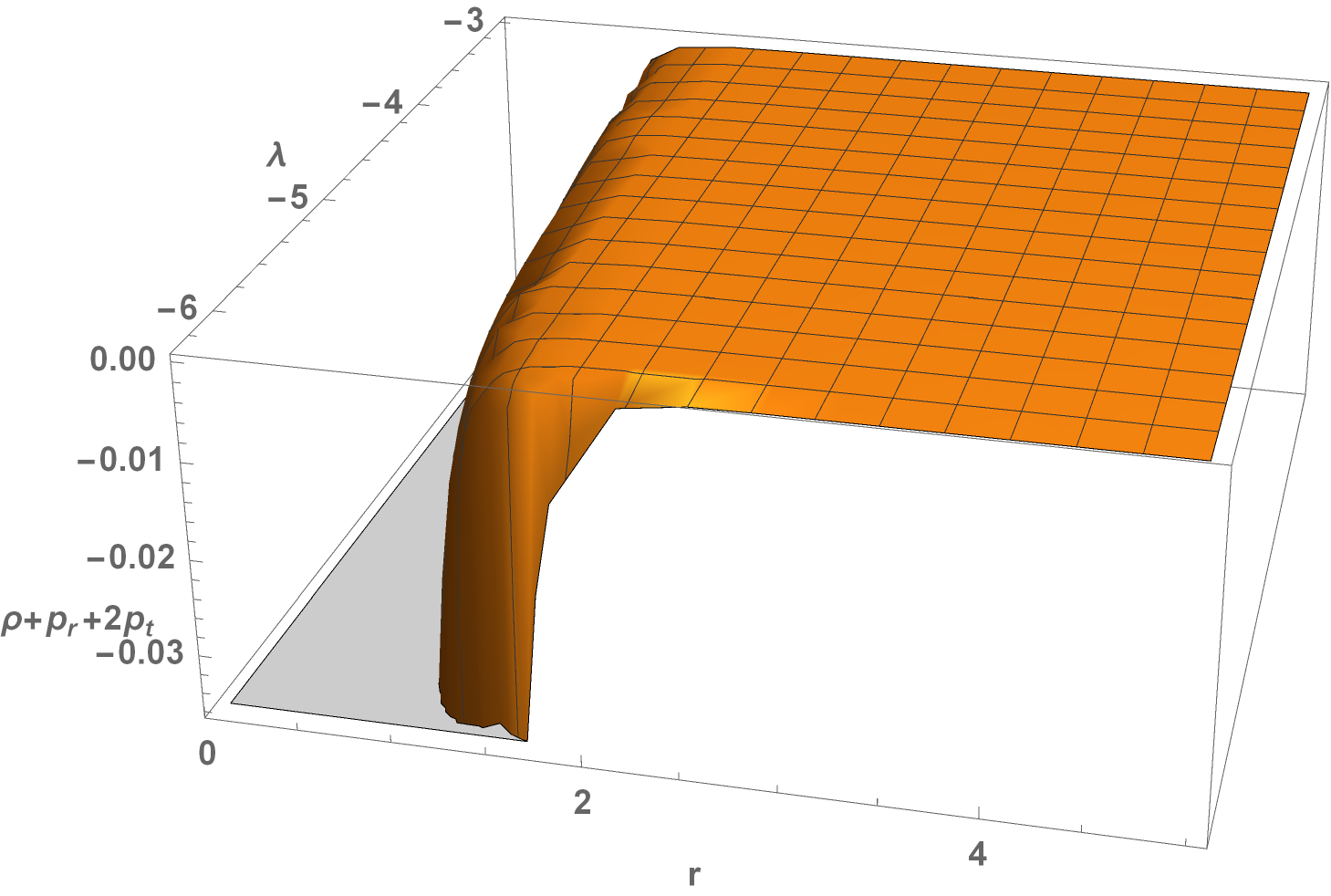}
  \caption{SEC, $\rho+p_r+2p_t\geq 0 $.}\label{ch7fig15}
\endminipage
\end{figure}

\begin{figure}[H]
\minipage{0.48\textwidth}
\includegraphics[width=75mm]{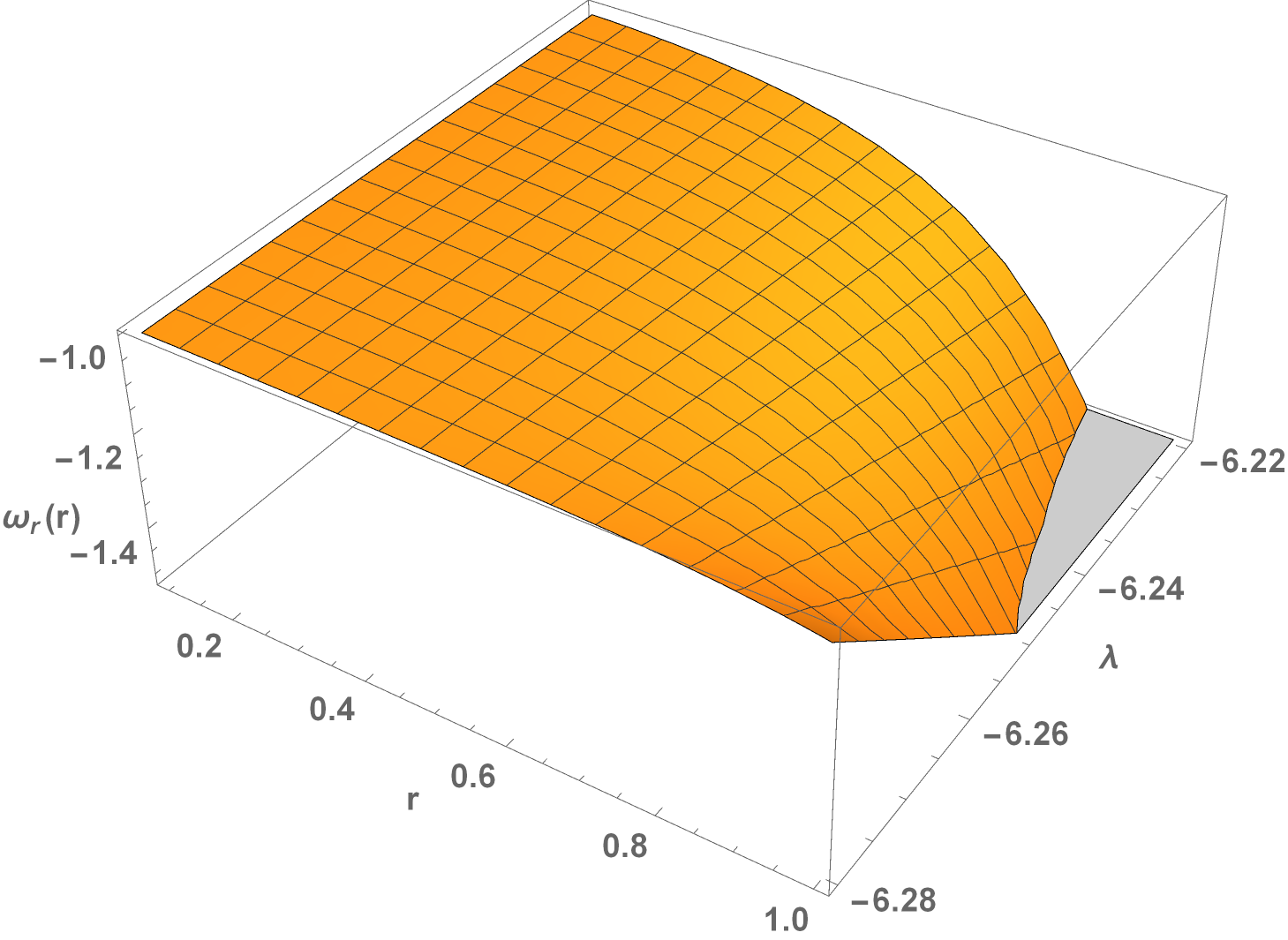}
  \caption{Radial EoS parameter $\omega_r$ as a function of $r$ and $\lambda$.}\label{ch7fig16}
\endminipage\hfill
\minipage{0.50\textwidth}
\includegraphics[width=75mm]{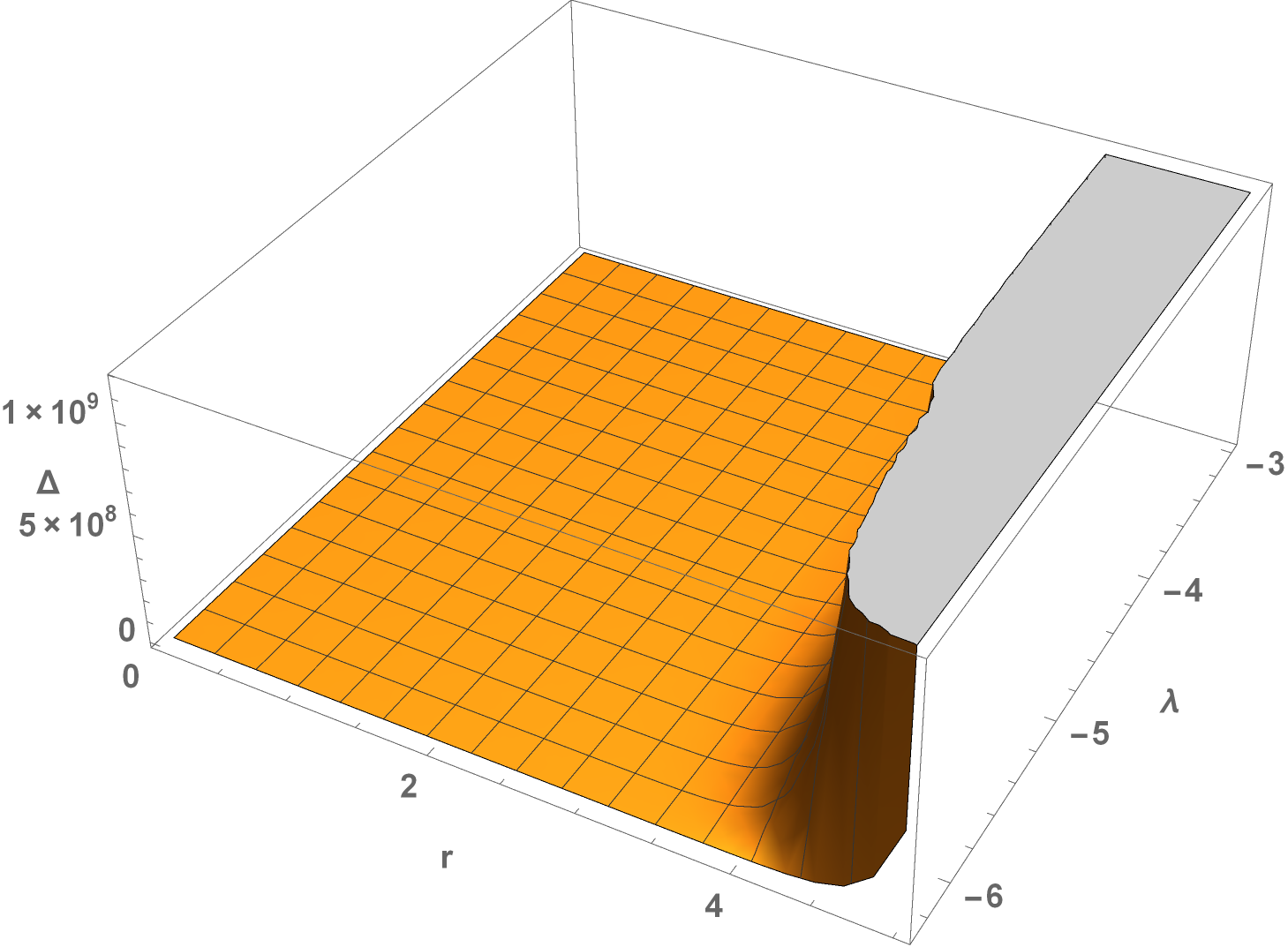}
  \caption{Dimensionless anisotropic parameter $\Delta$ as a function of $r$ and $\lambda$.}\label{ch7fig17}
\endminipage
\end{figure}

From the above quantities, the plot of the energy density as well as the ECs are shown in Fig.  \ref{ch7fig10} to Fig. \ref{ch7fig15}. In all figures we consider $m=2$ and $n=3$.
From Fig. \ref{ch7fig11} it is observed that NEC for radial pressure is violated. DEC for radial pressure is valid for $r>0$ as shown in Fig. \ref{ch7fig13}. 
One can observe from Fig. \ref{ch7fig15} that SEC violates everywhere and tends to zero for large values of $r$.
The radial EoS parameter $\omega_r(r)$ is always $<-1$ in support of the violation of NEC for modified gravity as shown in Fig. \ref{ch7fig16}.
The dimensionless anisotropic parameter is depicted in Fig. \ref{ch7fig17} for this case.
\section{Conclusion}\label{ch7conclusion}
In this chapter the WH solutions are deeply analyzed within the $f(R,T)$ gravity as the background theory. The analysis of WHs in modified gravity theories are mainly originated by the possibility of obtaining material sector solutions that obey the energy conditions, so that they do not have to be referred to as ``exotic''. According to the $f(R,T)$ gravity authors, the $T$-dependence of the theory is motivated by the (possible) existence of imperfect fluids in the universe \cite{Harko11}. In this way, the study of WHs, whose matter content is anisotropically distributed, is well motivated in this theory.
We have substituted the Morris-Thorne WH metric (\ref{eqnwh}) in the field eqns. (\ref{ch76}) obtained for the choice of $f(R,T)=R+2\lambda T$ in the model. The material solutions are obtained from a hyperbolic shape function, such as in ref. \cite{Farook/2008}, for logarithmic \cite{Pavlovic/2015} and constant redshift functions \cite{cataldo/2011,Farook/2007}. 
Moreover, it is important to mark that departing from many references in the present literature, we have not assumed any EoS parameter rather, we have obtained it from the model. It is shown in two cases in Fig. \ref{ch7fig9} and Fig. \ref{ch7fig16} respectively.  There is a remarkable feature about $\omega_r$ that can be appreciated in those figures and is discussed below.

In both figures, representing non-constant and constant redshift functions, the radial EoS parameter is $<-1$ for approximately the entire space. This indicates that the concerned WHs are filled by a phantom fluid. Recall that a phantom fluid permeating the whole universe is an important alternative to explain the cosmic acceleration.
Furthermore, eqn. (\ref{ch7omega2}) can be rewritten as a power series as follows
\begin{equation}\label{ch7fe}
\omega_r=-1-\frac{9 (\lambda +2 \pi ) r^2}{\lambda +3 \pi }-\frac{81 (\lambda +2 \pi ) r^4}{5 (\lambda +3 \pi )}+O\left(r^5\right),
\end{equation}
so that $\omega_r \rightarrow -1$ as $r\rightarrow 0$. The above equation explicitly shows the phantom aspect of the EoS obtained for the WH. As mentioned above, $r\rightarrow0$, $\omega_r\rightarrow-1$. If that was the case for the whole WH, we would have a sort of ``dark energy wormhole''. However, as one gets away from $r=0$, the ``phantom contributions'' start to dominate and $\omega_r$ decreases its values, characterizing a phantom WH.
WHs filled by phantom fluids have been analysed in the literature \cite{Lobo/2013}. In this chapter, a phantom fluid has shown to be responsible for supporting WHs with the geometrical features proposed in section \ref{ch7sec:whm} within $f(R,T)$ theory.

The results obtained from the ECs applications are quite interesting. One can observe that it is indeed possible to respect the ECs in the present theory. Apart from SEC, all the ECs presented in section \ref{ch7sss:ec1}, from a non-constant redshift function, are respected, at least for a range of values of the radial coordinate. 
On the other hand, the ECs shown in section \ref{ch7sss:ec2}, for a constant redshift function, have the NEC $\rho+p_r\geq0$ and SEC disobeyed. This may be an important clue that constant redshift functions yield unsatisfactory results regarding ECs applications, so that this particular case could be discarded from further WH modelling. 

The dimensionless anisotropic parameter given in eqn. (\ref{ch7an}) is depicted in Fig.  \ref{ch7fig9a} and Fig. \ref{ch7fig17}. This quantity is deeply approached in ref. \cite{cattoen/2005}. In ref. \cite{Lobo/2013}, an EoS is given in terms of $\Delta$. From Fig. \ref{ch7fig9a} and Fig. \ref{ch7fig17} it is clear that $\Delta>0$, which implies that the geometry is repulsive in both models due to the anisotropy of the system. One can conclude that, in principle, the repulsive character of the anisotropy compensates the attractive nature of gravity for a range of the parameters of the WH models.

At the end, it is important to remark that in order to get the WH solutions we need not use eqn. (\ref{ch710.1}). The system of eqns. (\eqref{ch7rho} - \eqref{ch7pt}) has shown to be soluble from the assumptions in section \ref{ch7sec:whm}. In this way, in the present approach, the equation for the non-conservation of the energy-momentum tensor in $f(R,T)$ gravity merely puts a stricter bound in the values of the free parameter $\lambda$. The solutions for $\rho$, $p_r$ and $p_t$ satisfy eqn. (\ref{ch710.1}) for $-4\leq\lambda\leq-3$ in the logarithmic redshift function case (section \ref{ch7ss:lrf}) and for $\lambda\sim-6.275$ in the constant redshift case (section  \ref{ch7ss:crf}). It is to be noted that these values for $\lambda$ are in agreement with the ECs.


\chapter{Final remarks and future perspectives} 

\label{Chapter8} 

\lhead{Chapter 8. \emph{Final remarks and future perspectives}} 

The results obtained in the whole study are summarized briefly in this chapter.  It is essential to know the origin, shape, structure formation, evolution and ultimate fate of our Universe. This can be effectively achieved by constructing mathematical models within different gravitational theories. Hence, the cosmological models obtained in this thesis in $f(R,T)$ gravity theory will expectedly, help in knowing our Universe in better way. The objective of this work has been to develop the framework of $f(R,T)$ gravity, and to investigate the present accelerated expansion of the universe through different cosmological model in it. This framework has been  used to study the ways in which $f(R,T)$ gravity can be considered to be special or unique.\\
 The first chapter of this thesis is completely introductory and motivational, as it is needed to prepare the ground to address the subsequent chapters. In all the chapters, we have focused on modified $f(R,T)$ gravity cosmological models in account of accelerated expansion of the universe. Chapter \ref{Chapter6} and chapter \ref{Chapter7} deal with few aspects of higher order curvature scalars in $f(R,T)$ gravity along with the phenomenology of astrophysical objects (e.g. WH solutions) and its cosmological viability.\\ 
Furthermore, the parametrization of a DP in terms of cosmic time and redshift, is considered as an important tool to obtain the exact solution of field equations. In fact, the evolution of DP with respect to time and redshift of each models are consistence with its observational aspects, which provides the observational viability of each theoretical models in this thesis. It is worth noting here that some particular parametric forms of DP like, time-varying DP, LVDP, and PVDP are used in obtaining the exact solutions. Moreover, some other aspects like, ECs, state-finder diagnosis parameter and more specifically evolution of EoS parameter are also made to get the models consistent with observational results.  This evolution of EoS parameter is always consistent with the current observations in all the models of each chapters. In particular, the transitional behavior of the universe from radiation era to DE era are smoothly depicted through this parameter, which are described in each chapter subsequently. In order to understand the homogeneous and anisotripic nature of the universe, we have considered Bianchi type III, VI$_0$ in chapter \ref{Chapter2}, and type I space-time in chapter \ref{Chapter3}, chapter \ref{Chapter4}, and first model in the chapter \ref{Chapter5}. The Bianchi type I space-time is considered as the most generalization of the FLRW space-time, which is flat and homogeneous, and  chosen in second model of chapter \ref{Chapter5}. In order to explain both the early and the late-time acceleration in a geometrical way, without invoking huge amount of DE, several combinations of curvature invariants, like,  $R_{\mu \nu}R^{\mu \nu}$, $R_{\mu \nu \sigma \gamma}R^{\mu \nu \sigma \gamma}$..., can be considered into the gravitational action. For this purpose, in chapter \ref{Chapter6}, we have obtained some cosmological models with higher order curvature scalar and its mixed form (both positive and negative)  within $f(R,T)$ formalism.    
In case of viability of cosmological model through ECs, it provides a strong evidence towards the positiveness of matter content. Hence, most of the cosmological models are constructed on the basis of validation of ECs. It would be interesting to investigate the other part of ECs, i.e. violation of ECs, which indicates the presence of some 
``exotic matter" in space-time. One of the best examples of violation of ECs is the existence of WH solutions. In this sense, we have obtained a WH solution model within $f(R,T)$ formalism in the second model of chapter \ref{Chapter6}. Furthermore, these exotic matters support phantom fluid distribution which can be depicted through the EoS parameter with $\omega<-1$. This phenomenons are covered in chapter \ref{Chapter7} along with the WH solution in two representative cases.   

In future, more work within this theory is required to fully understand  how these solutions should be interpreted and what their physical effects could be. There are several directions in which future
research on these subjects can proceed. For example, one can extend the $f(R,T)$ into more nonlinear cases to understand the details of matter energy coupling behavior. Also, one can focus on the generalization of model with several matter contents as a gravitational matter source, or, modify the gravitational actions by adding extra functional on the perturbation basis.  Furthermore, it would be possible to obtain the observational consistency of cosmological model within this theory through other constraints or free parameters. Then, we may focus on the generalization of the modified $f(R,T)$ gravity in several aspects, and extract reliable physical effects from these models which will be one of the most useful ways to study the evolution of universe in future generations.







\addtocontents{toc}{\vspace{2em}} 

\backmatter


\label{References}
\lhead{\emph{References}}

\cleardoublepage
\pagestyle{fancy}
\label{Publications}
\lhead{\emph{List of Publications}}
\chapter{List of Publications}
\textbf{Publication from thesis:}
\begin{enumerate}
\item P.K. Sahoo, B. Mishra, Parbati Sahoo, S.K.J. Pacif, \emph {Bianchi type string cosmological models in $f(R,T)$ gravity,} Eur. Phys. J. Plus, \textbf{131} (2016) 333, Springer, Impact factor 2.240, SCI, 
\url{https://doi.org/10.1140/epjp/i2016-16333-x}

\item P.K. Sahoo, Parbati Sahoo, B.K. Bishi, \emph {Anisotropic cosmological models in $f(R,T)$ gravity
with variable deceleration parameter,} Int. J. Geom. Methods Mod. Phys.,  \textbf{14} (2017) 1750097, World Scientific, Impact factor 1.068, SCI-E, 
\url{https://doi.org/10.1142/S0219887817500979}

\item P.K. Sahoo, Parbati Sahoo, B. K. Bishi, S. Ayg\"{u}n, \emph {Magnetized strange quark matter in $f(R,T)$ gravity with bilinear and special form of time varying  deceleration parameter,} New Astron., \textbf{60} (2018) 80-87, Elsevier, Impact factor 0.92, SCI,  
\url{https://doi.org/10.1016/j.newast.2017.10.010}

\item P.K. Sahoo, Parbati Sahoo, B.K. Bishi, S. Ayg\"{u}n, \emph {Magnetized strange quark model with Big Rip singularity in $f(R,T)$ gravity,} Mod. Phys. Lett. A \textbf{32} (2017) 1750105, World Scientific, Impact factor 1.308, SCI,   
\url{https://doi.org/10.1142/S021773231750105X}

\item P.K. Sahoo, S.K. Tripathy, Parbati Sahoo: \emph {A periodic varying deceleration in $f(R,T)$ gravity,}  Mod. Phys. Lett. A, \textbf{33} (2018) 1850193, World Scientific, Impact factor 1.308, SCI, 
\url{https://doi.org/10.1142/S0217732318501936}

\item P.K. Sahoo, P.H.R.S. Moraes, Parbati Sahoo, B. K. Bishi, \emph {$f(R,T)=f(R)+\lambda T$ gravity models as alternatives to cosmic acceleration,} Eur. Phys. J. C, \textbf{78} (2018) 736, Springer, Impact factor 5.172, SCI,   
\url{https://doi.org/10.1140/epjc/s10052-018-6211-4}

\item P.K. Sahoo, P.H.R.S. Moraes, Parbati Sahoo: \emph {Wormholes in $R^2$-gravity within the $f(R,T)$ formalism,} Eur. Phys. J. C, \textbf{78} (2018) 46, Springer, Impact factor 5.172, SCI,  
\url{https://doi.org/10.1140/epjc/s10052-018-5538-1}

\item P.K. Sahoo, P.H.R.S. Moraes, Parbati Sahoo, G. Ribeiro: \emph {Phantom fluid supporting traversable wormholes in alternative gravity with extra material terms,} Int. J.  Mod. Phys. D, \textbf{27}, (2018) 1950004, World Scientific, Impact factor 2.17, SCI, 
\url{https://doi.org/10.1142/S0218271819500044}
\end{enumerate}
\textbf{Other Publication:}
\begin{enumerate}
 \item Parbati Sahoo, Raghavender Reddy, \emph { LRS Bianchi type-I bulk viscous cosmological models in $f(R,T)$ gravity,} 
Astrophysics, \textbf{61}, 1 (2018), 153-162, Springer, Impact factor  0.755 \\
\url{https://doi.org/10.1007/s10511-018-9522-0}
 \item P. H. R. S. Moraes, P. K. Sahoo, Barkha Taori, Parbati Sahoo, \emph {Phantom energy dominated universe as a transient stage in $f(R)$ cosmology,} Int. J.  Mod. Phys. D, World Scientific, Impact factor 2.17, SCI, \url{https://doi.org/10.1142/S0218271819501244}
\end{enumerate}
\cleardoublepage

\pagestyle{fancy}
\lhead{\emph{Biography}}
\chapter{Biography}
\textbf{Brief Biography of Supervisor}\\
\textbf{Dr. Pradyumn Kumar Sahoo} is presently working as Associate Professor in Dept. of Mathematics, Birla Institute of Technology and Sciences, Pilani, Hyderabad Campus. He received his Ph.D. degree in Mathematics from Sambalpur University, Sambalpur, India in 2004. He has been actively involved in research, teaching, and academic administration for the past fifteen years. He has authored more than 80 research publications in various renowned national and international peer-reviewed journals. He has also presented several research works and delivered invited talks in national and international conferences held in India and aboard. He serves as an expert reviewer and editorial member for a number of reputed scientific research journals, and also Ph.D. examiner in several universities. He is also recipient of Science Academics Summer Research Fellowship, UGC Visiting fellow, Expert Reviewer, Physical Science Projects, SERB-DST, Govt. of India and lifetime membership of many scientific societies. He has successfully executed two projects from UGC, Govt. of India in 2014, and DAAD-RISE Worldwide in 2018. He has also visited counties like Canada, South Korea, Germany, Belgium, Japan, Poland, Switzerland, UK and presented his research work and delivered invited talks in different scientific events. \\ 

\textbf{Brief Biography of Co-supervisor}\\
\textbf{Dr. Bivudutta Mishra} is presently working as Associate Professor in Dept. of Mathematics, Birla Institute of Technology and Sciences, Pilani, Hyderabad Campus. He received his Ph.D. degree in Mathematics from Sambalpur University, Sambalpur, India in 2003. He has published around 80 research papers in reputed national and international peer-reviewed journals. He has also presented several research works and delivered invited talks in national and international conferences held in India and aboard. He has successfully completed two projects: one each from UGC, and SERB-DST. He has also reviewed several research papers of highly reputed journals, Ph.D examiner, BOS members of several universities. He has also visited counties like Canada, UAE, Germany, Republic of China, Malaysia, Italy, Czech Republic, Russia, Australia, Switzerland, Japan, and presented his research work in different scientific events. \\ 

\textbf{Brief Biography of Candidate}\\
\textbf{Ms. Parbati Sahoo} completed her Bachelor of Science from N. C. (Auto) College, Jajpur, Odisha, Master of Science and M.Phil. from Utkal University, Bhubaneswar, Odisha. She has about one year of academic experience in teaching at EATM, Bhubaneswar, Odisha from 2012-2013. She has been appointed as full time Ph.D. scholar in Dept. of Mathematics, Birla Institute of Technology and Sciences, Pilani, Hyderabad Campus from 2015-2019 under the supervision of Dr. Pradyumn Kumar Sahoo and co-supervision of Dr. Bivudutta Mishra. She has published 10 scientific publications in reputed international journals. She has presented her research papers in several national and international conferences held in India and aboard. She has received international travel grant from DST, Govt. of India to present her part of research work as poster presentation at the 
3rd HEL.A.S. Summer School and DAAD School on Neutron Stars and Gravitational Waves held at Department of Physics, Aristotle University of Thessalonki, Greece, from 8-12 October 2018.

\end{document}